# Simulating Nonlinear Neutrino Oscillations on Next-Generation Many-Core Architectures

by

## Vahid Noormofidi

B.S., Computer Science, Amirkabir University of Technology
(Tehran Polytechnic), 2006
M.S., Computer Science, University of Tehran, 2008

DISSERTATION

Submitted in Partial Fulfillment of the
Requirements for the Degree of

Doctor of Philosophy
Computer Engineering

The University of New Mexico

Albuquerque, New Mexico

November, 2015

# Simulating Nonlinear Neutrino Oscillations on Next-Generation Many-Core Architectures

by

**Vahid Noormofidi**

B.S., Computer Science, Amirkabir University of Technology
(Tehran Polytechnic), 2006
M.S., Computer Science, University of Tehran, 2008

Ph.D., Computer Engineering, University of New Mexico, 2015

## Abstract

In this work an astrophysical simulation code, XFLAT, is developed to study neutrino oscillations in supernovae. XFLAT is designed to utilize multiple levels of parallelism through MPI, OpenMP, and SIMD instructions (vectorization). It can run on both the CPU and the Xeon Phi co-processor, the latter of which is based on the Intel Many Integrated Core Architecture (MIC). The performance of XFLAT on configurations and scenarios has been analyzed. In addition, the impact of I/O and the multi-node configuration on the Xeon Phi-equipped heterogeneous supercomputers such as Stampede at the Texas Advanced Computing Center (TACC) was investigated.

# Contents

*Contents*

*Contents*

*Contents*

*Contents*

# Glossary

Accelerator  A processor that typically is installed as a card on motherboard. The main processor can offload its tasks on co-processors.

API  Application Programming Interface (API) is a set of functions and protocols to build software applications.

Compute node  An independent machine in which there one or more CPUs are installed. Some of them may be equipped with co-processors as well. Compute nodes can be linked together via high-performance network links.

Co-processor  A processor that typically is installed as a card on motherboard. The main processor can offload its tasks on co-processors. Co-processors
can also be employed independently.

CPU  A Central Processing Unit is an integrated circuit chip that executes instructions of computer program by performing the arithmetic, log-
ical, and control operations specified by the instructions.

GPU  Graphics Processing Unit. Traditionally, the are utilized for computer graphics tasks. By improving their performance, GPUs can also be employed for high-performance computing applications.

*Glossary*

HDD  Hard Disk Drive. Is a type of computer storage in which data is stored on one or more rapidly rotating metal disks.

| | |
|---|---|
| Library | A collection of pre-written codes, procedures, classes, or values that may be employed for developing softwares. |
| MPI | Message Passing Interface is a portable and standardized message-passing system that can be exploit for inter-node communications. |
| Node | An independent machine in which there one or more CPUs are installed. Some of them may be equipped with co-processors as well. |
| | Nodes can be linked together via high-performance network links. |
| OpenMP | Open Multi-Processing is an interface for parallel programming on shared memory multiprocessing environments. |
| PCI Express | Peripheral Component Interconnect Express. A high-speed serial bus that typically connects accelerators to processors. |
| RAM | Random-Access Memory. It is a type of memory that allows data elements to be accessed irrespectable of their physical location inside |
| | the memory. RAM is normally associated with volatile types of memory in which stored data is lost is the power is removed. Most modern computers' main memory is RAM. |
| Run Time | Run time or runtime is the time during which an application is executing. |
| SIMD | Single Instruction, Multiple Data is a parallel programming approach in which machines exploit data level parallelism by issuing a single instruction on multiple data simultaneously. |
| Supercomputer | A powerful computer containing many individual computing nodes that are connected via a high-speed interconnection link. |

*Glossary*

Thread         A lightweight process that may share some part of its memory with the other threads.

# Chapter 1

# Introduction

## 1.1    Astrophysics, Supernovae, and Neutrinos

Human being has always been curious about the universe. Astrophysics is the branch of science that attempts to understand the universe and the human future in it. The goal of astrophysics is "to ascertain the nature of the heavenly bodies, rather than their positions or motions in space." [Keeler, 1897]. There are a broad range of objects that are studied in astrophysics such as: extra-solar moons and planets, stars, supernovae, white dwarfs, neutron stars, black holes, galaxies, quasars, dark matter, dark energy, and the cosmic microwave background. Astrophysics address fundamental questions including the origin and evolution of stars, galaxies, and the universe.

At the end of its life, a massive star explodes as a *supernova* and at the same time its core collapses under its own gravity into a neutron star, which is the remnant of the core containing mostly neutrons [Woosley and Janka, 2005]. Therefore, during the explosion supernova can briefly outshine the entire galaxy. In a few seconds a supernova emits as much energy as the Sun emits over its entire life span [Giacobbe, 2005].

The explosion expels almost all of a star's material at a relativistic speed, driving a shock wave into the surrounding interstellar medium [Schawinski et al., 2008]. Supernovae are crucial to the chemical evolution of the universe which had only hydrogen and helium immediately after the Big Bang. All other heavier elements are born inside stars, during supernovae, or potentially during other dramatic astrophysical events such as neutron star mergers. Therefore, without supernovae there may have never been each of us who are made up from heavy elements that are generated or distributed by supernovae.

It turns out that around 99% of the total energy of a supernova is carried away by about $10^{58}$ particles called neutrinos within a minute of core collapse. Neutrinos are extremely difficult to detect as they have no electric charge and they interact with matter only through the weak interaction. In fact, about 65 billion solar neutrinos per second pass through every square centimeter on the surface of the Earth [Bahcall et al., 2005]. There are three types or *flavors* of neutrinos in the particle physics Standard Model, electron neutrino, mu neutrino, and tau neutrino ($\nu_e$, $\nu_\mu$, and $\nu_\tau$). Neutrinos play critical role in nucleosynthesis processes in supernovae. In fact the electron neutrino can affect the number density of protons and neutrons:

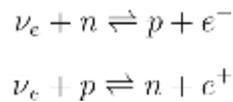

$$\nu_e + n \rightleftharpoons p + e^-$$
$$\nu_e + p \rightleftharpoons n + e^+$$

where $n$, $p$, $e^-$, and $e^+$ are neutron, proton, electron and positron, respectively. Therefore, neutrinos can influence the production of heavy elements in the supernovae ejecta by changing the proton to neutron ratio. In addition, since the supernova envelope is transparent to neutrinos, by investigating the neutrinos that are emitted from supernovae astrophysicists can probe into supernovae.

One of the most important breakthroughs in particle physics in the last twenty years was the discovery of neutrino (flavor) oscillations in which neutrinos of different flavors change to another during propagation [Olive et al., 2014]. In supernovae neu-

*1.1. ASTROPHYSICS, SUPERNOVAE, AND NEUTRINOS*

trinos of different flavors can have different luminosities. Besides, they have different average energies which are determined by the energy spectral functions. The energy spectrum of the neutrino is proportional to the number of neutrinos as a function of the neutrino's energy. Because, only the electron-flavor neutrinos and anti-neutrinos play important roles outside the neutron star and because neutrinos can change their flavors, neutrino oscillations are important to supernova dynamics and the origin of the heavy elements.

There are two complementary ways to study neutrino oscillations in supernovae. The first approach is to detect neutrinos emitted from supernova. However, supernovae are rare events, and on average only one supernova occurs in a century in our home galaxy [Hirata et al., 1987, Bionta et al., 1987]. The other approach is to study them via simulations and to model neutrino flavor oscillation in supernovae.

In principle, the flavor quantum state of neutrinos emitted from a proto-neutron star in a supernova depends on seven parameters: time ($t$), distance ($r$), emitting position on the surface ($\Theta,\Phi$), energy ($E$), and trajectory direction ($\vartheta,\phi$). The solution of the complete simulation of neutrino evolution in supernovae depending on all seven variables and cannot be solved analytically or computationally at present. In order to study neutrino oscillations simpler models are always employed [Duan et al., 2006]. For example, the *bulb model* is the simplest stationary model in which the spherical symmetry was assumed [Duan et al., 2006], thus the only parameters are $r$, $E$, and $\vartheta$. Even in this simple model, thousands of angle trajectories and hundreds and discretized energy bins are required for achieving good numerical resolution. In more complicated models, orders of magnitude more trajectories are required for each new parameter added back to the model. In addition, it is essential for a complete simulation to store the flavor quantum state of the many interacting particles in memory. Therefore, the very large number of neutrino trajectories in the system as well as the complexity of the geometry and environment makes these simulations very challeng-ing. Therefore, the next-generation of supercomputers can be very helpful for the study.

One might wonder if within a few years this complexity and high computational demand could be largely addressed by Moore's law. Moore stated back in 1965 (reviewed in [Moore, 2006]) that the number of transistors in a dense integrated circuit doubles approximately every year (or doubles every two years in the 1975 revised version of Moore's law [Moore et al., 1975]); consequently microprocessor and compute capabilities correspondingly double at the same rate. However, while computing capability has indeed been growing exponentially since then (see Fig. 1.1), adding even one more parameter to a simulation can cause computational complexity to grow by a few orders of magnitude (see Fig. 1.2). In order to perform a complete neutrino oscillations simulation, a large number of neutrino trajectories may be required. This requirement dictating the simulations' fidelity may have to be balanced against the ability to perform a feasible simulation on the current generation of supercomputers.

## 1.2    The History of Supercomputers

A supercomputer is a computer with a high-level computational capacity. The performance of a supercomputer is measured in *floating point operations per second* (FLOPS). Modern supercomputers can perform over quadrillions of floating point operations per second and it is expected that within a decade their performance reach up to exaflops ($10^{18}$ floating point operations per second) [Wikipedia, 2015g].

The history of supercomputers goes back into 1960s when Seymour Cray designed the first commercial supercomputer, the CDC 6600, with nearly 1 megaflops performance [Wikipedia, 2015b]. The CDC 6600 *Central Processing Unit* (CPU) was dedicated solely to computations rather than to handle all tasks such as memory and I/O. This was the first example of what later came to be called *reduced*

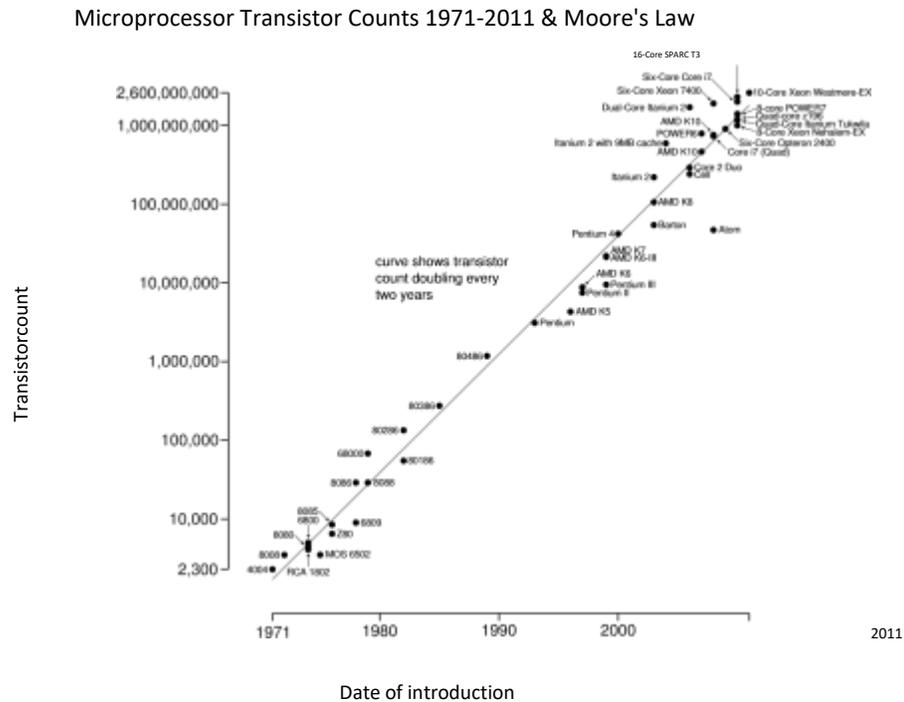

Figure 1.1:                    CPU transistor counts as a function of date of introduction [Wikipedia, 2015i].

*instruction set computer* (RISC) design. CDC 6600 was succeeded by CDC 7600, which could deliver ten times more performance [Wikipedia, 2015c]. For CDC 7600 Cray focused on the concept of an instruction pipeline. A pipeline improves the performance by feeding in the next instruction before the first has completed (similar to assembly line in a manufacturing process), thereby having each unit effectively work in parallel, as well as the machine as a whole (see Fig. 1.3). Cray continued his domination in supercomputing in the 1970s by introducing Cray-1 which could deliver ten times more performance than the previous record holder by utilizing vector processors (see Fig. 1.3) [Wikipedia, 2015e]. Compare to scalar processors, whose instructions operate on single data items, a vector processor is a CPU that implements an instruction set containing instructions that operate on an array of data called vector. Cray-1 was succeeded in 1982 by the 800 megaflops Cray X-MP, the first Cray shared-memory parallel vector processor in which the memory was shared among two processors [Wikipedia, 2015f]. In late 1985 the nCUBE 10 was released, which was based on a set of custom chips. In each compute node of nCUBE10 there was a processor chip with a 32-bit ALU, a 64-bit FPU [Wikipedia, 2015j]. Also, in 1985 the very advanced Cray-2, the first gigaflops capable supercomputer capable of 1.9 gigaflops peak performance, succeeded the first two models as the world fastest supercomputer [Wikipedia, 2015e]. During 1980s there were also other designs such as Thinking Machines CM-1 and CM-2 in which hypercubic design were employed [Wikipedia, 2015d]. Each CM-1 microprocessor had its own 4 kbits of RAM, and the hypercubic array of them was designed to performed the same operation on multiple data points simultaneously, *i.e.*, to execute tasks in the single instruction, multiple data (SIMD) fashion. The supercomputing trend continued in 1990s with the introduction of massively parallel supercomputers such as Thinking Machines CM-5/1024, the first supercomputer with more than one thousand processors [Wikipedia, 2015d]. In 1996, ASCI Red became the first supercomputer to break the 1 teraflops performance barrier. The ASCI Red computer was the first large scale supercomputer to be built entirely of common commercially available components [Wikipedia, 2015a]. Prior to 2000, nearly all commercial CPU architectures were of single-core, singlethreaded design (Single Instruction Multiple Data instruction (SIMD) set was not even available in commodity CPUs before 1997 [Peleg et al., 1997]). During the 2000s the use of commercial CPUs in supercomputers accelerated. It was in 2008 that IBM Roadrunner became the first supercomputer to perform more than 1 petaflops of computations [Wikipedia, 2015h] using Cell processors which were also used in the PlayStation 3 game console [Costigan and Scott, 2007]. Roadrunner was also the first hybrid supercomputer. Previous supercomputers only used one type of processor, *i.e.* CPU, Roadrunner, however, employed TriBlades in which two dual-core Opterons and four PowerXCell 8i CPUs interconnected to each other [Wikipedia, 2015h]. The Roadrunner can be considered as an Opteron cluster with Cell accelerators, with each node consisting of a Cell attached to each Opteron core and the Opterons to each other. In the last decade mainstream microchip companies such as Intel and AMD, have started packing more cores on a single die area [Geer, 2005], adding multi-threading (having multiple tasks run on a single core simultaneity) functionality onto each core [Magro et al., 2002], and increasing the length of SIMD registers [Thakkur and Huff, 1999, Firasta et al., 2008] by several folds. Consequently, hierarchically-parallelized softwares started to play more important role in the mainstream desktop applications such as web-browsers, video games and office applications.

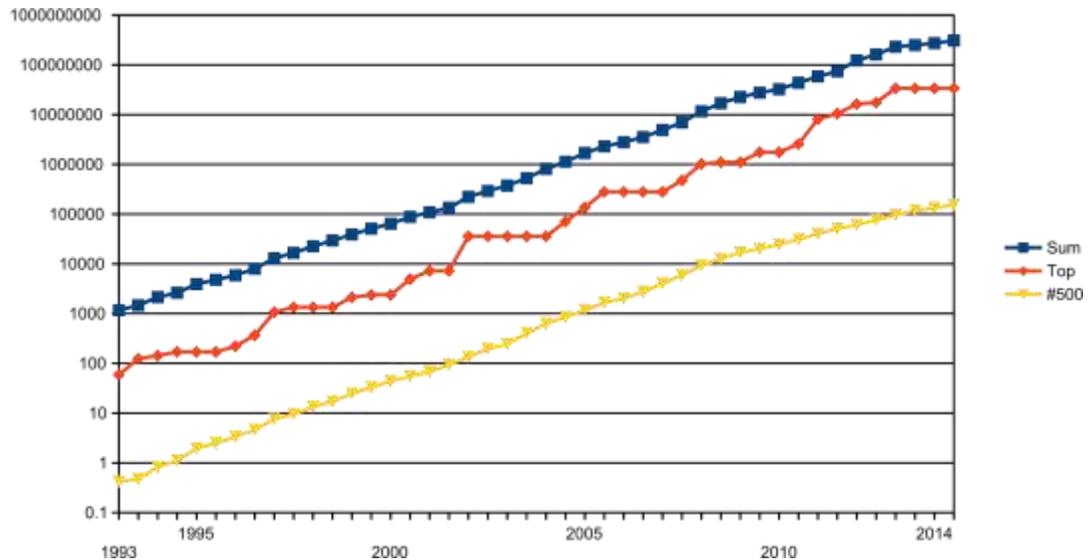

Figure 1.2: The combined performance of 500 largest supercomputers (blue), the fastest supercomputer (red), and the supercomputer on 500th place (yellow) as functions of time [Wikipedia, 2015k]. Data based on Top500 (http://http://top500.org/)

In recent years, High-Performance Computing (HPC) community has adopted the concept of heterogeneous computing and has shifted from employing only tra-ditional CPUs to also incorporating accelerators and co-processors such as nVIDIA Tesla Graphics Processing Unit (GPU) [Kirk, 2007] and the Intel Xeon Phi with the Intel Many Integrated Core Architecture (Intel MIC) [Chrysos and Engineer, 2012]. Today, most petaflop class supercomputers have hybrid or heterogeneous designs meaning that they are equipped with accelerators, such as GPU and Xeon Phi, in addition to traditional CPUs. These accelerators and co-processors, which are extension cards installed on computers' motherboard, are designed for fine-grained, massively parallel computation and possess great computing capabilities. Modern supercomputers typically exploit several levels of parallelism. As illustrated in Fig. 1.4, a modern supercomputer may contain many compute node which are connected to each other via high-speed interconnect links. Each compute node may have a few co-processors and CPUs. In each CPU or co-processor, there are a few to many cores available. Each of the cores may be able to handle a few hardware-enabled threads. Finally, depending on the processors' architecture, the execution units of each core may be equipped with vectorized units, which can execute a single instruction on multiple data. Hybrid codes are required to fully utilize all of the different levels of parallelism on a modern supercomputer. Hybrid codes can utilize the vectorization units to process arrays of data, different cores and processors and multiple compute nodes to process different chunks of data. Therefore, multiple order of magnitudes speedup can be expected when a fully optimized and parallelized code runs on a modern supercomputer.

| Instr. No. | Pipeline Stage | | | | | | |
|---|---|---|---|---|---|---|---|
| 1 | IF | ID | EX | MEM | WB | | |
| 2 | | IF | ID | EX | MEM | WB | |
| 3 | | | IF | ID | EX | MEM | WB |
| 4 | | | | IF | ID | EX | MEM |
| 5 | | | | | IF | ID | EX |
| Clock Cycle | 1 | 2 | 3 | 4 | 5 | 6 | 7 |

Figure 1.3: Basic five-stage pipeline (IF = Instruction Fetch, ID = Instruction Decode, EX = Execute, MEM = Memory access, WB = Register write back). In the fourth clock cycle (the green column), the earliest instruction is in MEM stage, and the latest instruction has not yet entered the pipeline [Wikimedia, 2015].

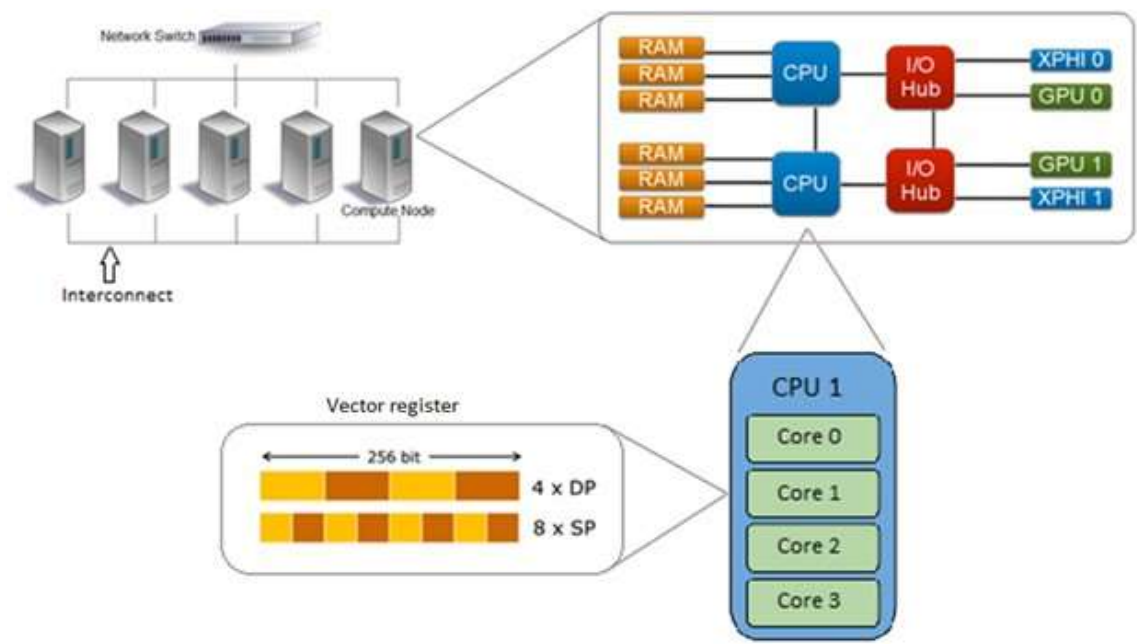

Figure 1.4: Cluster nodes and internal architecture of each node on a heterogeneous supercomputer. At the highest level compute nodes are interconnected via high-speed interconnect links. Inside each node, there may be multiple CPUs as well as accelerator cards. Typically, inside each processor there are a few to many cores. Finally, within each core, a vector instruction can be issued on multiple data simultaneously.

## 1.3    Accelerators

In a heterogeneous supercomputers, accelerators such as Graphics Processing Units (GPU) and Intel Xeon Phi are employed to boost the computing capability and improve the concurrency of each node. Initially, GPUs were introduced for computer graphics tasks, especially video games computations, so that CPU can be freed up for other tasks. By increasing the complexity of video games, the demand for more powerful GPUs drove the evolution of GPUs to the point where GPUs were flexible enough for have general computations other than graphics [Mark et al., 2003]. The trend of generating higher quality graphics made GPUs even more powerful than CPUs [Kirk, 2007].

The latest generation of mainstream CPU architectures have typically less than ten cores available on the die area [Wikipedia, 2014] and large part of the die area is dedicated to cache memory and control unit [Magro et al., 2002]. Therefore, the execution units occupy only small part of the die area. On the other hand, accelerators and co-processors usually have many smaller and simpler cores than CPU such that they can handle a larger number of light in-order hardware-enabled threads (see Fig. 1.5), and most of the die area is dedicated to Floating-Point Units (FPU) and Arithmetic Logic Units (ALU), which are for instruction execution.

Despite the fact that GPUs can accelerate scientific computations, the programming model of GPUs can be challenging. Traditional CPU codes are not able to run on GPUs without modifications in their data structures, algorithms, and memory models. Therefore, new codes must be developed to run on GPU. In addition, one also should have low-level knowledge of GPU architecture including GPUs' memory hierarchy and the new parallel programming paradigm in order to harness the power
of GPU efficiently.

Having failed with their experimental Larrabee project (Intel GPU chip code-name for GPU) [Seiler et al., 2008], Intel introduced the Many Integrated Core Architecture (MIC) in Xeon Phi co-processors in 2012 [Chrysos and Engineer, 2012]. Due to difficulties of GPU programming, Intel decided to stick with its established x86 microprocessor architecture for its teraflops co-processor. The Xeon Phi can be installed on a compute node as an extension card (similar to GPU cards), and it runs an embedded micro-kernel Linux Operating System (OS) in memory. Intel MIC is principally a many-core x86-64 microchip with several extensions [Seiler et al., 2008, Chrysos and Engineer, 2012]. As a result, legacy codes may be compiled and run on the Xeon Phi without significant modifications. Being able to write a parallel code once and run it on both CPUs and MIC architectures is a great opportunity for developers, thus a great advantage for Intel MIC over GPU programming. However, many applications do not exploit many levels of parallelism entirely. As a result, code optimization is necessary in order for legacy codes to run efficiently on the Intel MIC.

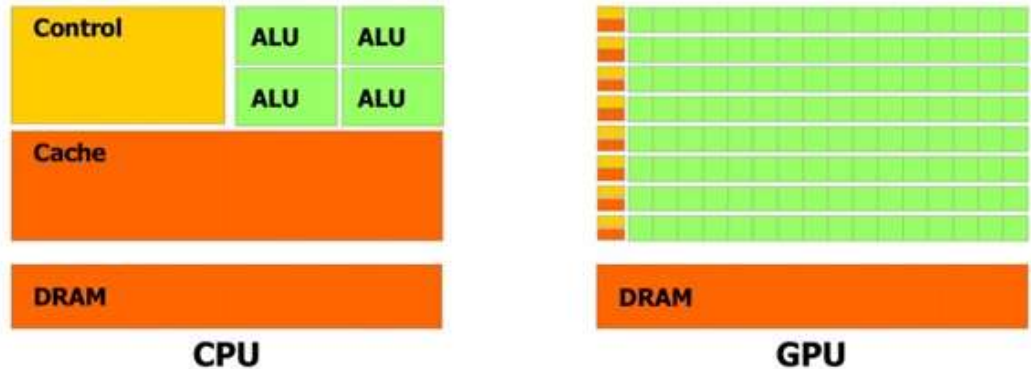

Figure 1.5: Comparison of the CPU and GPU die area. The CPU has more on-chip memory (shown in orange) and larger control unit (shown in yellow) than GPU. However, the GPU devotes more transistors to data processing (shown in green) than CPU [NVIDIA, 2012a].

Since there are still several hardware differences between the Intel MIC and traditional CPU, code optimizations are inevitable to make the chip appropriate for high-performance parallel applications. For example, the chip has a smaller cache area to make room for higher number of cores. In fact it completely removed the Level 3 cache [Chrysos and Engineer, 2012]. Additionally, in comparison to traditional out-of-order x86 CPU cores, each core on Xeon Phi has simpler in-order design [Chrysos and Engineer, 2012] and is capable of handling four hardware threads in contrast to two on CPU. Moreover, each thread on the Xeon Phi have access to wider Single Instruction Multiple Data (SIMD) vector registers [Chrysos and Engineer, 2012]. Nevertheless, by taking the advantage of the same optimization techniques for both the CPU and the MIC, performance improvement can be obtained on both architectures. However, to truly unleash the power of Xeon Phi, a computer code has to be designed carefully to utilize the special features of Xeon Phi such as 512-bit-long registers and 200+ hardware threads.

The goal of this work was to design and create a new high-performance software framework for astrophysics simulations of neutrino oscillations in supernovae which can run on the next-generation many-core architectures hardware as well as existing hardware components. The new code, XFLAT, is capable of employing several levels of parallelism and scales very well on multi-node systems. It is designed with a hybrid architecture in order to exploit all of the parallelism technologies available on hardware simultaneously. It is also designed to be modular so that new physical simulations can be performed by simply swapping in new modules into the code.

# Chapter 2

# Accelerators and Co-processors

The trend of employing non-general purpose processors for scientific computations has been boosting in recent years. The introductions of General Purpose Graphics Processioning Units (GPGPU) and Intel Many Integrated Core (MIC) architecture can be seen as two pivot points in high-performance computing trend in recent years.

## 2.1    GPUs

Despite the fact that CPUs have been evolving and adding more parallelism capabilities over a long period, GPUs were invented exclusively to perform the same tasks on a large set of pixels in parallel. As a reuslt, there was commonly more than one logical and computational unit in GPUs architecture design. Hence the GPU design was parallelized from the start. However, their architecture was fixedfunction pipeline for many years until 2001 that programmable pipeline is introduced [Lindholm et al., 2001]. A fixed-function pipeline contains a set of configurable processing state that were accessible by a set of callable functions [OpenGL, 2015a]. A graphics processing unit that contains the fixed-function pipeline, exposes a differ-ent predefined rendering pipeline based on user-provided configuration rather that user-provided programs [OpenGL, 2015a]. Therefore, based on the rendering criteria, user can decide which one of the predefined rendering pipeline configuration have to be loaded on the hardware (see Fig. 2.1).

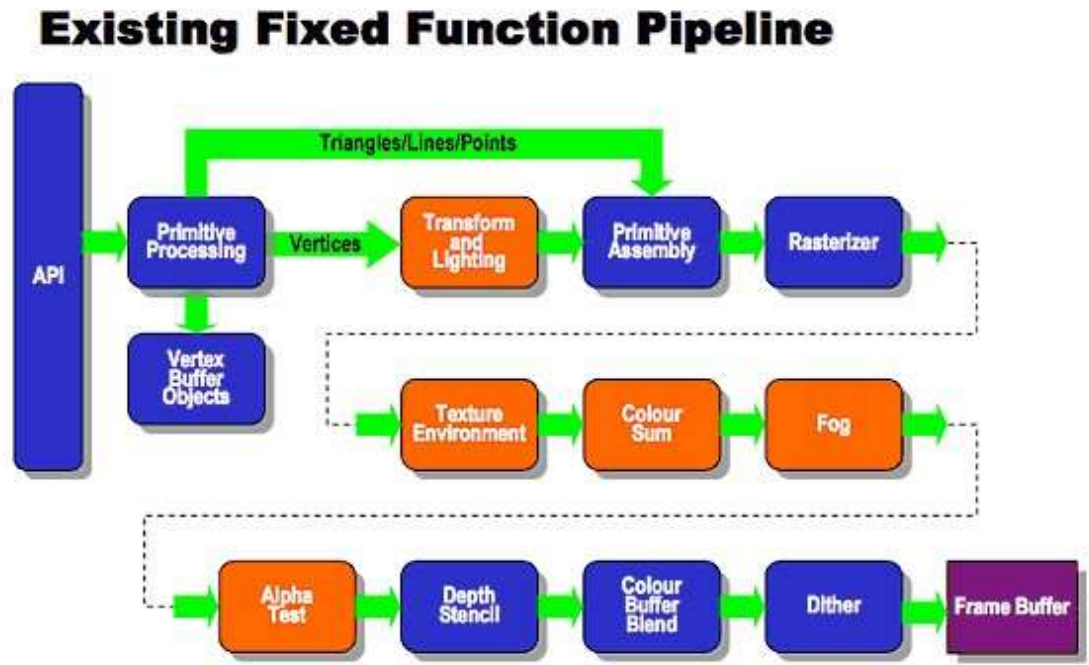

Figure 2.1: Existing fixed function pipeline hardware before the introduction of programmable pipelines in GPUs. Each box in orange has a fixed and non-programmable functionality. After the introduction of programmable shaders, those fixed-function units have replaced with vertex and fragment shader units. The boxes in blue remain identical before and after the introduction of programmable pipelines. (khronos.org 2015).

In 2001 nVIDIA introduced its programmable pixel *shaders* pipeline architecture in their GPUs [Lindholm et al., 2001], which allows programmers to develop customized shading programs on GPU for graphics and image processing fields [Purcell et al., 2002, Sugita et al., 2003] and other scientific research branches and purposes [Thompson et al., 2002, Mark et al., 2003, Kru¨ger and Westermann, 2003, Moreland and Angel, 2003, Atanassov et al., 2003]. A shader program is a piece of code which performs graphics shading and special color effects on objects. This transition from fixed-function pipeline, which is non-flexible pipeline for performing fixed tasks, to programmable pipeline which is shader based pipeline, introduced General Purpose computing on GPUs (GPGPU) and opened a new door for high performance computing as well [1].

For a long time after the introduction of programmable shaders, developers had to write their own shaders in GPU's assembly language [2]. Later, several higher level languages were introduced which made the GPU programming easier. One of the first high level GPU programming languages was nVIDIA Cg (C for graphics) [Mark et al., 2003]. Cg was a flexible C-like language that was designed for GPU shader

---

[1] Previously the entire GPU computation powers was only available for video games and rendering applications, even though GPUs can be capable of doing more single precision floating point operations per second (FLOPS) than CPUs [Nyland et al., 2007]. Before the introduction of programmable pipelines in GPUs, in order to use GPUs for scientific researches, scientists had to learn about the rendering pipeline deeply. Furthermore, most of the time there were no suitable way to convert a scientific problem in a way that fit into the fixed-function pipeline, thus the usage of GPUs were limited [Swanson, 1995]

[2] After the introduction of programmable shaders for the rendering pipeline in GPUs (The rendering pipeline is the sequence of steps that are taken when rendering objects), scientists started to study the appropriateness of GPUs in other research fields. In early generation of programmable GPUs, it was possible to write simple programs for *Vertex shader* and *Pixel shader* units of GPU. Since, the basic polygon for representing a graphical scene is triangle (in fact, all the geometrical characters in a scene are split down to triangles), a special hardware unit is dedicated for processing triangles' vertices. Therefore, the Vertex Shader is the programmable shader stage in the rendering pipeline that handles the processing of individual vertices [OpenGL, 2015c]. Likewise, the other important stage of the rendering pipeline relates to the processing of individual pixels in which most of the GPU computations happen. Hence, the Pixel Shader or Fragment Shader is a usersupplied program that performs pixel processing such as setting of colors and depth values [OpenGL, 2015b]

programming. It allowed graphics programmers to configure the vertex and pixel shader units and, therefore, change the behavior of the GPU rendering pipeline without dealing with the low-level assembly language. Using Cg programmers can write high-level structures and algorithms with the use of for-loops, if-else statements, functions etc. Because each vertex is addressed by using three numbers (for x, y, and z direction) and each pixel contains multiple color channels (red, green, blue, and alpha blending), a functionality similar to SIMD was present in Cg which together with multiple ALUs on vertex and pixel shader units provides additional parallelism
for the programmers to utilize at a high level.

The GPGPU trend continued further with the introduction of the Compute Unified Device Architecture (CUDA) by nVIDIA in 2007 [Kirk, 2007, Buck, 2007]. The CUDA platform enable developers to harness GPU power in high-performance computations in scientific researches. Since then, the number of cores on GPUs have increased dramatically, from tens in the first generation CUDA-enabled GPUs to thousands in the GPUs in Kepler and Maxwell generation [NVIDIA, 2012b]. As shown in Fig. 2.2, there are many simple cores in one GPU block in contrast to a few number of complex cores in a typical CPU. In high-end GPUs such as the GK110
illustrated in Fig. 2.3, there are more than one core blocks (a.k.a. Streaming Multiprocessors) on the die area. The parallel nature of the GPU itself plays an essential role in scientific computations. Thus, by employing CUDA or other GPU programming languages such as Open Computing Language (OpenCL) [Stone et al., 2010] or C++ Accelerated Massive Parallelism (C++ AMP) [Gregory and Miller, 2012], scientists can write general applications which can run on GPU without knowing
about computer graphics or the graphics pipeline.

The adaptation of GPU in scientific researches has increased dramatically in the past decade [Ryoo et al., 2008, Nickolls et al., 2008]. Porting legacy codes to GPU have helped researchers to accelerate many applications. For example, molecular dynamics research codes LAMMPS [Brown et al., 2011, Brown et al., 2012], NAMD [Phillips et al., 2008], and GROMACS [Hess et al., 2008] have enhanced their computation speed greatly by using GPUs. GPUs have also been utilized in particle simu-lations, such as cosmological simulations [Hamada et al., 2009, Spurzem et al., 2009, Belleman et al., 2008, Nyland et al., 2007], artificial neural networks [Jang et al., 2008], graph theory [Harish and Narayanan, 2007, Vineet and Narayanan, 2008], Fast Fourier transform [Nukada et al., 2008], bioinformatics [Ligowski and Rudnicki, 2009], weather prediction [Michalakes and Vachharajani, 2008].

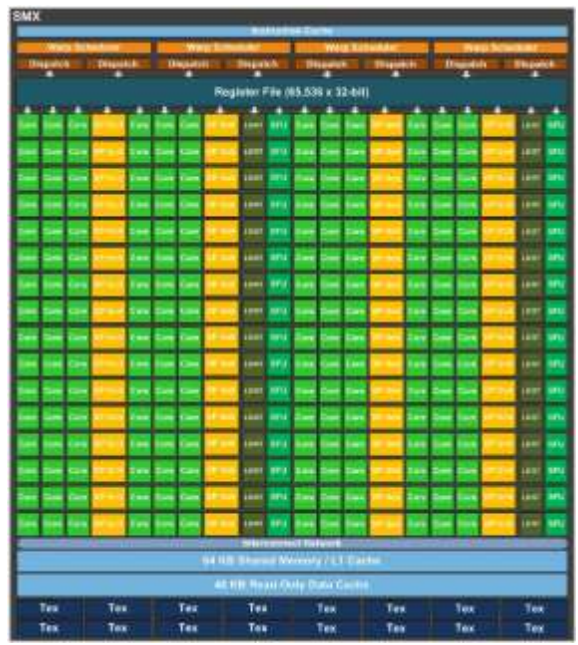

Figure 2.2: A cluster of tiny cores for the nVIDIA Kepler GPU architecture. There are hundreds of single floating point units (shown in light green) as well as double precision units (shown in yellow). The other important units are special function units (shown in green) which are responsible for the computations of transcendental functions (nVIDIA 2012).

The wide adoption of GPGPU in the past few years is largely due to the CUDA parallel programming model. CUDA is accessible to software developers via an extension over standard languages such as C, C++ and FORTRAN. There are three abstractions in CUDA. A hierarchy of thread groups, shared memories, and barrier synchronization. These layers of abstraction provides fine-grained data parallelism and thread parallelism, nested within coarse-grained data parallelism and task parallelism. As a result, programmers can partition a problem into coarse sub-problems which can be solved independently in parallel by blocks of threads. Each sub-problem can then be partitioned into finer pieces which can be solved cooperatively in parallel by all threads within the block. At the same time, scalability is preserved since each block of threads can be scheduled on any of the available multiprocessors within a GPU, in any order, concurrently or sequentially.

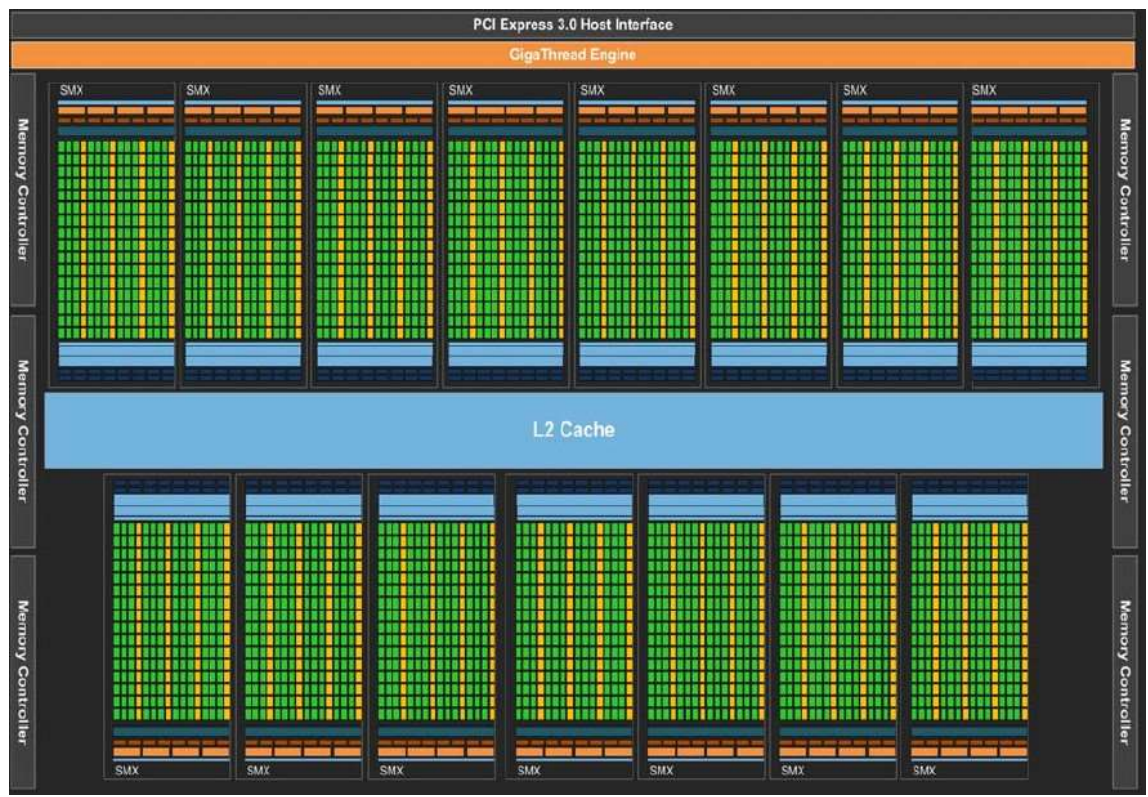

Figure 2.3: Full chip block diagram of nVIDIA GK110 architecture containing clusters (Streaming Multiprocessors) of CUDA core. Blocks can share data using the shared L2 cache unit (shown in light blue in the middle), and each one can access the global memory outside the chip through memory controllers (shown in grey on sides) (nVIDIA 2012).

In order to complete a computation task on GPU (or "Device"), programmers define special functions, or "kernels" which, when called from the CPU (or "Host"), are executed in parallel on streaming multiprocessors. A kernel contains a grid of blocks of threads which have equal number of threads (see Fig. 2.4).

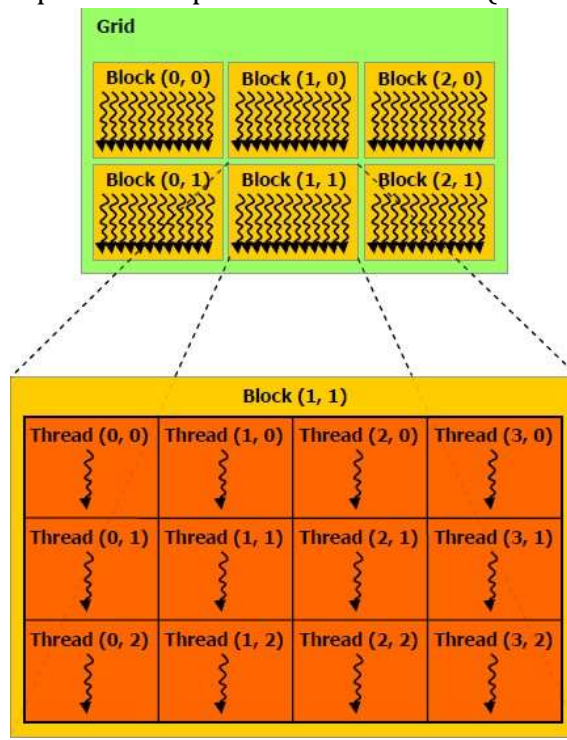

Figure 2.4: Grid of thread blocks. A grid (show in green) may contain a multidimension array of blocks (shown in yellow). Within each block there is an array of threads (shown in orange) (nVIDIA 2014).

CUDA threads may access data from multiple memory regions as illustrated by Fig. 2.5. Each thread possess a private local memory where local variables and data stored. Similar to registers, access time to this type of memory is short. However, if the programmer allocates too much of such memory, the allocated memory will reside on the GPU global memory (the off-chip memory). Each block has a shared memory space visible to all threads of the block within the lifetime of the block. The shared memory is also an on-chip memory and the access time is short. In addition, all threads have access to the same global memory, which resides on graphics card's RAM. There are also two read-only memory spaces accessible by all threads: the constant and texture memory spaces, which are optimized for different memory usages. As a result, programmers must understand the advantages and disadvantages of each memory in order to write codes which can run efficiently on GPU.

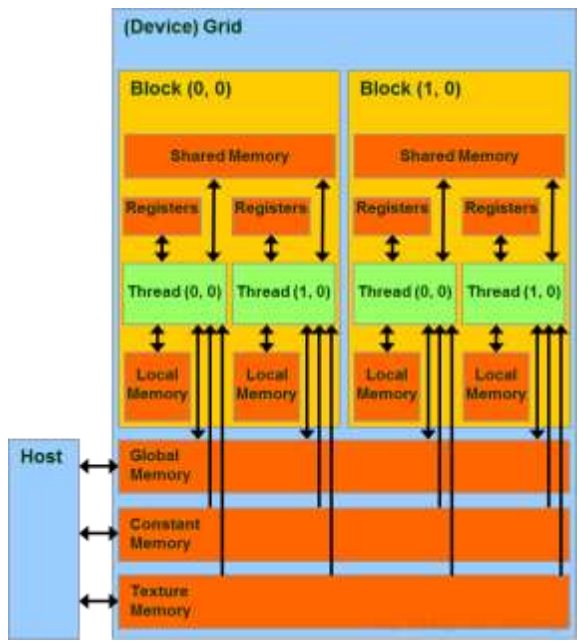

Figure 2.5: CUDA memory hierarchy. Threads (shown in green) can access to different memory locations each with their own advantages and disadvantages (shown in orange) (nVIDIA 2014).

When a kernel is launched from a C code on the CPU (host), only the grid of thread blocks and their associated memory reside on the GPU (device) and the rest of the program remains on the CPU. Therefore, a host program can continue its executions while the kernels are run on the GPU (see Fig. 2.6). The CUDA programming model assumes that both the host and the device maintain their own separate memory spaces in their own RAM, referred to as the host memory and device memory, respectively. A program on the host manages the global, constant, and texture memory spaces visible to device kernels through function calls to the CUDA runtime system. This includes device memory allocation and deallocation as well as data transfer between the host and device memories.

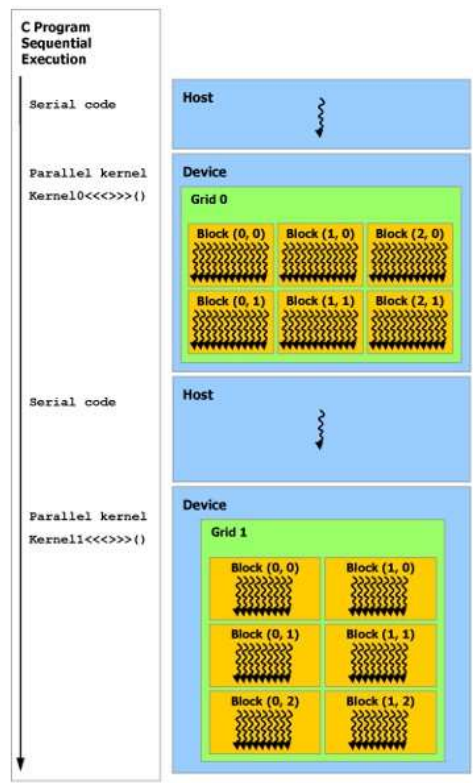

Figure 2.6: Launch of kernels from host. Host and device can have their own executions (shown in blue). Multiple grids (shown in green) can be launched from the host. (nVIDIA 2014).

There are a number of challenges involving in the usage of GPUs for generalpurpose programming. First, traditional C/C++ codes cannot run on GPUs without extensive modifications. In order to port the traditional codes to GPU, programmers are required to learn a new language, new techniques, and adopt new strategies. Second, in order to use CUDA or the other GPGPU programming platforms efficiently, developers have to deal with many low-level details such as micromanaging the shared memory, controlling memory flow between CPU and GPU, and with new version of old concepts such as grids, blocks, and threads. For these reasons, programmers frequently have to restructure their codes and data structures and come up with more optimized algorithms, as well as rewrite many parts of their code so as they can run on GPUs efficiently.

## 2.2  MICs

### 2.2.1  The architecture of the MIC

Intel had a GPU project in 2008 that eventually failed [Seiler et al., 2008]. After failing with their experimental GPU, they introduced the Many Integrated Core Architecture (MIC) in Xeon Phi co-processors in 2012 [Chrysos and Engineer, 2012]. Unlike the architectures of GPUs which may change with each generation, Intel decided to stick with their well established x86 architecture for its teraflops co-processor.

The architecture of Intel Xeon Phi is shown in Fig. 2.7. There are about 60 simple x86 cores on the microchip die area. Each core is connected to the other cores via a high-speed bi-directional ring interconnect, which results in full coherency of the Level 2 cache units. Therefore, despite the fact that there is no Level 3 cache for the Xeon Phi, each core can access the L2 cache associated with other cores via the ring interconnect in a constant access time that is not influenced by the location of the cores [Fang et al., 2014]. Thus the entire set of L2 cache units can act in a fashion similar to a distributed L3 cache [Fang et al., 2014]. The type of memory chip on the card is Graphics Double Data Rate (GDDR) memory, the same memory used in contemporary GPU cards. GDDR memory provides higher bandwidth and lower access time in comparison to the main memory for CPUs [Chrysos and Engineer, 2012].

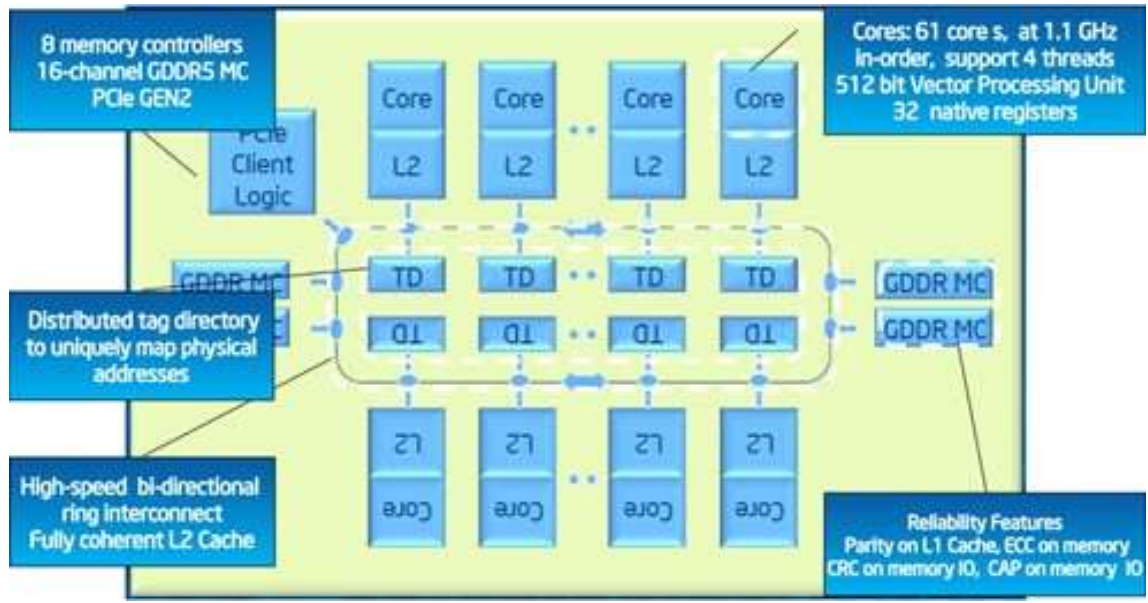

Figure 2.7: Intel Xeon Phi (MIC) architecture (Intel 2013).

Nevertheless, the Intel MIC sees this memory in exactly the same way that the CPU sees the main memory. All caching tasks are performed in the background and are transparent to the programmer (*i.e.*, the data is fetched from main memory to the L2 cache, then to the L1 cache inside the x86 code and into scalar or vector registers automatically if necessary). In this way, developers do not need to be concerned about understanding different types of memory in the hierarchy or micro-managing the underlying memory caches.

There are two pipelines inside each Xeon Phi core as illustrated by Fig. 2.8. Therefore, the core itself is dual issue per cycle (one for scalar and one for vector) [Jeffers and Reinders, 2013, Intel, 2013] and four hardware threads are available on each core. Most of the vector instructions have a 4-clock latency, which can be hidden by round-robin scheduling of multiple threads on a core. Therefore, the effective pipeline throughput is one-per-clock [Jeffers and Reinders, 2013, Intel, 2013]. Since the instruction decoding unit has a 2-cycle latency, no back-to-back cycle issue is possible from the same thread; therefore, at least two hardware threads are needed to fully utilize each core [Jeffers and Reinders, 2013, Intel, 2013]. Unlike Intel's previous vector instruction sets such as 64-bit MMX (for integers only), 128-bit Streaming SIMD Extensions (SSE), and 256-bit Advanced Vector Extensions (AVX) instruction sets, the Intel MIC has a new vector processing unit with 512-bit registers (see Fig. 2.9). As a result, in each cycle the Xeon Phi can execute an instruction on twice or more data than CPU does.

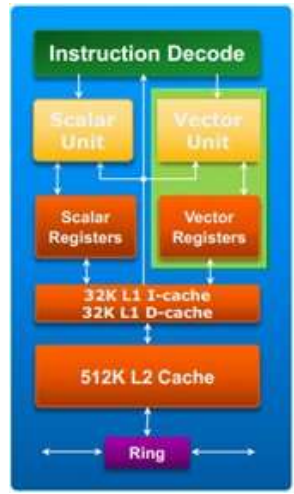

Figure 2.8: Intel Xeon Phi core architecture. There are two execution units per core, the scalar and vector units (shown in yellow). Each unit has access to several memory regions (shown in orange). (Intel 2013).

There are a few differences between the Intel Xeon Phi processor and traditional CPU. First, because the clock rate of the Xeon Phi is lower than CPU, it is not optimized to run serial code as it designed for parallelized/vectorized code. Second, the co-processor usually ships with lower memory than the host and less is available to programs [Intel, 2013]; thus, programs that require a larger amount of data may not be able to run on the co-processor. Third, compared to the out-of-order CPU cores, the MIC cores are simpler and cannot efficiently handle complicated code paths

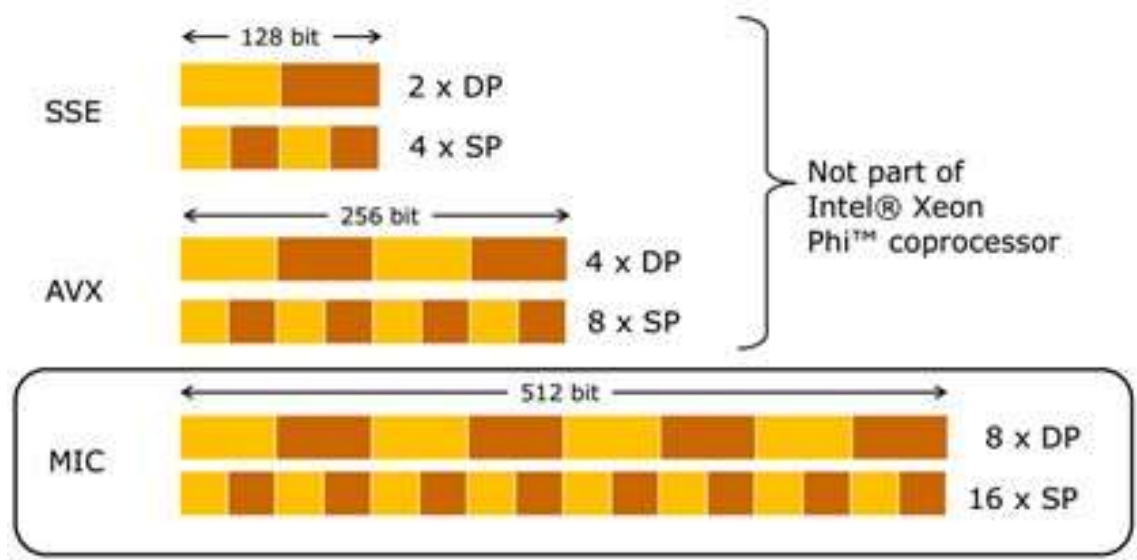

Figure 2.9: Intel Xeon Phi vector processing unit can operate on wider registers compare to traditional CPU's vector units. It can issue instructions on 8 double precision (DP) or 16 single precision (SP) numbers simultaneously (Intel 2013).

with many out-of-order jumps or nested conditional blocks.

There are several ways to utilize the vetorization unit on the Intel MIC: Intel Cilk Plus, Intel MKL, compiler auto-vectorization, and SIMD compiler directives [Jeffers and Reinders, 2013]. The Intel Cilk Plus is an extension of the C/C++ language which makes programming on the Xeon Phi easier for developers. However, it is still new, not portable and only available on Intel compilers. The Intel MKL is the Intel's math libraries that can be called within codes. It is optimized and easy to call, thus no code changes are required to scale forward. However, MKL is only math libraries which cannot be employed for general purpose solutions. In addition, it has been shown that the Intel MKL is not reliable on non-Intel x86 CPUs (*i.e.* AMD) and causes a drop in the performance of codes [Fog, 2015]. Although, the auto-vectorization technique only requires minor changes in codes, programmers need to make loops as simplest as possible so the compiler can vectorize them well [Jeffers and Reinders, 2013]. Furthermore, the functionality is limited by language and compiler technology and is not portable. The vectorization with SIMD directives is a reliable and predictable technique and is standardized in OpenMP 4.0 [OpenMP4.0, 2015]. The SIMD directives are powerful, yet need to be used carefully
to avoid changing the meaning of the program [Jeffers and Reinders, 2013].

Likewise, various techniques can be employed for parallelising code on the Intel MIC. The Intel Cilk Plus, Intel TBB, and OpenMP are among them. As mentioned earlier, Intel Cilk Plus extends C++ for parallelism. The parallelism can be achieved by employing special keywords, attributes, directives, and runtime libraries [Jeffers and Reinders, 2013]. The Intel Threading Building Blocks (TBB) is a compiler independent C++ template library that extends C++ for rich parallelism via C++ template libraries [Jeffers and Reinders, 2013]. Another technique is employing OpenMP for parallelisation. OpenMP works with compiler directives and runtime library which supports parallel loops, tasking model, and portable locks. Due to the standard nature of OpenMP and its portability, in XFLAT, OpenMP was employed
for intra-processor vectorizations and parallelisations.

There are considerable differences between the MIC and the GPU from the perspective of architectural design. First, in contrast to the GPU, Intel MIC is not optimized for concurrent out-of-cache random memory access by large numbers of threads [Intel, 2013]. Furthermore, the Intel MIC architecture has a traditional coherentcache architecture, whereas GPUs have a memory architecture specialized for localized shared memory processing [Intel, 2013]. Moreover, "threads" and "cores" mean something very different, as GPU versions are limited and lighter-weight [Intel, 2013]. Furthermore, on GPU, cores are limited to perform simple floating point operations and a number of threads can act as SIMD unit. Since the Intel MIC architecture is similar to traditional CPUs, it is a better choice for accelerating highly vectorized

and parallelized codes. They can be run on both CPU and the MIC without any

```
#pragma offload target(mic : target_id) \ in(data_in : length(SIZE) \ inout(data_inout:
length(SIZE)
for (int i = 0; i < SIZE; ++i)
{
     /// calculations here!
     data_inout[i] += data_in[i];
}
```

Figure 2.10: Offloading a loop to co-processor. The offload programming model of Xeon Phi is similar to the GPU programming model.

further modification simultaneously.

Typically, programs run on accelerators such as GPUs via *offload* mode in which the main program still continues its run on the CPU. The computational portions of the code and the required data are only transferred to the GPU. After completing the calculations, the results are transferred back to the main program on the CPU. Hence the non-computational parts such as input/output, MPI calls and other network communication remain on the CPU. However, on the Intel MIC there are two methods to run programs on the co-processor. The first method is to run codes in offload mode in a way that is similar to GPU. This method is compatible with the GPU programming style, thus codes that are written for GPU can run on the MIC with subtle modifications. Fig. 2.10 illustrates the offload programming model. The other approach is to run codes in *symmetric* mode which implies an entirely independent program executing on the MIC co-processor [Jeffers and Reinders, 2013]. In this mode, programs run on the Xeon Phi precisely in the way that they run on the CPU. Consequently, the Intel MIC can be seen as an MPI node at runtime, so that all the communication and I/O tasks are performed without interrupting the CPU. Hence, it is possible to have another MPI process running on the CPU simultaneously. Fig. 2.11 depicted different programming models on the Xeon Phi.

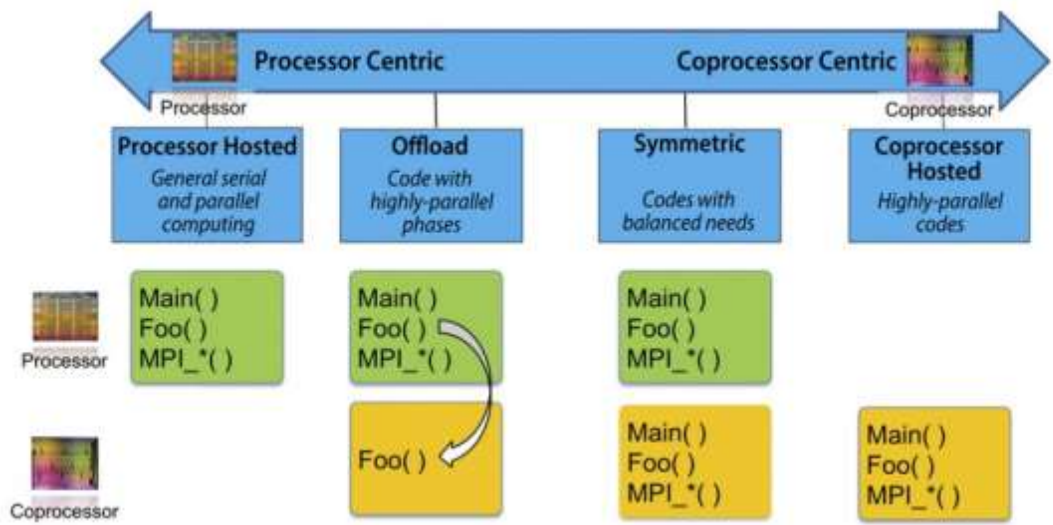

Figure 2.11: Xeon Phi programming models. A code can run on either the CPU (shown in green) or the MIC (shown in yellow) or both. [Jeffers and Reinders, 2013].

In addition, from the perspective of the software stack, there are a number of important differences between GPUs and the Xeon Phi: the former is controlled by a driver software, yet the latter is controlled by a micro-kernel Linux operating system [Jeffers and Reinders, 2013, Intel, 2013]. As a result, the Linux Standard Base core libraries are available on the Xeon Phi (see Table 2.1). On accelerators such as GPU, a portion of the driver resides on the CPU (a.k.a. Host) and controls the accelerator via the PCI Express bus. The driver is responsible for managing accelerators, providing low level software stack for data transfer, and running kernel codes on them by providing the runtime environment for programs. By having a Linux operating system running on the Xeon Phi, there will be no need to have the co-processor being controlled from the Host. As the co-processor is Internet Protocol (IP) addressable, communicating with the MIC can be directly completed without routing via the Host. Additionally, because there is a Linux OS on the co-processor, various I/O, MPI and other third-party libraries can be compiled, built, and run on the Xeon Phi as well. Therefore, in contrast to GPU, there is no need to write the Table 2.1:    Linux Standard Base (LSB) core libraries on the Intel MIC [Jeffers and Reinders, 2013].

| Component | Description |
|---|---|
| glibc | the GNU C standard library |
| libc | the C standard library |
| libm | the math library |
| libdl | programmatic interface to the dynamic linking loader |
| librt | POSIX real-time library |
| libcrypt | password and data encryption library |
| libutil | library of utility functions |
| libstdc++ | the GNU C++ standard library |
| libgcc s | a low-level runtime library |
| libz | a lossless data compression library |
| libcurses | a terminal-independent method of updating character screens |
| libpam | the Pluggable Authentication Module (PAM) interfaces |

required libraries for the Xeon Phi from scratch.

Since both the Xeon Phi and CPU share the same underlying architecture (x8664 instruction set), it has been demonstrated that optimizing code for the Intel Xeon

Phi, can result in more optimized code for Xeon CPU as well [Calvin et al., 2013, Jeffers and Reinders, 2013]. Consequently, programmers need to not maintain two separate code bases each with different data structures, algorithms, and optimization paths. Understandably, this is a huge advantage over GPU programming for developers, which they are required to maintain multiple programs for multiple targets.

## 2.2.2    The raw performance of the MIC

Although, the MIC is equipped with more cores and lengthier vector registers, its performance advantages over the CPU depends on the codes that run on it, dramatically. Therefore, prior to any real-world benchmarks, the Intel Xeon Phi should be benchmarked for basic arithmetic operations (*i.e.* additions, multiplications and transcendental functions) in order to find the processor's capabilities and limitations for the basic and most utilized mathematical functions in the developed code.

First the raw performance of the CPU and Xeon Phi on Stampede was benchmarked using a code adapted from [Jeffers and Reinders, 2013] with the structure illustrated in Fig. 2.12 (see Appendix B). In the innermost loop of this code simple floating point operations (*i.e.* additions and multiplications) were performed on an array or vector of double-precision (DP) floating-point numbers. The widths of the vectors were taken to be a multiple of that of the SIMD registers of the computing component (256 bits or 4 DP for the Xeon CPU, and 512 bits or 8 DP for the Xeon Phi). The same vector operations were repeated 10 million times in the middle loop to maintain data locality. In the outermost loop all of the hardware threads were utilized to achieve the best performance. The results of these benchmarks with different vector widths in the innermost loop are shown in Fig. 2.13. These results show that the floating point performance of the Xeon Phi is highly sensitive to the width of the vector, while the performance of the CPU is relatively stable. The floating point performance of the Xeon Phi is best when the width of the DP vectors is 64; in that test, the Xeon Phi ran 10 times as fast as the CPU. However, the performance of the Xeon Phi degrades substantially as the vector width increases.

Since XFLAT employs transcendental functions such as sin() and exp(), the transcendental function performance of the CPU and Xeon Phi were benchmarked using a code similar to that of Fig. 2.12 with the simple floating point operations replaced by a pair of sin() and cos() functions in one series of tests, and exp() in the other (see Fig. 2.13). These results suggest that the transcendental function performance on the Xeon Phi is relatively stable against the width of the vectors. For the tests on sin() and cos() the Xeon Phi ran 6–8 times as fast as the CPU, but for exp() the Xeon Phi is only 3–4 times better.

The double-precision (DP) floating-point benchmarks showed that the Xeon Phi ///
8/244 OpenMP threads for CPU/Xeon Phi.

```
#pragma omp parallel for for (t:
NUM_THREADS)
{
  /// Repeat 10 million times.
  for (itr: LOOP_COUNT)
  {
    /// VECTOR_WIDTH is a multiple of 4/8 /// for
    CPU/Xeon Phi. #pragma simd for (s:
    VECTOR_WIDTH)
    {
      /// floating point operations ...
    }
  }
```

}

Figure 2.12: High-level structure of the benchmark code for floating point performance.

can outperform CPU by ten times only when the vector of double-precision numbers is short. However, by increasing the length of the vector beyond 64 double-precision numbers, the performance of the Xeon Phi drops dramatically and becomes comparable to that of the CPU. From the transcendental functions benchmarks it can be concluded that the performance of transcendental functions on the ˜60 core MIC is a few times better than the 8-core CPU. For the short vector of numbers, the sin()/cos() and exp() on Xeon Phi ran 8 times and 4 times as fast as the CPU, respectively. However, by increasing the vector length beyond 64 double-precision numbers, the sin()/cos() on Xeon Phi ran only 6 as fast as the CPU, and the exp() on 244 threads of the Xeon Phi was slightly higher than 3 times as fast the CPU's 8 threads.

As a result, the performance of Xeon Phi is dependent on the loop structures of applications and the length of vectorized loops. Thus, it can be varied dramatically by a slight changes in the size of data. It also directly depends on the type of the employed operations and functions in applications. Therefore, the decision employing Xeon Phi in real world applications depends on the application type and may or may not be economically beneficial.

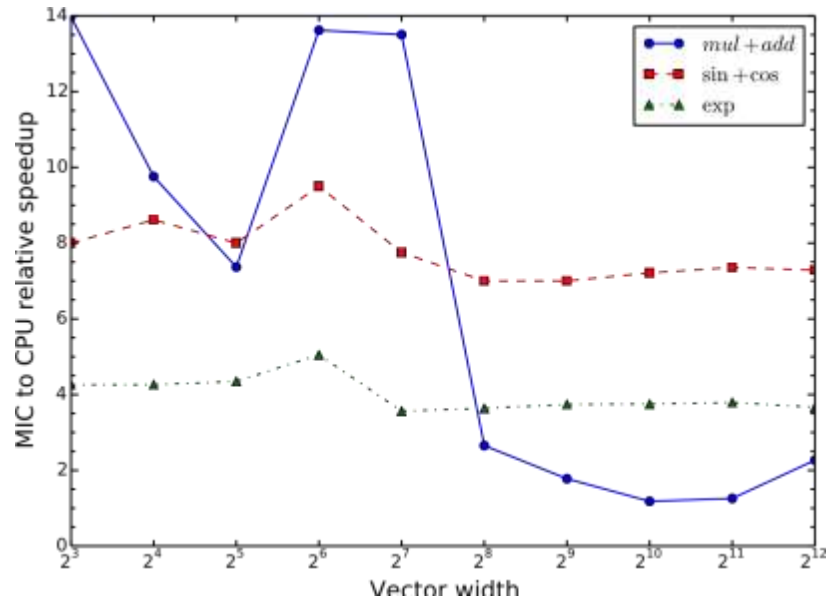

Figure 2.13: The relative speedup of the floating point and transcendental functions for the CPU and Xeon Phi. The width of vector is the length of the innermost loop (vectorized loop) in the kernel.

### 2.2.3    Previous works on the MIC

After the introduction of the Xeon Phi, many researchers started to explore the capabilities of the new co-processor. Many researches haven been performed on the Xeon Phi via benchmarking kernels in order to study the advantages and disadvantages of the co-processor. In addition, several real-world applications were developed to harness to power of the Xeon Phi for scientific researches. As a result, Xeon Phi was utilized in many scientific research fields.

[Pennycook et al., 2013] employed Xeon Phi for molecular dynamics. They employed intrinsic functions, which are mapped by the compiler directly onto the corresponding assembly instruction, as well as the compiler #pragma in order to exploit SIMD registers. They code on the MIC achieved 1–1.4× performance improvement
over a single 8-core Xeon E5-2660 CPU.

[Crimi et al., 2013] ported a Lattice Boltzmann code on the Xeon Phi. They used the offload programming approach. For vectorization they employed the intrinsic functions. Their code on the Xeon Phi achieved 1.12× higher performance than a 16-core system.

[Leang et al., 2014] utilized Xeon Phi for quantum chemical calculations. They compared the Xeon Phi with a system that were equipped with two 8-core Xeon E5-2650. For all Phi runs, the MIC KMP ₋AFFINITY environmental variable for the co-processor was set to compact as this was found to offer better overall performance relative to the default affinity setting of scatter. To improve data transfer performance on the Phi, the environmental variable MIC ₋USE 2MB BUFFERS=2M was set to turn on the use of 2 MB page buffers. They achieved 1.5–2× improvement over the
host.

[Apra` et al., 2014] reported an implementation of many-body quantum chemical methods on the Xeon Phi. The auto-vectorization and offload model were chosen for vectorization and design strategies, respectively. Moreover, performance benefit was observed by setting the environment variable MIC USE 2MB _BUFFERS=16K. They reported 2.1× improvement over a Xeon E5-2670 CPU.

[Teodoro et al., 2014] employed the Intel MIC for microscopy image analysis (digital pathology). Digital Pathology involves the analysis of images obtained from whole slide tissue specimens using microscopy image scanners. They employed the offload programming approach and utilized the compiler #pragma in order to vectorize their code. Their Xeon Phi implementation achieved 0.4 – 1.9× performance
over a system with two 8-core Xeon E5-2680 processors.

[Wende and Steinke, 2013] was exploited the Intel MIC in Swendsen-Wang multicluster algorithm for the 2D/3D Ising model (ferromagnetism in statistical mechanics). They utilized the intrinsic functions for vectorizations and the environmental variable KMP _AFFINITY=balanced,granularity=fine for threads pinning. Speedups up to a factor 3 over a single 8-core Xeon E5-2680 was observed, however, for the smaller problem size the performance of the CPU and MIC were comparable.

[Liu and Schmidt, 2014] utilized Intel MIC in biological sequence database search. Their code was written in for Xeon Phi with a set of SIMD intrinsics. They investigated the offload model to coordinate multiple co-processors to perform sequence alignments. Their Xeon Phi code achieved 1.19 – 2.49× higher performance than
implementations on a single 8-core CPU.

[Park et al., 2013] developed a code for efficient backprojection-based synthetic aperture radar computation using Intel MIC. They reported a 1.9× improvement over two 8-core Xeon E5-2670 CPUs.

[Halyo et al., 2014] employed the Intel MIC for real time particle tracking based on Hough transform at the Large Hadron Collider (LHC). In order to utilize the co-processor, data was moved from the CPU to the co-processor using the offload functionality. In addition, the auto-vectorization functionality of the compiler was employed for utilizing SIMD units. The reported Xeon Phi's performance was 0.3× of two 12-core E5-2697v2 CPUs (Xeon Phi was three times slower).

[Knoll et al., 2013] exploited the Xeon Phi for an application of ray tracing and volume rendering large molecular data. They employed SPMD language (special languages that generates vectorized code) for code vectorizations. They reported about 2× performance improvement on the Xeon Phi over 8-core Xeon E5-2680 CPU.

Table 2.2: Summary of the implementation approaches on the Intel MIC for various researches. The '-' means the approach did not mention in the paper. For the Multinode column, N/A means that the code was run on single node, thus there was no multi-node support.

|  | Exe. mode | Multi-node | Multi-thread | SIMD |
|---|---|---|---|---|
| [Pennycook et al., 2013] | - | MPI | - | Intrinsic |
| [Crimi et al., 2013] | Offload | N/A | - | Intrinsic |
| [Leang et al., 2014] | Offload | N/A | OpenMP | - |
| [Apra` et al., 2014] | Offload | - | OpenMP | auto-vec. |
| [Teodoro et al., 2014] | Offload | - | OpenMP | auto-vec. |
| [Wende and Steinke, 2013] | Native | MPI | OpenMP | Intrinsic |
| [Liu and Schmidt, 2014] | Offload | - | OpenMP | Intrinsic |
| [Park et al., 2013] | Offload | MPI | OpenMP | Intrinsic |
| [Halyo et al., 2014] | Offload | N/A | OpenMP | Intrinsic |
| [Knoll et al., 2013] | - | MPI | OpenMP | SPMD |

| [Kulikov et al., 2015] | Offload | MPI | OpenMP | - |

[Katja Malvoni and Solar Designer and Josip Knezovic, 2014] utilized the Intel MIC for cracking a password hashing scheme based on the Blowfish block cipher. Their application's performance on the Xeon Phi were comparable to the performance of a 4-core i7-4770K CPU.

[Kulikov et al., 2015] developed an astrophysics code on Xeon Phi and they reported over 10× improvement over a single-core of an E5-2690 CPU (Around 1.2× improvement over an 8-core CPU).

The implementation methods of the mentioned researches on the Intel MIC is summarized in Tab. 2.2

All the reported speedup can be normalized to a reference hardware. Since several works (including the current work) employed Stampede supercomputer, all the hardware were normalized to the Stampede's MIC and CPU:

$$\frac{\dfrac{MIC\# \ of \ cores}{Stampede's \ MIC\# \ of \ cores} \times \dfrac{MIC's \ freq.}{Stampede's \ MIC's \ freq.}}{\dfrac{CPU\# \ of \ cores}{Stampede's \ CPU\# \ of \ cores} \times \dfrac{CPU's \ freq.}{Stampede's \ CPU's \ freq.}} \qquad (2.1)$$

Fig. 2.14, depicts the relative speedup of each works normalized to the Stampede's hardware. As it is observable, the expected speedup for a full fledged application is at most around four times on a lower-end CPU and by increasing the CPU power it decreases and to be around $1 - 1.5\times$ for the current generation of the Xeon Phi. XFLAT could achieve up to three times speed up on the Stampede's MICs and up to four times speed up on Bahcall's machine. The specification of the machines are shown in Tab.5.2 and Tab.5.3.

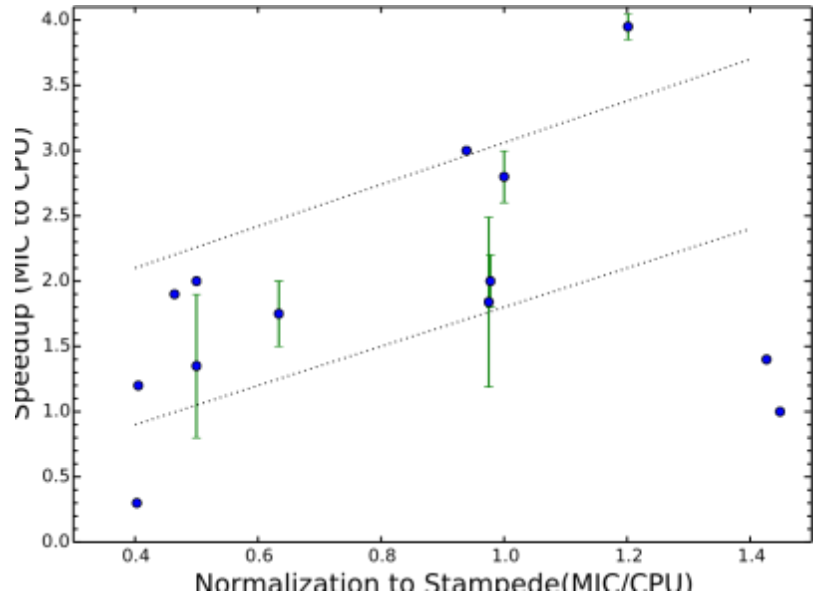

Figure 2.14: Illustration of the relative MIC to CPU speedup of different applications. The x-axis is the *MIC⁰s cores×MIC⁰s freq. / CPU⁰s cores×CPU⁰s freq.* normalized to the Stampede's hardware.

# Chapter 3

# Neutrino Oscillations

## 3.1        Physics of Neutrino Oscillations in Supernova

As discussed earlier, neutrino oscillations in which neutrinos can change their flavor state is a quantum mechanical phenomenon [Barger et al., 2012]. In neutrino oscillations phenomenon a neutrino created with a specific flavor ($\nu_e$, $\nu_\mu$ or $\nu_\tau$) can later have a different flavor. In neutrino oscillations the probability of measuring a particular flavor for a neutrino varies periodically as it propagates through space. Neutrino oscillations implies that the neutrino has a non-zero mass, which was not included as part of the original Standard Model of particle physics [Barger et al., 2012].

Neutrino oscillations evidences can be collected from multiple sources: Solar neutrino oscillations [Davis et al., 1968] (in which the neutrinos that are produced in the Sun are studied), atmospheric neutrino oscillations [Fukuda et al., 1998] (when a cosmic ray hits a nucleus in upper atmosphere, neutrinos may be created during the process), reactor neutrino oscillations [An et al., 2012] (neutrinos may be created in nuclear reactors' core), and Beam neutrino oscillations [Agafonova et al., 2010] (neutrinos may be created in particle accelerators).

In a supernova, it has been shown that, because there are so many neutrinos emitted from the neutron star within such a short time, flavor oscillations of neutrinos with different energies that are propagating along different trajectories are not independent of each other but are correlated [Duan et al., 2010]. As illustrated in Fig. 3.1, following the flavor evolution of neutrino $v_p$ would entail knowing the flavor states of all neutrinos on which it forward scatters (only the trajectories of neutrinos along the escaping directions are accounted), such as $v_q$ and $v_k$ propagating on trajectories which intersect $v_p$'s trajectory at points $P$ and $Q$, respectively. Note that the flavor histories of $v_q$ and $v_k$ are not independent, as they have undergone a forward scattering at point $K$ in the past. This phenomenon is known as *collective*

*neutrino oscillation*. Since it is impossible to solve the supernova neutrino oscillation equations analytically, and it is very challenging to follow collective oscillations of supernova neutrinos numerically, the only model in which collective neutrino oscillations have been solved numerically so far is the *neutrino bulb model* in which the physical conditions are assumed to be exactly spherically symmetric for which the azimuthal symmetry around any radial direction [Duan et al., 2006].

Even in this seemingly simplistic model, as calculations of the oscillations of the neutrinos of different energies and emission angles are highly correlated, millions of coupled, nonlinear differential equations have to be solved simultaneously. Hence, with a huge number of neutrinos emitting from the surface, tracing the neutrino-neutrino evolution history quickly become an intractable simulation problem.

Not only solving each equation requires many computationally expensive functions (*i.e.* sin, cos, and exp) to be employed, yet still the whole model only works for the neutrino-neutrino forward scattering interactions, which means it does not even include the interaction of neutrinos that scatter along backward direction. Since for each step and each particle's beam so many quantum mechanical equations have to be solved, and there might be millions of steps and millions of neutrino beams in the system, the entire simulation requires vast computational power which is only available on very high-end supercomputers.

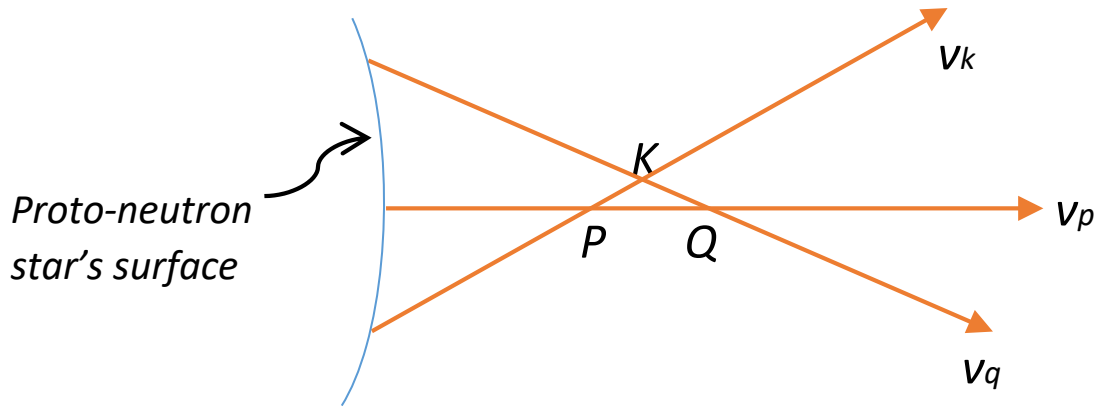

Figure 3.1: Illustration of non-linear problem for intersecting neutrinos trajectories. Points $P$, $Q$, and $K$ are intersections of the world lines for three neutrino beams, $\nu_p$, $\nu_q$, and $\nu_k$. Hence, the flavor evolution histories of these neutrinos will quantum mechanically coupled.

As previously mentioned, one of the most common ways to simplify the simulation is to reduce the dimensionality by studying simpler models such as isotropically spherical neutrino bulb model. The assumed spherical symmetry in the bulb model reduces the required calculations dramatically, as neutrino trajectories with the same emission angle become equivalent. Thus, neutrinos emitted in the same flavor and energy state and with the same emission angle have identical flavor evolution history [Duan et al., 2006].

As mentioned earlier there are three kinds of neutrinos–the $\nu_e$, $\nu_\mu$, and $\nu_\tau$–in the particle physics Standard Model. Each particle has an anti-particle ($\bar{\nu}_e$, $\bar{\nu}_\mu$, $\bar{\nu}_\tau$). The fluxes and energy spectra may be different for each neutrino and antineutrino. As a result, the flavor content of the neutrino field above the proto-neutron star surface and its evolution history in time and space can be important, since many environments associated with compact objects and the very early universe are dominated by neutrinos and their interactions. The objective of nonlinear neutrino oscillation simulations in supernovae is to study and understand the flavor evolution of the neutrino field when $\nu_e$ and $\bar{\nu}_e$ mix with neutrinos and anti-neutrinos of other active flavors. Hence, the quantum state of the system must be studied.

The quantum state of a particle can be described by the *wavefunction*. The wavefunction of the flavor of a neutrino in the two flavor system is written as:

$$\dot{\psi}_\nu \equiv \left( \begin{array}{c} a \\ b \end{array} \right)$$

$$(3.1)$$

where $a$, and $b$ are the amplitudes for neutrino to be in the $\nu_e$, $\bar{\nu}_e$, and $\nu_x$ ($\bar{\nu}_x$) ($x$ represents either the $\mu$ or $\tau$ neutrino or a linear combination of these) flavor states, respectively.

The development in time and space of the complex amplitudes is followed that describes the flavor states of neutrinos, and the full quantum kinetic equations reduce to a Schrödinger equation. The Schrödinger equation determines how the wave function evolves over time; i.e., the wavefunction is the solution of the Schrödinger
equation.

For a given neutrino trajectory, with parameter $l$, which is the traveled length of a neutrino (as the speed of neutrinos is assumed to be nearly the same as the speed of light, thus $l$ and time $t$ are identical and interchangeable), the Schrödinger can be written as: d

$$i \frac{}{dl} \psi_\alpha (\mathbf{\hat{v}},E;l) = (H_0 + H_m + H_{\nu\nu}) \cdot \psi_\alpha (\mathbf{\hat{v}},E;l), \qquad (3.2)$$

where $\alpha$, $\mathbf{\hat{v}}$ and $E$ are the initial flavor, (the unit vector of) the propagation direction, and the energy of the neutrino, respectively. $H_0$ is the Hamiltonian in vacuum, $H_m$ is the matter potential, and $H_{\nu\nu}$ is the neutrino potential because of the ambient neutrinos. The propagation direction $\mathbf{\hat{v}}$ is fully described by the polar angle $\vartheta$ between $\mathbf{\hat{v}}$ and the radial direction when axial symmetry about the radial direction is imposed. In the neutrino bulb model the $\mathbf{\hat{v}}$ is determined by $\vartheta$. As depicted in Fig. 3.2, at any given radius $r$ all the geometric properties of a neutrino beam can be calculated using only $r$ and $\vartheta_0$. Hence, $\vartheta$ and $\Theta$ are related to $\vartheta_0$ through:

$$\frac{\sin \vartheta}{R_\nu} \equiv \frac{\sin \Theta}{l - l_0} \equiv \frac{\sin \vartheta_0}{r}, \qquad (3.3)$$

where $l \equiv r \cos\vartheta$ and $l_0 \equiv R_\nu \cos\vartheta_0$.

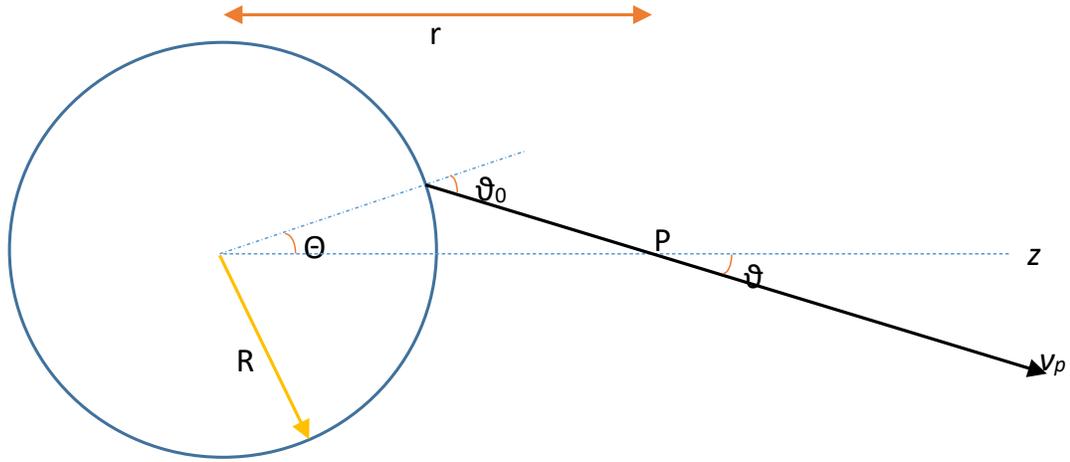

Figure 3.2: Geometry of the neutrino bulb model. The neutrino beam $v_p$ is emitted from a point on the proto-neutron sphere with radius $R_\nu$, with polar angle $\Theta$ and intersect the $z$-axis at point $P$ with angle $\vartheta$. All the geometry properties can be calculated from the radius $r$ and the emission angle $\theta_0$, which is defined with respect to the normal direction at the point of emission on the star ($\vartheta_0 = \Theta + \vartheta$).

However, in other models such as the *extended bulb model* [Mirizzi, 2013] where axial symmetry is not imposed, $\hat{\mathbf{v}}$ is determined by both $\vartheta$ and the azimuthal angle $\phi$ about the radial axis. Thus, the $\hat{\mathbf{v}}$ is directly related to the geometry of the model. For $n$ neutrino flavors, $\psi$ is a vector of $n$ complex variables, and $\mathsf{H}_0$ and $\mathsf{H}_{\nu\nu}$ are both $n \times n$ Hermitian matrices. Since the neutrino potential $\mathsf{H}_{\nu\nu}$ (ambient neutrinos forward-scattering) contributes to the Hamiltonian, a serious complication arises and renders the problem nonlinear, as the interactions which dictate flavor transformation amplitudes are themselves dependent on the neutrino flavor states. Neutrino modes are binned in each flavor by the energy function and emission angle.

As a result, the flavor evolution of the neutrino wavefunction can be derived from Eq. 3.2:

$$\frac{\mathrm{d}}{l}\psi_\alpha\left(\hat{\mathbf{v}}, E; l\right) = \left(\mathsf{H}_0 + \mathsf{H}_m + \mathsf{H}_{\nu\nu}\right) \cdot \psi_\alpha\left(\hat{\mathbf{v}}, E \quad \text{i}\_\_ \quad ; l\right) \mathrm{d}$$ where $\theta$ is the vacuum

mixing angle,

$$= \frac{1}{2} \begin{pmatrix} -\Delta\cos 2\theta + A + B & \Delta\sin 2\theta + B_{e\tau} \\ \Delta\sin 2\theta + B_{e\tau}^* & \Delta\cos 2\theta - A - B \end{pmatrix} \psi_\alpha\left(\hat{\mathbf{v}}, E; l\right), \quad (3.4)$$

and $\Delta$, $A$, and

$B_{e\tau}$ are the vacuum potentials induced by neutrino mass difference, the matter, and the background neutrinos, respectively. The appropriate Hamiltonian for anti-neutrinos is obtained by making

the transformation:

$$A \to -A, \quad B \to -B, \quad B_{e\tau} \to -B_{e\tau}^*$$

In Eq. 3.4, the $H_0$ is the vacuum oscillations. Since neutrinos have non-zero mass they may change their flavors by only propagating through vacuum. Therefore, the vacuum oscillations term is:

$$i \frac{d}{dl} \psi_\alpha(\mathbf{v},E;l) = -H_0 \psi_\alpha(\mathbf{v},E;l), \quad \frac{d}{dl} \frac{1}{2} \begin{pmatrix} -\Delta\cos2\theta & \Delta\sin2\theta \\ \Delta\sin2\theta & \Delta\cos2\theta \end{pmatrix} \quad (3.5)$$

where $\Delta$ is the vacuum oscillations frequency and is defined as:

$$\Delta \equiv \frac{\delta m^2}{2E_\nu}, \quad (3.6)$$

where $\delta m^2$ is the neutrino mass-squared difference, and $E_\nu$ is the energy of the neutrino. The mass-squared difference can be defined in terms of the appropriate neutrino mass eigenvalues. Although, the absolute value of the the mass-squared difference ($\delta m^2_{23}$) can be determined experimentally, it is still not known if the masssquared difference is a positive or a negative value. Therefore, simulations are run with both the positive and the negative values.

Electrons interact with neutrinos via weak force. As a result, the matter density above the neutron star plays an important role as well. The matter potential ($H_m$) in Eq. 3.4 is defined as:

$$A = \sqrt{2}G_F n_e = \sqrt{2}G_F Y_e n_b, \quad (3.7)$$

where $G_F$ is the Fermi coupling constant, $n_e$ is the net electron number density that is related to baron density with $n_e = Y_e n_b$, where $Y_e$ is the electron fraction, and $n_b$ is the baryon density.

The difficulty in solving Eq. 3.4 stems from the neutrino potential ($H_{\nu\nu}$), as shown:

$$\mathsf{H}^{\nu\nu} = \sum_{\alpha'} \int dE' \int d\hat{\mathbf{v}}' \, (1 - \hat{\mathbf{v}} \cdot \hat{\mathbf{v}}') \left[ \varrho'_{\nu_\alpha} f_{\nu_\alpha}(E') \frac{L_{\nu_\alpha}}{\langle E_{\nu_\alpha} \rangle} - \varrho'^{*}_{\bar{\nu}_\alpha} f_{\bar{\nu}_\alpha}(E') \frac{L_{\bar{\nu}_\alpha}}{\langle E_{\bar{\nu}_\alpha} \rangle} \right] \quad (3.8)$$

where the quantities with primes are associated with the ambient neutrinos, the quantities with bars are associated with anti-neutrinos. $L_{\nu\alpha}$, h$E_{\nu\alpha}$i, and $f_{\nu\alpha}(E^0)$ are the energy luminosity, average energy and normalized energy distribution function of $\nu_\alpha$, respectively.

The $\%^0$ in Eq. 3.8 is called the density matrix and can be calculated directly from $\psi_{\alpha 0}(\mathbf{v}^{\wedge 0}, E^0; l)$ as is shown:

$$\varrho \equiv \begin{pmatrix} |a|^2 - |b|^2 & 2ab^* \\ 2a^*b & -|a|^2 + |b| \end{pmatrix}, \tag{3.9}$$

where $a$ and $b$ are the components of the wavefunction ($\psi_\alpha$) in Eq. 3.1.

For a small distance, Eq. 3.4 can be solved numerically as:

$$\psi_\nu\left(l+\delta l\right) \simeq \exp\left(-i\mathsf{H}\delta l\right)\psi_\nu\left(l\right.$$

$$= \frac{1}{\lambda}\begin{pmatrix} \lambda\cos(\lambda\delta l)-ih_{11}\sin(\lambda\delta l) & -ih_{12}\sin(\lambda\delta l) \\ -ih_{12}^*\sin(\lambda\delta l) & \lambda\cos(\lambda\delta l)+ih_{11}\sin(\lambda\delta l) \end{pmatrix}, \quad (3.11)$$

$$(3.10)$$

where $h_{11}$ and $h_{12}$ are the diagonal and off-diagonal elements of the Hamiltonian $H$, and $\lambda$ is defined as:

$$\lambda \equiv \sqrt{h_{11}^2+|h_{12}|^2} \tag{3.12}$$

## 3.2 Implemented Physics Modules

### 3.2.1 Bulb Model

As previously discussed, the implementation of this project was intended to support a variety of geometries, and the "neutrino bulb model" is only one of them. In the neutrino bulb model, there are multiple neutrino trajectories along different zenith directions (the $\vartheta$ angles). The emission from each point on the star's surface was assumed to be identical, thus the study of a single emission point is sufficient. However, azimuthal symmetry was assumed in the bulb model. Azimuthal symmetry means that all the trajectories along a particular zenith angle but along different azimuthal directions (the $\phi$ angles) was assumed to be identical. Fig. 3.3 illustrates the neutrino bulb model. There are different trajectories along zenith direction emitted from a point on the surface, and every point on the surface was assumed to be equivalent.

Consequently, Eq. 3.8 can be rewritten for the bulb model as:

$$H_{\nu\nu} = \sum_{\alpha'} \int dE' \int (1 - \cos\vartheta\cos\vartheta') \left[ \varrho'_{\nu_\alpha} f_{\nu_\alpha}(E') \frac{L_{\nu_\alpha}}{\langle E_{\nu_\alpha} \rangle} - \varrho'^*_{\bar\nu_\alpha} f_{\bar\nu_\alpha}(E') \frac{L_{\bar\nu_\alpha}}{\langle E_{\bar\nu_\alpha} \rangle} \right] d(\cos\vartheta')$$

(3.13)

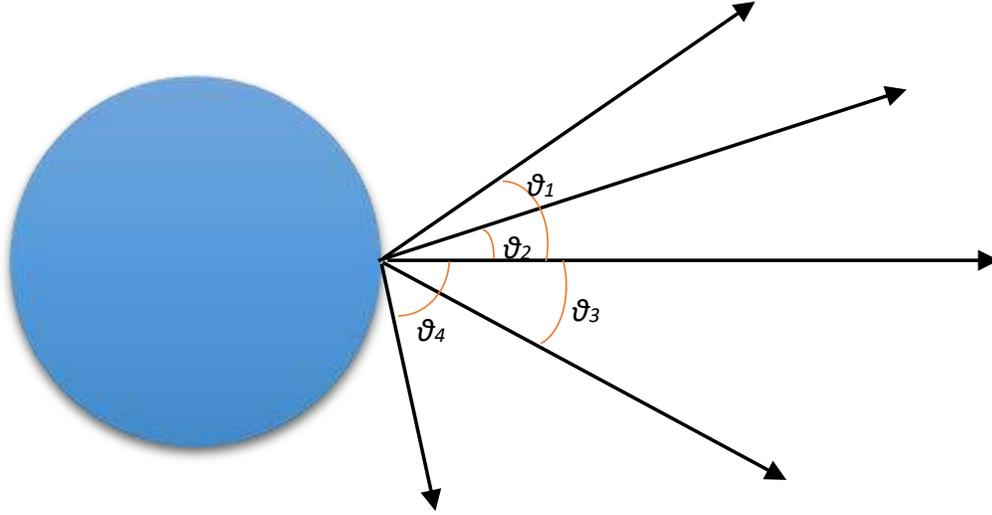

Figure 3.3: Illustration of multi-zenith angle model. There are different trajectories along zenith direction ($\vartheta_1$, $\vartheta_2$, $\vartheta_3$, $\vartheta_4$) emitted from a single point on the star's surface (shown in blue), and every point on the surface is assumed to be equivalent.

where $\vartheta$ is the polar (zenith) angle of the neutrino of interest and $\vartheta'$ is the polar (zenith) angle of the background neutrino. At each point the $\cos\vartheta'$ can be calculated as:

$$\cos\vartheta' = \sqrt{1 - \left(\frac{R}{r}\right)^2 (1 - \cos^2\vartheta'_0)}$$

(3.14)

where $R$ is the proto-neutron star's radius, $r$ is the distance from the center of the proto-neutron star to the current point and $\vartheta'_0$ are the initial angles at the surface (see Fig. 3.2) for which the angle bins are calculated. From Eq. 3.14, $d\cos\vartheta'$ can be computed as:

$$d\cos\vartheta' = \frac{1}{2}\left(\frac{R}{r}\right)^2 \frac{d\cos^2\vartheta'_0}{\cos\vartheta'}$$

(3.15)

where $\vartheta_0$ angles are only need to be calculated once at the proto-neutron star's surface.

## 3.2.2    Single-angle Model

This model is the simplified version of the bulb model. In fact, it is the simplest implemented geometry in XFLAT. Similar to the bulb model, every points on surface of the proto-neutron star emits neutrinos identically, however, there is a single identical emitting angle for all of the points on the surface. Therefore, the required amount of computations is a few orders of magnitude less than the bulb model since there are no multi-zenith angle trajectories for this model. As a result, Eq. 3.13 can be simplified further as:

$$\mathsf{H}^\nu = D(r/R_\nu)\sum_\alpha \int \left[\varrho'_{\nu_\alpha}f_{\nu_\alpha}(E')\frac{L_{\nu_\alpha}}{\langle E_{\nu_\alpha}\rangle} - \varrho'^*_{\bar\nu_\alpha}f_{\bar\nu_\alpha}(E')\frac{L_{\bar\nu_\alpha}}{\langle E_{\bar\nu_\alpha}\rangle}\right]\mathrm{d}E',$$

(3.16)

where $r$ is the current radius, $R_\nu$ is the proto-neutron star's radius, and the geometric factor $D(r/R_\nu)$ is defined as:

$$D(r/R_\nu) \equiv \frac{1}{2}\left[1 - \sqrt{1 - \left(\frac{R_\nu}{r}\right)^2}\right]^2$$

(3.17)

Furthermore, due to simplicity of this model and presenting a single trajectory, only the vector registers within a single core can be employed for computations. Therefore, multi-core or multi-node run is not supported for this model. As a result, a run on Xeon Phi can not accelerate the computations.

### 3.2.3    Extended Bulb Model

The extended bulb model is essentially the same as the neutrino bulb model with the inclusion of azimuth trajectories (the $\phi$ angles). Fig 3.4 illustrates the extended bulb model including the azimuth angles. For each zenith angle ($\vartheta$), there is a cone of azimuth angle trajectories. The parameter $r$ is the distance from the center of the proto-neutron star to the current calculation point and $R$ is the neutron star's radius. Moreover, minor perturbation may be added to azimuth beams to distinguish them from each other. However, this model is not truly a representation of neutrino supernova model anymore. Breaking the azimuthal symmetry (by adding beams along the azimuth direction), causes breaking the spherical symmetry. Therefore, the model is not self-consistent anymore. The simulation of this model is merely to study behavior of neutrinos in a multi-azimuth experimental model, and to help the prediction of their evolution history in the more complicated environments.

Therefore, Eq. 3.8 can be rephrase as:

$$H^{\nu\nu} = \sum_{\alpha} \int \mathrm{d}E' \int (1 - \cos\gamma) \left[ \varrho'_{\nu_\alpha} f_{\nu_\alpha}(E') \frac{L_{\nu_\alpha}}{\langle E_{\nu_\alpha} \rangle} - \varrho'^*_{\bar{\nu}_\alpha} f_{\bar{\nu}_\alpha}(E') \frac{L_{\bar{\nu}_\alpha}}{\langle E_{\bar{\nu}_\alpha} \rangle} \right] \mathrm{d}\varphi' \mathrm{d}(\cos\vartheta')$$

$$\tag{3.18}$$

where $\cos\gamma = \cos\vartheta\cos\vartheta^0 + \sin\vartheta\sin\vartheta^0 \cos(\phi - \phi^0)$.

The new model requires enormously higher amount of computations, since for each zenith angle trajectory, there may be hundreds to thousands of azimuth angles. Therefore, the entire calculations are increased by at least two order of magnitudes. As a result, employing supercomputers that are equipped with accelerators are unavoidable. In addition, since the memory requirement is at least two order of magnitudes higher, it may not be possible to run the code on a single processor or on a single Xeon Phi co-processor. As a result, multi-node high-performance machines are necessary for performing the computations of this module.

### 3.2.4   Plane/Point Model

Another one of the geometries to study the neutrino oscillation is the plane geometry. The recent progresses in instability calculations have shown that if the azimuthal asymmetries are taken into account, in linearized stability calculations, a new kind of instability the multi-azimuth angle appears, which happens at smaller radii. Since, adding the azimuth angles to the bulb model breaks the spherical symmetry, this

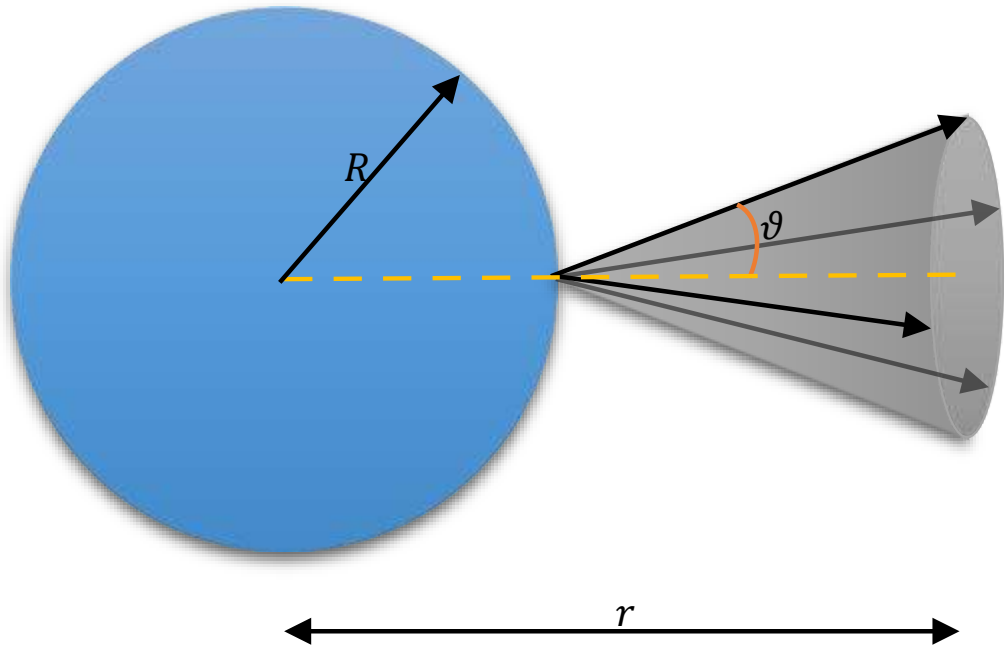

Figure 3.4: Illustration of the extended bulb model including the azimuth angles. For each zenith angle ($\vartheta$), there is a cone (shown in grey) of azimuth angle trajectories. The parameter $r$ is the distance from the center of the proto-neutron star (shown in blue) to the current calculation point and $R$ is the neutron star's radius. model tries to restore the broken symmetry in order to make the model self-consistent. As a result, all the emitting points are assumed to be on a plane surface, therefore there is no breaking in the symmetry anymore. As depicted in Fig. 3.5 all the points on the plane surface are assumed to be the equivalent, hence a particular neutrino trajectory emitting along $(\vartheta, \phi)$ direction from point $p_0$ would have identical property and history as another neutrino trajectory emitting along $(\vartheta, \phi)$ direction from another point $p_1$. For that reason, all neutrino trajectories alongside the same direction $(\vartheta, \phi)$, and emitted from different points on the plane surface, would have precisely the equivalent evolution history.

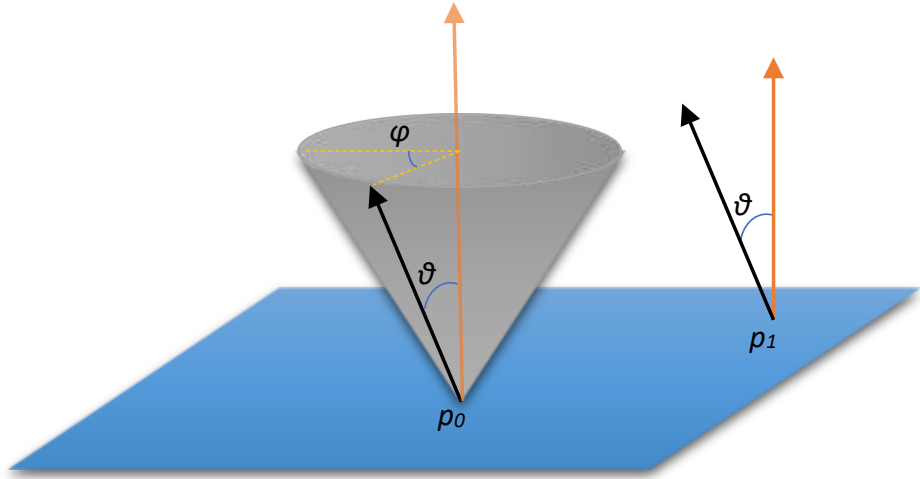

Figure 3.5: Demonstration of the plane geometry. The point $p0$ and $p1$ have identical emission. For every zenith angle $\vartheta$, there are multiple neutrino trajectories along different $\phi$ angles (shown as the grey cone).

Hence, Eq. 3.8 can be recast as:

$$\mathsf{H}^{\nu\nu} = \sum_\alpha \int \mathrm{d}E' \int (1 - \cos\gamma)$$

$$\left[\varrho'_{\nu_\alpha} f_{\nu_\alpha}(E') j_{\nu_\alpha}(\varphi', E') - \varrho'^*_{\bar\nu_\alpha} f_{\bar\nu_\alpha}(E') j^*_{\bar\nu_\alpha}(\varphi', E')\right] \mathrm{d}\varphi' \mathrm{d}\left(\cos\vartheta'\right), \qquad (3.19)$$

where $\cos\gamma$ is defined as $\cos\gamma = \cos\vartheta\cos\vartheta^0 + \sin\vartheta\sin\vartheta^0\cos(\phi - \phi^0)$, and $j_{\nu\alpha}(\phi^0, E^0)$ is the number flux of $\nu_\alpha$ with energy $E^0$ emitted in $\phi^0$ direction.

## 3.2.5    Cylinder/Line Model

It is of interest to understand the behavior of neutrinos when there are multiple emitting points from a surface. However, there is still no complete model available for the multiple emitting point simulation of supernova. As an alternative to the neutrino bulb model, in the cylinder model an infinitely long cylinder emits neutrinos from multiple points. To make the model more straightforward neutrino trajectories are all perpendicular to the axis of the cylinder. As illustrated in Fig. 3.6, neutrinos that are emitted from points $p_1$, $p_2$, $p_3$, and $p_4$ are different. The points that located along the $z$ axis are assumed to be all identical. Therefore, the instability comes from different points on a slice of the cylinder, emitting different neutrino beams at the same time.

In this model, Eq. 3.8 can be written as:

$$H_{\Phi\vartheta} = \int \mathrm{d}E' \int_{-\pi/2}^{\pi/2} (\varrho'_{\nu_\alpha}(\Phi', \vartheta') - \varrho'^*_{\bar\nu_\alpha}(\Phi', \vartheta')) \frac{\cos\vartheta'}{\sqrt{1 - \left(\frac{R_\nu}{r}\right)^2 (1 - \cos^2\vartheta')}}$$
$$\left[1 - \cos\left(\arcsin\left(\left(\frac{R_\nu}{r}\right)\sin\vartheta\right) - \arcsin\left(\left(\frac{R_\nu}{r}\right)\sin\vartheta'\right)\right)\right] \mathrm{d}\vartheta', \qquad (3.20)$$

where $\Phi$ is the latitude angle of the neutrino of interest and $\Phi^0$ is the latitude angle of the background neutrino, and $\Phi' = \Phi + \vartheta' - \vartheta + \arcsin\left(\left(\frac{R_\nu}{r}\right)\sin\vartheta\right) - \arcsin\left(\left(\frac{R_\nu}{r}\right)\sin\vartheta'\right)$.

Because there may be hundreds to thousands of different emitting points, the complete calculations require multiple orders of magnitudes more computational power than the bulb model. As a result, the required computations for this model as well as the other models beyond the neutrino bulb model will require multi-node machines and accelerators, thus employing high-performance hybrid codes in order to harness the power of new heterogeneous supercomputers are mandatory.

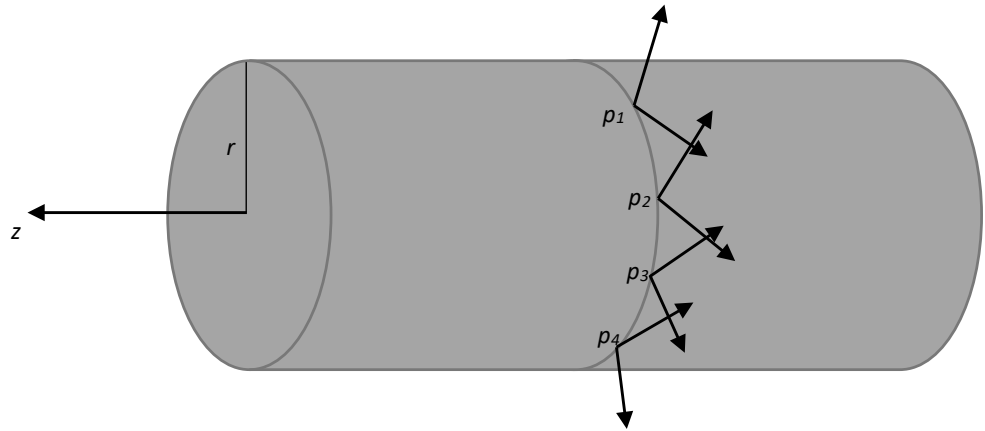

Figure 3.6: Illustration of the cylinder model and its emitting points' trajectories. Neutrinos emitting from points $p1$, $p2$, $p3$, and $p4$ are different, however, since the azimuth symmetry is assumed, the points that located along the $z$ axis are all identical. $r$ is the radius of the cylinder.

# Chapter 4

# Design, Implementation, and Validation of the XFLAT code

## 4.1 Numerical Implementation

It has been shown that the neutrino oscillations in the bulb model can be solved numerically using the previously developed code, FLAT [Duan et al., 2008]. In FLAT the quantum flavor states of neutrinos, $\psi_\alpha(\vartheta,E;r)$, at a given radius $r$ were described by a multi-dimensional array psi alpha[theta, E]. At each radius $r$, a summation over the elements of this array were performed to obtain the neutrino potential $H_\nu$. The numerical algorithm similar to the midpoint method was subsequently employed to solve the Schr̈odinger equation and to evolve $\psi_\alpha(\vartheta,E;r)$ one radial step further. In a typical run 100 − 1,000 discrete energy bins may be required in order to achieve the desired resolution for the energy spectra function. In addition, over 1,000 polar angle ($\vartheta$) bins may be required to achieve numerical convergence. In other words, millions of nonlinear differential equations have to be solved simultaneously for the computations of the neutrinos flavor evolution in the bulb model. More

*4.1. NUMERICAL IMPLEMENTATION*

realistic models, in which the geometry is more complex, can have significantly larger problem sizes. For example, the inclusion of the azimuth ($\phi$) dimension in the extended bulb model increases the problem size by a at least a couple of orders of magnitude, depending on the number of $\phi$ beams. As a result, in order to accelerate the computations and explore the new models, employing modern supercomputers seems inevitable.

One of the computationally time consuming sections is the calculation of the $H_\nu$. In order to accelerate the computations, it is desirable that those calculations run on more than one compute nodes. At the same time, the number of nodes should be kept minimum in order to minimize the overhead of the inter-node communications. Furthermore, the larger the problem's size is (for more complex supernova models), the higher computational power is reuired. The balance between the two factors can be achieved by employing accelerators or co-processors such as Graphics Processing Units (GPU) and the Xeon Phi. In principle the computational power of accelerators is higher than traditional CPUs.

As mentioned in previous chapters, in order to maintain the same code for both accelerators and CPUs, the Intel Xeon Phi co-processor (which is based on the Intel x86 architecture) was chosen for the XFLAT framework development.

XFLAT was written in C++ and contains about ten thousand lines of code. It employs an algorithm similar to FLAT for solving the Schr¨odinger equations. However, unlike FLAT, which exploited only MPI, XFLAT implemented and utilized three levels of parallelism. In Fig. 4.1 the high level code structure of XFLAT is shown for one of the the extended neutrino bulb models. At the top level, the $\vartheta$ angle bins were distributed among compute nodes (either a CPU or a Xeon Phi) and nodes communications were handled via MPI. At the middle level (*i.e.*, on a CPU or Xeon Phi) $\vartheta$ angle bins were further dispatched to individual cores via OpenMP. At the bottom level (within a thread), the energy bins' loops were performed via Sin-

*4.1. NUMERICAL IMPLEMENTATION*

```
/// Each node receives array of neutrino beams init_NBeam(psi[]);
/// The main neutrinos' evolution loop while
(!termination_conditions) { ...
```

```
/// Divides local polar bins among OpenMP threads
#pragma omp parallel for for (theta:
POLAR_ANGLE_COUNT) { ... for (phi:
AZIMUTH_ANGLE_COUNT) { ...
    /// Distributes energy bins over SIMD
    #pragma simd for (E:
    ENERGY_BINS)
    {
       /// wavefunction computations of neutrino beams(theta,phi,E)
       psi[theta,phi,E].calc(...);
    }
  }
}
/// Inter-node MPI communication; nodes exchange the partial result
MPI_AllReduce(...);
}
```

Figure 4.1: High-level structure of XFLAT parallelism for the extended supernova model. The first loop is over the polar angles ($\vartheta$), the loop middle is over the azimuth angles ($\phi$), and the most inner loop is over energy bins. gle Instruction with Multiple Data (SIMD) paradigm. SIMD units cannot directly be utilized by C++ complex numbers. Therefore, in order to use SIMD unit efficiently, double-precision floating-point arrays (*i.e.*, ar alpha[E], ai alpha[E], *etc.*) represent the real and imaginary components of the variables in complex vectors $\psi_\alpha(\vartheta,\phi,E;r)$ for a single $\phi$ and $\vartheta$, and at a given $r$. These arrays are then grouped into an object element of neutrino beam array NBeam[theta,phi] for a given $\phi$ and $\vartheta$.

## 4.2 Architecture

As formerly stated, there are two methods for running applications on the Xeon Phi: the offload mode and the symmetric mode. However, the offload mode entails several compromises. For instance, programs have to use one less core, since the idle core is responsible for code and data transfers to/from the MIC [Jeffers and Reinders, 2013]. Thus, there is always less processing power available in offload mode. Furthermore, the offload mode is not suitable for very short tasks. Since the overhead of data transfers as well as thread creation may suppress the computational segment performance. On the other hand, in symmetric mode, there is no code or data transfer during run time except the inter-node MPI communication. Therefore, all of the hardware resources such as cores and hardware threads are available for computation. In addition, in symmetric mode the CPU does not have to be attached to the codes that run on the MIC, hence it is feasible to launch another process on the CPU as an additional MPI task. As a result, in this project the symmetric mode was chosen for XFLAT implementation. Because the offload mode requires maintaining two separated codes for CPU and MIC, and in comparison to the symmetric mode it has higher overhead, the symmetric mode is preferable.

There are several factors that affect the overall design of the code. For instance, the architecture of XFLAT should be as flexible as possible so that it can perform the simulation of diverse geometries and scenarios. Furthermore, since XFLAT is a high-performance code, it is supposed to be able to exploit all of the available parallelism levels and hardware resources (see Fig. 4.2).

In order to increase the flexibility of the code, XFLAT was designed to be modular. This modularity can save scientists and developers' time and effort by allowing them to change a single physics module without any requirement of knowledge about the internal structures of the other modules. Since various geometries, different par-ticle types in the system, diverse energy spectra functions, ... might be required, a module can simply be switched with another module, thus only a code rebuilding is necessary before running XFLAT.

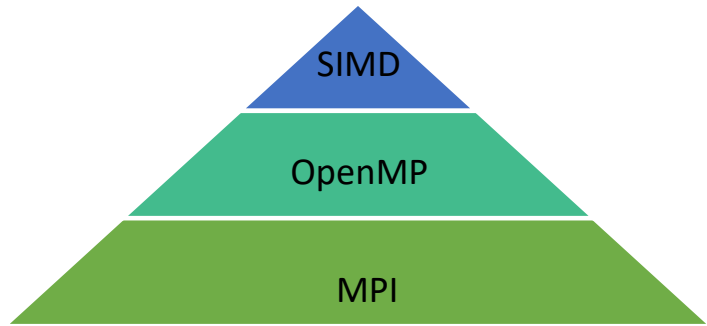

Figure 4.2: Parallelism hierarchy of XFLAT. MPI (show on the bottom) was employed for inter-node parallelism, OpenMP (shown in the middle) was exploit for multi-thread parallelism on each processor, and SIMD vectorization (shown on the top) was utilized for the last level of parallelism within each core.

As illustrated in Fig. 4.3, the software framework architecture is a dual-layer design. The upper layer is responsible for general portions of the code including the numerical, physics, and I/O modules. The lower layer is responsible for per particle calculations such as calculations of the Hamiltonians, computing neutrino's flavor evolution. In this approach, if the particles in the simulation are required to be replaced by other particle's categories, only the lower module has to be swapped with a replacement module. In addition, if the geometry and physics of the simulation are required to be changed, the only affected module will be the physics module in the upper layer. The modularity of the code dictates that each module encapsulates all of its internal data, structures, and functionalities within the module, and only communicates with other modules via their exposed Application Programming Interface (API) functions. This requirement guarantees that by swapping modules the functionality of the rest of the framework will not be affected.

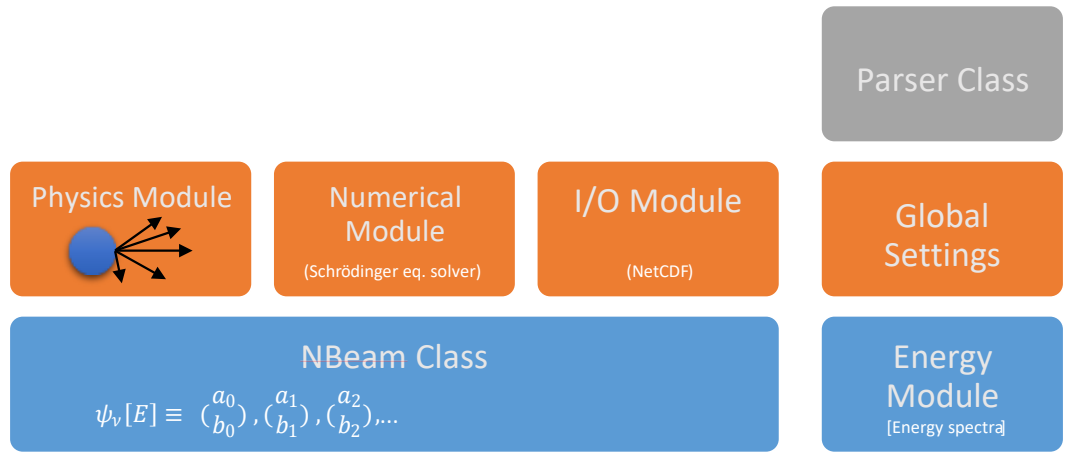

Figure 4.3: High level hierarchical illustration of XFLAT modules. The bottom layer (shown in blue) is responsible for per particle calculations. The upper layer (shown in orange) is responsible for the general functionalities such as geometrical and I/O methods. Other modules (shown in grey) function as helper modules.

Together with the data structure, the algorithm was designed so that the overhead of the inter-node communications, data movement within memory, and I/O operations are minimized. As illustrated by Fig. 4.4, the application starts from the Parser class, which receives a configuration file from the command prompt console. The Parser extracts the variables from the configuration file and stores them into the Utility module. Afterwards, the Parser starts reading the configuration file line by line, if it encounters character '#', the line is commented out and should be skipped. Otherwise, the Parser scans each line from left to right. The configuration file's tokens begin from the beginning of each line and are in TokenID= <value> format. The Parser class maintains a list of valid tokens (i.e. 'TokenID='), thus tries to match the read token with one of the pre-defined tokens. Whenever a matching is found, the <value> next to the token is read and stored onto the corresponding token's variable. The rest of each line may contain a comment, which starts with a '#', as well. The Parser class continues scanning all of the lines of the configuration file and stores their value onto the Utility module for later usage.

Subsequently, the application continues within the top-layer modules by calling the initialization functions of every module in a precise order. Initially, the global module calls the init() functions of the NBeam module, the Physics module, the I/O module, and the Matter profile module respectively.

## 4.2.1    NBeam module

The NBeam module is at the lowest level of the hierarchy that implements the NBeam class. Since each instance of the NBeam class represents a single neutrino beam, it contains an array of $\psi_\alpha(\sim v, E)$ in which each element represents a neutrino with a particular energy. As a result, the NBeam's init() function is responsible for memory allocation of the energy bins as well as invoking the init() function of the Energy module in which the energy spectra function initializes. For each component of the wavefunction, there is an array of double precision numbers inside the NBeam class. For instance, the wavefunction of the two-flavor system contains two complex numbers ($a,b$) and each complex number can be represented by two double precision floating point numbers, one for the real part and the other for the imaginary part (see Fig. 4.2.1). As a result, four double numbers can describe the flavor state of a neutrino at a given energy ($\psi_\alpha(a_r, a_i; b_r, b_i)$). Consequently, for a given range of energy, there can be four arrays $a_r[E], a_i[E], b_r[E], b_i[E]$ and each element of those arrays corresponds to a flavor state for a particular energy bin $[E - \frac{1}{2}\Delta E, E + \frac{1}{2}\Delta E]$.

The NBeam class encapsulates several private (internal) methods. Those methods including the density method in which the density matrix calculations are performed (see Eq. 3.9), and the neutrino evolution method in which the corresponding flavor state $\psi_\alpha(\sim v, E)$ evolves one step further for a given Hamiltonian $\hat{H}$ (see Eq. 3.11). In addition, the NBeam class contains several public (external) methods that are accessible from the other modules. One of the important interface functions is the getESum() in which the numerical integration over all energy bins are performed (the first integral in Eq. 3.8). The other important function is the evolveBins(...) function that loops over all energy bins and evolves each corresponding flavor state with a given Hamiltonian (see Eq. 3.11). Another NBeam's public method is the addAvg(NBeam&) method in which the average of the flavor states between two NBeam objects is calculated (one object is passed by reference and the other is the this pointer). Additionally, in order to calculate the error between two neutrino beams, another method (the calcErr(NBeam&)) is provided in which it receives an

```
class NBeam { private:

    // Each array's range is [E_min:E_max] double ar[], ai[],
br[], bi[]; // density matrix method void density(ar[E], ai[E],
br[E], bi[E], ...); // single energy flavor state evolution void
U(Hamiltonian H[], ...); ... public:
    // calculates the integral over energy bins void
    getESum(...);
    // loops over energy range and evolve all of the flavor states void
    evolveBins(Hamiltonian H[], ...);
    // computes the average of the flavor states of two NBeam objects void
    addAvg(NBeam& beam);
    // calculates the error of flavor states double
    calcErr(NBeam& beam); ...
};
```

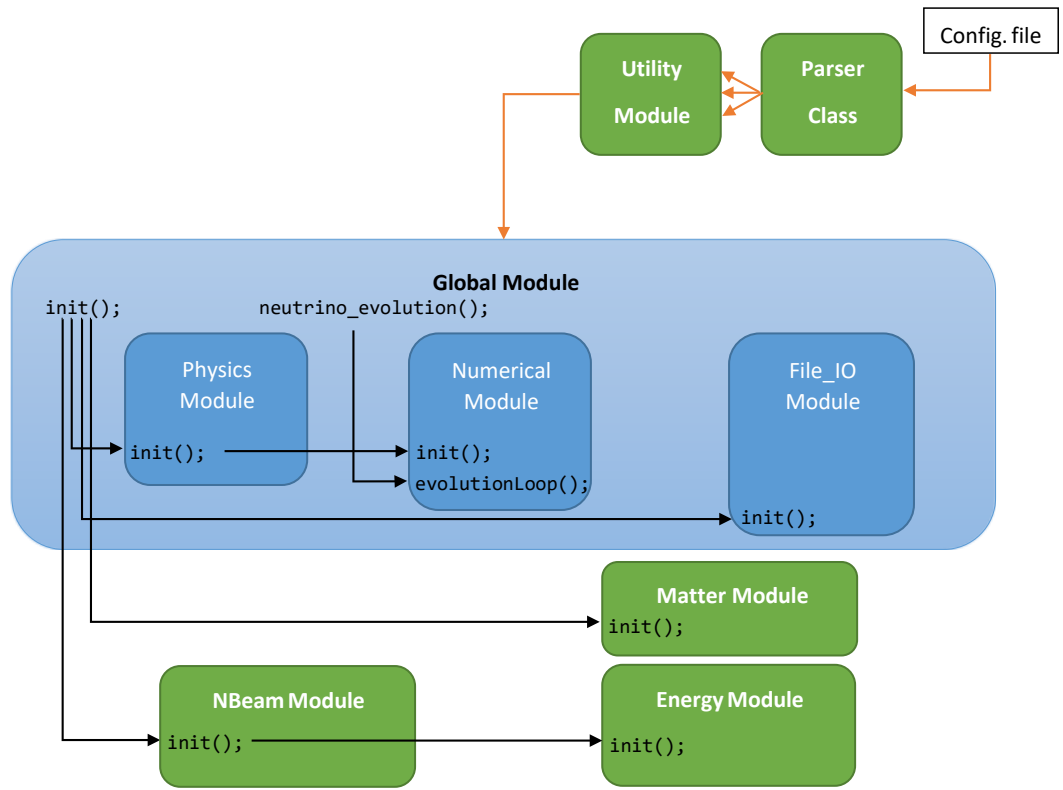

Figure 4.4: High-level illustration of XFLAT initialization. The Global module (shown in light blue) in which the upper layer's modules are encapsulated, begin initialization of each module in a particular order.

Figure 4.5: NBeam class overall structure and its private and public functions, for the two particle system. object of NBeam as an argument and loops over energy bins to calculate the error between the flavor states of the current object, the this, and the received object. All of the public functions in the NBeam class exploit SIMD instructions for computations of $a_r[E], a_i[E], b_r[E], b_i[E]$ components for the $\psi_\alpha(\sim v, E)$.

Since NBeam objects were distributed over processors' cores and the NBeam module is located on the bottom layer of XFLAT architecture, one NBeam instance can entirely exploit the first layer of parallelism within each core (SIMD) for the flavor state calculations (see Fig. 4.6).

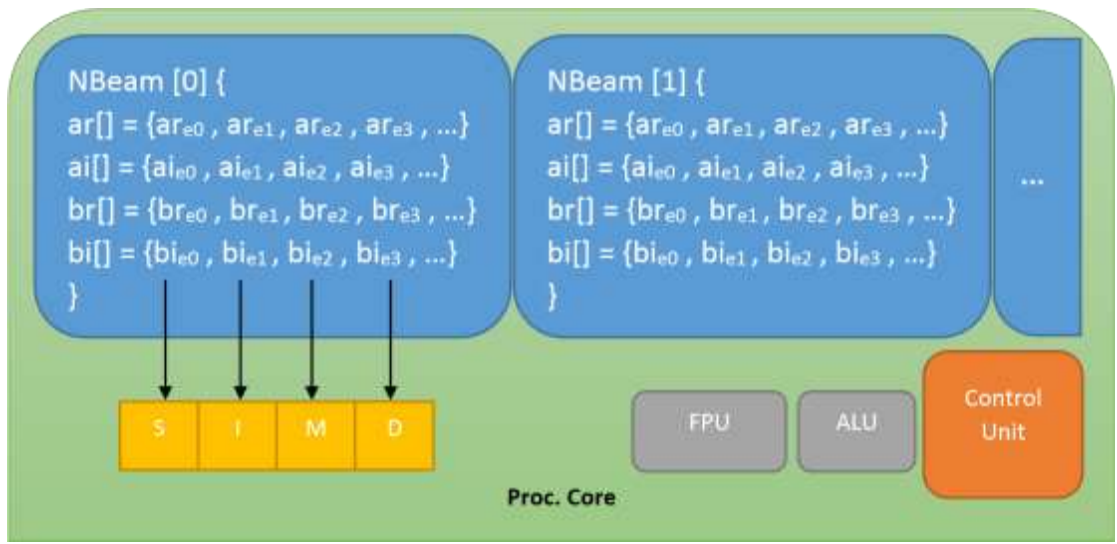

Figure 4.6: NBeam objects can employ the lowest level of parallelism, SIMD unit (shown in yellow). Each NBeam instance (shown in blue) issue vector instructions in order to utilize SIMD unit.

## 4.2.2   Physics module

The Physics module is the next module from which the init() function is called from the global module. The Physics module is responsible for the implementation of geometrical related functions as well as the computations of the Hamiltonian. The module encapsulates functions in which computations of the arrays of NBeam objects are performed via OpenMP threads (see Fig. 4.7). The instructions within those functions are based on the plugged-in geometry module and may be different between modules. The Physics module's init() function has two major responsibilities. First, it is responsible for allocating the angle dependent arrays (such as $\cos(\vartheta)$ bins). Furthermore, it calls the init(int) method of the Numerical module to which an integer as the length parameter is passed. The length parameter is the number of all trajectories along all different directions, such as zenith direction ($\vartheta$) and azimuth ($\phi$) angles. Since, the Physics module maintains the information about the number of dimensions and the length of each, the Numerical module can receive the length parameter only from the Physics module.

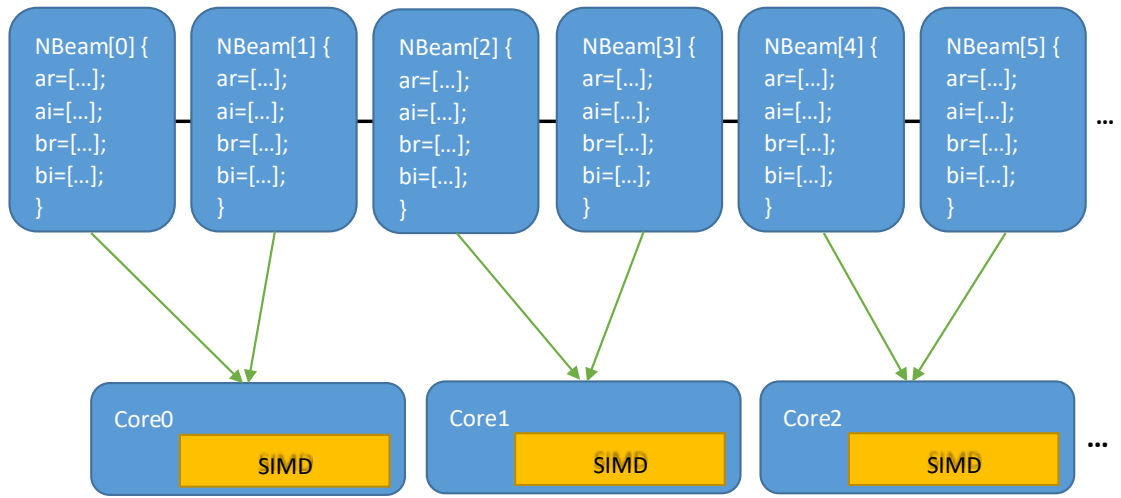

Figure 4.7: Arrays of NBeam objects are distributed over processor's cores via OpenMP. Thus, each core may be responsible for the calculations of several NBeam class instances.

Within the Physics module there are several methods that are accessible and utilized from the Numerical module. One of those methods is the initBeam(NBeam*) which receives an array of NBeam objects, loops over each element of the array and calls the constructor of each element. This function should be called before any function calls from the NBeam class in order to initialize and allocate memory for the wavefunction's components within each object. Thus, for each NBeam class array, the Numerical module calls the initBeam(NBeam*) method to initialize each array's elements. There are two more important methods within the Physics module that are utilized from the Numerical module. The first one is the newHvv(double*&) method in which the memory for an array of Hamiltonians are allocated (each particular neutrino beam trajectory has a particular Hamiltonian). The last method is the deleteHvv(double*&) in which the previously allocated memory for Hamiltonians is freed.

### 4.2.3    Numerical module

Upon calling the init(int) method of the Numerical module, depending on the implemented numerical algorithm, multiple arrays of NBeam instances may be required to be allocated and maintained onto the main memory. For each allocated NBeam array, the initBeam(NBeam*) method of the Physics module must be called to initialize the array. Hence, the evolutionLoop() method in the Numerical module is called from the global module to begin the neutrino flavor evolution calculations. It continues the neutrinos evolution calculations until one of the application criteria reaches to its maximum value. The application ending criteria parameters are: the maximum radius, the maximum allowable run time, or the maximum number of radial iterations. Before starting the evolution loop, the first step within the evolutionLoop() method is to perform memory allocation for the arrays of Hamiltonians. This can be performed by calling the newHvv(double*&) method from the Physics module. Afterwards, the neutrino evolution function is called from the global module.

    The flowchart of the neutrino evolution loop is illustrated by Fig. 4.8. The neutrino evolution function is responsible for solving the the the Schrödinger equation (Eq. 3.2) and evolving the flavor state of each neutrino beam accordingly. Fig. 4.9 illustrates the algorithm steps in order to solve the Schrödinger equation within the evolutionLoop() function. The blue squares represent wavefunction states of neutrinos, and the green squares represent the calculated Hamiltonian from the previous state functions. The green arrows represent the length and the direction of neutrino evolution by applying the Hamiltonian, the 'Op(Avg)' box represents averaging between two state functions, and the '?' box shows the comparison of two

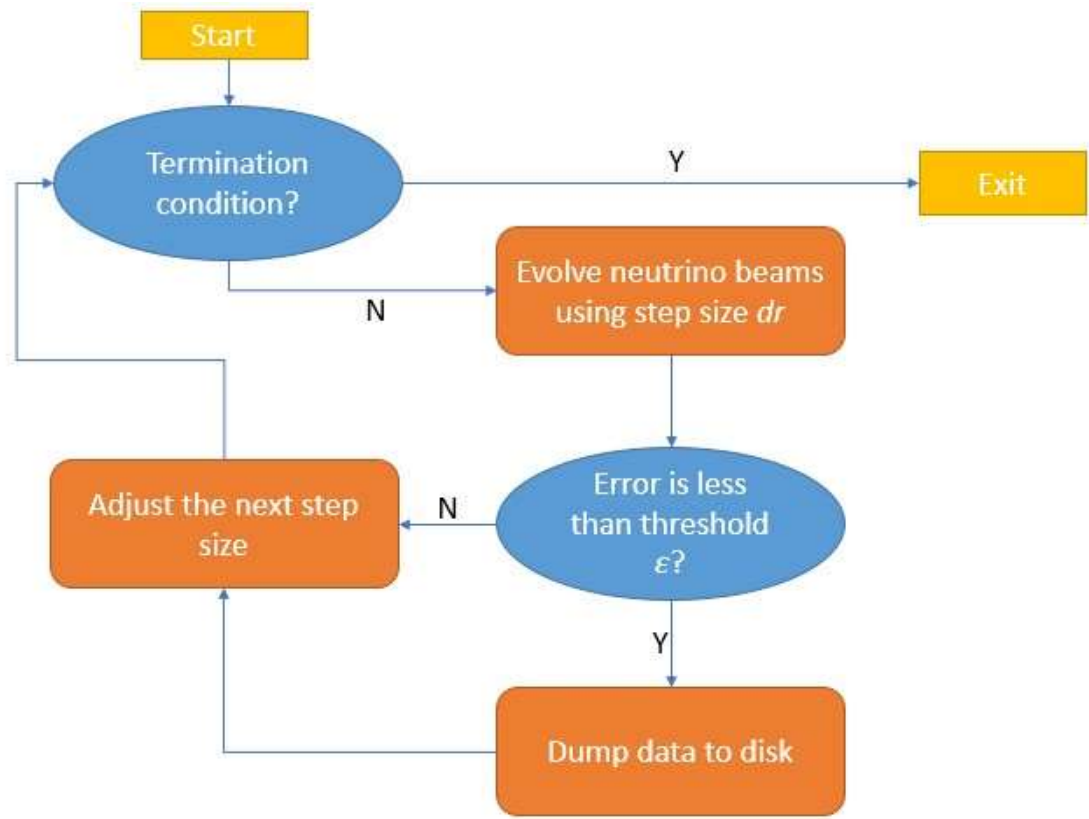

Figure 4.8: Illustration of the high-level neutrino evolution flowchart. The loop always starts by checking the termination conditions. The three major phases: the evolving of neutrino beams phase, the dumping data phase, and the adjusting step size phase (shown in orange) are performed accordingly.

state functions and if the error is less than a predefined threshold then saving the result.

As observable in Fig. 4.9, within the Numerical module at several points, the neutrino-neutrino Hamiltonian matrix (Eq. 3.8) has to be calculated based on the previously-calculated neutrinos' wavefunction. Each matrix calculation can be performed via a function call within the Physics module. The calc Hvv(...) method is responsible for computing the neutrino self-coupling Hamiltonian integral for which the loops and instructions within the function depend on the implemented geometry.

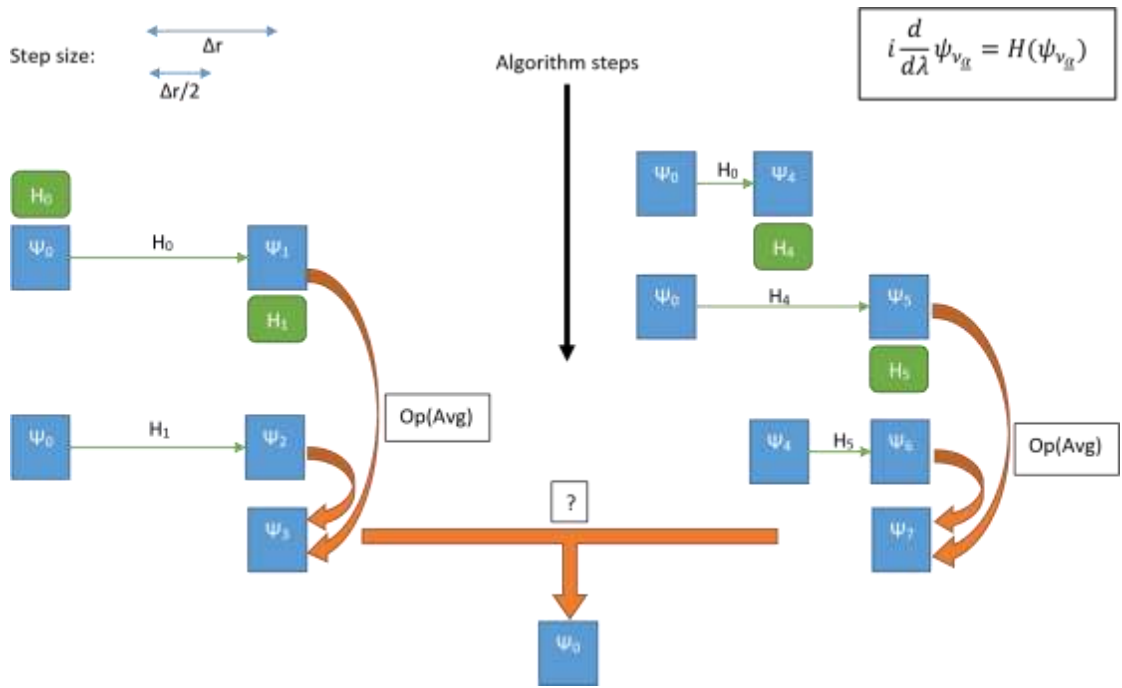

Figure 4.9: Illustration of the numerical algorithm for solving the Schro¨dinger equation. The algorithm starts from top to bottom, and left to right. The blue squares represent flavor states of neutrinos, the green squares represent the calculated Hamiltonian from the previous state functions, the green arrows represent the direction and step size of neutrino evolution using Hamiltonians, the 'Op(Avg)' boxes represent the averaging between two flavor states, and the '?' box shows the comparison of two flavor states (if the error is less than a predefined threshold then saving the result).

Once this method is called, it receives the current NBeam array's pointer as an argument and performs a summation over the array's elements via OpenMP threads. It results in computing the partial Hamiltonian integral summation (the pseudocode for the neutrino bulb model corresponding to Eq. 3.8 is illustrated in Fig. 4.10). Within the loop, by calling each NBeam object's NBeam::getESum() method, in which the SIMD unit is employed, the summation of wavefunctions over energy bins is calculated internally. If the code runs on multi-node configuration, there should be several data exchange points within the algorithm. Thus, utilizing the MPI functions are // parallel loop via OpenMP #pragma omp parallel for for (int ang = 0; ang < THETA_BINS; ++ang)

```
{
  // loop over neutrino flavors for (int n = 0;
  n < FLAVOURS; ++n)
  { ...
    // calculates Sum(E) of NBeam objects internally (SIMD) beam[neutrino_idx
      ].getESum(res_neu[] ); beam[anti_neutrino_idx].getESum(res_aneu[]);

    ...
  }
  // apply cos(t) and dcos(t) to the result angle_calc(result[], ...);
}
```

Figure 4.10: Illustration of the loop structure that is responsible for computing the partial Hamiltonian integral summation.

inevitable, since the NBeam arrays are distributed over multiple compute nodes (see Fig. 4.11). Consequently, to evaluate the final Hamiltonian integral summation's result, each MPI task exchanges its partially calculated integral result with all other tasks by employing the MPI reduction method. Afterwards, each node will have the final Hamiltonian integral value .

After the calculations of the Hamiltonian, the corresponding neutrino flavor state $|\psi(r)\rangle$ residing at the parameter value $r$ on a given trajectory must be evolved one step further for the given Hamiltonian $H$ and step size $\Delta r$. Therefore, the evolve(…) method within the Physics module is called in which it receives an array of NBeam objects and performs a loop over them via OpenMP threads. Within the loop, the NBeam's NBeam::evolveBins(…) method is called for each object. Within the NBeam::evolveBins(…) method, the SIMD unit is exploited to perform the neutrinos' flavor state evolution for all energy bins. The neutrinos' flavor state evolution can be performed by applying the calculated neutrino self-coupling Hamiltonian $H$ (the pseudocode corresponding to Eq. 3.11 is illustrated in Fig. 4.12) and updating the flavor states accordingly. The result of the evolved flavor state $|\psi(r)\rangle$ is saved onto another one-dimensional array of NBeam objects.

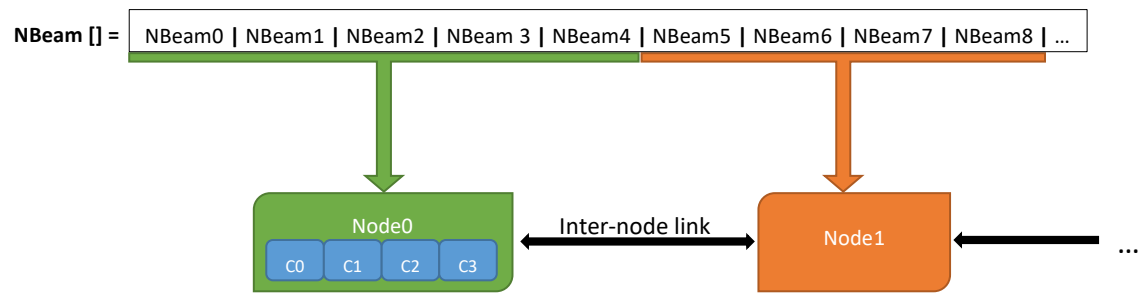

Figure 4.11: Distribution of NBeam objects over multiple nodes. Each node based on its computational capacity will receive a particular load. For instance, Node0 (shown in green) is more powerful than Node1 (shown in orange), thus it received more NBeam objects [NBeam0:Nbeam4] to process on, than Node1 that received only four objects [NBeam5:Nbeam8].

After evolving the neutrinos' flavor state, the next step is to compute the average of two calculated flavor states. The two flavor states were evolved separately using two different Hamiltonians and different step sizes. Therefore, the avgBeam(...) function from the Physics module is called to perform this task. This function receives two NBeam object arrays as the argument to calculate the average between their wavefunctions. The function employs OpenMP threads for looping over NBeam objects. OpenMP threads are responsible for calling each NBeam object's

NBeam::addAvg(NBeam&) method. The NBeam::addAvg(NBeam&) is accountable for computing the average of neutrino flavor states of all energy bins between two NBeam objects (via SIMD) (see Fig. 4.13). Since the evolving and averaging steps can be completed separately on each node, no MPI message-passing call is required for those tasks, thus there is no network communication overhead.

```
#pragma omp simd for (int e = 0; e <
NUM_of_EBINs; ++e)
{ ...
    // calculates the vacuum Hamiltonian getH0(e, h0);
    // add vaccum hamiltonian to hamilt[] ...
    double lambda = sqrt( hamilt00^2 + hamilt01^2 ); double ldr
    = lambda * dr; double cosCoef = cos(ldr); double sinCoef =
    sin(ldr) / lambda; ...
    // complex numbers multiplications for result[] ...
    // save the final values to neuBeam's components neuBeam2->ar[e] =
    result00_real + result01_real; neuBeam2->ai[e] = result00_iamg +
    result01_imag; neuBeam2->br[e] = result10_real + result11_real; neuBeam2-
    >bi[e] = result10_imag + result11_imag;
}
```

Figure 4.12: Evolving the neutrinos' flavor state for all of the energy bins within the NBeam class. Since the computations of different energy bins are all identical, the loop was vectorized via SIMD.

The last step of the neutrino evolution loop is to find the global maximum difference (error) between the flavor states of the two final evolved neutrino beams (the $\psi_3$ and $\psi_7$ in Fig. 4.9). Hence, there should be one more OpenMP-managed loop over the NBeam objects to find the maximum difference between the two arrays of the neutrinos' flavor state. Within the loop, for each element within the NBeam object's array, the `NBeam::calcErr(NBeam&)` method is called. This method calculates the error among all the energy bins via SIMD instructions. Afterwards, the global maximum error may be found by exchanging the results for the local maximum errors of nodes via MPI. Consequently, if the computed global error is less than a predefined threshold, the results of the evolved wavefunctions $|\psi(r)\rangle$ are accepted. Then, depending on the I/O module configurations, the wavefunctions may also be required to be saved onto a file. The Network Common Data Form (NetCDF) library was employed to I/O tasks (Further I/O APIs can be supported by implementing I/O modules in XFLAT). Afterwards, the next iteration continues with the final evolved neutrino's flavor states. On the other hand, if the global error is more than the threshold, the computed evolved flavor states are discarded, the next step size is adjusted accordingly, and the next iteration resumes with the previous flavor states.

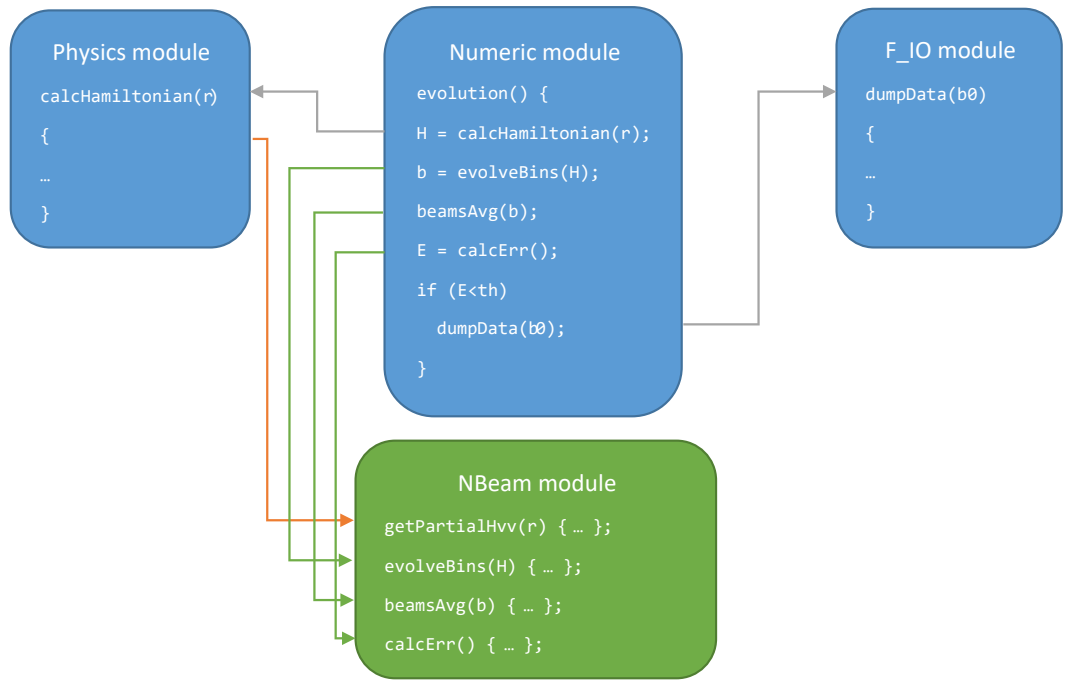

Figure 4.13: Function dependency between modules and the order of the function calls. Upper layer's modules are shown in blue and bottom layer module is shown in green.

## 4.2.4    I/O module

The I/O module is the next module that its init() function is called from the global initialization process. That function is responsible for initializing data files in which the results of computations are stored. NetCDF file format was chosen for dumping data on disk. Several dump modes may be provided via the configuration file's parameters. Thus, XFLAT may create several data files in which different calculations' result with different format that may be stored. The first implemented dump mode is responsible to take a snapshot of the neutrino flavor states by writing them onto a file. Hence, without any further process it write the raw values of neutrinos' wavefunction on disk. Later, that file may be used for resuming the computations from a particular radial point. However, if many snapshots are required during runtime, the size of the data file may expand. For instance, for the problem size of $1000(\vartheta) \times 100(\phi) \times 100(energy\ bins)$ the size of a single snapshot could be $1000 \times 100 \times 100 \times 4(particles) \times 4(wavefunction's\ components) \times 8(size\ of\ double) = 1.28\ GB$, thus if the code runs for 100 km and only one snapshot is required per kilometer, the total file size will be 128 GB! As a result, there may be issues in opening and manipulating the data file due to its size. On the other hand, if each snapshot is saved on a single file, the overhead of opening and closing files may affect the overall performance. Therefore, XFLAT can be provided a parameter via the configuration file that defines the maximum number of saved snapshots per file. When the maximum number of allowable snapshots is reached, XFLAT closes the current data file and creates a new file. As a result, the overhead of opening and closing a file per snapshot was eliminated and if the program crashes during runtime, the data are preserved onto the older files.

The second implemented dump mode stores less data by preprocessing data before saving it. Thus, instead of dumping the wavefunction's components for every beam it first calculates the average flavor states of all energy bins for a single neutrino's trajectory, then stores the result onto a file. Therefore, due to the averaging process there is no need to write the flavor state's results per energy bin individually. As a result, the data file becomes significantly smaller. The drawback is that the generated file cannot be used for XFLAT resumption since the information related to each individual energy bins were lost.

Similar to the other parts of XFLAT, the dump mode functionality is expandable as well. As a result, other dumping modes can be added in the future based on the requirement. XFLAT can simultaneously work with the combination of dump modes since the provided dump mode's code in the configuration file represent the bit pattern of the dump mode. Hence, each bit of the dump code in the configuration file represents a separate dump code. In the configuration file, the dump code 0 means no I/O tasks. The dump code 1 represents the first mode and the dump code 2 represents the second mode. However, the next dump mode may be represented by 4 instead of 3! Since the binary representation of 3 is 011, this code will enable both the first and the second dump modes simultaneously. Thus, the third dump mode is represented by binary number 100 (4 in decimal representation) and can be combined with the other modes as well. For instance, the dump code 5 enables the first and the third dump modes (101 in binary), the code 6 enables the second and the third dump modes (110 in binary), and the code 7 enables all of the three dump modes (111 in binary). Consequently, further dump modes may be added to XFLAT in future similar approaches.

As it will be discussed in the performance analysis section, the I/O performance on the the Xeon Phi is very poor, thus heavy I/O tasks should be avoided on the Xeon Phi, otherwise the co-processor's poor I/O performance can hurt the overall performance of the application. Therefore, as the solution to that issue, in XFLAT it is possible to redirect the Xeon Phi's I/O traffic to the corresponding CPU. Thus, instead of dumping data by Xeon Phis directly onto the disk, their processed data on the Xeon Phis may be written onto disk indirectly by CPUs. Prior to redirecting the I/O traffic, each instance of XFLAT has to know if it is running on the Xeon Phi or CPU. Nonetheless, when XFLAT runs on the Xeon Phi in symmetric mode, thus from the view point of the application there is no definite difference between a CPU and a Xeon Phi and there is no API that can be employed to detect the difference. By default, on the MIC the hostname is the same as its host (CPU) with the addition of '-mic' at the end. For instance, a host name might be hostid1, thus its first MIC's hostname will be hostid1-mic0 and its second MIC's hostname, if installed, will be hostid1-mic1. As a result, each instance of XFLAT can query their host name by calling MPI Get processor name() method and checking whether or not the host name includes '-mic' suffix. Afterwards, all the XFLAT instances exchange their hostname by calling MPI Allgather() method, in order to find their mate. For instance, on a compute node with two CPUs and two MICs, the first CPU rank is 0, the second CPU rank is 1, the first MIC rank (connected to CPU 0) is 2, and the second MIC rank (connected to CPU 1) is 3. Therefore, the first MIC (rank=2) must send its data to the first CPU (rank=0), and the second MIC (rank=3) must send its data to the second CPU (rank=1). If the XFLAT task runs on the MIC, it tries to find the CPU id (the MPI rank) to which it is connected. Since, every process maintains the list of processors ranks and host names, the XFLAT task that runs on the MIC starts from *rank* −1 and checks the host names. If it encounters anther MIC immediately, it means the current MIC is the second installed MIC, thus should connect to the second CPU. If by decreasing the rank a CPU is encountered first, that means the current MIC is the first installed MIC, so it continues rank decrements until it finds the first CPU of the current node (at this point by one more decrements the compute node may change and the host names change completely, *e.g.* hostid1
to hostid0, thus it is a sign of finding the first CPU rank on the current node).

As mentioned earlier, XFLAT can resume the computation from a previous run. In order to complete the resumption task, first XFLAT must check whether or not an input data file was provided to the application. If an input file was provided to XFLAT, the neutrinos' flavor state of NBeam objects must be initialized using the neutrinos' flavor state of the last snapshot that was stored in the file. Therefore, in contrast to the NBeam objects initialization from within the phys::initBeam(NBeam*) function, the objects are initialized with the final flavor states which were saved onto the file. As a result, the phys::initBeam(NBeam*) function may call the fillInitData(NBeam*) method from the I/O module in order to open the provided data file, read the stored neutrinos' flavor state, and initialize the NBeam objects with them.

In order to act properly, the I/O module requires information about the data that is going to be written onto the disk. Those information are the length and the number of dimensions as well as the name of each dimension. The I/O module can query those information from the Physics module, since the Physics module maintains the geometry related information. The first method is the phys::beamLen() that returns the length of the NBeam arrays. The second method is phys::getDim() which returns the number of dimensions in the geometry. The next method is getDimInfo(string str[]) which returns the name of each dimension so as to distinguish them in the file. The I/O module may use those information for formatting the saved data. The next two methods are the phys::startDim() and phys::countDim() that are utilized by the I/O module in order to receive an array of starting point and an array containing the length of each data dimension, respectively. The phys::startBeamIdx() and phys::endBeamIdx() methods return the index of the beginning and ending beam of the first dimension (Since the neutrino beams are distributed over compute nodes by the first dimension, the starting and ending beam indecis are required for nodes to function properly). For instance, if there are 1000 zenith angle beams and are distributed over ten identical nodes, the first node receives NBeam[0:99], the second node receives NBeam[100:199], *etc.*. Hence, there will be no conflict between nodes data. The last public method is the phys::firstDimLen() that returns the total length of the first dimensions on which the data is distributed over nodes (depending on the implemented geometry module, the first dimension length may be the size of $\vartheta$ or $\phi$ dimensions).

## 4.2.5    Matter module

In addition to the previous initialization function, another module in XFLAT from which its init() function is called by the global module is the Matter module. This module is responsible for implementing the matter profile.

## 4.2.6    Energy module

This module is responsible for the implementation of the energy spectra function for neutrinos. The NBeam module employs the functions of this module.

## 4.2.7    Utility module

Another important module in XFLAT is the Utility module in which the global variables that are fetched from the configuration file are stored as well as the application state and miscellaneous functions. Miscellaneous functions are the functions that do not belong to any modules, yet they are employed from several modules. For instance, the complex number multiplication mul _cmplx() and the norm2() methods are two important methods that are utilized by several other modules. Since the other modules have access to the Utility module's data, it contains the global state of the application that is used to indicate whether or not the application is in a specific state such as the benchmark state. The initialization part is different in the benchmark mode (*e.g.* there is no requirement for loading and initializing the I/O module during benchmarks), thus all modules must have access to the Utility module's information.

The XFLAT's last step after completing the neutrino evolution loop and storing the results, is the memory deallocation by calling the freemem() functions. Hence, the global module calls the I/O module's freemem() in which the NetCDF open files are closed. The next call is the Physics module's freemem() method in which the deconstructor of NBeam objects are called from within a loop. Hence, all the assigned memory is deallocated gradually before the application termination.

As discussed earlier, multiple copies of the NBeam arrays are maintained in memory in which the intermediate calculation results are maintained. Those arrays are among the data that are required to be transferred between XFLAT modules. Since they are all represented by one-dimensional arrays, modules can transfer them via their pointers (a single variable) without the requirement to pass further data. For instance, the number of required NBeam array is defined in the Numerical module, then their pointers are directed to the Physics module for memory allocation, and at the end of each iteration the pointer to the final result is forwarded to the I/O module for I/O tasks. As a result, the number of required NBeam arrays by the
solver does not place extra overhead on the inter-module communications, thus does not produce performance bottleneck.

## 4.3    High-level Parallelisation

As previously mentioned, XFLAT initially allocates one-dimensional NBeam's object arrays of neutrino's angle beams onto the main memory. For the reason that the number of neutrino beams can be too high in complicated geometries, the computations should be able to employ multi-node computing environments such as supercomputers. Therefore, XFLAT should dispatch the NBeam arrays over multiple nodes. For inter-node communications, the MPI functions were employed since MPI is traditionally the *de facto* standard for inter-node communications on supercomputers. Typically, one MPI process on each processor is sufficient. The processor can be either a CPU installed on a socket or the Xeon Phi card installed on a PCIe slot. Creating higher number of MPI processes per processor is supported as well, yet normally no significant improvement in the performance is seen. Besides, higher number of MPI tasks per processor result in increasing the MPI communications overhead dramatically by increasing the number of nodes in run time. In addition, the number of synchronization points in the code should be kept minimized to reduce the processors' idle time as well as the MPI overhead. Consequently, in XFLAT there are only three major MPI synchronization points in the main loop. At the first synchronization point, the root node broadcasts global variables such as the next calculated radius or the termination conditions to all of the nodes. At the second communication point, nodes exchange the result of the background neutrinoneutrino partial Hamiltonian in order to compute the final integral summation for every node. Finally at the third exchanging point, nodes exchange their local maximum error in order to find the global maximum error among all neutrino's beams. Fig. 4.14 depicted the location of those points inside the modules.

Modern CPUs as well as the Intel MIC contain multiple cores, thus within each MPI node there may be an additional level of parallelism. As a result, the NBeam object arrays are dispatched over all of the available cores and hardware threads via OpenMP. Since inside a processor or co-processor the memory model is shared, *i.e.* the entire memory is accessible by all cores; OpenMP is a very appropriate choice here. In this way, the threads' communications and synchronizations remains within the processor. Therefore, there will be no message outside a processor. The advantage of keeping threads' communications within a processor is that the communication between CPUs or or between the CPU and MIC is slow in comparison to intra-processor communications. The CPU-CPU communication is performed via

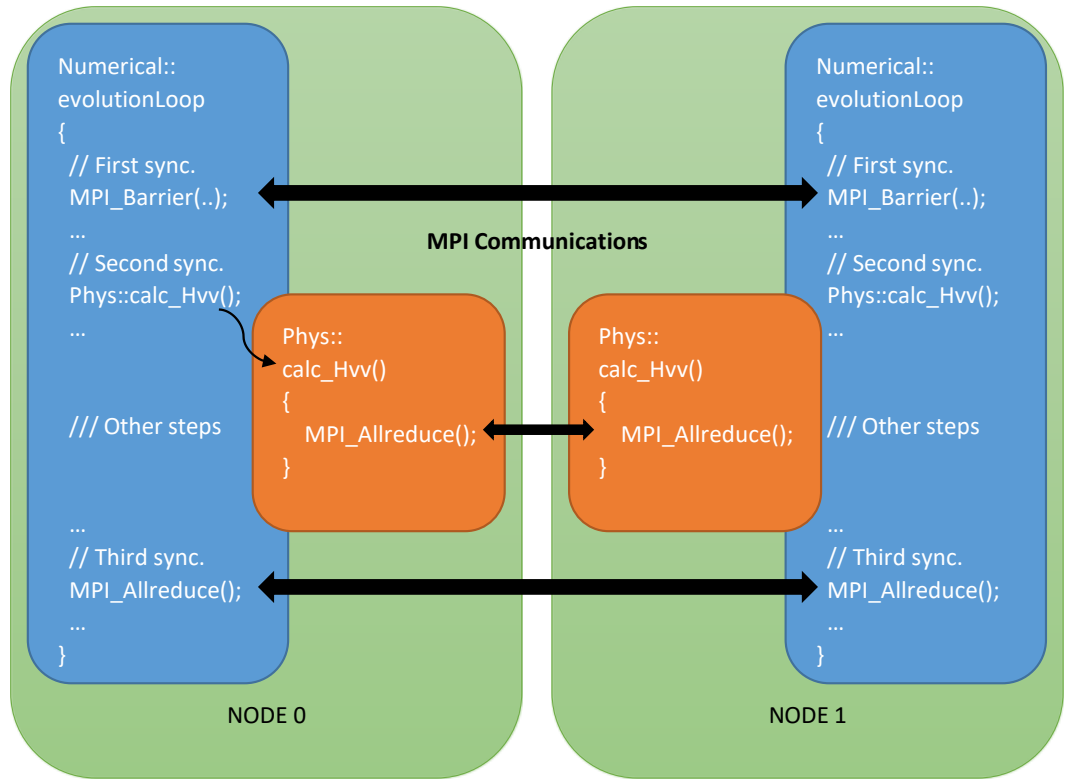

Figure 4.14: MPI communication points between nodes (shown in green). There are several MPI calls within the Numerical module (shown in blue) as well as MPI calls within the Physics module (shown in orange).

QuickPath Interconnect (QPI) bus, and the CPU-MIC communication is performed via the high latency PCIe bus. As a result, OpenMP is preferred over MPI for intraprocessor computations since cores are able to perform context switching between lighter-weight threads such as OpenMP threads (each hardware thread can handle one OpenMP thread efficiently) more efficiently. Placing multiple processes on a core decreases the performance due to process context switching overhead. Therefore, all
available hardware threads on the CPU and on the Xeon Phi (the number depends on the MIC's model, and ranges from 228 to 244 threads), can be exploited via OpenMP.

The last level of parallelism is the SIMD unit within each processors' core. As previously mentioned, neutrinos that propagate along one direction have different energies. Since the calculations on energy bins are identical, multiple energy bins' data may be packed together and the calculations can be performed simultaneously via SIMD. Hence, each OpenMP thread exploit SIMD instructions to perform calculations on the NBeam object's inner arrays in parallel. As a result, XFLAT can utilize all of the available levels of parallelism on a modern heterogeneous supercomputer.

## 4.4  Optimization

Although, the performance of XFLAT may be satisfactory on traditional CPUs, there are still several challenges in order to optimize it to run efficiently on the Intel MIC co-processors as well as the latest generation of CPUs. Even before the code development, the first step in the code optimization is to identify potential performance bottlenecks. Since bottlenecks can arise due to various factors including thread overhead, memory latency, cache line evictions, pipeline evictions, *etc.*, understanding their root causes can be beneficial in resolving bottlenecks more rapidly. The performance analysis can be performed by studying the behavior of the code as well as by employing analyzing tools. There are various tools available on the market to help programmers find code bottlenecks quickly. The analyzer that was utilized for a part of the XFLAT performance analysis is the Intel vTune. It can be exploited to perform code analysis on both Xeon CPUs and the Xeon Phi. Once the code is
analyzed, vTune provides numerous helpful information including memory latency, cache lines eviction rate, the processor's pipeline occupancy, the processor's cycles per instruction ratio, SIMD instruction unit occupancy, *etc.* categorized by function calls (see the example in Fig. 4.15). Consequently, it is easy to see which procedure on which processor's unit may be a potential cause of a bottleneck.

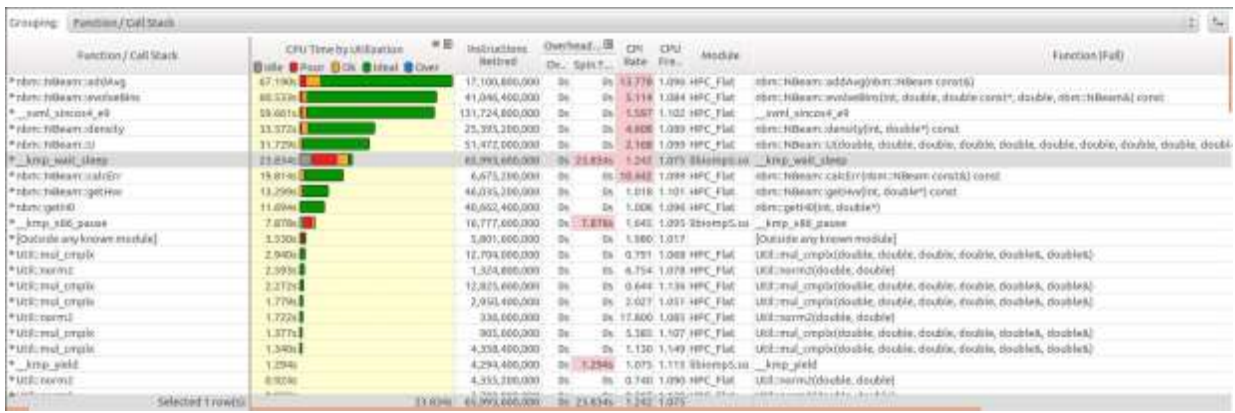

Figure 4.15: Intel vTune report section screenshot.

**Techniques and Tools**

The standard Object-Oriented Programming (OOP) approaches may not be helpful for high-performance computing codes as the first step. Since in general, the first goal of OOP is to compromise performance for the sake of *Rapid Application Development* (RAD) cycles. Therefore, for many applications typically maintaining clear class hierarchies, and utilizing virtual and multiple inheritances have higher priority than performance. On the other hand, typical high-performance implementation techniques result in having less complex data structures, thus less flexibility in code in order to gain more performance. For that reason, the implementation approach used in this project has been to attempt to compromise neither on flexibility nor on performance. As a result, the code architecture was designed to be a dual-layer and modular as much as possible. The priority of the internal structure of the lower layer is to gain high-performance computing capability on both the CPU and Xeon Phi, and the design goal of the upper layer is flexibility so that it is able to have expansion opportunity as well as supporting different modules. Hence, various challenges should be met to achieve this goal as well as several more improvements are required in the code to make sure that the implementation is HPC-ready. The following techniques and methods are among the most general and important optimization

techniques used in XFLAT implementation:

**Changing the Array of Structures (AoS) to the Structure of Arrays (SoA):** There are two high-level and general approaches for designing and implementing data structures. The simpler approach is to maintain an array of structures in which each structure encapsulates variables for a single particle. The other approach is to maintain a single structure in which arrays of data are allocated and maintained. Hence, each element in an array belongs to a single particle (see Fig. 4.16).

AoS approach is more simple for development and expansion since each particle is represented by an independent object. In addition, by accessing an object, all the related data is accessible. Nevertheless, AoS structures require gather/scatter methods in order to get/set similar fields of various objects. It may introduce extra latency for memory accesses as well, since in order to access data elements multiple jumps within memory space is unavoidable. Furthermore, non-continuous memory access may hurt the SIMD performance, since there must be several memory loads in order to fill up a SIMD register.

SoA approach can address performance issues related to AoS approach. By maintaining separated arrays each containing data element is possible to preserve contiguous memory access. Therefore, accessing the data elements in memory and streaming memory to SIMD units may remain continuous. As a result, with a single load it is possible to fill up a cache line or a SIMD register. Moreover, the problem of accessing an identical field for all objects is now the problem of accessing neighbor elements within a continuous array. As a result, the performance of SoA approach can be significantly higher than the performance of AoS approach. On the other hand, SoA approach may introduce several issues as well. For instance, choosing SoA approach may reduce the locality for accessing multiple fields of the original structure instance. However, the performance improvement of the SoA approach make it the preferable choice over AoS approach for HPC applications.

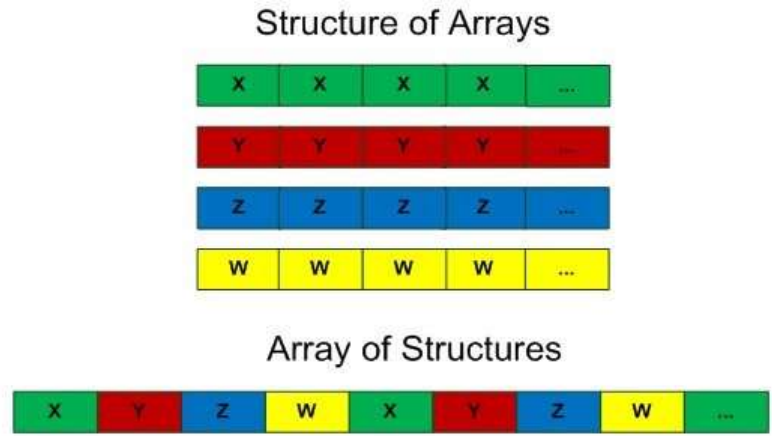

Figure 4.16: Array of Structures (AoS) and Structure of Arrays (SoA) representation inside memory [Intel, 2015b]. In AoS, elements related to a single data structure reside continuously onto memory. In SoA approach, for each field, there is an array in which data fields related to different structures reside continuously.

   In XFLAT, all of the lower modules' data are characterized as double precision arrays, where each element represents the wavefunction of a single neutrino. For instance, for the two-flavor system, since the neutrino's wavefunction has two complex components (3.1), there are four double precision arrays representing the four wave-function components. Therefore, there are two double precision arrays for the first complex number *a*, designated by ar[], ai[] (one for the real part, one for the imaginary part) and two double precision arrays for the second complex number *b*, designated by br[], bi[]. Accordingly, as opposed to creating one object per energy bin holding the double precision components, which would result in memory fragmentation and non-aligned memory access, only one object is assigned for the

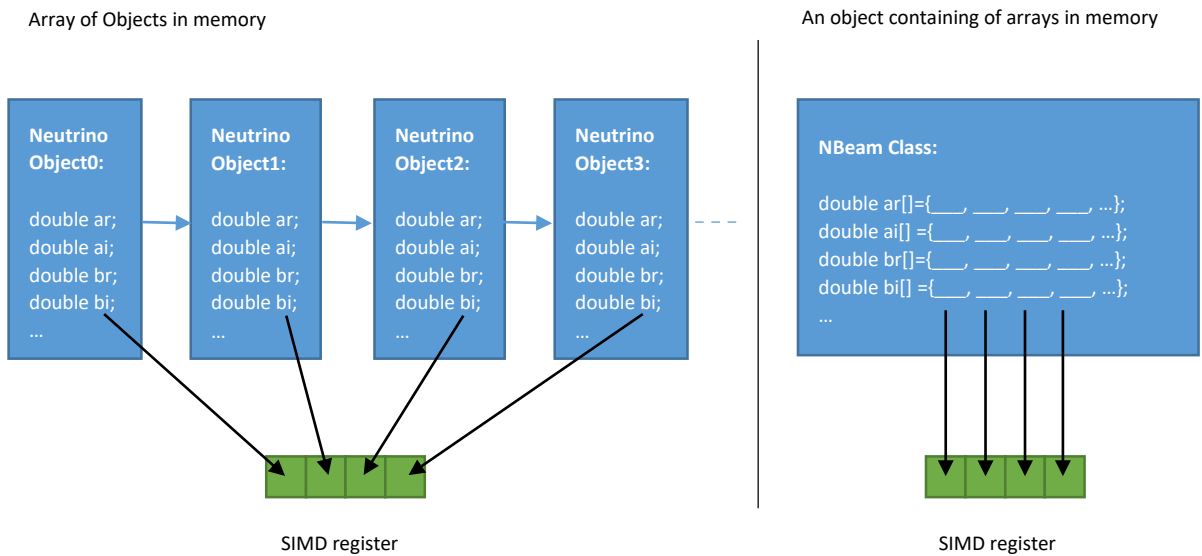

Figure 4.17: XFLAT data access pattern for AoS and SoA approaches. In the AoS approach (shown on the left), in order to fill up a SIMD register (shown in green), fragmented data must be extracted from multiple objects (shown in blue). On the other hand, in the SoA approach (shown on the right), with a single continuous fetch multiple data can be load onto a register simultaneously.

entire energy bins. Thus, contiguous aligned double precision data arrays are maintained for the wavefunctions' components, which result in performance improvement

(4.17).

Since, in XFLAT the upper layer modules commonly have no data to maintain (as they only encompass functions that perform computations on the lower layer's data), there is no requirement to instantiate the upper layer modules more than once. Consequently, they are not implemented as class or array of classes, instead their functionalities are encapsulated in C++ namespaces. Using namespace, has several advantages over class objects, as the flexibility and modularity is preserved and at the same time there would be no object creation overhead. Thus, there is neither memory fragmentation nor non-aligned memory issues for the upper layers sections.

**Data Alignment:** Fundamentally, data alignment denotes accessing the data at a memory offset equal to some multiple word size. Memory alignment has an important role in affecting both the cache hit rate and the SIMD instruction performance in most modern processor architectures. The memory cells are read and written word by word. Each word contains a few bytes, typically four (32 bits) or eight (64 bits) bytes. Hence, by accessing a memory cell, more than a single byte is accessible. As a result, the number of memory fetch may depend on the way that data reside onto memory. Fig. 4.18 illustrates the difference between non-aligned and aligned memory. As observable, when data is aligned to the word size by a single fetch, the data can be accessible, however, more than one fetch may be required to access non-aligned data.

The data alignment role is critical inside cache memory and vector units. If the data is non-aligned to the correct alignment length, which implies that the starting point of the data array is not a multiple of a predefined word size (64 bytes cache lines and vector registers), it can affect the application performance. The reason is that in order to fetch data onto cache lines, the compiler has to fetch data from memory more than once to fill up a cache line (see Fig. 4.18). Multiple data fetch from main memory onto cache is one of the causes of performance loss in processors due to slowness of the main memory compare to the cache memory. Likewise, in order to fill up the SIMD vector registers, if the data is not aligned to the vector registers' length, once more the compiler has to perform multiple fetches from memory to fill up vector registers, which results in a huge impact on the vectorization pipeline performance.

For the implementation of XFLAT, the 64 byte alignment was chosen on both the Xeon CPUs and the Intel MIC. There were a few motivations behind choosing the 64 byte alignment. First, the length of the Intel MIC vector registers are 512 bits

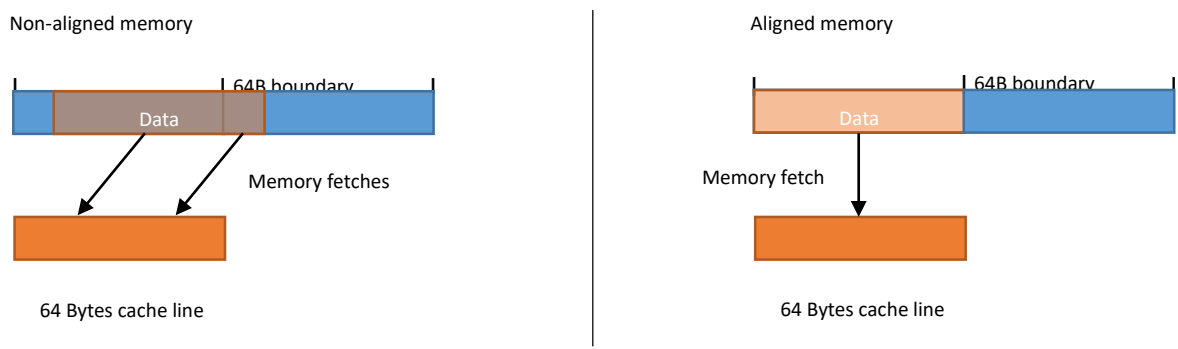

Figure 4.18: Difference between the non-aligned (the left side) and aligned memory (the right side) and their impact on performance. If the data is non-aligned to 64bytes memory boundaries (shown in blue), there must be more than one fetch in order to extract the data from the memory.

(64 bytes). Second, the cache line length on both the Xeon CPUs and the Intel MIC is 64 bytes. Therefore the choice of 64 byte alignment is the most appropriate on both chips. As a result, it tends to minimize memory fetch attempts and maximize the performance of the SIMD unit on both platforms.

**Inter-socket communication on Non-Uniform Memory Access domains:** This technique only has an impact on Non-Uniform Memory Access (NUMA) multisocket systems, as it is only associated to inter-socket memory access. Prior to NUMA architecture, the most common shared memory architecture was Uniform Memory

Access (UMA) in which all processors within a single node shared the physical memory uniformly. However, scaling the UMA architecture was hard and required complex hardwares and sophisticated softwares to control and manage the memory access. In contrast, modern multi-processor systems adopt NUMA architecture in order to simplify the hardware and software architecture (see Fig. 4.19).

Nonetheless, NUMA architecture has its own challenges. Virtually, all modern operating systems (OS) do not allocate memory when the allocation methods are called. In fact, operating systems allocate memory as soon as the first-touch hap-pens to the memory, *i.e.* when a thread attempts to touch a part of the memory (read/write) for the first time. At that time, the OS allocates the memory such that it resides as close as possible to that thread. Therefore, although in multi-socket machines single OpenMP process can access and manage all the hardware threads on all CPUs, if the memory allocation or the initialization section is completed by the master thread (which is typically the situation in many applications due to simplicity), all memory is allocated in such that the access time of the master thread to the allocated memory is minimized. Hence, during execution, the rest of the threads on other processors must access memory through the CPU-CPU QPI bus, thus resulting in higher memory latency. For instance in Fig. 4.20, if the thread on *CPU*0 tries to access to the memory that is controlled by *CPU*1, it has to access it through the QPI interconnection for which the latency is higher.

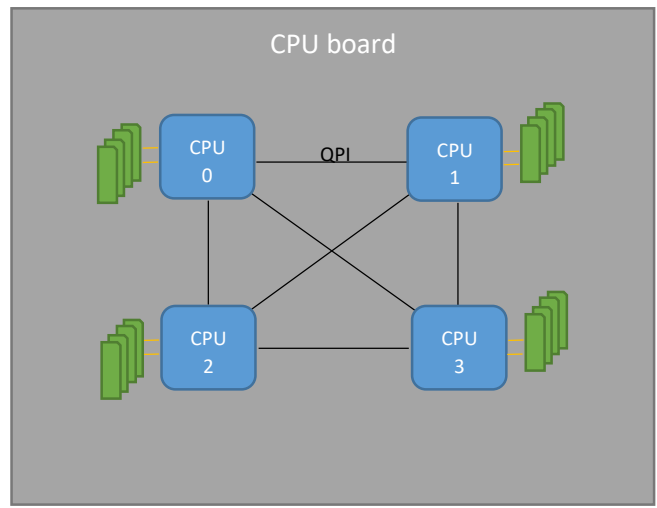

Figure 4.19: Non-Uniform Memory Access architecture. The memory (shown in green) connected to a particular CPU (shown in blue) is visible and accessible by threads on other CPUs through the QPI link.

There are multiple approaches to resolve this issue. The first method is attempt-ing to parallelize every first-touch memory access in the code by using multi-threaded loops, which may not be feasible in every application and situation due to code dependencies in the initialization functions. The second solution is to eliminate the issue completely by maintaining independent MPI tasks on each CPU socket.

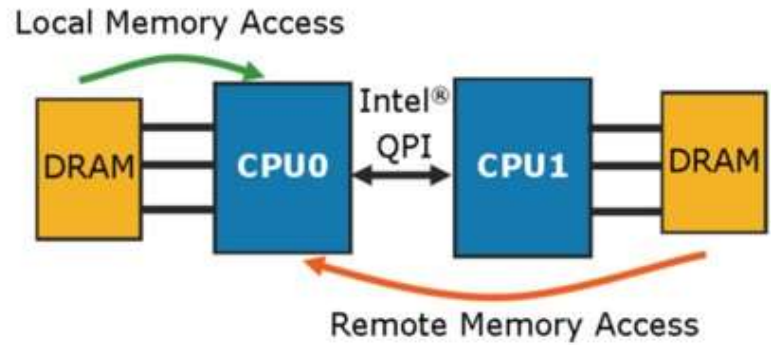

Figure 4.20: Illustration of accessing the memory (shown in yellow) in multi-socket CPU (shown in blue) architecture and the impact on the thread performance due to the route to the memory [SlideShare, 2015].

In XFLAT, the latter approach was adopted, therefore as an alternative of allowing OpenMP to manage an entire multi-socket workstation or CPU board, one MPI process run per socket. Thus, the communications of the OpenMP threads are internal to each CPU. Accordingly, inter-socket communications consist solely of the MPI messages that happen at only a few points per iteration. As a result, inter-socket OpenMP thread overhead was eliminated completely.

**Fusing Functions:** In modern processors, the role and performance of cache memory is important in overall application performance. As depicted by Fig. 4.21 the latency of cache memory is several orders of magnitude lower than the main memory (RAM), however, the capacity of this special memory is limited compared to the main memory. Therefore, when a function's instructions, which are being executed on the processor, try to access data, the data latency varies based on the data location. If the required data can be found within the cache memory, the execution performance will be higher due to the higher speed of the cache memory. On the other hand, if the data is not found onto the cache, the processor should fetch the data from the main memory, thus due to the main memory latency the CPU idle time may increases. Since, the amount of available cache memory is limited and less than the main memory, it is not possible to fetch the entire main memory onto the cache. Therefore, typically only a limited memory section can be fetched onto cache (see Fig. 4.22). The fetched memory is mostly related to the instructions that are being executed on the CPU. Consequently, when the function that is utilizing the CPU returns, its data is evicted from the cache to make room for new functions' data. Fetching memory onto cache lines is an expensive and time consuming task for CPU. The situation can become worse, if the consequent function tries to work on the same set of data. As a result, the CPU must fetch the identical data, for which it just evicted, from the main memory onto the cache again.

There are numerous sections in a typical application where different functions perform calculations on the same set of data, and the functions are called continuously. Assume the first function is called, thus upon the call the required data must be fetched from the main memory onto the processor's cache. If the data is a large array (most of the data in XFLAT are in this format), the data cannot stay inside the cache for a long period due to its size. Therefore, when the function complete its tasks and returns, the cache lines may be evicted. Afterwards, as soon as the second function is called, the same set of data must be fetched from the main memory once more. Obviously, this extra memory fetch may have a huge impact on the application's performance. In order to eliminate this bottleneck, functions that are working on the same set of data, if possible, should be fused together as one multipurpose function (see Fig. 4.23). In this manner, the required data for computations is only fetched once and all the function's instructions can perform calculations on the cached data.

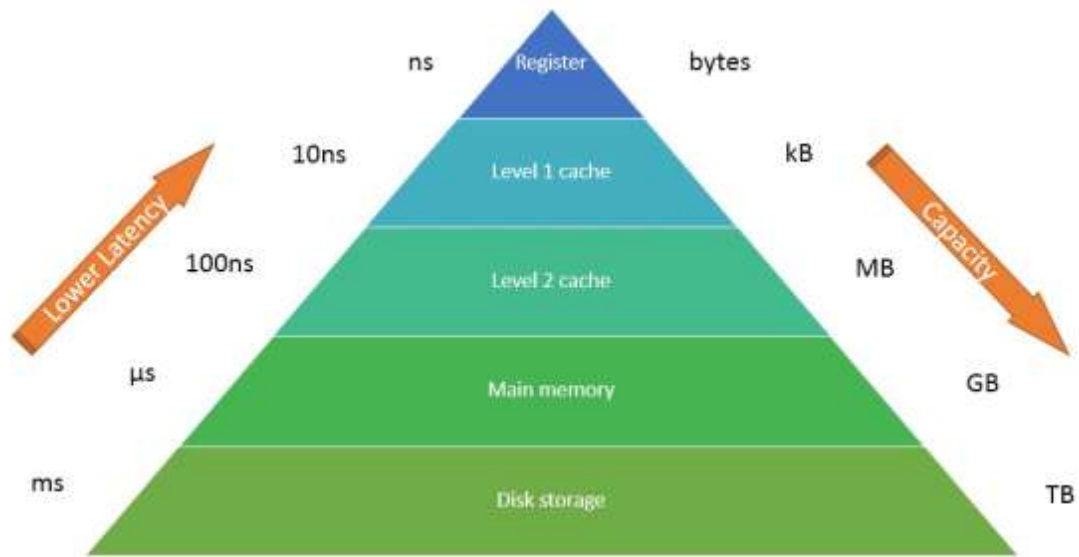

Figure 4.21: Illustration of memory hierarchy. CPU internal registers (shown on the top) have the highest speed and lowest capacity. The disk storage (shown on the bottom) has the highest capacity and slowest access time. All of the other memory levels reside in between the two memory holders.

In XFLAT, in order to eliminate unnecessary memory fetches and since several of the functions in the NBeam class work on the wavefunction components (the same data), several procedures of the NBeam class were fused together. Therefore, as an alternative to call several functions consequently, a single fused function may be called to perform the combined tasks. Nevertheless, this fusing technique should be used in parallel to the separated functions. Since, there might be circumstances where only one of the functionalities of the NBeam class is required, and in that situation calling a fused function will result in an unnecessary computational overhead. Hence, in XFLAT, multiple API functions are provided in order to support variety of situations including the fused functions that must be utilized only in the appropriate places.

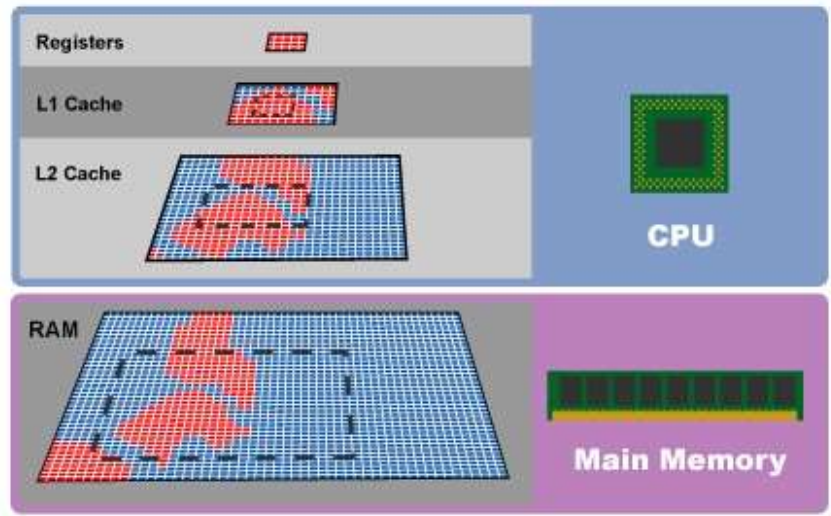

Figure 4.22: Caching pyramid illustration. For each memory fetch, the CPU fetches a region of memory containing the required data. If a particular memory location is referenced at a particular time, then it is likely that nearby memory locations will be referenced in the near future [ArsTechnica, 2015].

## 4.5    Code Validation

Prior to any performance analysis, XFLAT should be validated in several approaches and against the previous neutrino oscillations codes. Although XFLAT supports a wide range of physics and geometries, since the previous code developed only for the bulb model, the XFLAT validates the bulb model.

At the first step, it is possible to compare the results of the two canonical XFLAT's modules with each other, the bulb model and the extended bulb model. It is expected that the result of the multi-zenith supernova run (buld model) be similar to the result of the multi-azimuth multi-zenith run (extended bulb model), only for the inverted mass hierarchy, *i.e.*, $\delta m^2 < 0$. The survival probability results of a particular neutrino is expected to be similar. It indicates that the number of neutrinos that remained in their initial flavors at a particular step. Hence, an experiment can start with a pure beam of known flavor $v_x$, and at the end observe to see how many have changed their flavors, and how many remained in their initial flavors.

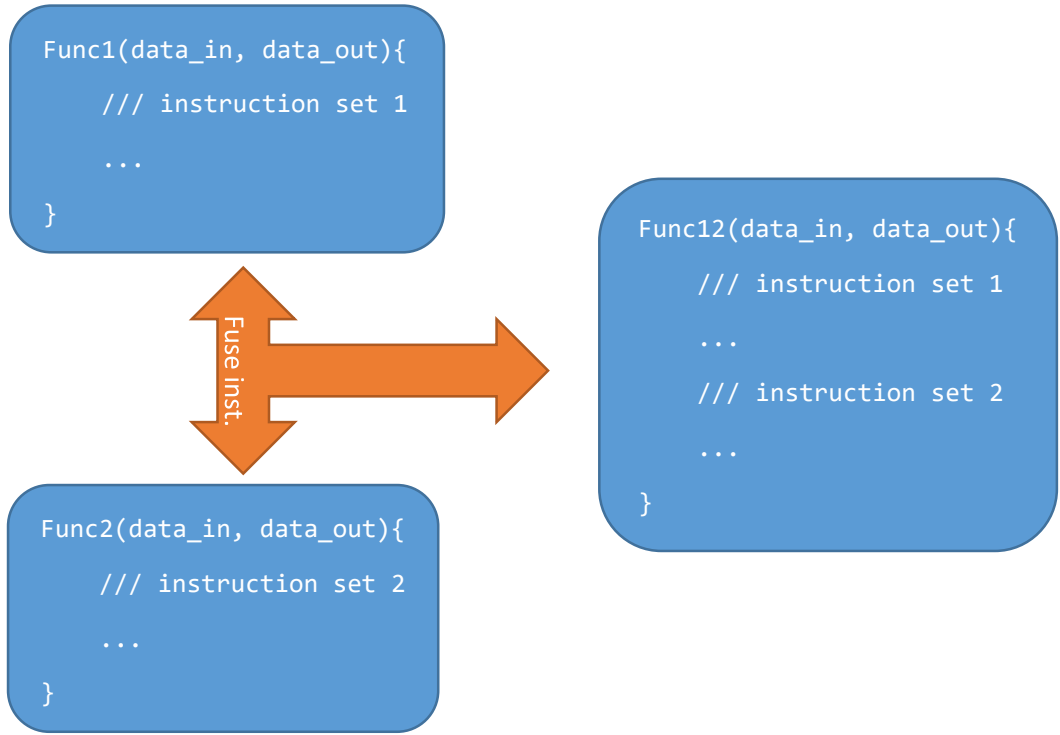

Figure 4.23: Fusing different functions that work on the same data together. The two functions on the left, in which the same set of data is processed within them, can be fused together to form the fused function on the right side.

XFLAT was run with the bulb model (Phy MA.cpp) module as well as the extended bulb model (Phy _MAA.cpp) module. For both runs the energy distribution function, $f_{v\alpha}(E^0)$, is taken to be of the Fermi-Dirac form with two parameters ($T_v, \eta_v$),

$$f_{\nu_\alpha}(E') \equiv \frac{1}{F_2(\eta_\nu)} \frac{1}{T_\nu^3} \frac{E^2}{\exp(E/T_\nu - \eta_\nu + 1)}, \tag{4.1}$$

where $\eta_\nu$ is the degeneracy parameter, $T_\nu$ is the neutrino temprature, and

$$F_k(\eta_\nu) \equiv \int_0^\infty \frac{x^k dx}{\exp(x - \eta) + 1}. \tag{4.2}$$

For all numerical calculations, unless it stated otherwise, $\delta m^2 = -3 \times 10^{-3}\, eV^2$, $\theta = 0.1$, $L_\nu = 10^{51}\, erg/s$, $n_{b0} = 1.63 \times 10^{36}\, cm^{-3}$, $L_{\nu e} = L_{\nu e^-} = L_{\nu x} = L_{\nu x} = 10^{51}\, erg/s$, h$E_{\nu e}$i = 11 $MeV$, h$E_{\nu e^-}$i = 16 $MeV$, h$E_{\nu x}$i = h$E_{\nu x}$i = 25 $MeV$, $\eta_e = \eta_{e^-} = \eta_x = \eta_{\bar{x}} = 3$. With these choices, $T_{\nu e}$' 2.76 $MeV$, $T_{\nu e^-}$' 4.01 $MeV$, $T_{\nu x} = T_{\nu x}$' 6.26 $MeV$. The baryon density is obtained as:

$$n_b \simeq \frac{2\pi^2}{45} g_s \left( \frac{M_{NS} m_N}{m_{PI}^2} \right)^3 S^{-4} r^{-3} \tag{4.3}$$
$$\simeq \left(4.2 \times 10^{30} cm^{-3}\right) g_s \left( \frac{M_{NS}}{1.4 M_0} \right)^3 \left( \frac{100}{S} \right)^4 \left( \frac{10 km}{r} \right)^3 ,$$

where $m_N$ is the mass of a nucleon, $m_{PI}$' 1.221 $\times 10^{22} MeV$ is the Plank mass, $S$ is the entropy per baryon that is set to $\frac{11}{2}$, $r$ is the distance from the center of the proto-neutron star, $M_0$ is the solar mass and $M_{NS} = 1.4 M_0$. In reality, however, the baryon density near the proto-neutron star is much higher than $n_b$. Indeed, the baryon density near the surface of the proto-neutron star is better represented by:

$$n_b' \simeq n_{b0} \exp\left( -\frac{r - R_\nu}{h_{NS}} \right), \tag{4.4}$$

where $h_{NS}$ is the scale height.

The number of angle beams were 800, the range of energy function and the number of energy bins were 0 – 80 $MeV$ and 160 bins, respectively. For the extended bulb model, each zenith angle had 100 azimuth beams as well. The computations started at 50 km, and the survival probabilities are shown at 250 km.

Fig. 4.24 depicts the survival probability of electron neutrino for the bulb model and extended bulb model runs. The vertical and horizontal axis show the zenithangle and energy, respectively. The color red indicates that the survival probability is 100% and blue indicates 0% of survival probability. Furthermore, in Fig. 4.25, the difference between the results of the bulb model and extended bulb model is depicted. As noticeable, the maximum difference is only 0.00029 (the values' range is [0, 1]).

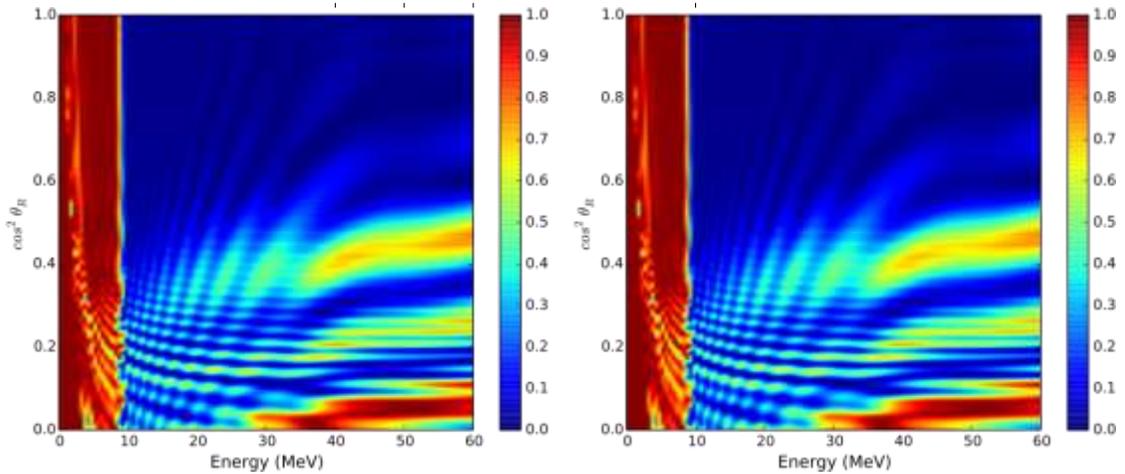

Figure 4.24: Survival probability of electron neutrino for the bulb model (left) and extended bulb model (right) runs for the inverted mass hierarchy. The vertical and horizontal axis show the zenith-angle and energy, respectively. The color red indicates that the survival probability is 100% and blue indicates 0% of survival probability.

In addition to the results for the electron neutrino, Fig. 4.26 depicts the survival probability of anti-electron neutrino for the bulb model and extended bulb model runs. In Fig. 4.27, the difference between the results of the bulb model and extended bulb model is depicted. As observable, the maximum difference is only 0.00049 (the values' range is [0, 1]).

As observable in the results, there is no observable differences between the two codes as they produce visually identical results. In addition, the maximum absolute difference between the two runs was always less than 0.05%.

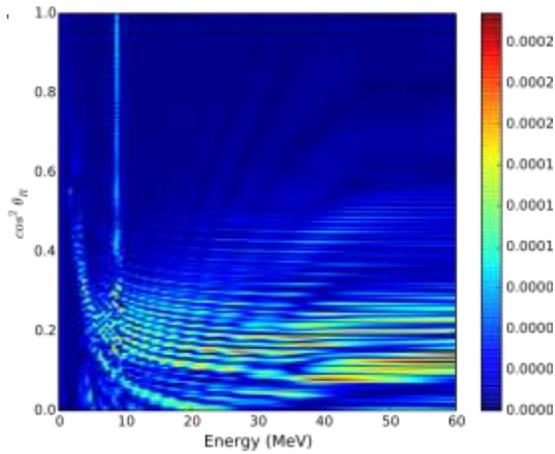

Nevertheless, this comparison can only be performed for the inverted mass hierFigure 4.25: The difference between the bulb model (left) and extended bulb model (right) runs for the results of the survival probability of electron neutrino for the inverted mass hierarchy. The vertical and horizontal axis show the zenith-angle and energy, respectively. The color red indicates that the survival probability is 100% and blue indicates 0% of survival probability.

archy, since the two geometries only produce similar results for the inverted mass hierarchy. For the normal mass hierarchy, as depicted by Fig. 4.28, the survival probability results of the two geometries, for the electron neutrino, are completely different (As shown in Fig. 4.29, the maximum absolute difference between the two geometries is 100%).

The next step in the validation process was to compare the result of XFLAT against a previously bulb model developed code [Duan and Shalgar, 2014]. Both codes are based on [Duan et al., 2006] research, yet the development path were isolated from each other. The [Duan and Shalgar, 2014] code was developed separately at Northwestern University. Thus, the algorithms, data structures, and numerical calculations are completely different between XFLAT and [Duan and Shalgar, 2014].

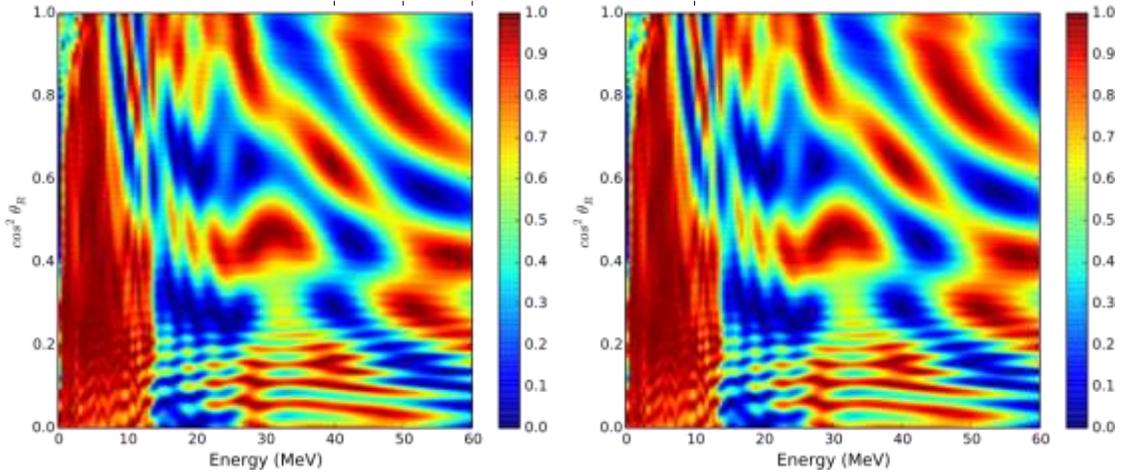

The simulation of the collective neutrino oscillations can be categorized into two classes, *i.e.*, the single-split and multi-split simulations. The most important obserFigure 4.26: Survival probability of anti-electron neutrino for the bulb model (left) and extended bulb model (right) runs for the inverted mass hierarchy. The vertical and horizontal axis show the zenith-angle and energy, respectively. The color red indicates that the survival probability is 100% and blue indicates 0% of survival probability.

vational consequence of the collective effects is an exchange of the $\nu_e$ ($\bar{\nu}_e$) spectrum with the $\nu_x$ ($\bar{\nu}_x$) spectrum in certain energy intervals. Such a flavor exchange is called a "swap", whereas "splits" are sharp boundary features at the edges of each swap interval [Dasgupta et al., 2009].

First, the results of the single-split runs were compared. In order to produce the following results, the previously mentioned parameters were employed, except the number of angle beams for which 2046 were chosen and the starting radius was at 20 km. The survival probabilities are shown at 250 km. Fig. 4.30, depicts the results of the two codes, and Fig. 4.31 illustrates the absolute difference between the results of the two codes. In addition, in Fig. 4.32 the initial and final spectra of the XFLAT's run (top), the absolute difference between XFLAT and [Duan and Shalgar, 2014] (bottom), are depicted. As observable, the absolute difference of spectra between the two codes is about $10^{-4}$, thus both results are virtually identical.

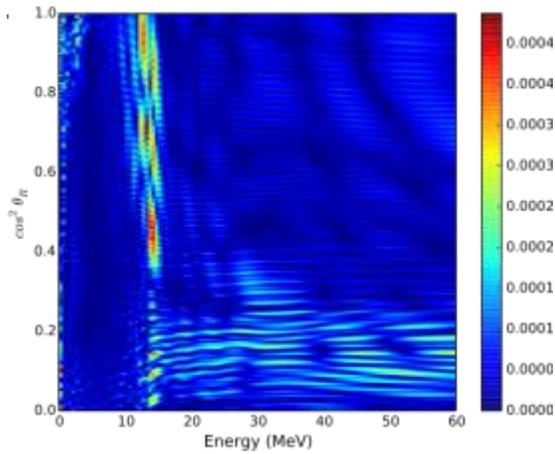

Figure 4.27: The difference between the bulb model (left) and extended bulb model (right) runs for the results of the survival probability of anti-electron neutrino for the inverted mass hierarchy. The vertical and horizontal axis show the zenith-angle and energy, respectively. The color red indicates that the survival probability is 100% and blue indicates 0% of survival probability.

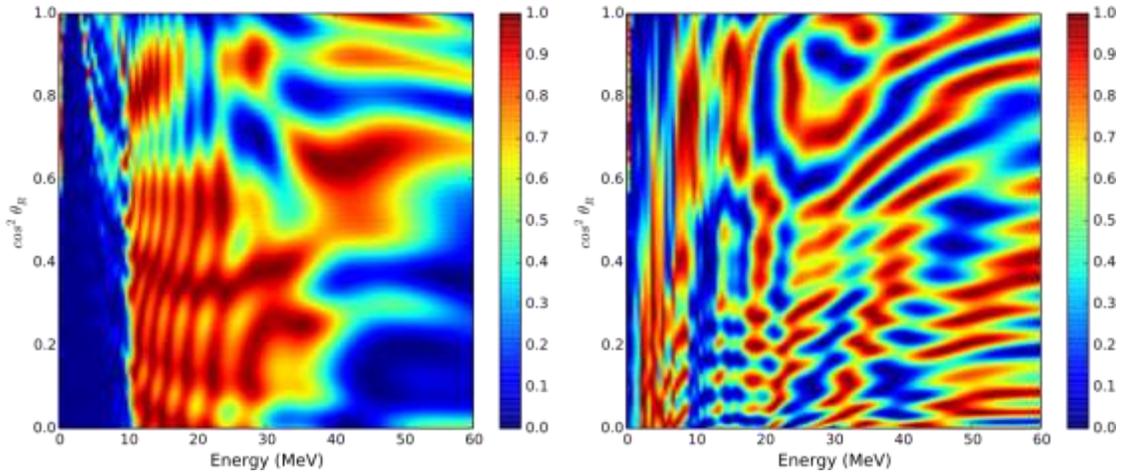

Figure 4.28: Survival probability of electron neutrino for the bulb model (left) and extended bulb model (right) runs for the normal mass hierarchy. The vertical and horizontal axis show the zenith-angle and energy, respectively. The color red indicates that the survival probability is 100% and blue indicates 0% of survival probability.

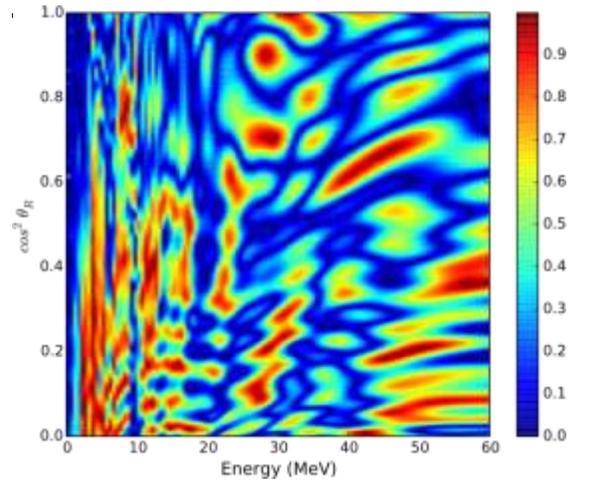

Figure 4.29: The difference between the bulb model (left) and extended bulb model (right) runs for the results of the survival probability of anti-electron neutrino for the normal mass hierarchy. The vertical and horizontal axis show the zenith-angle and energy, respectively. The color red indicates that the survival probability is 100% and blue indicates 0% of survival probability.

The previous comparisons used single-split energy spectrum. However, the singlesplit energy spectrum simulation is more stable than the multi-split energy spectrum simulation. Therefore, XFLAT and the [Duan and Shalgar, 2014] code were bench-marked for a multi-split spectrum run as well.

As illustrated by Fig. 4.33, there is an agreement between the results of XFLAT (left) and [Duan and Shalgar, 2014] (right) in the more unstable multi-split run. In addition, in Fig. 4.34 the change of the energy spectra of neutrino (left) and antineutrino(right) for the multi-split spectrum runs are depicted. The dashed and dotdashed lines are the initial spectra of the electron and tau neutrinos, respectively. As observable, the absolute difference of spectra at the final radius ($r$ = 400 km), is on the order of $10^{-4}$.

For all runs the parameters in Eq. 4.1 were taken as follow: $L_{ve}$ = $4.1 \times 10^{51}$ *erg/s*,

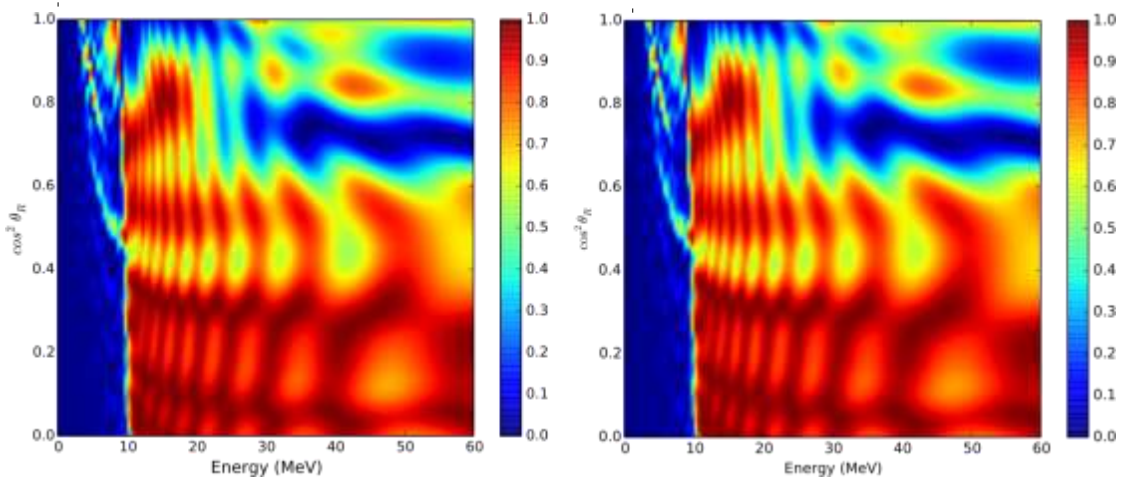

$L_{\nu_{e^-}}$ = 4.3 × $10^{51}$ *erg/s*, $L_{\nu_x}$ = $L_{\nu_{x^-}}$ = 4.1 × $10^{51}$ *erg/s*, $\eta_e$ = 3.9, $\eta_{e^-}$ = 2.3, $\eta_x$ = $\eta_{x^-}$ = Figure 4.30: Survival probability of electron neutrino for XFLAT (left) and [Duan and Shalgar, 2014] (right). The vertical and horizontal axis show the zenithangle beams and energy bins, respectively. The color red indicates that the survival probability is 100% and blue indicates 0% of survival probability.

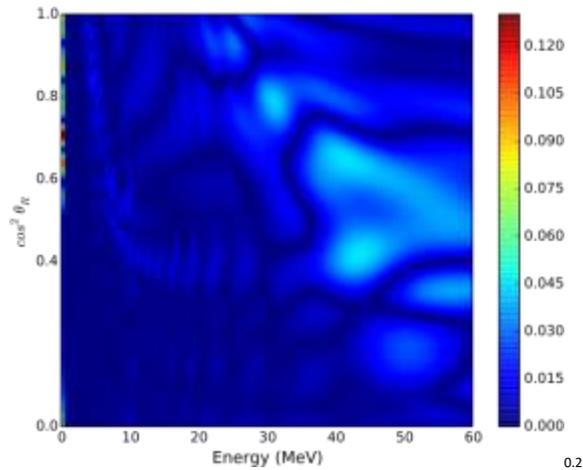

Figure 4.31: The absolute difference of the survival probability of electron neutrino between two XFLAT runs (left) and [Duan and Shalgar, 2014] (right) for the

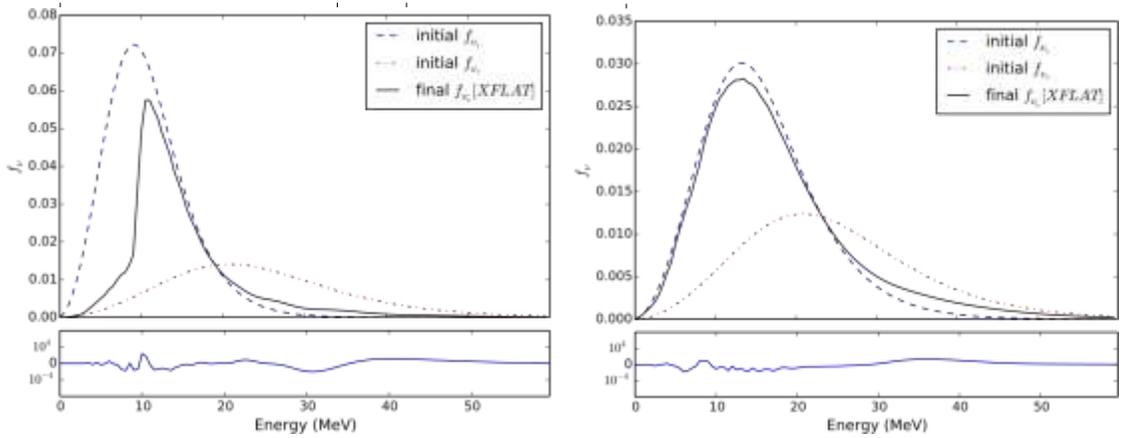

singlesplit simulation. The vertical and horizontal axis show the zenith-angle beams and energy bins, respectively.

Figure 4.32: Change of energy spectra of neutrino (left) and anti-neutrino(right) for the single-split spectrum runs. The dashed and dot-dashed lines are the initial spectra of the electron and tau neutrinos, respectively. The solid line are the corresponding spectra at $r$ = 250 km for XFLAT's result. The bottom panel illustrate the absolute differences between the XFLAT and [Duan and Shalgar, 2014] runs for the final spectra.

2.1, $T_{\nu e}$ ' 2.1 $MeV$, $T_{\nu \bar{e}}$ ' 3.4 $MeV$, $T_{\nu x} = T_{\nu \bar{x}}$ ' 4.4 $MeV$, the final radius was at 400 km and the number of zenith angles were set at 10000. For XFLAT the error tolerance was set to $10^{-10}$ and for the [Duan and Shalgar, 2014] code, in order to make the code converge, the error tolerance was set to $10^{-11}$. Nevertheless, the definition of error tolerance between the two codes are completely different. As a result, it is difficult to set the error tolerance to a particular number for both codes. For XFLAT choosing lower error tolerance produced satisfactory results as well (see Fig. 4.35).

As depicted in Fig. 4.36, the absolute difference between the XFLAT runs (with $10^{-8}$ and $10^{-10}$ error tolerance) (left plot) is higher than the absolute difference between the XFLAT run (with $10^{-10}$ error tolerance) and [Duan and Shalgar, 2014] run (with $10^{-11}$ error tolerance) (right plot). It is observable that the maximum absolute difference between the two XFLAT results is around 0.06 and the difference between XFLAT and [Duan and Shalgar, 2014] results is around 0.04.

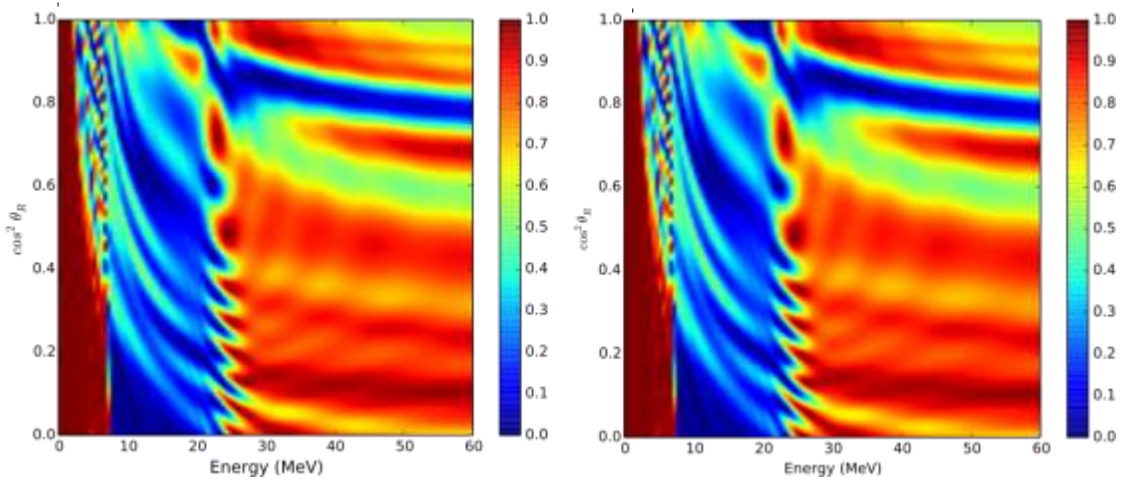

Figure 4.33: Survival probability of electron neutrino for XFLAT (left) and [Duan and Shalgar, 2014] (right) for the multi-split spectrum runs. The vertical and horizontal axis show the zenith-angle beams and energy bins, respectively. The color red indicates that the survival probability is 100% and blue indicates 0% of survival probability.

Furthermore, Fig. 4.33 depicts the survival probability for anti-electron neutrino for XFLAT (left) and [Duan and Shalgar, 2014] (right). Similar to the electron neutrino's plots, once more the results of the two codes are similar. In addition, as shown in Fig. 4.38, the absolute difference between two XFLAT results with $10^{-8}$ and $10^{-10}$ error tolerance (left plot) is similar to the difference between the results of XFLAT with $10^{-10}$ and [Duan and Shalgar, 2014] with $10^{-11}$ error tolerances.

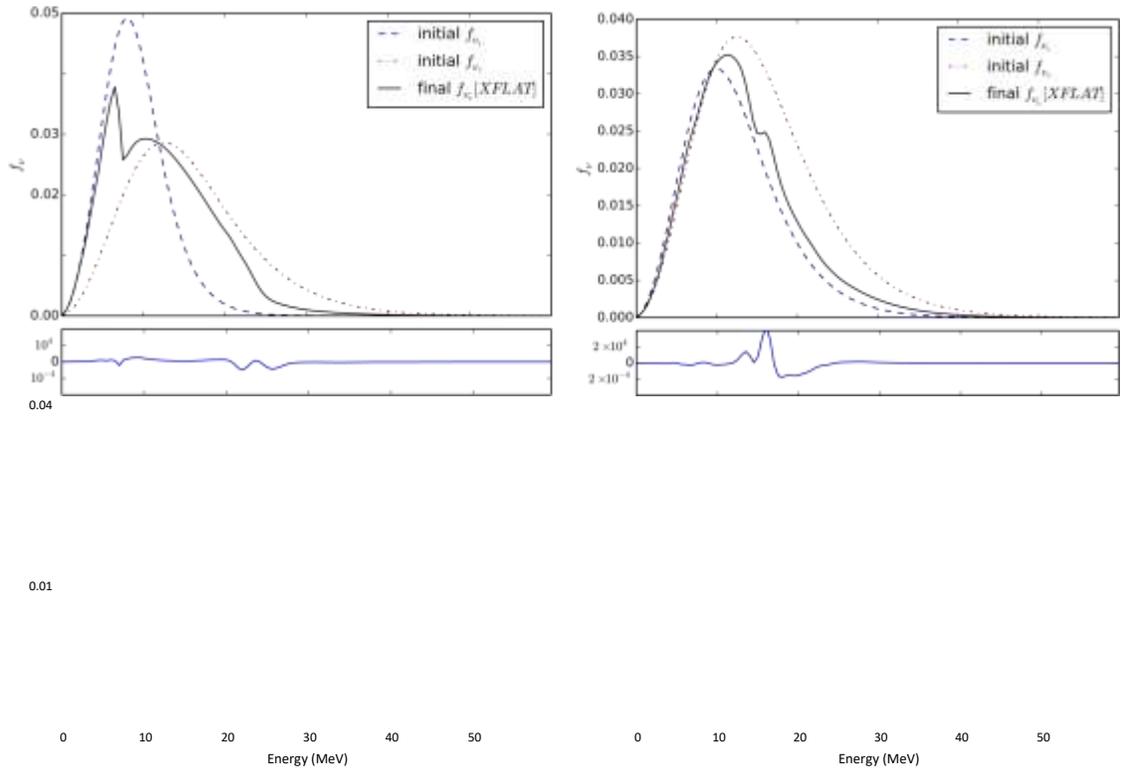

Figure 4.34: Change of energy spectra of neutrino (left) and anti-neutrino(right) for the multi-split spectrum runs. The dashed and dot-dashed lines are the initial spectra of the electron and tau neutrinos, respectively. The solid line are the corresponding spectra at *r* = 400 km for XFLAT's result. The bottom panel illustrate the absolute differences between the XFLAT and [Duan and Shalgar, 2014] runs for the final spectra.

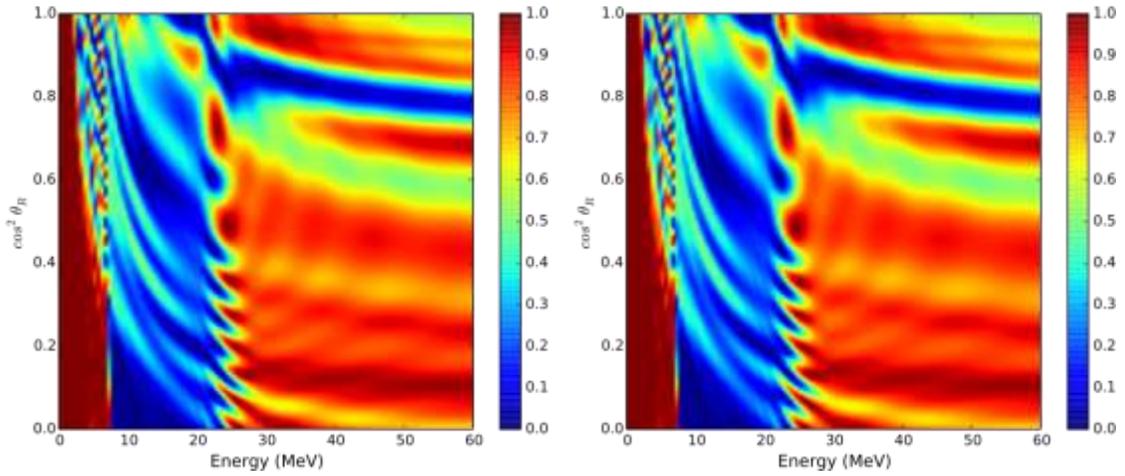

Figure 4.35: Survival probability of electron neutrino for XFLAT from multi-split spectrum runs with $10^{-8}$ (left) and $10^{-10}$ (right) error tolerance. The vertical and horizontal axis show the zenith-angle beams and energy bins, respectively. The color red indicates that the survival probability is 100% and blue indicates 0% of survival probability.

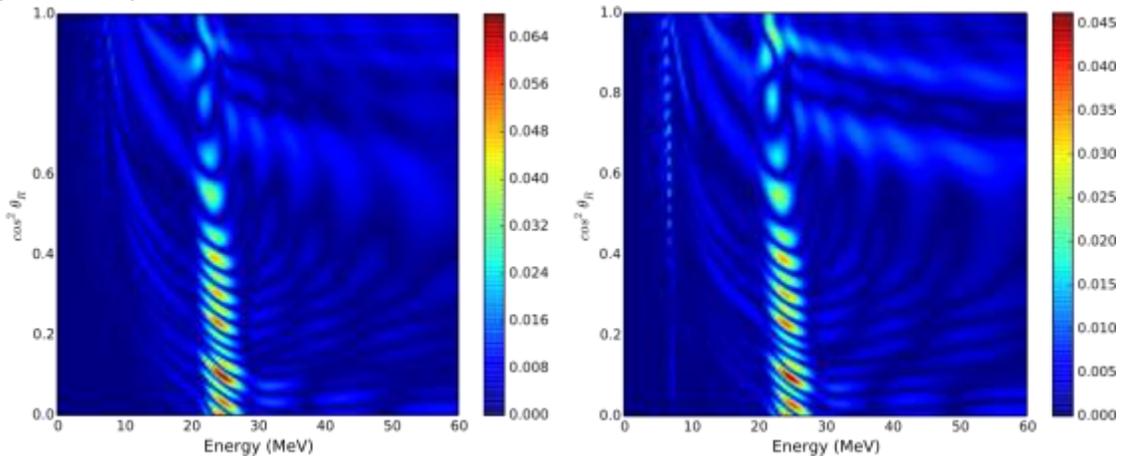

Figure 4.36: The absolute difference of the survival probability of electron neutrino between two XFLAT runs (left), with $10^{-8}$ and $10^{-10}$ error tolerance, and between XFLAT with $10^{-10}$ error tolerance and [Duan and Shalgar, 2014] with $10^{-11}$ error tolerance (right). The vertical and horizontal axis show the zenith-angle beams and energy bins, respectively.

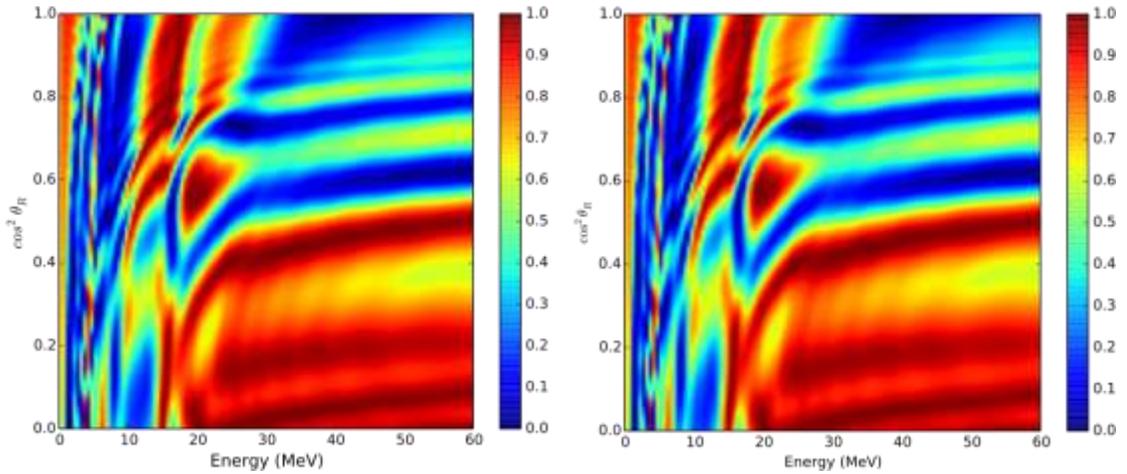

Figure 4.37: Survival probability of anti-electron neutrino for XFLAT (left) and [Duan and Shalgar, 2014] (right) from multi-split spectrum runs. The vertical and horizontal axis show the zenith-angle beams and energy bins, respectively. The color red indicates that the survival probability is 100% and blue indicates 0% of survival probability.

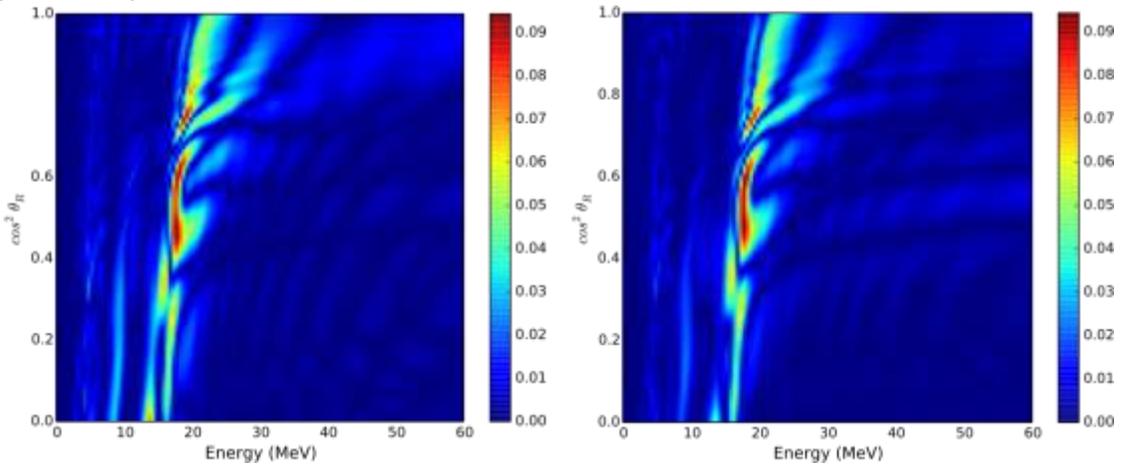

Figure 4.38: The absolute difference of the survival probability of anti-electron neutrino between two XFLAT runs (left), with $10^{-8}$ and $10^{-10}$ error tolerance, and between XFLAT with $10^{-10}$ error tolerance and [Duan and Shalgar, 2014] with $10^{-11}$ error tolerance (right). The vertical and horizontal axis show the zenith-angle beams and energy bins, respectively.

# Chapter 5

# XFLAT Performance

## 5.1    Measurement of Performance

The performance of a parallel code on a new architecture can be studied in two approaches. The first approach is to create isolated benchmarks (kernels) in order to understand the bottlenecks of the new platform and code. The second approach is to study the overall performance of the newly developed code (performance analysis). By benchmarking the standalone kernels, instead of the entire XFLAT, the source of the performance improvements and bottlenecks can be highlighted. Thus, XFLAT benchmarks can be used to identify the performance bottlenecks of the internal structures and algorithms of the code. Finding the root source of performance improvements or bottlenecks plays a major role in a parallel code's efficiency.

The significant factor in parallel code development was the amount of speedup that may be obtained by running the code on parallel systems. In simple words, *speedup* was defined as the relative performance improvement when running a task. This relative improvement typically is shown as instructions per cycle (IPC). IPC is the average number of instructions executed per each processor's clock cycle. Instructions include the set of operation codes (opcodes) for a particular processor. The operation codes are those instructions that identifies the operation to be performed on processor (*e.g.* multiplications and additions). Another way to define the speedup is the convention of cycles per instruction (CPI) that is the length of time between successive instruction completions. Hence, the speedup is defined as:

$$S = \frac{T_{old}}{T_{new}},$$
(5.1)

where $S$ is the resultant speedup, $T_{old}$ is the old execution time without improvement, and $T_{new}$ is the new execution time with improvement.

Consequently, *linear speedup* or *ideal speedup* is obtained when the speedup for $p$ processors ($S_p$) are equal to $p$ which means that the scalability is perfect. For example, this implies that by doubling the number of processors, the speed doubles as well. The other important metric that can be derived from speedup is *efficiency* that is defined as:

$$E_p = \frac{S_p}{p},$$
(5.2)

where $E_p$, the efficiency, is defined as the speedup divided by number of processors.

Typically, the efficiency value is between zero and one, approximating how wellutilized a processor is in solving a particular problem, compared to how much the time of a processor is wasted in communication and synchronization.

Unfortunately, the way that the speedup and efficiency were defined make them inappropriate for measuring the efficiency of codes on heterogeneous muti-node systems. On a homogeneous system, there is only one type of processor as well as one type of available memory. However, those definitions are not directly applicable on a multi-node heterogeneous system, which is equipped with both CPUs and MICs. On heterogeneous system, there are different processor types each with different clock frequency and cache, various memory levels each with different bandwidth and latency, and multiple buses and I/O routes available. Hence, there may be numerous factors that can affect the overall performance of the system. As shown in this chapter, the performance of XFLAT depends on the MIC to CPU load ratios as well as the distribution way of the load on the MIC's threads. When the optimum number of nodes is not known in advance, the prediction of the speedup becomes perplexing. For instance, Fig. 5.1 illustrated the multi-node XFLAT speedup relative to a single node (CPU only). On each node, there were two CPUs and two MICs which the benchmark measured the number of calculated radial steps for about 100 seconds. As observable, when the desired number of nodes was less than 10, the 3:1 MIC to CPU load ratio resulted in the best performance. For instance by employing 7 nodes and 3:1 load ratio, the speedup was about 22 times higher than the single node's result. However, when the number of nodes were chosen to be 14 in advance, the optimum load ratio was 2:1 and the relative speedup was 32 time higher than the single node's result. Hence, by doubling the number of nodes, the speedup may not be doubled since on heterogeneous systems the performance depends on many other factors including the load distributions and load ratios.

Likewise, the efficiency metric cannot be applied since multiple types of processors with different capability were available during run time. Thus, the efficiency metric definitions should be modified accordingly.

As a result, the speedup and efficiency metrics that were traditionally used to describe the scalibity of a code, cannot be employed for performance analysis of hybrid codes on multi-node heterogeneous systems. On heterogeneous systems, several factors can affect the overall performance of a code, hence the code scalibility prediction cannot be achieved by exploiting traditional metrics, thus new metrics should be explored in order to describe the behavior of a code on multi-node heterogeneous systems.

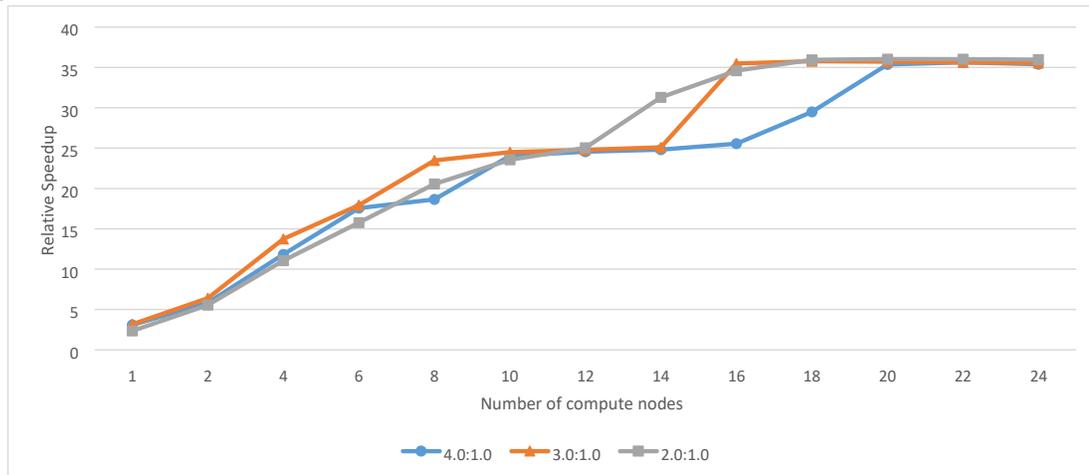

Figure 5.1: Illustration of XFLAT speedup relative to single node (CPU only) on a multi-node environment. The blue, orange, and grey curves show the MIC to CPU load ratios of 4:1, 3:1, and 2:1, respectively.

## 5.2 Kernel Benchmarks

Kernels are typically small codes that developed with the purpose of testing and analyzing a piece of software or hardware. The performance analysis of large softwares (*e.g.* XFLAT) may be hard, due to having many modules and parts. Each software modules may have its own bottleneck that can affect the overall performance dramatically. Kernels can analysis each functionality or module by isolating and testing a particular functionality. Hence, employing kernels for performance analysis may be vital for analysis of large softwares.

As previously mentioned, there are several techniques to improve the performance of an HPC code including data alignment, fusing functions, reducing inter-node communications, and changing loops' structure. In addition, studying the performance of low level details such as dereferencing pointers and transcendental functions can be helpful since XFLAT utilized them frequently. In this section, the performance impact of those techniques will be studied as independent kernels.

### 5.2.1 Floating point and transcendental functions performance

One of the most important criteria of any performance benchmarks is measuring the raw performance of each processor. Hence, kernels are required to measure the performance of the floating point operations as well as transcendental functions. XFLAT only utilizes double precision floating point operations and it heavily employs transcendental functions as well. As shown in this chapter, all of the double precision version of the transcendental functions were implemented in software.

They exploit SIMD units for internal calculations, thus the knowledge of their performance plays an important role in the performance analysis of XFLAT or any HPC code that ex// Two transcendental operations per iteration

```
// Outer loop for
(...) {
    // Vectorized loop #pragma omp
    simd for (int i : ARRAY_LENGTH)
    {
        A[i] = sin(A[i]);
        B[i] = cos(B[i]);
    }
}
```

Figure 5.2: Illustration of the vectorized loop's structure for transcendental functions benchamrk.

ploits double precision calculations. In XFLAT, since the majority of the innermost loops were vectorized, all of the following benchmarks were performed inside vectorized loops (see Fig. 5.2.1). The widths of the vectors were taken to be a multiple of the SIMD registers width (256 bits or 4 DP for the Xeon CPU, and 512 bits or 8 DP for the Xeon Phi). In order to maintain data locality, the same vector operations were repeated ten million times inside the middle loop. Furthermore, in the outermost loop all of the hardware threads were utilized to achieve the best performance.

Three similar kernels were prepared for different operations. The only variable in kernels was the length of the vectorized loop that was 8–4096 double precision elements. The first kernel only benchmarked simple floating point operations (additions and multiplications), thus one double precision addition and one double precision multiplication were performed per iteration. In the next kernel, one *sin* and one *cos* calculations in double precision were performed per loop iteration. The last kernel benchmarked the performance of the double precision version of *exp* function.

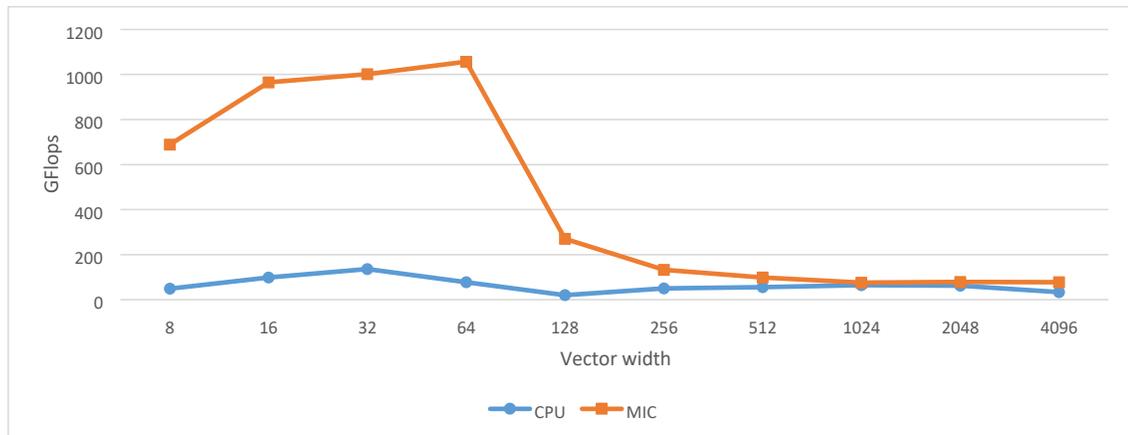

Figure 5.3: Floating point performance of CPU and MIC (Xeon Phi) for various vector width. The performance of the Xeon Phi (orange curve) degraded by increasing the vector width beyond 64, however, the CPU's performance (blue curve) remained steady.

As illustrated in Fig. 5.3, the Xeon Phi can reach over 1 TeraFlops (1000 GigaFlops) as advertised [Jeffers and Reinders, 2013]. However, the performance of the Xeon Phi is highly sensitive to the width of the vector. On the other hand the performance of the CPU is relatively stable. The floating point performance of the Xeon Phi is best when the width of the DP vectors is 64; thus, the Xeon Phi ran 10 times as fast as the CPU. However, the performance of the Xeon Phi degraded substantially as the vector width increased.

The results of transcendental functions benchmarks are shown in Fig. 5.4. These results suggest that the transcendental function performance on the Xeon Phi is relatively stable against the width of the vectors. For the tests on *sin*() and *cos*() the Xeon Phi ran 6–8 times as fast as the CPU, but for *exp*() the Xeon Phi is only 3–4 times better.

The performance drop may be due to the loop structure as well. As a result, alternative benchmarks with different loop structure were performed on Xeon Phi.

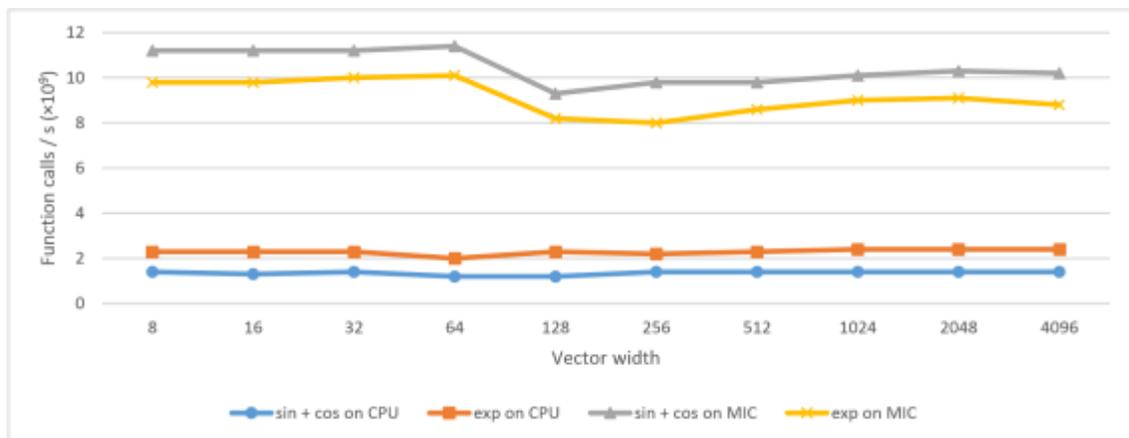

Figure 5.4: Transcendental function performance of CPU and MIC (Xeon Phi). For the tests on sin() and cos() the Xeon Phi (grey curve) ran 6–8 times as fast as the CPU (blue curve), but for exp() the Xeon Phi (yellow curve) is only 3–4 times better (orange curve).

In addition, the compiler was upgraded from Intel C++ 13.1 to recently installed Intel C++ 15.0 on Stampede. Fig. 5.6 illustrates the performance of floating point operations (additions and multiplications) on Stampede's Xeon Phi. The difference between the new and old kernels is that unlike the previous kernel, within the new kernel's loop, there were two and four independent array's operations per iteration, respectively (see Fig. 5.2.1). The results suggest that by going beyond the length of the 64 double precision numbers, the performance of the Xeon Phi drops dramatically. The issue may be due to the current generation of Xeon Phi's software or hardware
architecture.

Fig. 5.7 illustrates the performance of *sin*() and *cos*() when there were four arrays' operations within the vectorized loop. As noticeable, by changing the compiler version and loop structure, the overall Xeon Phi's performance was still for transcen-dental functions.

The performance of the transcendental function within parallelized and vectorized

```
// Two transcendental operations per iteration

// Outer loop for
(…) {
    // Vectorized loop #pragma omp
    simd for (int i : ARRAY_LENGTH)
    {
        A[i] += x1 * B[i];
        C[i] += x2 * D[i];
        E[i] += x3 * F[i];
        G[i] += x4 * H[i];
    }
}
```

Figure 5.5: Illustration of the vectorized loop's structure with four array's operations per iteration.

loops may depend on another factor as well. As previously mentioned, since all of the double precision version of the transcendental functions were implemented in the software, SIMD units are utilized internally for transcendental calculations. This means that the benchmarks' loop as well as the functions within those loops utilize SIMD units. Depending on the compiler flags, for transcendental functions the internal employed function call may be different, which may affect the final result.

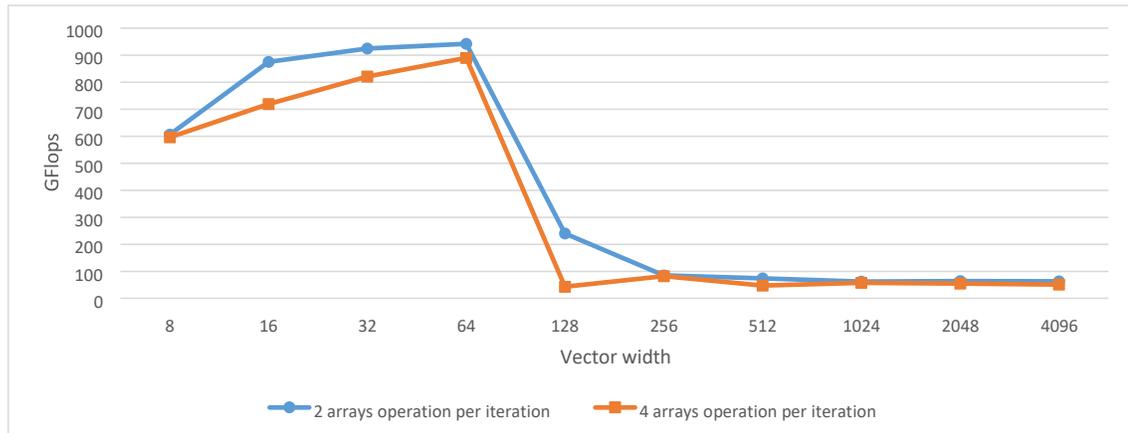

Figure 5.6: The performance of floating point operations on the CPU and MIC (Xeon Phi). The blue and orange curves illustrate benchmarks with two and four independent lines of arrays' operations, respectively.

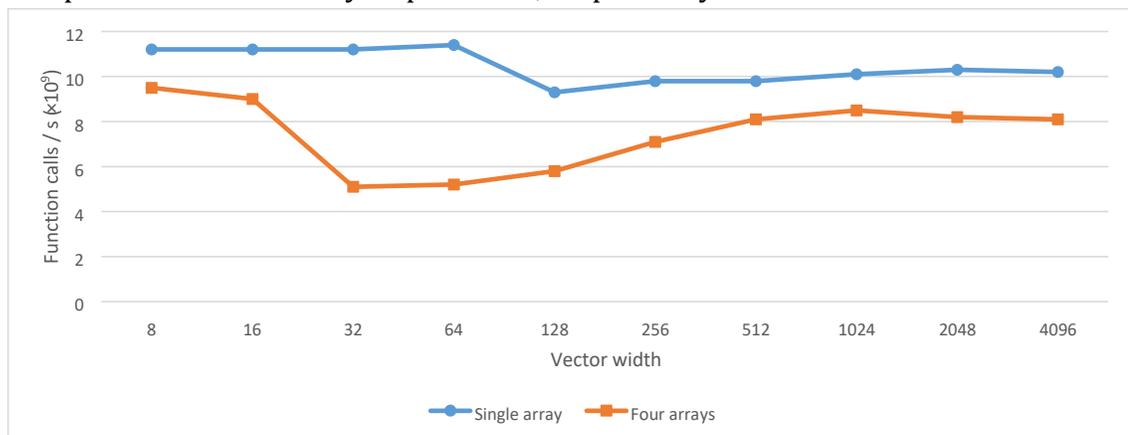

Figure 5.7: Transcendental function performance of CPU and MIC (Xeon Phi). The blue and orange curves illustrate benchmarks with one and four independent lines of arrays' operations, respectively.

This fact can be confirmed by looking into the generated assembly code of a built kernel using different flags. For instance, there can be multiple internal function calls depending on the compiler flags for the kernel that utilized the sin and cos functions. Fig. 5.2.1, Fig. 5.2.1, Fig. 5.2.1, and Fig. 5.2.1 show the internal called functions that were employed when the code was compiled with -S -no-vec, -S -no-vec -xHOST, -S, and -S -xHOST flags respectively. The compiler -S flag generates the assembly file output, the -no-vec flag prevents loops to be vectorized, and the -xHOST flag force the compiler to generate AVX vector specific instructions. Unlike SSE instructions (and its later versions SSE2, SSE3, SSSE3, SSE4, *etc.*) on the previous generations of CPU on which two double precision floating point numbers per instruction can be handled, the AVX instructions can perform a single instruction on four double

..B1.33:

```
        movsd       40(%rsp), %xmm0
        movl        %r12d, %r14d
        addsd       .L_2il0floatpacket.88(%rip), %xmm0
        movsd       %xmm0, 40(%rsp)
        call        __libm_sse2_sincos
```

Figure 5.8: The code is compiled with -no-vec flag. The employed transcendental function is libm sse2 sincos in which one sin and one cos are calculated.

precision floating point numbers simultaneously. As it can be seen in the generated assembly codes, even when the compiler is forced not to vectorize the loop, it still calls SSE2 version of a function to calculate sin and cos simultaneously. Even the -S - xHOST flag cannot change the utilized function when the compiler is forced not to generate vectorized loop. However, by lifting the no-vectorization restriction, normally the compiler pack a sin and cos functions together and computes them together. Only by forcing the compiler to utilize AVX registers, it can be seen that the sin and cos functions are employed separately to pack four elements together and issue the similar instruction on them simultaneously.

As a result, in a vectorized loop, the transcendental functions should always be employed with caution, since depending on the loop's structure and compiler flags the underlying function call may be different.

## 5.2.2 Structure of Arrays (SoA) vs. Array of Structures (AoS)

At mentioned earlier, there are two general approaches for developing codes in which computations are performed on arrays of data. The first and more convenient approach is building a structure that encapsulates a group of variables, afterwards

```
        vmovsd      40(%rsp), %xmm0
```

..B1.33:

```
movl        %r12d, %r14d
vaddsd      .L_2il0floatpacket.88(%rip), %xmm0, %xmm0
vmovsd      %xmm0, 40(%rsp)
call        __libm_sse2_sincos
```

Figure 5.9: The code is compiled with -no-vec and -xHOST flags. The employed transcendental function is libm sse2 sincos in which one sin and one cos are calculated.

..B1.33:

```
xorl %r12d, %r12d addsd
    %xmm8, %xmm9 movaps
    %xmm9, %xmm0 unpcklpd
%xmm0, %xmm0 call
    __svml_sincos2
```

Figure 5.10: The code is compiled without using any of -no-vec and -xHOST flags. The employed transcendental function is svml sincos2 in which two sin and two cos are calculated.

allocating memory for an array of those structures. For example one can design a neutrino class that encapsulates the flavor state of a single neutrino. Next, for each energy and angle beam, one instance of the object is allocated. This approach is more straightforward to implement and easier to expand. However, for performing the same operations on components of all neutrinos, data must be fetched from noncontinuous and sparse locations onto vector registers (see Fig. 4.17). As a result, the amount of memory fetch and the latency increase per each cycle.

On the other hand, if an structure can encapsulate a continuous range of data, the compiler can provide continuous streams of data in order to utilize SIMD registers efficiently. As an example, in XFLAT an instance of the neutrino class encapsulates

..B1.33:

arrays of neutrino flavor states for which each element represents a particular energy. xorl   %ebx, %ebx

```
        vaddsd     %xmm8, %xmm9, %xmm9
        vmovddup %xmm9, %xmm1 vinsertf128 $1,
        %xmm1, %ymm1, %ymm11 vmovaps
            %ymm11, %ymm0 call
            __svml_cos4
```

..B1.57:
```
        vmovapd    %ymm0, %ymm10
        vmovaps    %ymm11, %ymm0
        call  __svml_sin4
```

Figure 5.11: The code is compiled with only -xHOST flag. The employed transcendental functions are svml cos4 and ⎽⎽svml sin4 in which four sin and four cos are calculated independently.

Therefore, an object encompasses a range of neutrino flavor states. As a result, in order to perform the same computations on the neutrino wavefunction's components, they can be fetched and stored onto SIMD registers continuously.

In order to study the performance impact of SoA and AoS approaches, two kernels were developed. The first one contained a class in which four arrays were defined, thus the design approach was SoA. The length of the arrays were 100 double precision numbers, and 1000 instances of the class were created. For the second kernel, a single class was developed in which four single variables were encapsulated, thus the design approach was AoS. Afterwards, 100k instances of the class were allocated. Simple floating point calculations were performed on both kernels' data for 1M iteration count. The results of the benchmark on a single CPU are shown in Fig. 5.12.

..B1.33:

One can see that even for simple floating point operations, the performance of the SoA approach is about twice higher than the performance of the AoS approach. Although, the performance gap may vary based on the loops' structure and length of arrays, for high-performance application the SoA approach is recommended. Since,

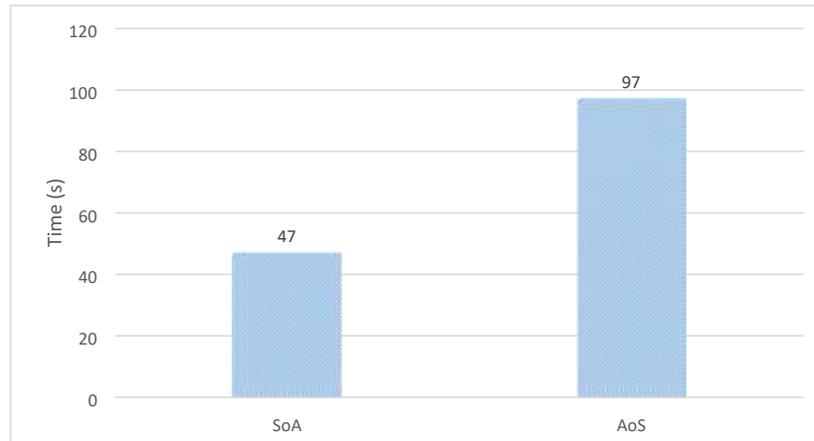

Figure 5.12: Performance comparison of two general designs: Structure of Arrays (SoA) vs. Array of Structures (AoS).

the AoS is more convenient and easier to implement, it can be employed for general applications.

## 5.2.3    Dereferencing pointers inside vectorized loops

During the development of XFLAT, several performance fluctuations were observed on the Xeon Phi platform. One of the strange issues was related to dereferencing pointers within a vectorized loop and the way that arguments were passed to a function. The NBeam class contains several vectorized loop inside which function calls are performed. Thus, pointers and data are passed to the called functions. One important function, which is called frequently, is the neutrino evolution function that loops over neutrino bins, evolves neutrino flavor states, and saves the result onto another neutrino array. Another frequently called function is the density function that receives an array of neutrino flavor states and computes the density matrix for each wavefunction. Both functions are called within vectorized loops and the loops performed reduction operation on arrays.

```
// The input is passed to the class by object reference // The output
is saved onto the class array member for (index: ARRAY_LENGTH) {
      //1) both input and output are explicitly passed as arguments
      evolve1(input[index], output[index], ...);

      //2) the output is an array's member. // thus, it is
      accessible within the function evolve2(index,
      input[index], ...);

      //3) the input is passed via a reference to the object's function // within the
      function the input arrays are accessed directly evolve3(&object_input, index, ...);

      //4) similar to 3) but first the input array is dereferenced // thus, the input data
      access are performed via simple pointers evolve4(&object_input, index, ...);

      //5) similar to 4) but first the output arrays are dereferenced // so member' array
      are accessed via simple pointers evolve5(&object_input, index, ...);
}
```

Figure 5.13: Different approaches for passing arguments and calling the evolution function. Within the function elements are accessed from inside a vectorized loop.

In order to find the root cause of the performance issue, two kernels with similar behavior to XFLAT were developed. The first kernel contained a class that implements a function in which the same instructions as the evolution function in XFLAT were utilized. The second kernel's class had a function similar to the density method in XFLAT with the same set of instructions. There were five ways to call, pass arguments, and utilize the evolution function as shown in Fig. 5.13. Likewise, there were four ways to call, pass arguments, and utilize the density function as illustrated by Fig. 5.14.

```
// The arrays are member of the class for
(index: ARRAY_LENGTH) {
      //1) the data arrays are dereferenced and used within the method density1(index,
      ...);

      //2) the arrays' elements are accessed directly for calculations
      // without any dereferencing in advanced density2(index, ...);

      //3) the data arrays are passed directly to the function density3(output[index], ...);

      //4) the data arrays are dereferenced before calling the function
      density4(output[index], ...);
}
```

Figure 5.14: Different approaches for passing arguments and calling the density function. Within the function elements are accessed from inside a vectorized reduction loop.

Fig. 5.15 illustrates the performance of the five mentioned evolution functions on the CPU as well as the MIC. As it can be seen, the performance on the MIC varied by a large margin. The performance of the first method on the MIC was more than twice higher than the performance of the fifth method. In addition, Fig. 5.16 depicts the performance of the density functions on the CPU and the MIC. Once more, the performance on the MIC fluctuated by a large margin.

As a result, the best performance was achievable when the pointers were dereferenced before each loop. Therefore, the compiler can generate more optimized code when the pointers are dereferenced before starting a vectorized loop. However, on the CPU, this behavior was not observed. In XFLAT, in order to achieve the highest possible performance, all the pointers were chosen to be dereferenced before vectorized loops.

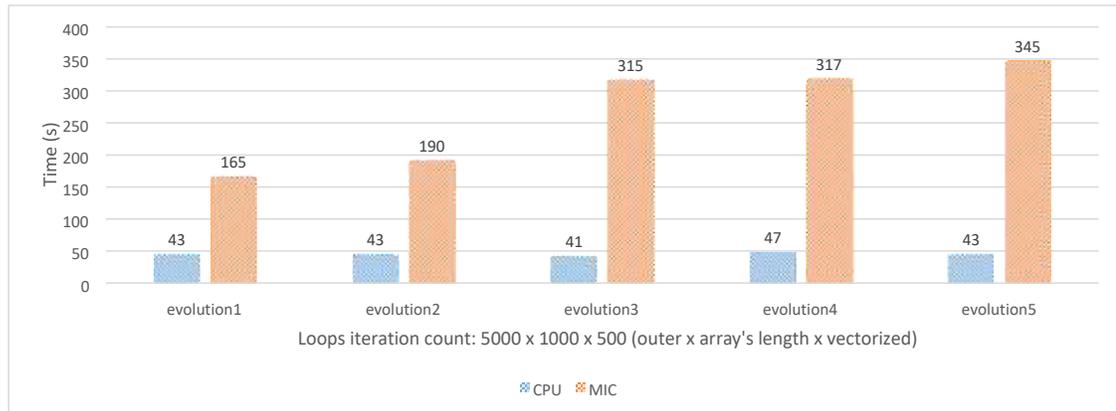

Figure 5.15: Performance comparison of the evolution functions on the CPU and the MIC. On the MIC, the results (orange bars) fluctuated, whereas on the CPU, the results (blue bars) were steady.

## 5.2.4    Calling functions and their arguments

In the previous section, the body of the functions contained complex instructions (similar to XFLAT code). Performing complex instruction can affect the overall

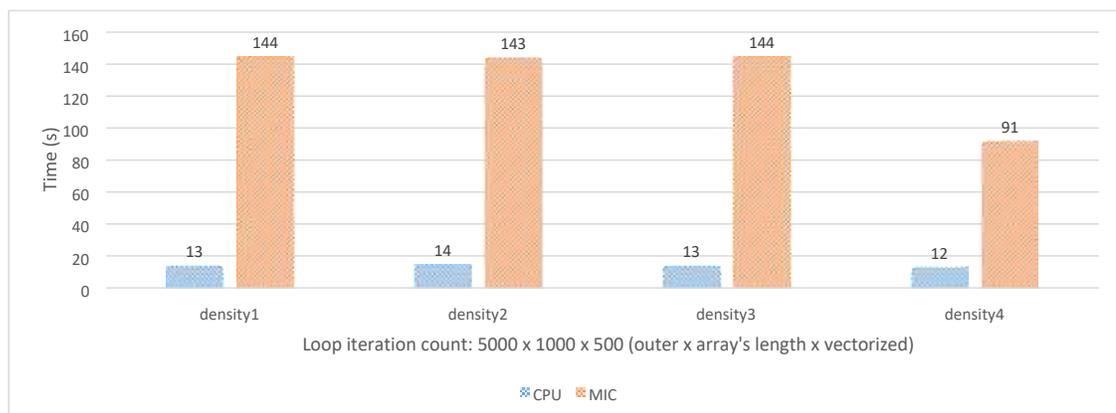

Figure 5.16: Performance comparison of the density functions on the CPU and the MIC. On the MIC, the results (orange bars) fluctuated, whereas on the CPU, the results (blue bars) were steady.

```
#pragma omp simd for (int index :
ARRAY_LENGTH)
{
    /// Only one of the following is called each time!
    func1(index, &results); func2(ar[index], ai[index], br[index], bi[index],
    &results);
}
```

Figure 5.17: Two different approaches of passing arrays' elements to a function. In the first approach, the function only the received the index of elements, whereas in the second approach each element was passed to the function independently.

timing. Therefore, a simpler kernel was designed in order to study the effects of different ways of sending arguments to functions. Within the kernel a single and simple function was called from inside a vectorized loop in two approaches. Fig. 5.2.4 shows the two approaches a function can access elements of an array. The first method was passing the element's index to the function. Within the functions, the received index was employed to look up arrays' elements. The second method was passing each element as a separated argument. In principle, the performance of the two methods should not differ much. Nevertheless, the compiler could generate more optimized code for the second approach.

The designed kernel had two nested loops. The outer loop which only repeated the inner loop. And the inner loop was responsible to loop over array elements and call the functions with proper arguments. The first function only received an index as an argument (implicit method) and the second function received each arrays' elements as an argument (explicit method). If the previously observed issue was not related to the code's instructions, the behavior was expected to be observed again. Fig. 5.18 illustrates the performance of the kernels on the CPU and MIC. Overall, 10 billion arrays' elements were accessed in three different scenarios. The length of arrays were 10000, 2000, and 1000 elements for the first, second, and third scenarios, respectively. The outer loop iteration count was 1 million, 5 million, and 10 million for the first, second, and third approaches, respectively.

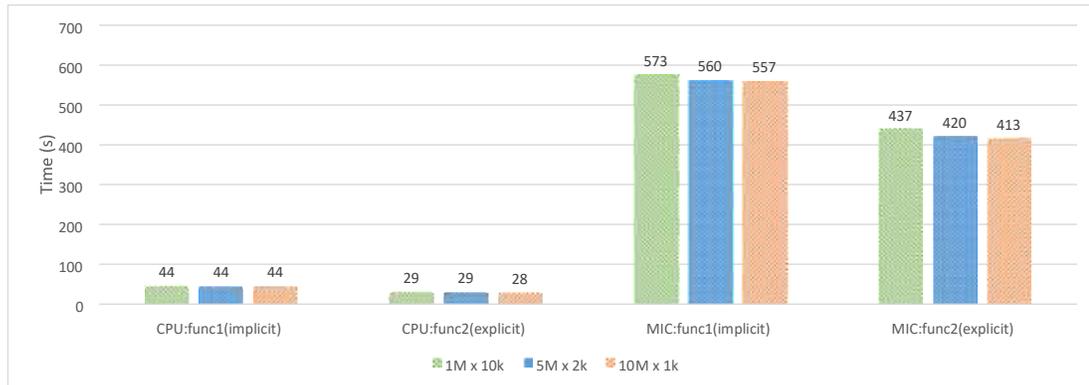

Figure 5.18: Performance of the CPU and MIC in two scenarios (implicit and explicit) for different loop lengths (outer loop len × inner ₋loop len). The kernel was benchmarked for three loop lengths: $1M × 10k$ (green bars), $5M × 2k$ (blue bars), and $10M × 1k$ (orange bars).

As illustrated in Fig. 5.18, the performance of the explicit method on the CPU and the MIC was 35% and 25% higher than the implicit method, respectively. Another noticeable difference is the performance difference between the CPU and the Xeon Phi. This difference, which is about 12–15 folds, shows that calling a function on the Xeon Phi is an expensive task compared to the CPU. It is still not clear that whether or not the issue is due to a bug in compiler or a limitation in hardware.

## 5.2.5    OpenMP parallel for loops

There are several approaches to turn a serial region into parallel via OpenMP. Implementing parallel regions and nested loops together can cause performance bottleneck if programmers do not take into account the effects of the nested parallel loops. The situation can become more complicated inside hybrid (MPI+OpenMP) codes such as XFLAT in which depending on the location of the synchronization points the performance of parallel regions may be different. XFLAT has an outer loop that continues during the life time of the application. Within the loop, there are several regions that were parallelized using OpenMP. There are four possible approaches that a parallel region and a parallel loop can be implemented in XFLAT. Fig. 5.19 illustrates the first approach on which a parallel region encloses the outer loop as well as multiple parallel for loops. Note that there are single threaded sections before each for loop. The next approach, as illustrated by Fig. 5.20, is to enclose the internal region of the outer loop only. Thus, the single threaded regions as well as the inner for loops are enclosed with OpenMP pragma. Fig. 5.21 depicted another approach to parallelize a region, which is to have the parallel region enclose everything and adding the nowait pragma to the inner for loops. In this way every time that a thread completes its for loop computations, it does not stay idle at the end of the loop and continues outside the for loop. The last method is to parallelize only the inner for loops using OpenMP pragma. Hence, there is no need to define the outer parallel region or to have single threaded regions inside the outer loop (see Fig. 5.22).

Several kernels were required in order to benchmark the performance of the mentioned approaches. In all of the benchmarks, the outer loop contained four parallel regions and four single threaded regions. Each single threaded region resided exactly before one for loop. The computations within each region depends on the previous region's result to make sure that the compiler did not remove and optimize out any part of the kernels.

Since the amount of computations did not change between kernels, any performance difference was due to the different parallelism approaches. The kernels were benchmarked using three different for loops' length. For the first run, the outer loop iteration count was set to 100k and every inner loop iteration count was set to 10k. The results of the MIC and CPU for the four approaches are illustrated in Fig. 5.23. #pragma omp parallel

```
{ Loop(termination_conditions)
    {
        #pragma omp single {
            /// single-threaded code
        }
        #pragma omp for for
        (int i : index)
        {
            /// multi-threaded code
        } ...
    }
}
```

Figure 5.19: First approach for parallelizing a region via a parallel region that encloses everything and single regions within the loop.

On the CPU there was no visible performance difference between those methods and on the MIC the maximum difference was about 10 seconds for 100k iteration count. For the next run, the outer loop iteration count was set to 500k and every inner loop iteration count was set to 2k. As shown by Fig. 5.24, on the CPU there was almost no performance difference between different approaches, however, on the MIC the maximum performance gap was about 40 seconds. For the last run, the outer loop iteration count was set to 1M and each inner loop ieration count was set to 1000. This time as depicted in Fig. 5.25, the CPU performance fluctuation was about 7 seconds, however, on the MIC the maximum performance gap was increased to about 100 seconds. The performance gap on the MIC may be due to MIC's simpler core architecture and lower clock rate.

Over 1 million iterations the performance difference between the mentioned approaches was negligible on the CPU. On the MIC the performance was below 100

```
Loop(termination_conditions)
{
    #pragma omp parallel
    {
        #pragma omp single {
            /// single-threaded code
        }
        #pragma omp for for
        (int i : index)
        {
            /// multi-threaded code
        } ...
    }
}
```

Figure 5.20: Second approach for parallelizing a region via a parallel region inside the main loop that encloses everything.

seconds. Note that in the real applications the 1 million iterations of the outer loop may take hours or days to complete, therefore 100 seconds of difference is still negligible on the MIC.

For XFLAT implementing the fourth method was chosen. There were two main reasons for that. First of all, the performance of the third and fourth method were always the best, and the second and more important factor was its simplicity. The simplicity of the fourth method comes from the fact that there is no need to define the OpenMP parallel region to enclose the inner parallel for loops. Furthermore, defining the single threaded regions is not required as well. Consequently, MPI functions can be put after each for loops without requiring to treat them as special lines of code inside OpenMP parallel regions. As a result, the implementation became simpler, the maintenance became easier, and the debugging phase became less complicated. #pragma omp parallel

```
{ Loop(termination_conditions)
    {
        #pragma omp single {
            /// single-threaded code
        }
        #pragma omp for nowait for
        (int i : index)
        {
            /// multi-threaded code
        } ...
    }
}
```

Figure 5.21: Third approach for parallelizing a region via a parallel region that encloses everything and single regions within the loop. Threads at the end of parallel for loop does not wait for the other threads.

```
Loop(termination_conditions)
{
```

```
/// single-threaded code …
#pragma omp parallel for for (int i
: index)
{
      /// multi-threaded code
} …
}
```

Figure 5.22: Fourth approach for parallelizing a region via separated parallel for regions.

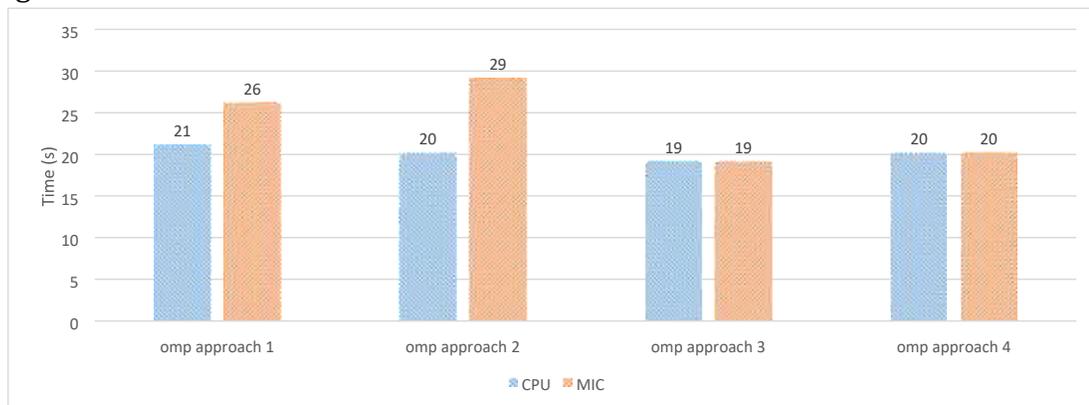

Figure 5.23: Illustration of CPU (blue bars) and MIC (orange bars) OpenMP performance for 100k × 10k loop configuration (outer loop len × inner loop len).

## 5.2.6   The effects of NUMA on multi-socket code performance

On multi-socket systems, multiple processors may be installed on a single motherboard and managed by a single image of an OS. Therefore, the OS can manage all of the available cores. Furthermore, the OS manages all of the available memory, although each RAM module is directly connected to only one CPU socket. As a result, one single instance of an application can employ every accessible core as well as the entire amount of RAM. On multi-threaded codes, depending on the application objective, at some points every thread may require to access a specific memory block. Since each block of memory resides on a unique location on physical RAM, threads that do not reside on the corresponding socket, can access the memory block only by going through extra buses. For example, in Fig. 4.19, if a thread on the CPU1 attempts to access a memory block that resides on the RAM module (that is connected to CPU2), it has to go through one more bus (QPI) in order to access the memory. It is not hard to imagine the performance impact of a situation that many threads on different CPUs attempt to access scattered memory blocks on a multi-socket system. The situation will even worsen if the application repeats the memory access within a loop for every iteration.

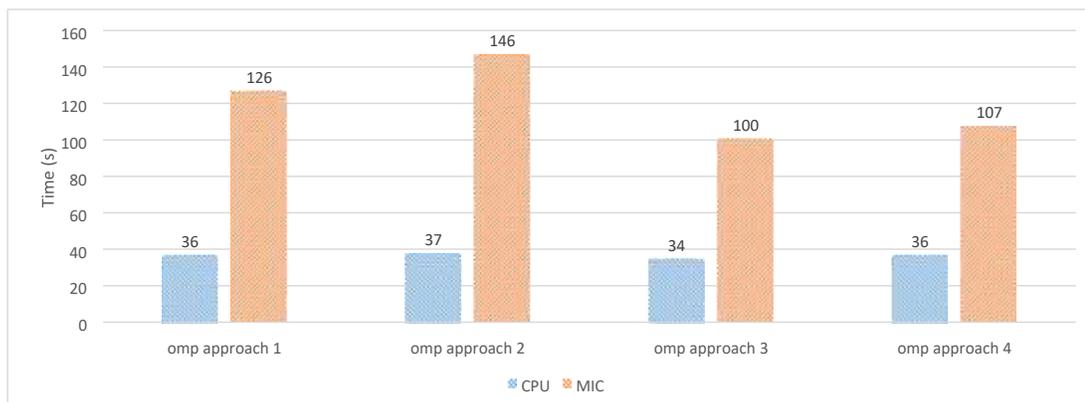

Figure 5.24: Illustration of CPU (blue bars) and MIC (orange bars) OpenMP performance for 500k × 2k loop iteration counts (outer loop len × inner loop len).

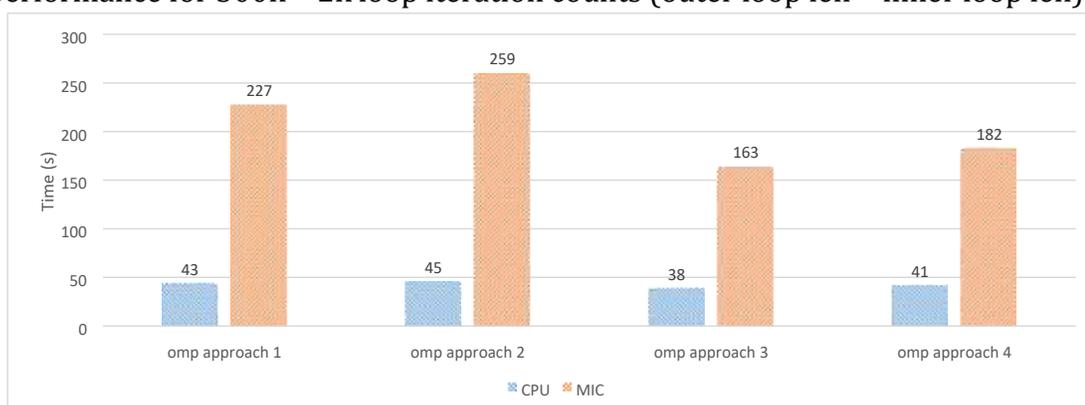

Figure 5.25: Illustration of CPU (blue bars) and MIC (orange bars) OpenMP performance for 1M × 1k loop iteration counts (outer loop len × inner loop len.

Two possible approaches can address the mentioned issue. The first approach is to make the memory initialization code multi threaded in the same way that the rest of the code is multi-threaded. Modern OSes do not allocate the requested memory before the first touch, *i.e.* the first attempt to read from or write to the memory. // By adding the OpenMP parallel for pragma, memory initialization
// become multi-threaded. As a result, the allocated memory resides
// near the thread who initialized it

```
#pragma omp parallel for for
(index: ARRAY_LENGTH)
Beam[i].init();
```

```
// The computational part is always multi-threaded
#pragma omp parallel for for
(index: ARRAY_LENGTH)
Beam[i].calc(...);
```

Figure 5.26: Parallel initialization of the memory as one of the solutions for the first touch memory issue.

Therefore, the allocated memory will reside as nearest as possible to the thread who first attempts to touch it. If the memory initialization code is multi-threaded in the same way as the rest of the code, it is highly probable that the initialized memory by a thread and the working memory of the thread remain identical (see Fig. 5.26). The second method is to eliminate the issue completely by running two separated instance of the code, each on a single socket. However, the second method requires multi-node support by employing the MPI communications.

XFLAT's functions fetch large amount of data per iteration. Therefore, if an instance of XFLAT runs on a multi-socket system, there might be a serious performance bottleneck in run time. In order to measure the performance impact of the first touch issue, XFLAT was benchmarked for three different scenarios on Stampede. For the first run, a single instance of XFLAT utilized the entire dual-socket compute node. For the second run, two separated instances of XFLAT, each employed one socket of a dual-node compute node. The communication between the two instances were performed via MPI and on the hardware side via QPI bus. For the third run, a single instance of XFLAT utilized an entire dual-socket compute node similar to the first run, however, the memory initialization section was parallelized via OpenMP. All runs were continued for about 100 seconds, the number of neutrino beams was 1200×100×80, and the number of computed radial steps was measured. As depicted in Fig. 5.27, the performance difference of the first run was half of the other runs, indicating that the inter-socket communications can impact on the performance of high-performance applications. As a result, the amount of inter-socket communications should remain minimum in order to achieve the maximum performance of
applications.

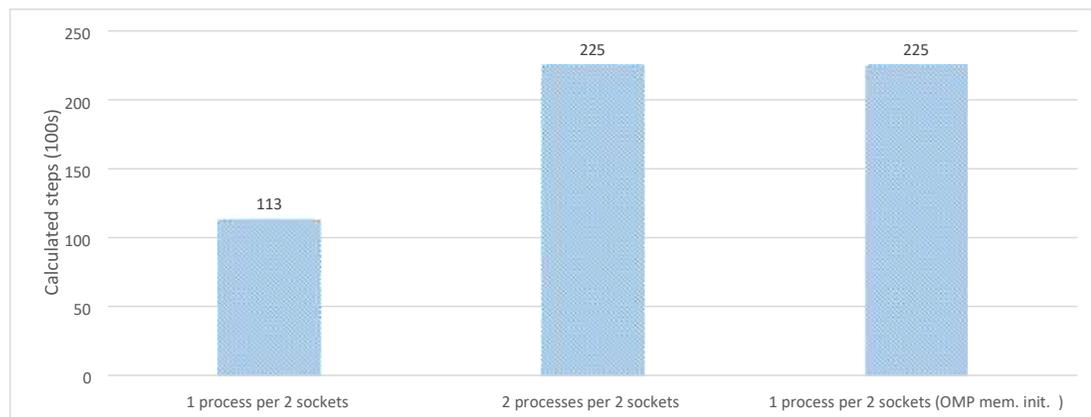

Figure 5.27: Performance of three methods of running XFLAT on a NUMA system. The calculated radial steps were measured for 100-second runs. From left to right, one XFLAT instance employed the entire dual-socket node, two XFLAT instances each employed a single CPU socket, and one XFLAT instance initialized memory in parallel and employed the entire dual-socket node.

## 5.2.7 Fusing functions

XFLAT contains several functions in which different instructions are performed on identical data. By calling each function, the required data must be fetched from the main memory. The fetched data is stored within the processor's fast memory, *i.e.* the cache memory. After the function returns to the calling point, its data that resides on the processor's cache memory may be evicted to make space for new functions' data. However, the new function may perform different calculations on the same set of data and there is no guarantee that the old data, within the cache, can be reused. Since the fetching time is on the order of microseconds and the execution time is on the order of nanoseconds (The clock frequency of the modern processors is typically above 1 GHz), reducing the amount of memory fetch can boost the overall performance dramatically.

In order to improve the performance, XFLAT provides fused functions, which were created by merging two functions that belong to one module together. The functions can be fused if and only if they process on the same set of data. The method of fusing functions is only applicable when the data remain unchanged between the two function calls.

As an example of implemented fused function, within NBeam class, the function that calculates the neutrinos' evolution can be fused with the function that calculates the partial summation over neutrinos' wavefuncion, since both functions may be called back to back and require the same set of data. In NBeam class, there is another fused function similar to the mentioned fused version that receives an array of neutrino beams as an extra argument. Thus, it can calculate the average of neutrinos' flavor states between the computed neutrino beams and the transferred as an input neutrino beam array (three functionalities fused together).

As an another example, at the end of the evolution loop, within the numerical module, there is a function that computes the maximum global error. Moreover, there is a function at the beginning of the evolution loop in which partial summation over neutrino's energy bins is performed. Since those functions perform calculations on the same set of data and the data remain unchanged between the two calls, they safely can be fused together. Nevertheless, the original version of those functions is performance were 15% and 30% higher on the MIC and CPU, respectively when the fused functions were employed (blue bars) compare to when the non-fused functions were employed (orange bars).

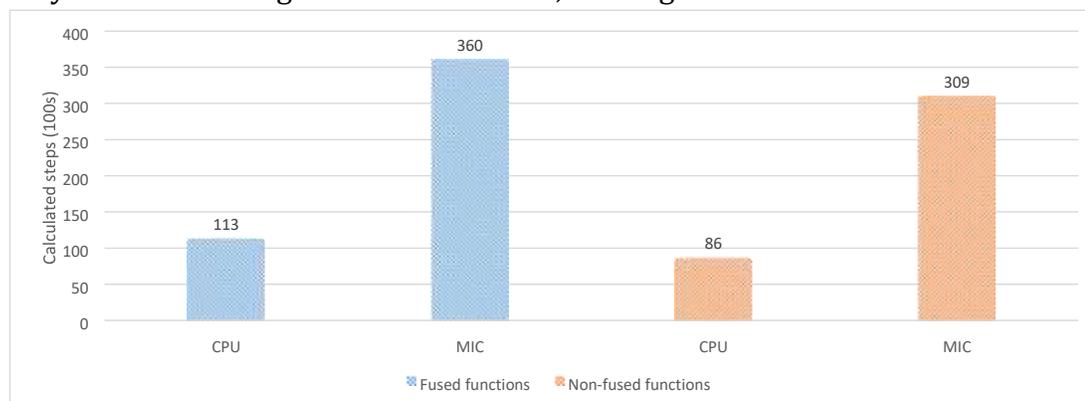

Figure 5.28: Performance of fused and non-fused functions on the CPU and the MIC. The calculated radial steps were measured for 100-second runs. The XFLAT

available as well, since depending on the algorithm the fused versions might not be applicable everywhere.

In order to study the effect of function fusing, XFLAT was benchmarked on Stampede in two approaches. The first run employed the plain non-fused functions, and the second run exploited the fused functions for calculations. Fig. 5.28 illustrates the performance of XFLAT for the problem size of 1200×100×80 (*θ*×*φ*× *Ebins*) on both the CPU and MIC. Every run continued for about 100 seconds and the number of calculated radial steps were measured. The XFLAT performance on the MIC improved by 15% when fused functions were used. On the CPU, the improvement was around 30% due to its larger cache size, in which more data can be maintained, thus the required memory fetches were reduced.

Although, the fusing function optimization was meant for the Xeon Phi, the CPU can gain even more benefit from the method since the cache size on the CPU is larger.

## 5.2.8    I/O performance

Every high-performance computing application requires to save the result of computations by writing data before termination. Since I/O tasks are slower than the rest of an application, there may be bottleneck in the I/O section of an HPC code. On the heterogeneous systems the performance hit may be even higher since there are different types of processors as well as multiple data buses involved. Depending on the configuration, XFLAT may require to write gigabytes of data onto the disk. Since, XFLAT employs NetCDF for saving data, the performance of NetCDF plays critical role in XFLAT performance. There are several approaches to write multi-dimensional arrays of data using NetCDF, thus several kernels with similar structure to the XFLAT's I/O module, were developed in order to find the most optimum approach.

On XFLAT, there are three methods to save multi-dimensional data on disk. The first method is to write data within nested loops as illustrated in Fig. 5.29. Nested loops may have more overhead than a single loop. Therefore, the second method is to implement a single loop in which the 1D data array is written onto disk (see Fig. 5.30). The third approach is similar to the second approach, however, instead of calling NetCDF function within the loop, the NetCDF function is only called once after the loop is completed. Thus within the loop, data is extracted from neutrino beams and saved onto a buffer. After the completion of the loop, the entire buffer is passed to the NetCDF function to be written onto disk.

In order to study the efficiency of the mentioned approaches, the I/O module of XFLAT was isolated and benchmarked as a separated kernel. As a result, the performance of the other calculations in other modules did not interfere with the performance of the I/O methods. The kernel was benchmarked on Stampede for both the CPU and MIC. Each kernel's snapshot was 128 MB and the total of 100 for (int theta : Theta_Angles) for (int phi : Phi_Angles) for (int p : Particle_Num) for (int c : WF_Components) for (int e : Energy_Bins) nc_put_var…(…, data[theta][phi][p][c][e]);

Figure 5.29: Saving data via NetCDF within nested loops.

```
for (int i : Neu_Beams_TotLen)
{ int index = calc_index(i); nc_put_var…(…, start[index], count[index],
      data);
}
```

Figure 5.30: Saving data via NetCDF within a single loop.

snapshots (12.8 GB) were written on Stampede's $SCRATCH disk. Fig. 5.32 shows the performance of each approach on the CPU and MIC.

As observable from Fig. 5.32, the Xeon Phi has a very poor I/O performance that may result in major bottleneck in run time. Therefore, direct I/O tasks should be avoided on Xeon Phi. Since, I/O tasks are inevitable in production environments, another I/O module was added to XFLAT. In the new module, since the I/O per-

```
for (int i : Neu_Beams_TotLen)
{ int index = calc_index(i); buffer[index]
    = data[index];
}
nc_put_var...(..., buffer);
```

Figure 5.31: Saving data via NetCDF after completion of a single loop. Within the loop data may be extracted from NBeam objects and store onto a buffer.

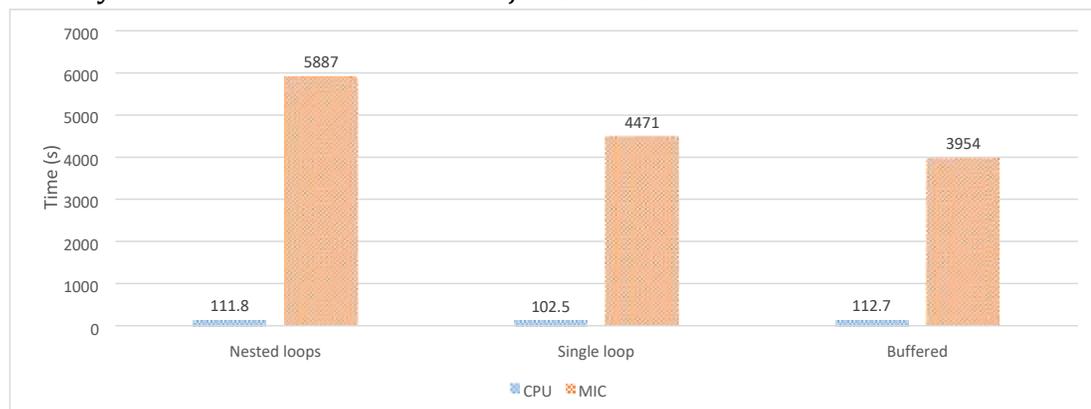

Figure 5.32: XFLAT's I/O performance on the CPU and MIC for three different approaches. The performance of the MIC (orange bars) was 35 to 53 times worse than the performance of CPU (blue bars).

formance of the CPU was satisfactory, the Xeon Phi sends its own data to its CPU mate instead of writing the data onto the disk directly. Afterwards, the CPU is responsible to save its own data as well as the Xeon Phi's data. As a result, in the new indirect module the Xeon Phi extracts data from NBeam objects and saves them onto a buffer, which later sends the buffer to the CPU. Consequently, CPU is the only processor type that is responsible to perform I/O tasks and writes data on disk. Fig. 5.33 illustrates the performance of the direct module in which each processor is responsible to write their own data on disk, and the indirect module in which the MIC sends its data to CPU for dumping on disk.

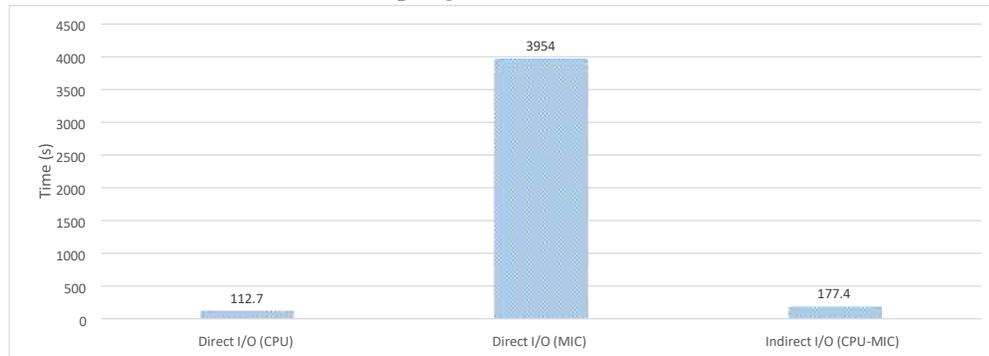

Figure 5.33: XFLAT's I/O performance on the CPU and MIC. From left to right direct I/O on the CPU, direct I/O on the MIC, and indirect I/O from MIC to CPU.

Table 5.1: Bethe workstation specifications.

| CPU | Intel Xeon CPU 6Core/12Thread E5-2620 @ 2.0 GHz |
|-----|--------------------------------------------------|
| Mem | 8x 4096 MB DDR3 @ 1333 MHz |
| MIC | Intel Xeon Phi 5110P 60Core/240Thread @ 1.053 GHz - 8 GB GDDR5 |

Table 5.2: Bahcall workstation specifications.

| CPU | 2x Intel Xeon CPU 6Core/12Thread E5-2620 v2 @ 2.1 GHz |
|-----|--------------------------------------------------------|
| Mem | 4x 16 GB DDR3 @ 1600 MHz |
| MIC | 2x Intel Xeon Phi 3120A 57Core/228Thread @ 1.1 GHz - 6 GB GDDR5 |

# 5.3 Performance Analysis

In order to analyze the performance of XFLAT, the code should be benchmarked on single-node and multi-node as well as homogeneous and heterogeneous environments. It is expected that the performance of the code scales linearly or near linearly when the load increases on processors. As a result, the code was benchmarked on three different available machines: Bethe (Tab. 5.1), installed at the Center for Advanced Research Computing (CARC) at the University of New Mexico; Bahcall (Tab. 5.2), located at Physics and Astronomy department of the University of New Mexico; and Stampede supercomputer (Tab. 5.3), installed at the Texas Advanced Computing Center (TACC). The TACC Stampede system is a 10 PFLOPS (PF) Dell Linux Cluster based on 6400+ Dell PowerEdge server nodes, each outfitted with 2 Intel Xeon E5 (Sandy Bridge) processors and an Intel Xeon Phi Coprocessor (MIC Architecture). The aggregate peak performance of the Xeon E5 processors is 2+PF, while the Xeon Phi processors deliver an additional aggregate peak performance of 7+PF [Sta, 2015].

As observable from the Stampede's specifications, Hyper-Threading (HT) was disabled on Stampede's compute nodes. It was confirmed by one of the TACC Stampede's administrators via email. According to them:

Table 5.3: Stampede Dell PowerEdge C8220z compute node specifications.

| CPU | 2x Intel Xeon CPU 8Core/8Thread E5-2680 @ 2.7 GHz |
|-----|----------------------------------------------------|

| Mem | 8x 4096 MB DDR3 @ 1600 MHz |
|-----|---------------------------|
| MIC | 2x Intel Xeon Phi SE10P 61Core/244Thread @ 1.1 GHz - 8 GB GDDR5 |

"HPC workloads are typically parallelized in a homogeneous fashion, and this is not optimal for HyperThreading – if all the threads are going after the same resources, then performance is not likely to improve much, and will often decrease due to contention for cache resources and DRAM banks. The primary reason we have disabled HyperThreading on all of our production systems is that the performance degradation due to incorrect assignment of processes/threads to logical cores can easily outweigh the (modest) benefits that HyperThreading might provide. The variability in performance due to incorrect assignment of processes/threads to logical cores would almost certainly confuse users and significantly increase our support

workload."

However, as depicted in Fig. 5.34, when utilizing Hyper-Threading, for XFLAT, there was still small performance gain. In the upcoming plots, an instance of XFLAT (extended supernova module), was run on each compute node. Unless stated otherwise, in all of the following benchmarks the number of energy bins was set to 100, the number of azimuth angles ($\phi$ bins) was set to 100, and only the number of zenith angles ($\vartheta$ bins) varied. There may be thousands of neutrino zenith angles in a typical problem size, and the number of zenith angles can be increased until it hits the maximum available memory on each node.

The starting point of problem size for the following benchmarks was $1000 \times 100 \times 100$ ($\Theta$ *angles* $\times \Phi$ *angles* $\times$ *Energy bins*) beams and the number of zenith angles increased by 500 bins until it hit $T \times 100 \times 100$ neutrino beams in which the parameter $T$ depended on the available memory for the CPU or MIC. All the Benchmarks continued for 100 radial steps.

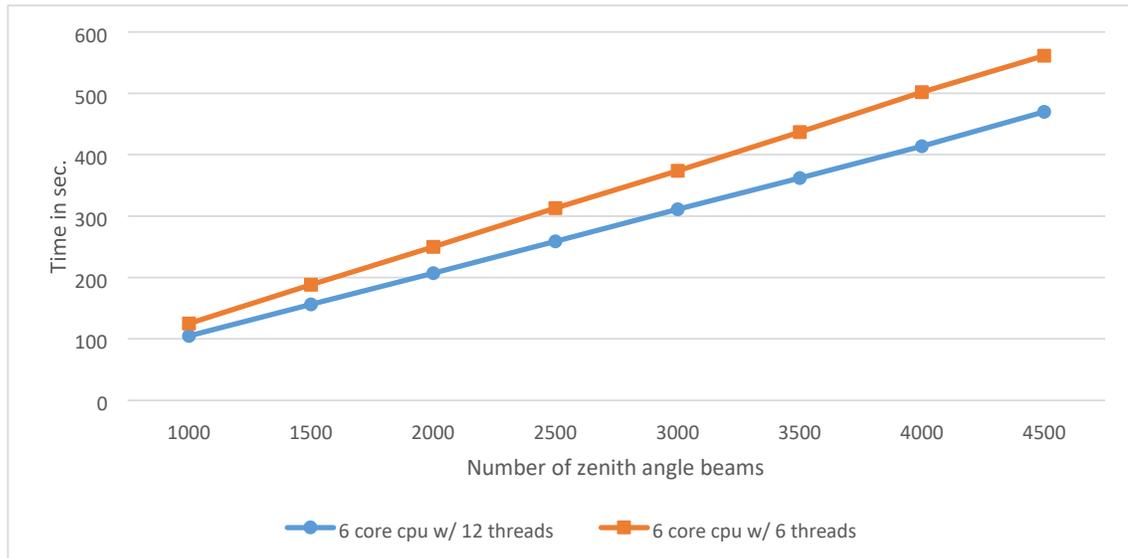

Figure 5.34: CPU benchmark with and without Hyper-Threading on Bahcall. The blue line shows a run by utilizing all threads on the processors and orange line shows a run by utilizing single thread per core on the CPU.

## 5.3.1    XFLAT single processor scalability

The first set of benchmarks was designed to experiment XFLAT scalability as they measured XFLAT timing when the load increased on a single processor. As shown in Fig. 5.35 on Bethe, by increasing the load on the CPU, the timing increased linearly. This shows the code can utilize the entire processor, since by increasing the load the performance did not decreased. However, surprisingly on the dual-socket Bahcall, at some point the timing started to improve!

This behavior was unexpected since typically by increasing the amount of load on a node, the performance behaves linearly or decreases after reaching a particular point. The first step to investigate more about the result was to take into account the hardware differences between the two machines. Bahcall equipped with more memory, thus it was possible to continue the benchmark using higher number of neutrino beams. However, the performance of the Bethe's CPU was about 30%40% better than the Bahcall's, although both CPUs are the same model and even Bahcall has the newer CPU revision (Bahcall's CPU is E5-2620v2 and Bethe's CPU is E5-2620). Hence, two unexplained behaviors were seen in Fig. 5.35. First, the improvement of XFLAT timing when the load increased on Bahcall. Second, the Bethe's CPU performed better than the CPU on Bahcall, although Bahcall was equipped with the newer generation of CPU.

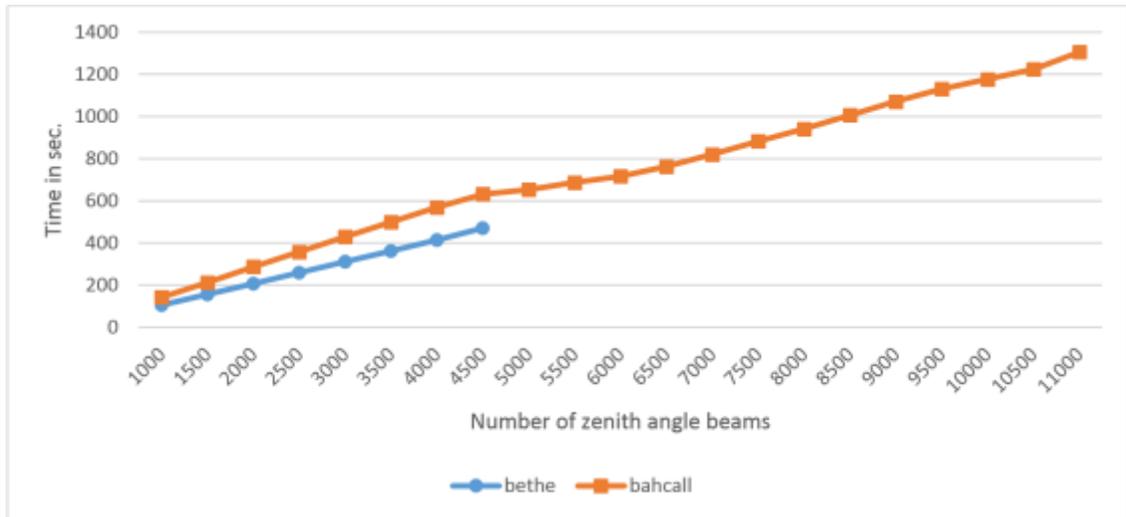

Figure 5.35: XFLAT performance on Bethe and Bahcall machines using one CPU. Horizontal line illustrates the number of zenith angles, the orange curve shows the run time on Bahcall machine and the blue line shows the run time on Bethe machine. Due to memory limitation the run could not be continued on Bethe.

By comparing the details of both machines' hardware, one major hardware configuration difference between Bahcall and Bethe is noticeable. On Bethe, the RAM modules were all connected directly to the single available CPU, yet on Bahcall half of the RAM modules were installed on the other zone, thus connected to the other CPU socket. On Bahcall the CPU accesses to the second half of the RAM via QPI bus (Fig. 5.36).

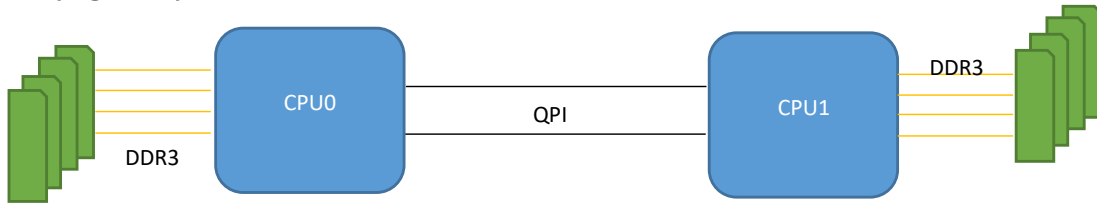

Figure 5.36: Illustration of the memory configuration in NUMA architecture machines, such as Bahcall. CPUs are connected via Intel QPI link and DDR3 is the link between the memory and processor.

For that reason, the timing improvement may be due to the fact that on Bahcall, when too many beams are allocated, some of them are actually allocated on the other memory zone, which is connected to the second CPU socket. Therefore, since accessing the RAM modules on the second zone was provided via QPI bus, the first CPU gained extra memory bandwidth. In other words, the CPU can access to the first part of the memory via its direct memory bus and access to the second part via QPI bus simultaneously.

The first step to confirm this hypothesis was to find out the physical location of the allocated memory by consulting the operating system. On most of the Linux systems this can be confirmed by probing the /proc/buddyinfo file. Bahcall's OS is RedHat based and according to RedHat:

"This file is used primarily for diagnosing memory fragmentation issues. Using the buddy algorithm, each column represents the number of pages of a certain order (a certain size) that are available at any given time. For example, for zone DMA (direct memory access), there are 90 of $2^{0*PAGE\_SIZE}$ chunks of memory. Similarly, there are 6 of $2^{1*PAGE\ SIZE}$ chunks, and 2 of $2^{2*PAGESIZE}$ chunks of memory available. The **DMA** row references the first 16 MB on a system, the **HighMem** row references all memory greater than 4 GB on a system, and the **Normal** row references all memory in between."

```
[vahid@bahcall ~]$ cat /proc/buddyinfo
Node 0, zone      DMA      2      0      1      1      1      0      1      0      1      1      3
Node 0, zone    DMA32  32349  22319  13269   5379   1716    570    268    181     87     59     32
Node 0, zone   Normal 113900  80322  53593  70524  36653  11679    900    435    289    343    777
Node 1, zone   Normal  17569  11022   7608   6106   5825   4750   2758   1514   1056   1273   1532
[vahid@bahcall ~]$
```

Figure 5.37: /proc/buddyinfo file content prior running XFLAT on Bahcall.

PAGE SIZE is a fixed-length continues block of memory that is the smallest unit of data for memory management on operating systems. Normally, PAGE SIZE is 4096 bytes. During huge memory allocation, the number of available pages decrease accordingly. Therefore, it is possible that if the memory, which is directly connected to the first CPU socket is exhausted, then the rest of the allocation is performed on the second memory zone that is connected to the second socket. In Fig. 5.35, the Bahcall's performance was virtually linear up to 4500 zenith angle beams, afterwards there was a sudden change when the number of allocated zenith angle beams was set to 5000. By probing the /proc/buddyinfo file during the allocation, it was possible to figure out if the performance gain was due to the memory allocations on both sockets that resulted in providing more memory bandwidth to the processor.

Fig. 5.37 illustrates the /proc/buddyinfo contents prior to XFLAT run on Bahcall machine. As noticeable, there were plenty of pages available on both sockets (Node 0 and Node 1).

For a problem size below the 5000 zenith angle beams, the required memory for XFLAT could fit onto one memory zone, therefore the code did not utilize more memory bandwidth. As a result, it scaled linearly. For example, as shown in Fig. 5.38 for a run with 2000 zenith angles, the entire required memory was only allocated onto Node 0.

However, as soon as the number of zenith angle beams reached to about 5000, the entire allocation could not fit onto one zone, thus several memory pages were allocated on the other memory zone. In this way, extra buses became available for transferring memory from/to CPU via QPI. Fig. 5.39 illustrates a run with 5000 zenith angle beams. In contrast to the previous screenshots, the available pages (especially large chunk size pages) were depleted on both zones (sockets), which confirm the memory distribution over both sockets.

Figure 5.38: /proc/buddyinfo file content when XFLAT instance fit onto one memory zone on Bahcall.

Figure 5.39: /proc/buddyinfo file content when XFLAT instance does not fit onto one zone on Bahcall.

Moreover, in order to confirm the result from the software point of view, XFLAT was benchmarked for two different scenarios. One benchmark was designed in the way that at the first run, the neutrino's beams were distributed evenly on both memory zones using all the available threads on both CPU sockets, but the calculations were performed by employing only the first CPU's threads. In this way, similar to the previous benchmarks, only one CPU was employed to process the data, yet half of the memory was connected directly to the CPU and the rest of the memory was connected to the CPU via QPI. Thus, it resulted in providing more aggregate memory bandwidth. In order to reach this goal, OpenMP threads were utilized. First, for the memory initialization all the available threads were used on Bahcall (12 threads per each CPU), as illustrated by Fig. 5.40. Hence, memory uniformly distributed over #pragma omp parallel for num_threads(24) for (int i = 0; i < neutrino_beams; ++i) new (&beam[i]) NBeam();

Figure 5.40: Parallel neutrino's beams memory allocation using all the available threads on Bahcall.

```
#pragma omp parallel for num_threads(12) for (int i =
0; i < neutrino_beams; ++i) exec(beam[i]);
```

Figure 5.41: Parallel execution of neutrino's beams using half of the available threads on Bahcall.

two sockets.

After performing memory initialization, the rest of the computational loops were performed by only the OpenMP threads on the first CPU. Therefore, the number of threads were adjusted to 12 (Fig. 5.41), since each CPU on Bahcall has 12 hardware threads.

Nevertheless, enabling only half of the available threads on Bahcall was not enough, since one has to make sure that only threads in the first CPU were participated in computational loops. This can be achieved by invoking OpenMP functions within the program or by setting the environment variables from shell. The settings and commands may vary between compilers and shells. For all of the benchmarks the Intel compiler was chosen and OpenMP threads were located as closest as possible to each other (export KMP AFFINITY=compact). As a result, utilizing the first CPU with 6cores/12threads became possible in run time.

Fig. 5.42 illustrates that by applying those changes, the code scaled linearly on Bahcall and the result was similar to Bethe's result. Furthermore, the performance of the Bahcall's CPU was very similar to Bethe's CPU due to providing more memory channels.

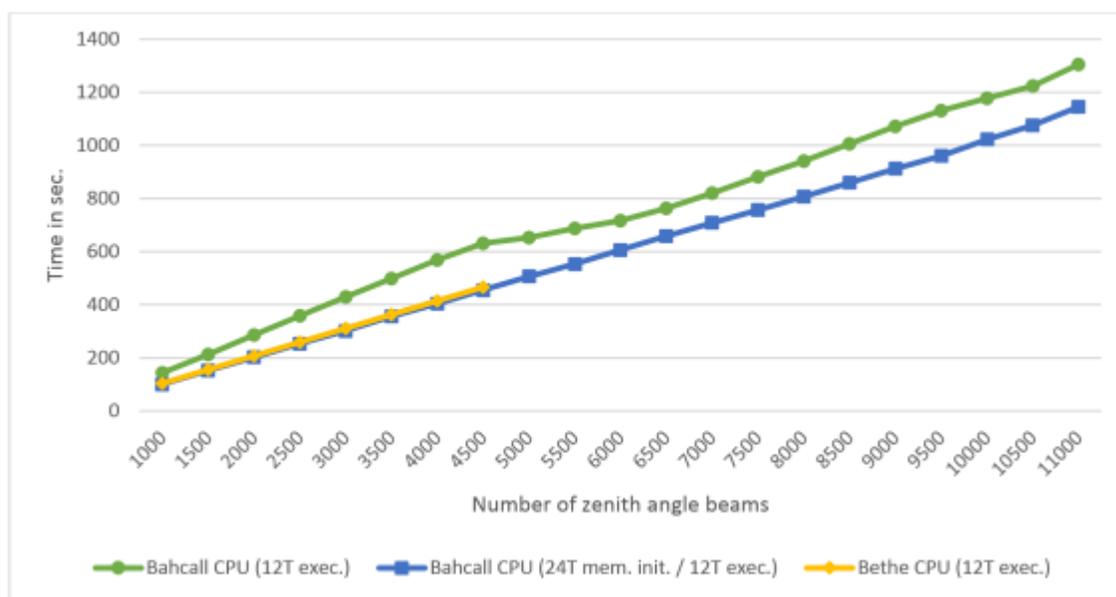

Figure 5.42: Illustration of XFLAT perofrmance on bahcall when different number of threads were utilized for memory initialization and execution. Green curve illustrates the performance when a single Bahcall CPU were utilized for memory initialization and execution. Blue curve illustrates the performance when both Bahcall CPUs were utilized for memory initialization but only a single CPU were employed for execution. Yellow curve illustrates the performance of the single-CPU Bethe machine.

As a result, the performance difference between the two machines was related to the memory bottleneck on Bahcall. Two RAM modules (each 16 GB) were connected to each CPU on Bahcall, however, on Bethe eight RAM modules (each 4 GB) were connected to the CPU. Since, the Bahcall's CPU could not reach the full capability due to the bandwidth bottleneck, the RAM configuration on the machine was changed to provide higher memory bandwidth. Previously, Bahcall was equipped with four 16 GB RAM modules, two installed on each zone (connected to each CPU). However, the Bahcall's CPU (Xeon E5-2620v2) can support quadruple-channel ar-chitecture, which increases data transfer rate by adding four communication channels between CPU and RAM modules.

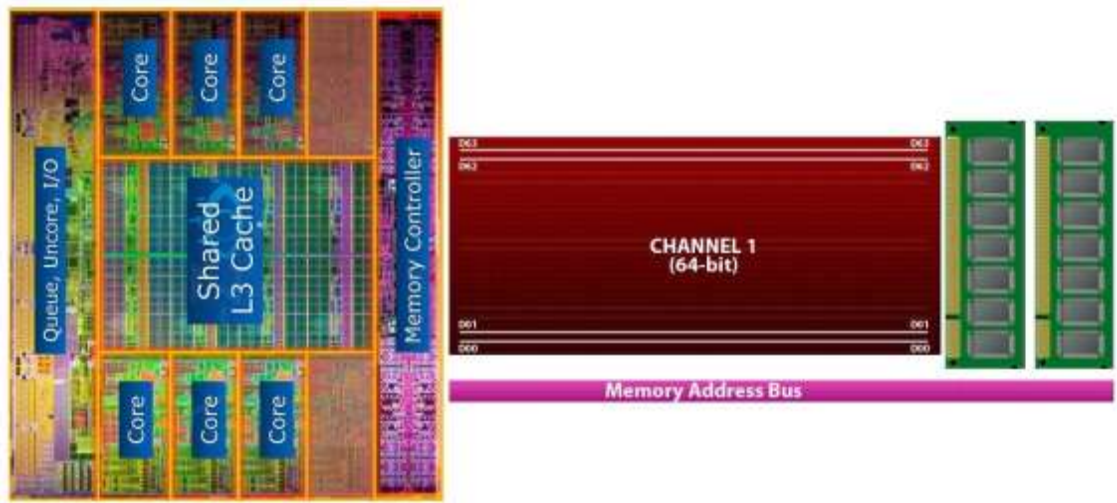

Figure 5.43: Illustration of single-channel memory configuration between Intel CPU and DDR3 memory modules [www.gamersnexus.net, 2015]. One 64-bit channel is available for transferind data to/from CPU.

The multi-channel technology effectively multiplies data throughput from RAM to the CPU's memory controller by providing multiple channels. In the single-channel configuration there is only one 64-bit memory channel to transfer data between the RAM and CPU (see Fig. 5.43). In the dual-channel configurations, two channels are available between the selected RAM modules and the memory controller, thus the effective data channel is equal to a 128-bit channel (see Fig. 5.44). The same is true for triple and quadruple channel configurations that are only available on highend CPUs such as Bahcall's CPUs. The multi-channel technology requires identical RAM modules. For instance, for the quadruple-channel configuration four identical RAM modules should be installed on specific slots on motherboard. The RAM modules should be identical in capacity, speed, latency, number of memory chips, and matching size of rows and columns of memory cells.

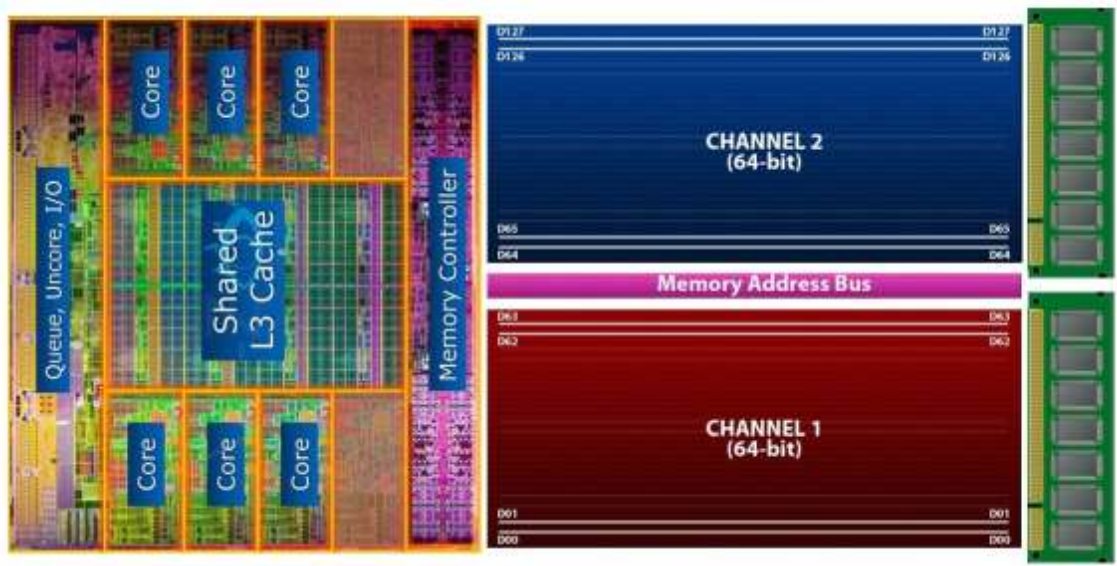

Figure 5.44: Illustration of dual-channel memory configuration between Intel CPU and DDR3 memory modules [www.gamersnexus.net, 2015]. Two 64-bit channels (one 128-bit aggregate channel) are available for transferind data to/from CPU.

Consequently, the RAM configuration on Bahcall's machine was changed and eight 8 GB RAM modules were installed on it, four modules on each CPU zone. The rest of the memory's specifications remained identical as before. Fig. 5.45 depicts the improvement for the single-CPU benchmark and Fig. 5.46 illustrates the improvement for the dual-CPU benchmark on Bahcall.

The next performance benchmark for the scaling of XFLAT was the performance benchmark on multi-socket systems. Bahcall and a single Stampede's compute node were utilized for that purpose. On Bahcall, Hyper-Threading was enabled, thus up to 24 threads could be employed for the benchmark. Since each node of Stampede supercomputer contains two 8-core CPUs, and Hyper-Threading was disabled on Stampede's CPUs, it was possible to utilize up to 16 threads per Stampede's node. Two separate XFLAT benchmarks were performed on a single Stampede's node. The first one launched one MPI task per socket (illustrated in Fig. 5.47), and the other one launched only one MPI task per entire compute node (one MPI task per two CPUs) as shown in Fig. 5.48.

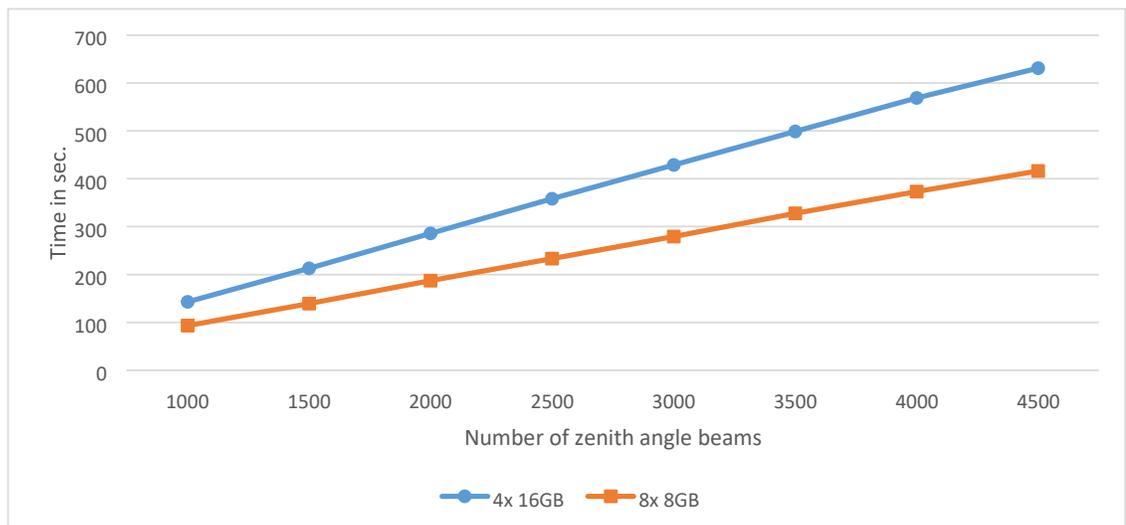

Figure 5.45: Illustration of Bahcall's single CPU timing when 4 RAM modules were installed (blue) and when 8 RAM modules were installed (orange).

Since XFLAT can uniformly distribute memory allocation on multi-socket nodes and since the QPI buses between CPUs on Stampede have enough bandwidth (Dual QPI, 8 GT/s per QPI bus), the performance difference between the two benchmarks is negligible as depicted in Fig. 5.49.

The technical reasons of the performance similarity between the two Stampede's benchmarks is due to the way that the code is implemented. There are two main causes that keep the performance satisfactory on the multi-socket benchmarks. First of all, during the memory initialization (first touch section), all of the threads contribute in the initialization section. This indicates that the memory section related to each thread resides as nearest as possible to the thread. Therefore, threads on one CPU do not have to try to access the other CPU's memory zone, thus the communication overhead between sockets reduce.

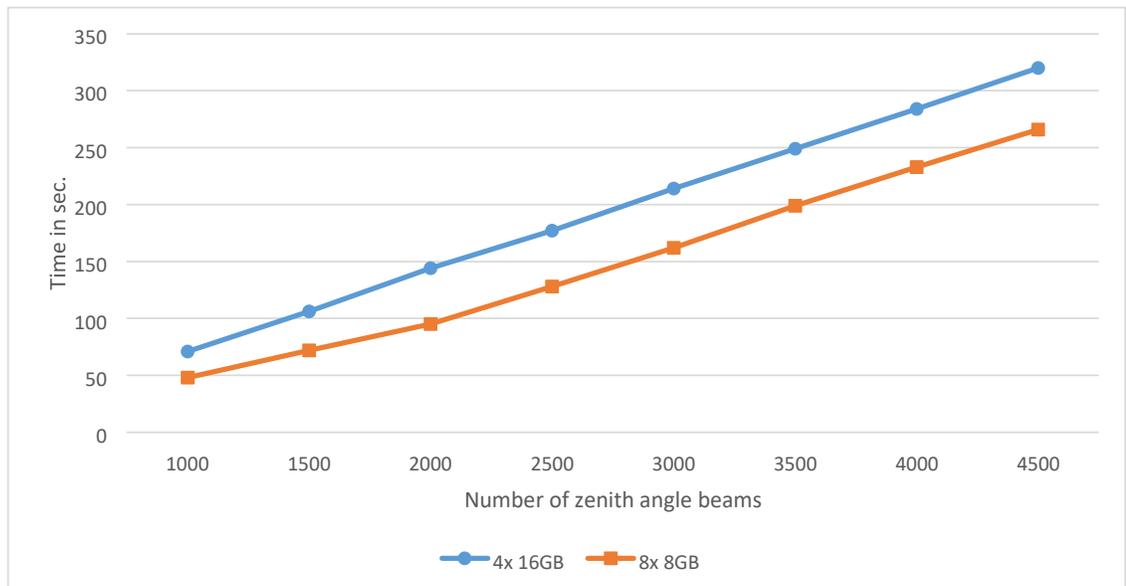

Figure 5.46: Illustration of Bahcall's dual CPU timing when 4 RAM modules were installed (blue) and when 8 RAM modules were installed (orange).

Another reason is because of the selected approach for implementing the Hamiltonian. In order to compute the total Hamiltonian, which depends on the plugged in module, one or more nested summation loops have to be computed. According to Eq. 3.8, the value of the integral can be calculated by performing a loop over all the neutrino's angle beams. Since, the neutrino beams are distributed over compute nodes, a bad implementation would transfer a lot of data between nodes in order to calculate the integral. Nevertheless, in XFLAT, each node can go through an independent loop for calculating the partial integral over their own set of beams. At the end of the loops, only the final value is exchanged between nodes via MPI Allreduce() function. Therefore, every node at the end of the Hamiltonian function has the final value of the Hamiltonian integral. In this way, only a few bytes are required to be exchanged after the completion of the nested loops. Within the neutrino evolution loop there are not many exchange points, thus there is no need for large communication bandwidth. As a result, it is expected that by increasing the number of nodes, the performance of XFLAT does not drop and scales linearly.

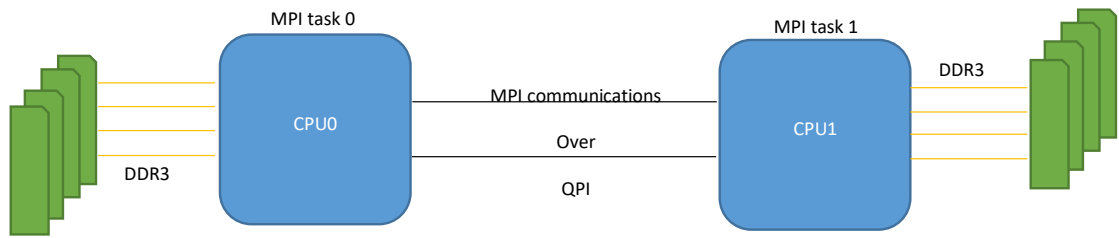

Figure 5.47: A single MPI task per socket that only employs a single CPU on NUMA architecture. Blue squares are CPUs and green recangles are RAM modules.

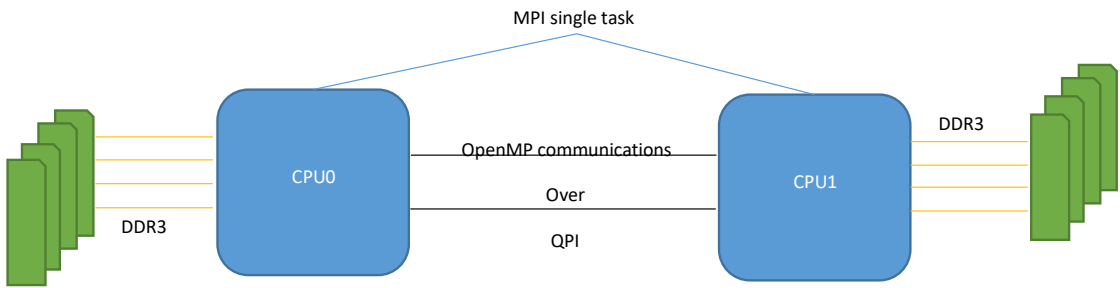

Figure 5.48: A single MPI task per node that employs entire system including all the available processors on NUMA architecture. Blue squares are CPUs and green recangles are RAM modules.

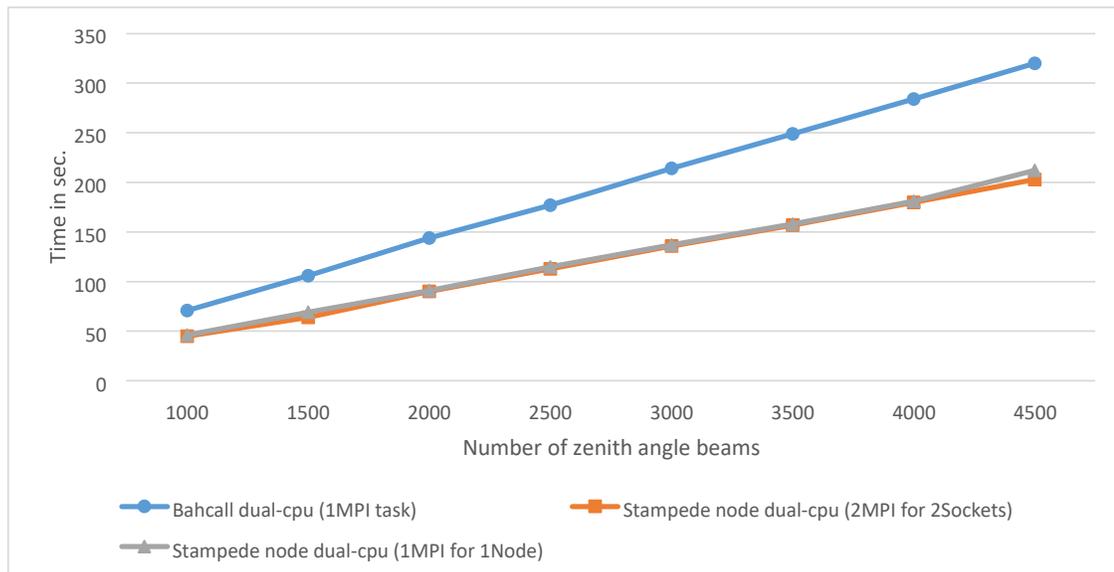

Figure 5.49: XFLAT benchmark utilizing different process configurations. Blue curve illustrates the performance on dual-socket Bahcall machine by employing only one MPI task. Orange curve illustrates the performance on dual-socket Stampede's node by employing one MPI task per socket. Grey curve illustrates the performance on Stampede's node using one MPI task per entire node.

Prior to studying the multi-node benchmarks, XFLAT was benchmarked on single CPU, dual CPU, single MIC, and dual MIC configurations on a single node. As depicted by Fig. 5.50 the XFLAT was benchmarked on the Stampede node for 1000 radial steps. The number of zenith angles started from 200 and scaled up to 1200. It was not possible to fit more beams onto the MIC's memory due to its limitation.

As observable, the MIC performance was about three times higher than the CPU on the Stampede. In addition, the performance of the dual-processor benchmarks was twice higher than the single-processor benchmarks.

## 5.3.2 XFLAT perofrmance on heterogeneous environments

Since XFLAT should be able to utilize both the MIC and CPU simultaneously, the next stage of XFLAT performance analysis was studying the code performance on heterogeneous (MIC and CPU together) environments. As the first step on a heterogeneous system, such as the Stampede supercomputer, the way of distributing data across different processors is required to be determined. Since on a heterogeneous environment the computational capability of each processor may differ across nodes, there should be a precise mechanism to decide what portion of data need to be placed on each processor based on their computational capabilities. If the code runs on a heterogeneous cluster with a load-imbalanced configuration, the MPI communication cost will be high. Due to load imbalance on nodes, the waiting time on synchronization points increases. That means a few tasks may finish their executions earlier and remain idle. Therefore, they have to wait for the other tasks to reach the same synchronization point. In addition to the idle nodes there may be oversubscribed nodes in run time.

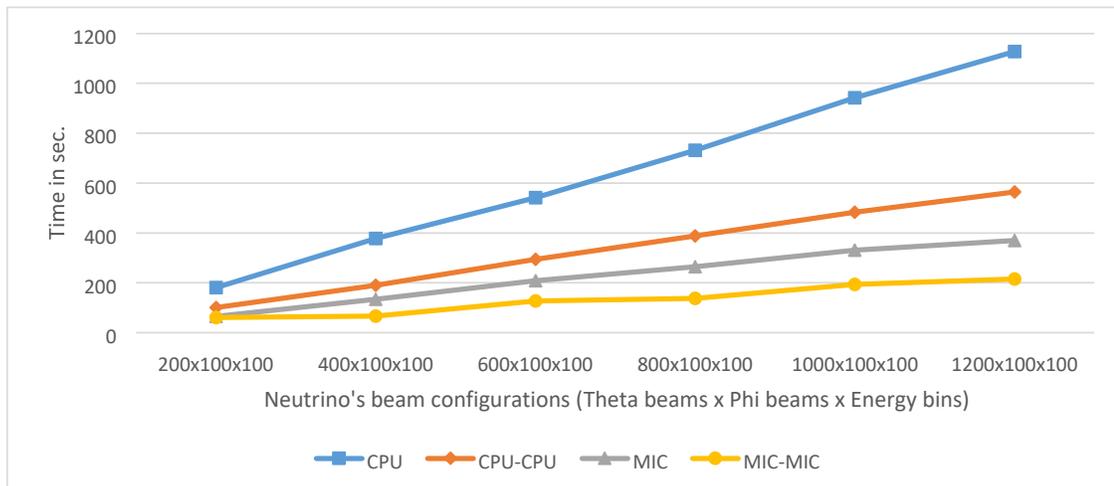

Figure 5.50: Illustration of XFLAT performance on Stampede's node. Blue line illustrates the single CPU performance. Orange line illustrates the dual CPU performance. Grey line illustrates the single MIC performance. Yellow line illustrates the dual MIC performance. The timing was measured for 1000 radial steps.

The first heterogeneous XFLAT benchmark was designed to find the sweet spot for the CPU and MIC on Stampede. Thus, the benchmark could help find the best MIC to CPU load ratio. The motivation for this benchmark was that the raw performance of a processor, the FLOPS capability, and the actual performance, which may be different based on the application's instructions, are not necessarily identical. Therefore, even when the raw performance of CPU and MIC is known in advance, the XFLAT performance may vary depending on the details of the employed modules and the problem size. In addition, due to the low-level differences in the underlying hardwares, the performance of a processor might be satisfactory for many instructions but not satisfactory for the other instructions. Therefore, depending on the employed instructions in a code, the relative performance (CPU vs. MIC) may vary. Although, the instruction set is similar for the CPU and MIC (both are x86), there are still several factors that can affect the final performance of the code. Hardware design factors such as the size and number of levels of cache memory, core frequency, memory bus frequency, number of hardware threads per core, memory bandwidth, length of the vector registers, can affect final performance dramatically. Therefore, both CPU and MIC must be benchmarked using the actual code in order to find the sweet spot for XFLAT code.

For the timing benchmark the number of neutrino beams were manually chosen to be $1000 \times 100 \times 100$ (Θ *angles* × Φ *angles* × *Energy bins*), and the number of total radial steps for the calculation was fixed at 100. The number of neutrino beams assigned to the MIC and CPU varied unless the optimal load ratio for the overall node was found. As depicted by Fig. 5.51, the execution time was shown for various MIC:CPU load ratios and the optimum point was in [2.7–3.0] range. The benchmark was also repeated for another problem size with different number of beams. Fig. 5.52 illustrates the benchmark with different problem configuration. This second benchmark was performed with $10000 \times 10 \times 100$ (Θ *angles* × Φ *angles* × *Energy bins*) beams. As observable the optimum point was still in the same range as the previous benchmark.

In Fig. 5.53 and Fig. 5.54, the maximum spent time within MPI functions were shown. The maximum MPI time was the maximum time that either the CPU or MIC spent in MPI functions, since the timing fluctuated on processors, only the maximum MPI times were measured among each processor's type. When the spent time of communication increases, the performance is supposed to decrease, thus the wasted time increases. The benchmark repeated for various MIC:CPU load ratios.

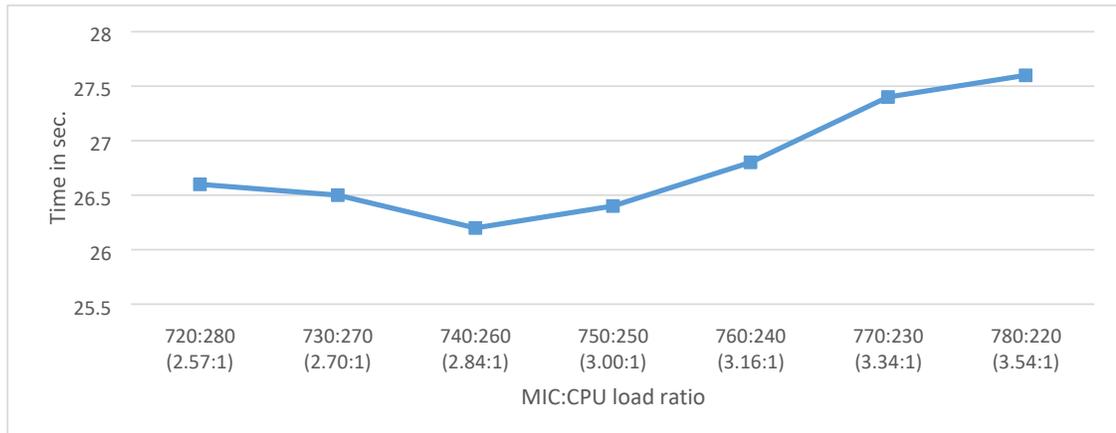

Figure 5.51: The timing benchmark of XFLAT on single-node heterogeneous environment. The execution time was shown for different MIC:CPU load ratios using 1000 zenith trajectories. Each point represents [number of theta angles on ˍMIC : number of theta ˍangles on CPU] (the normalized load ratio is shown in parenthesis).

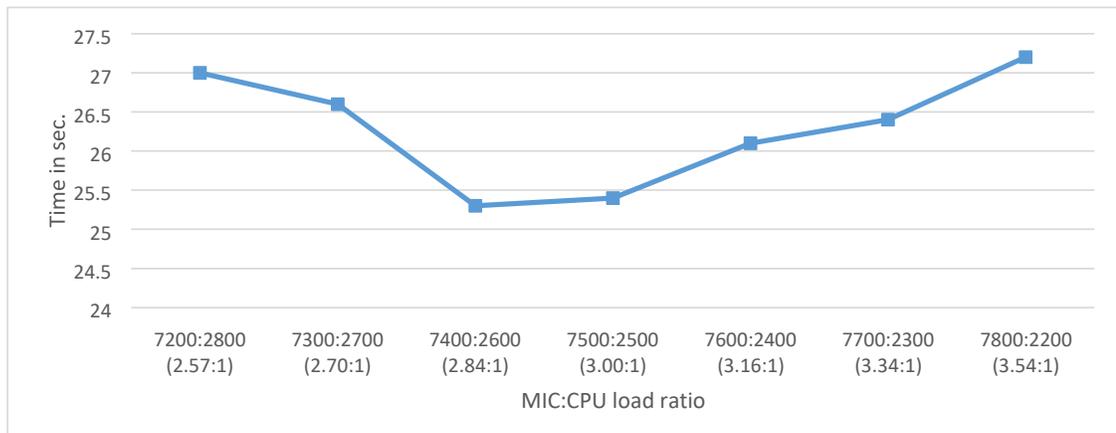

Figure 5.52: The timing benchmark of MPI times for XFLAT on single-node heterogeneous environment. The execution time is shown for different MIC:CPU load ratios using 10k zenith trajectories. Each point represents [number of theta angles on MIC : number of theta angles on CPU] (the normalized load ratio is shown in parenthesis).

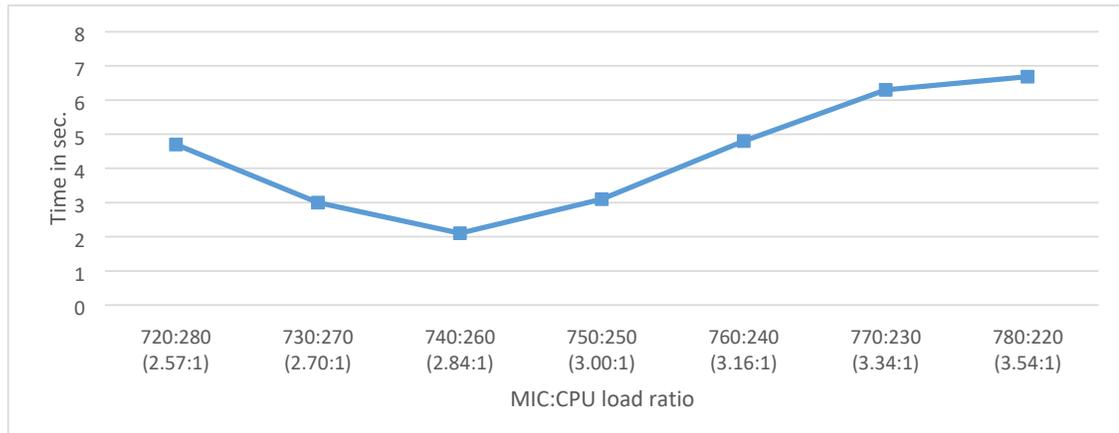

Figure 5.53: The timing benchmark of XFLAT on single-node heterogeneous environment. The spent time for MPI functions is shown for different MIC:CPU load ratios by employing 1000 theta beams. Each point represents [number of theta angles on MIC : number of theta angles on CPU] (the normalized load ratio is shown in parenthesis).

A minimum (optimum) region is observable around the similar range as the previous results. On both sides of the minimum region the time increases. This increase in timing was due to the fact that at least one processor may remain idle and wait for other tasks to reach the synchronization point (which is due to load imbalance between processors).

As observable, the best result was achieved when the load was distributed across both MIC and CPU based on their computational capability. Consequently, the MPI wait time was also minimized at the same point. The illustrated benchmarks show that on the Stampede nodes, the optimal MIC:CPU load ratio range should be in the range of 2.8:1 to 3.0:1. By finding the range of optimum load ratios, the next step can be benchmarking the code on heterogeneous multi-node environments.

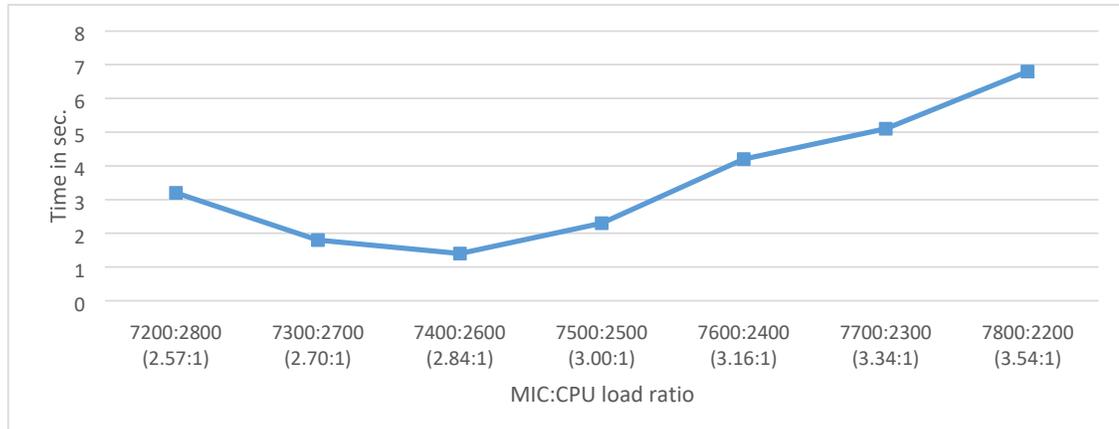

Figure 5.54: The timing benchmark of MPI times for XFLAT on single-node heterogeneous environment. The spent time for MPI functions is shown for different MIC:CPU load ratios by employing 10k theta beams. Each point represents [number of theta angles on MIC : number of theta angles on CPU] (the normalized load ratio is shown in parenthesis).

## 5.3.3 XFLAT performance on multi-node systems

As a result of previous benchmarks, in the multi-node benchmarks the MIC:CPU was set to be 2.9:1 to maintain a minimal processors' idle time.

The next benchmark on Stampede employed one CPU and one MIC per node. The problem size was set to 10000 × 10 × 100 beams and the number of processed radial steps was measured over a 100 second period of run time. The number of nodes varied from 1 to 16 nodes (see Fig. 5.55).

As observable, the result's behavior was virtually linear from one node up to 16 nodes. The 16-node benchmark utilized 32 MPI tasks overall, 16 of them employed 16 Xeon CPUs and the other 16 tasks employed the 16 Xeon Phis. The maximum spent time in MPI functions among all nodes was recorded, and the results are illustrated in Fig. 5.56.

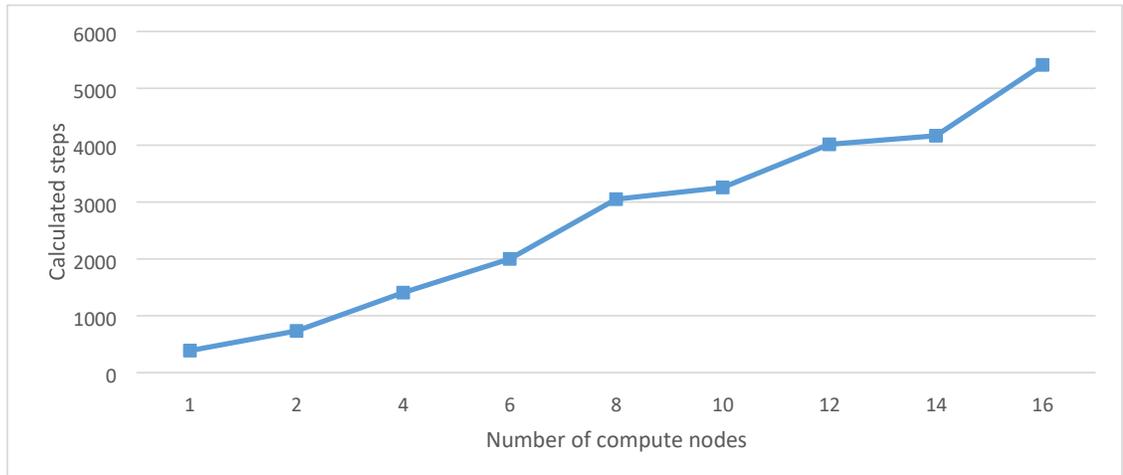

Figure 5.55: XFLAT heterogeneous multi-node performance on Stampede (one CPU and one MIC per node) for various number of compute nodes. The calculated radial steps were measured for about 100 second run.

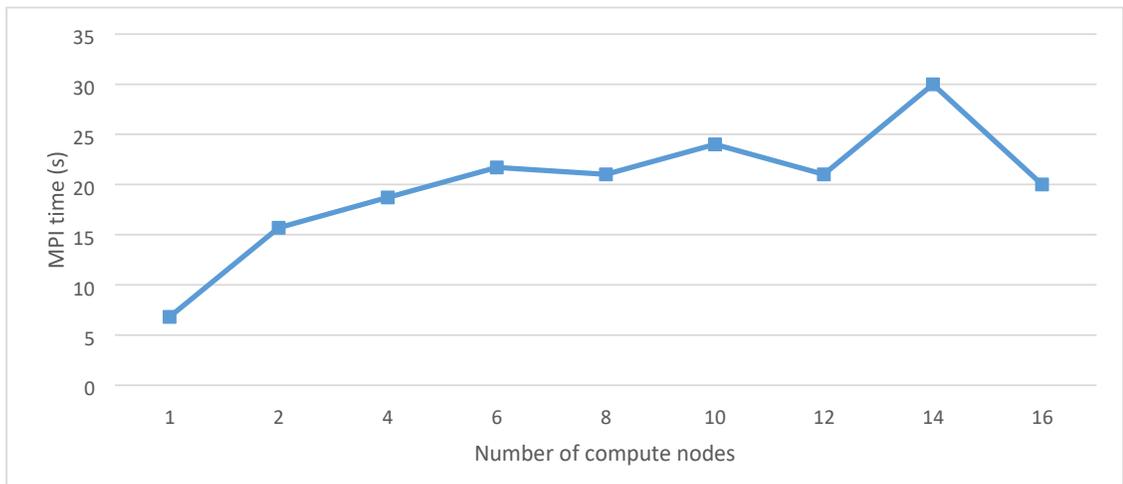

Figure 5.56: The maximum spent time within MPI functions for various number of nodes on Stampede (using one CPU and one MIC per node). The problem size was set to 10000×10×100 beams and the number of processed radial steps was measured over 100 second period of run time.

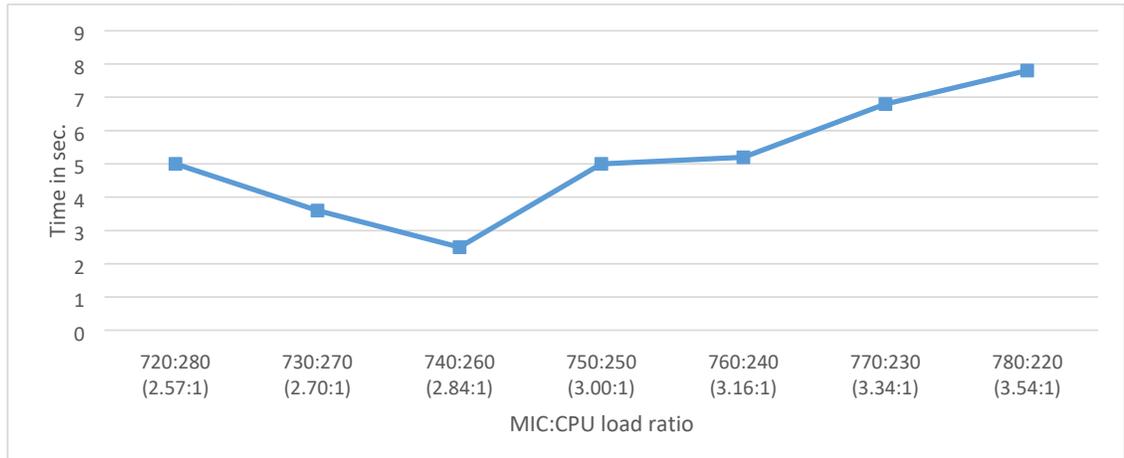

Figure 5.57: The timing benchmark of MPI times for XFLAT on single-node (dualCPU and dual-MIC) heterogeneous environment. The spent time for MPI functions is shown for different MIC:CPU load ratios by employing 1000 theta beams. Each point represents [number of theta angles on ̱MIC : number of ̱theta angles on CPU] (the normalized load ratio is shown in parenthesis).

The similar benchmarks were performed on Stamped by employing dual-CPU and dual-MIC nodes. On each node there were two MPI tasks on CPUs (one per CPU) and two MPI tasks on MICs (one per MIC).

Similar to the single processor benchmarks, the MIC:CPU ratio was set to 2.9:1. Fig. 5.57 and Fig. 5.58 illustrate the maximum time of MPI functions for the 1000× 100 × 100 and 10000 × 10 × 100 beam configurations, respectively.

Afterwards, a set of new benchmarks were performed using identical problem size in the single processor benchmarks, for various number of nodes (1 to 16 compute nodes). Consequently, the single-node benchmark utilized four MPI tasks (two on the CPUS and two on the MICs), thus the 16-node benchmark utilized 64 MPI tasks. The results are presented in Fig. 5.59.

Surprisingly, by employing beyond 8 nodes, the performance did not improve. In fact, there were no significant improvements until the number of nodes increased to

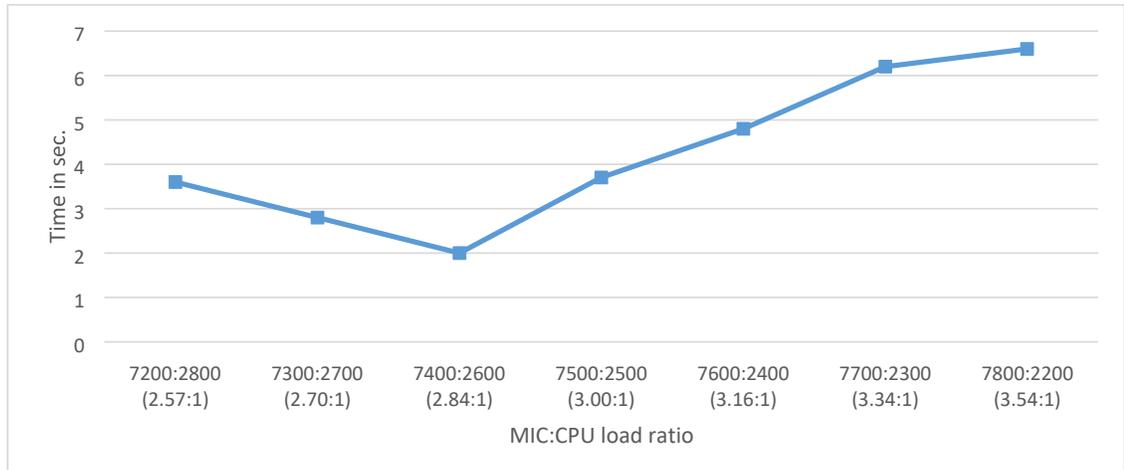

Figure 5.58: The timing benchmark of MPI times for XFLAT on single-node (dualCPU and dual-MIC) heterogeneous environment. The spent time for MPI functions is shown for different MIC:CPU load ratios by employing 10k theta beams. Each point represents [number of theta angles on _MIC : number of _theta angles on CPU] (the normalized load ratio is shown in parenthesis).

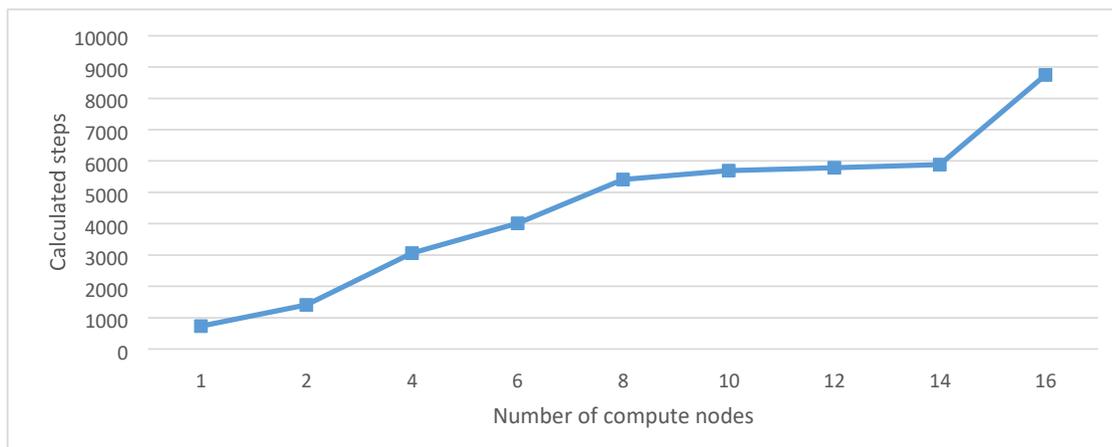

Figure 5.59: XFLAT heterogeneous multi-node performance on Stampede (two CPUs and two MICs per node) for various number of compute nodes. The calculated radial steps were measured for about 100 second run.

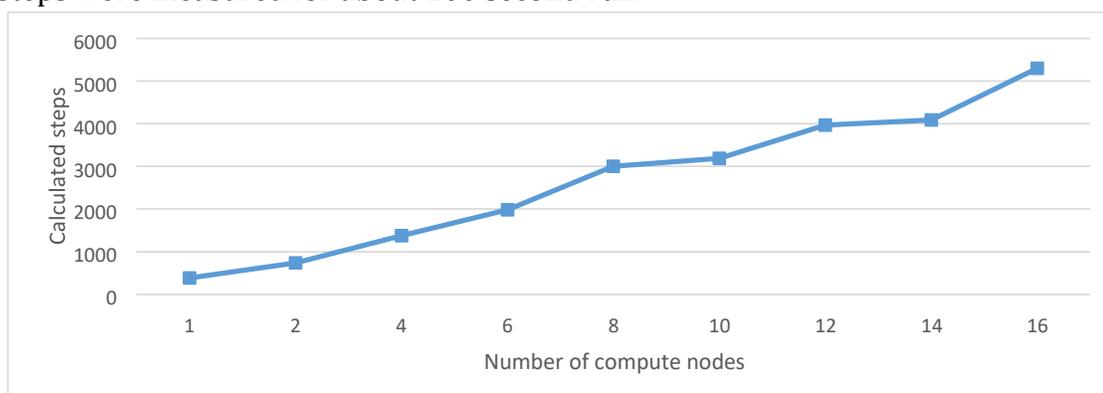

Figure 5.60: XFLAT heterogeneous multi-node performance on Stampede (two CPUs and two MICs per node) by employing double amount of load on each processor. The calculated radial steps were measured for about 100 second run.

16. By employing 16 nodes, an unexpected jump in performance is observable.

No improvement region in the result may be due to insufficient load on Xeon Phi processors. In order to confirm it, a benchmark was designed in which a double amount of load were placed on each processor. As depicted by Fig. 5.60, after placing twice more load on each component, the performance of the dual-processor benchmark appeared to be similar to the single-processor benchmark.

In addition, as an another test, the single-processor benchmark was continued beyond 16 compute nodes. Consequently, the load on each processor eventually became similar to the load on the dual-processors benchmark (For instance, for the 32-node single-processsor benchmark, the load on a CPU was identical to the load on a CPU in the dual-processors benchmark using 16 nodes). As illustrated by Fig. 5.61, the similar flat region that was observable in the dual-processor benchmark appeared
again by going beyond 16 nodes.

As a result, it can be concluded that the flat region in the performance result was not due to the number of employed nodes in the environment. One possibility for that behavior was the fact that by increasing the number of nodes, the overhead of sending and receiving MPI messages may affect the overall performance.

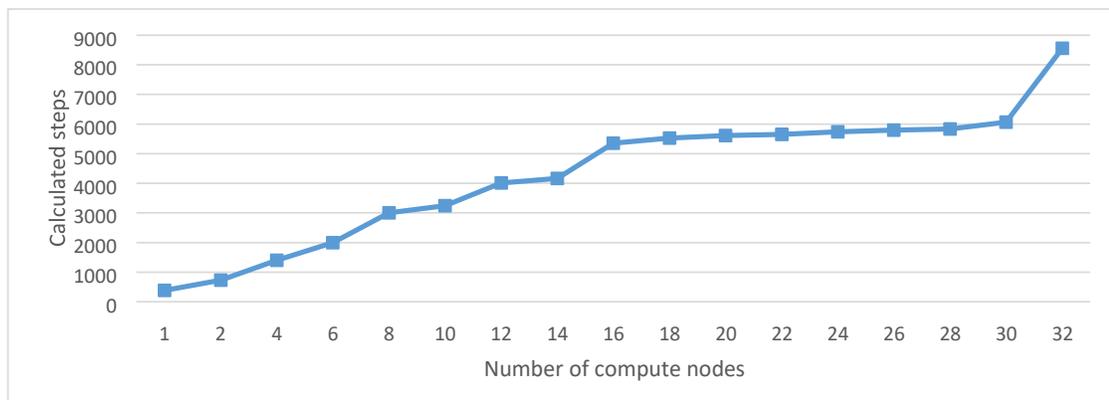

Figure 5.61: XFLAT heterogeneous multi-node performance on Stampede (one CPU and one MIC per node) by employing up to 32 nodes. The calculated radial steps were measured for about 100 second run.

Consequently, in order to validate the hypothesis the net amount of MPI messages' overhead should be extracted from the processors' idle time to figure out whether or not the overhead can affect the entire XFLAT performance. Thus, detailed studies on MPI tasks were required to understand the role of the MPI overhead in the XFLAT performance.

## 5.3.4    MPI communications' overhead

In order to extract and measure the overhead of MPI communications from the idle (waiting) time at synchronization points, two sets of benchmarks were designed. In the first set, barriers were placed prior to each MPI synchronization point, thus the timer started only after the barrier line. Hence, only the time of the MPI communication section (the MPI overhead) was measured (see Fig. 5.62). As a result, all of the processes reached the synchronization point virtually at the same time.

```
while (conditions)
{
     … // Computations!

     // All of the process are synchronized at this point MPI_barrier(…);

     timer.Start();
     MPI_Send(…);
     timer.Stop(); …
}
```

Figure 5.62: Pseudocode illustrating the measurement of MPI communication time excluding the idle time by placing barrier prior to the communication line.

In the second set of benchmarks the barrier line was removed, thus the idle time of each process was included in the overall time as well (see Fig. 5.63).

Consequently, following the removal of the barrier, due to the load imbalance between the MIC and CPU, one of them reached the synchronization point earlier. Therefore, it resulted in increasing the MPI time for that process. However, when

```
while (conditions)
{
    ... // Computations!

    timer.Start();
    MPI_Send(...);
    timer.Stop(); ...
}
```

Figure 5.63: Pseudocode illustrating the measurement of total MPI communication time including the idle time of processes.

there was a barrier exactly prior to the synchronization point, since both processes on the MIC and CPU reached the synchronization point virtually at the same time, the overhead of the waiting time was eliminated from the MPI functions timing. Therefore, during applications run time when processes on different nodes reached the same synchronization point simultaneously, the processes' idle time remained minimized. As a result, by minimizing the idle time more time could be dedicated for computations.

The designed benchmarks were performed for both blocking and non-blocking MPI calls. Blocking function are those invokes in which all of the processes must reach the same point before leaving the function, however, in the non-blocking versions, MPI functions can return immediately. In all of the following benchmarks the message size was equal to 8 double precision numbers, and the loop iteration count was 10k. In order to break the symmetry and to make sure that one process always arrived later, the computational section prior to the synchronization point was about three times higher on the Xeon Phi, hence the process on the CPU always reached the communication points earlier during the benchmarks without barriers (by at least tens of seconds). The result of the MPI send/receive are shown in Tab. 5.4 in which a barrier was presented before MPI invokes, and in Tab. 5.5 in which there was no barrier prior to MPI invokes. Furthermore, a ping-pong benchmark was performed in which the first process sent data to the second process, afterwards the second process sent data to the first one. Tab. 5.6 and Tab. 5.7 show the result for the ping-pong benchmarks when barriers were presented and when barriers were removed before MPI invokes, respectively.

There are several interesting observations in the results. First of all, it can be understood that the overhead of the MPI methods when initiated on the MIC side was higher (*i.e.*, when CPU attempted to send data to MIC, the overhead of the MPI method was lower than the case when the MIC attempted to send data to CPU).

Table 5.4: Time measurement for MPI Send/Receive between two nodes when a barrier was presented prior to MPI invokes. The first column shows the direction of the message flow. The second and third columns show the communication time for the MIC and CPU, respectively.

| Send/Recv | MIC comm. time | CPU comm. time |
|---|---|---|
| Blocking Sends | | |
| CPU->MIC | 2.2 | 0.05 |
| MIC->CPU | 0.7 | 0.8 |
| Non-blocking Sends | | |
| CPU->MIC | 2.0 | 0.06 |
| MIC->CPU | 0.8 | 1.1 |

Table 5.5: Time measurement for MPI Send/Receive between two nodes when there was no barrier prior to MPI invokes. The first column shows the direction of the message flow. The second and third columns show the communication time for the MIC and CPU, respectively.

| Send/Recv | MIC comm. time | CPU comm. time |
|-----------|----------------|----------------|
| Blocking Sends | | |
| CPU->MIC | 0.9 | 37 |
| MIC->CPU | 2.1 | 39 |
| Non-blocking Sends | | |
| CPU->MIC | 0.4 | 0.03 |
| MIC->CPU | 2.2 | 39 |

Furthermore, the task to which the data was sent (the receiver task) via the PCI Express bus had more wasted (idle) time (see Tab. 5.4 and Tab. 5.6).

Similar to the send and receive benchmarks, the next benchmarks measured MPI communication time for the MPI broadcast messages. The broadcast benchmarks showed that for the blocking calls the node on which the message was initialized, wasted slightly less time for sending messages (Tab. 5.8). However, for the benchmark without barrier, when MIC was the receiver its time at the synchronization point was less than the time when the MIC was a sender (Tab. 5.9). Note that the non-blocking broadcast data are not usable on the receiver side without making sure that Table 5.6: Time measurement for double MPI Send/Receive (ping-pong) between two nodes when a barrier was presented prior to MPI invokes. The first column shows the direction of the message flow. The second and third columns show the communication time for the MIC and CPU, respectively.

| Send/Recv | MIC comm. time | CPU comm. time |
|-----------|----------------|----------------|
| Blocking Sends | | |
| CPU->MIC+MIC->CPU | 2.8 | 2.7 |
| MIC->CPU+CPU->MIC | 3.6 | 0.8 |
| Non-blocking Sends | | |
| CPU->MIC+MIC->CPU | 2.6 | 2.9 |
| MIC->CPU+CPU->MIC | 3.6 | 1.1 |

Table 5.7: Time measurement for double MPI Send/Receive (ping-pong) between two nodes when there was no barrier prior to MPI invokes. The first column shows

the direction of the message flow. The second and third columns show the communication time for the MIC and CPU, respectively.

| Send/Recv | MIC comm. time | CPU comm. time |
|---|---|---|
| Blocking Sends | | |
| CPU->MIC+MIC->CPU | 3 | 40.7 |
| MIC->CPU+CPU->MIC | 5.5 | 41 |
| Non-blocking Sends | | |
| CPU->MIC+MIC->CPU | 3.1 | 40.2 |
| MIC->CPU+CPU->MIC | 5.5 | 41.1 |

the message has been arrived completely before trying to touch the buffer.

For the MPI reduction functions there was no non-blocking benchmark, since the nature of MPI reduction function is blocking. As shown in Tab. 5.10 and Tab. 5.11, for the MPI all reduce functions, there was no significant difference between the reduction with summation and reduction with maximum. Similar to the previous results, the removing of the barrier caused slightly increase in the timing on the MIC side.

In all of the kernel benchmarks, the MPI invokes overhead was subtle. ConseTable 5.8: Time measurement for MPI broadcast between two nodes when a barrier was presented prior to MPI invokes. The first column shows the direction of the message flow. The second and third columns show the communication time for the MIC and CPU, respectively.

| Broadcast | MIC comm. time | CPU comm. time |
|---|---|---|
| Blocking Broadcasts | | |
| CPU->MIC | 3.2 | 0.05 |
| MIC->CPU | 0.8 | 0.8 |
| Non-blocking Broadcasts | | |
| CPU->MIC | 0.2 | 0.00 |
| MIC->CPU | 0.9 | 0.01 |

Table 5.9: Time measurement for MPI broadcast between two nodes when there was no barrier prior to MPI invokes. The first column shows the direction of the message

flow. The second and third columns show the communication time for the MIC and CPU, respectively.

| Broadcast | MIC comm. time | CPU comm. time |
|-----------|----------------|----------------|
| Blocking Broadcasts | | |
| CPU->MIC | 1.0 | 37.2 |
| MIC->CPU | 2.1 | 38.1 |
| Non-blocking Broadcasts | | |
| CPU->MIC | 0.4 | 0.04 |
| MIC->CPU | 1.6 | 0.02 |

quently, in addition to those benchmarks the actual XFLAT code, with the actual instructions and messages, was benchmarked on Bahcall. The problem size was set to $555 \times 10 \times 100$ ($\vartheta \times \phi \times E$) neutrino beams. The time for completing 10k radial steps was measured and the benchmark was repeated for various MIC:CPU load ratios. The results of the benchmarks in which a barriers were presented before each MPI synchronization and when the barriers were removed, are shown in Tab. 5.12 and Tab. 5.13, respectively.

As shown in Tab. 5.12, the time of the MPI communications in XFLAT was not minimized at any point and fluctuated by varying the load ratio. That result was not Table 5.10: Time measurement for MPI reduction between two nodes when a barrier was presented prior to MPI invokes. The first column shows the direction of the message flow. The second and third columns show the communication time for the MIC and CPU, respectively.

| All reduction | MIC comm. time | CPU comm. time |
|---------------|----------------|----------------|
| All reduce(SUM) | 2.5 | 2.0 |
| All reduce(MAX) | 2.3 | 2.2 |

Table 5.11: Time measurement for MPI reduction between two nodes when there was no barrier prior to MPI invokes. The first column shows the direction of the message flow. The second and third columns show the communication time for the MIC and CPU, respectively.

| All reduction | MIC comm. time | CPU comm. time |
|---------------|----------------|----------------|
| All reduce(SUM) | 3.2 | 39 |

| All reduce(MAX) | 3.4 | 40.1 |
|---|---|---|

expected since a barrier was placed before each MPI communication points. Thus, the MPI communication time should not be the function of load ratio. According to the MPI standard: the MPI Barrier() function only guarantees that at some point all the processes are within the barrier function, however, it does not guarantee that all of them exit the barrier function at the same time [LLNL, 2015]. Therefore, the way that the barrier function was implemented may affect the result. Hence, the fluctuations may be due to the utilized MPI library on the Bahcall machine, which was MPICH library. Although, the MPICH library is one of the most popular and widely used implementation of the MPI standards, the HPC design was not the highest priority for its implementation, unlike the other MPI implementation such as MVAPICH [MVAPICH, 2015] or Intel MPI [Intel, 2015a] libraries (another widely used MPI library was Open MPI [MPI, 2015] which was not ready at the time of analysis for the Intel MIC architecture).

Since the Intel MPI library is not available as a free software and the MVAPICH is the MPI library over InfiniBand and 10GigE links, it was not possible to exploit them Table 5.12: XFLAT performance benchmark on Bahcall (using MPICH library) when barriers were placed prior to each MPI synchronization. The first column shows the load ratios of MIC to CPU (number of beams on the MIC and CPU). The second and third columns show the communication time on MIC and CPU, respectively. The fourth column illustrates the overall run time.

| MIC:CPU (beams#) | MIC comm. time | CPU comm. time | Overall time |
|---|---|---|---|
| 5.0:1 (463:092) | 17.7 | 13.0 | 254.3 |
| 4.5:1 (454:101) | 16.8 | 12.5 | 217.6 |
| 4.0:1 (444:111) | 16.5 | 13.3 | 215.3 |
| 3.5:1 (432:123) | 16.7 | 15.5 | 214.7 |
| 3.0:1 (416:139) | 14.1 | 14.1 | 209.0 |
| 2.5:1 (397:158) | 12.0 | 18.3 | 205.8 |
| 2.0:1 (370:185) | 11.3 | 20.9 | 245.2 |

| 1.5:1 (333:222) | 11.3 | 23.8 | 262.6 |
| 1.0:1 (278:277) | 11.4 | 21.3 | 332.1 |

on Bahcall machine (Bahcall was equipped with neither the InfiniBand cards nor the 10GigE cards). Fortunately, the Stampede supercomputer utilizes the Intel MPI library. Consequently, the XFLAT code and the Intel MPI library were benchmarked on one of the Stampede compute nodes. The benchmarks were repeated with the same neutrino beams configuration and the same iteration count. As observable in Tab. 5.14 and Tab. 5.15, unlike the previous results, the MPI communications time for benchmarks with barriers, remained constant. Therefore, the previously observed fluctuations in the benchmark results (with barriers) were only due to the MPICH library design and implementation. By switching over the Intel MPI library the expected results were achieved. In addition, the optimum MIC:CPU load ratio appeared to be around 3:1, which was in agreement with the results of the previous benchmarks.

Nevertheless, if the number of beams on the MIC is not enough, the distribution way of beams over threads may affect the result. Thus, it may affect the previous results. As a result, the previous benchmark was repeated by using ten times more Table 5.13: XFLAT performance benchmark on Bahcall (using MPICH library) when barriers were remove prior to each MPI synchronization. The first column shows the load ratios of MIC to CPU (number of beams on the MIC and CPU). The second and third columns show the communication time on MIC and CPU, respectively. The fourth column illustrates the overall run time.

| MIC:CPU (beams#) | MIC comm. time | CPU comm. time | Overall time |
| --- | --- | --- | --- |
| 5.0:1 (463:092) | 19.8 | 134.1 | 232.8 |
| 4.5:1 (454:101) | 19.4 | 90.1 | 199 |
| 4.0:1 (444:111) | 19.1 | 77.1 | 197.3 |
| 3.5:1 (432:123) | 19.3 | 65.5 | 195.7 |
| 3.0:1 (416:139) | 18.0 | 48.5 | 191.4 |
| 2.5:1 (397:158) | 26.3 | 24 | 187.1 |
| 2.0:1 (370:185) | 63.9 | 14.8 | 219.8 |
| 1.5:1 (333:222) | 89.3 | 12.7 | 236.7 |
| 1.0:1 (278:277) | 171.8 | 13 | 310 |

load on each processors (see Tab. 5.16 and Tab. 5.17).

In all of the previous benchmarks, the optimum MIC:CPU load ratio was around 3:1. This fact indicates that on Stampede, the optimum load on the MIC should be about three times higher than the CPU load in order to achieve the optimum performance. The results appeared to be in agreement with the previous finding of the sweet spot on Stampede.

Nevertheless, in order to find a more precise number for the sweet spot, a more refined range for the load ratios should be searched. From the previous results, it could be concluded that the refined range of 2.5–3.5 load ratios can be searched so as to find the sweet spot. Therefore, two benchmarks were performed on the refined range to find the precise sweet spot. The first benchmark utilized $555 \times 10 \times 100$ ($\vartheta \times \phi \times E$) number of neutrino beams similar to the previous benchmark (see Tab. 5.18), and the second benchmark performed by employing different neutrino beam number ($12000 \times 10 \times 100$). Consequently, the number of neutrino beams on each processor differed from the first benchmark (see Tab. 5.19). The timing were measured for Table 5.14: XFLAT performance benchmark on Stampede (using Intel MPI library) when barriers were placed prior to each MPI synchronization. The first column shows the load ratios of MIC to CPU (number of beams on each side). The second and third columns show the communication time on MIC, CPU, respectively. The fourth column illustrates the overall run time.

| MIC:CPU (beams#) | MIC comm. time | CPU comm. time | Overall time |
|---|---|---|---|
| 5.0:1 (463:092) | 3 | 3.5 | 168.9 |
| 4.5:1 (454:101) | 2.9 | 3.3 | 166.2 |
| 4.0:1 (444:111) | 2.9 | 3.3 | 164.1 |

| 3.5:1 (432:123) | 2.9 | 3.3 | 163.7 |
| 3.0:1 (416:139) | 3 | 3.3 | 163 |
| 2.5:1 (397:158) | 3 | 3.2 | 165.1 |
| 2.0:1 (370:185) | 3 | 3 | 186 |
| 1.5:1 (333:222) | 2.9 | 2.8 | 215.9 |
| 1.0:1 (278:277) | 2.9 | 2.9 | 265.2 |

calculating 1000 radial steps. The result showed that the more accurate optimum MIC:CPU load ratio was around 2.9:1.0. Fig. 5.64 depicted the normalized XFLAT performance for the two benchmarks. Similar to the previous findings, by placing enough amount of load on the MIC and CPU, the single-node MIC:CPU sweet load ratio is about 2.9:1, on Stampede.

In all of the benchmarks with a barrier, always about 3–4 seconds of overall MPI overhead was observable. That overhead was due to invoking MPI functions alone, and it was not possible to exclude it from XFLAT runs on multi-node environments. The overhead was still small enough that could not cause any major performance hit. Therefore, none of the benchmarks could explain the reason for previously observed flat region on heterogeneous multi-node benchmarks.

On a single node the MPI overhead was not problematic, however, for the multinode benchmarks the MPI communications were involved heavily. As a result, XFLAT multi-node performance benchmarks, without MPI overhead, should be performed.

Table 5.15: XFLAT performance benchmark on Stampede (using Intel MPI library library) when barriers were removed prior to each MPI synchronization. The first column shows the load ratios of MIC to CPU (number of beams on each side). The second and third columns show the communication time on MIC, CPU, respectively. The fourth column illustrates the overall run time.

| MIC:CPU (beams#) | MIC comm. time | CPU comm. time | Overall time |
| --- | --- | --- | --- |
| 5.0:1 (463:092) | 6 | 80.8 | 166.9 |
| 4.5:1 (454:101) | 5.8 | 71 | 165 |
| 4.0:1 (444:111) | 5.7 | 60.9 | 163.6 |
| 3.5:1 (432:123) | 5.7 | 39.7 | 162.6 |
| 3.0:1 (416:139) | 6.1 | 26.7 | 160.8 |
| 2.5:1 (397:158) | 10.1 | 15.3 | 162.4 |
| 2.0:1 (370:185) | 30 | 6.8 | 181 |
| 1.5:1 (333:222) | 64.1 | 6.6 | 211.4 |
| 1.0:1 (278:277) | 118.2 | 6.1 | 260.4 |

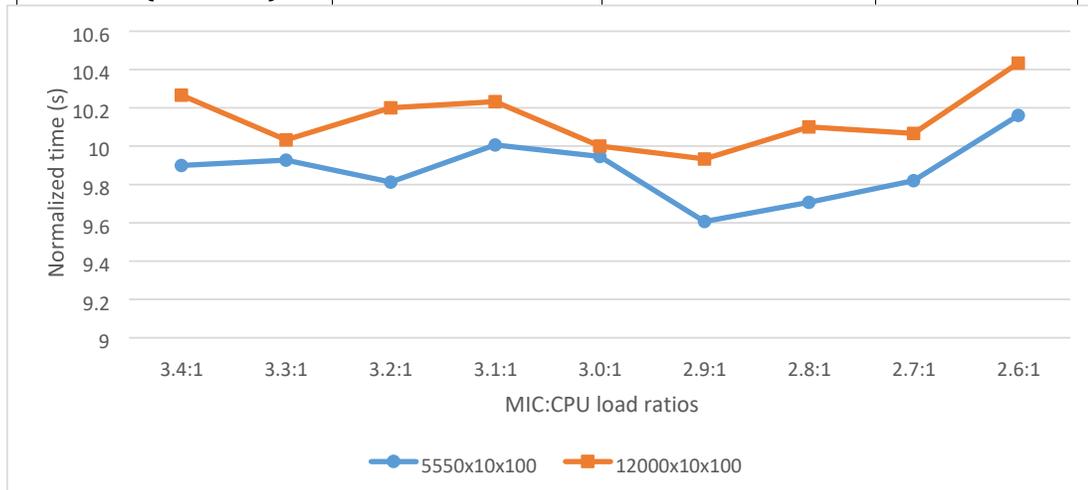

Figure 5.64: Stampede single-node benchmark for two different neutrino beam configurations, 5550×10×100 and 12000×10×100, for various MIC:CPU load ratios. Table 5.16: XFLAT performance benchmark, by employing ten times more load on each processor, on Stampede (using Intel MPI library) when barriers were placed prior to each MPI synchronization. The first column shows the load ratios of MIC to CPU (number of beams on each side). The second and third columns show the communication time on MIC, CPU, respectively. The fourth column illustrates the overall run time.

| MIC:CPU (beams#) | MIC comm. time | CPU comm. time | Overall time |
|---|---|---|---|
| 5.0:1 (4630:0920) | 3.5 | 3.9 | 1576 |
| 4.5:1 (4540:1010) | 3.5 | 3.9 | 1584 |
| 4.0:1 (4440:1110) | 3.4 | 3.9 | 1544 |
| 3.5:1 (4320:1230) | 3.6 | 4.1 | 1488 |
| 3.0:1 (4160:1390) | 3.6 | 3.8 | 1435 |
| 2.5:1 (3970:1580) | 3.6 | 3.7 | 1514 |
| 2.0:1 (3700:1850) | 3.4 | 3.5 | 1742 |
| 1.5:1 (3330:2220) | 3.5 | 3.5 | 2103 |
| 1.0:1 (2780:2770) | 3.4 | 3.5 | 2562 |

## 5.3.5 Eliminating inter-node message passing

It is possible to design a benchmark to simulate a multi-node run without any internode communication on a single node. This can be achieved by placing the amount of load on a single machine as equal as the load on a node on multi-node environment. For a fixed problem size, in a multi-node environment, by increasing the number of compute nodes, the amount of load on a single node decreases. As a result, the multi-node simulation on a single node becomes possible by adjusting the amount of load on a single-CPU and single-MIC configuration, accordingly.

Therefore, for benchmarking XFLAT, for each step the placed load was equal as the load on a machine in multi-node environment. Consequently, the inter-node communication was eliminated entirely and only the performance of processors were benchmarked. Therefore, a simulated benchmark utilizing a single node became possible. The initial problem size was set to 10000 × 10 × 100 neutrino beams for a single node. At each point, the load on the single compute node was adjusted Table 5.17: XFLAT performance benchmark, by employing ten times more load on each processor, on Stampede (using Intel MPI library) (using ten times more load on each processor) when barriers were removed prior to each MPI synchronization. The first column shows the load ratios of MIC to CPU (number of beams on each side). The second and third columns show the communication time on MIC, CPU, respectively. The fourth column illustrates the overall run time.

| MIC:CPU (beams#) | MIC comm. time | CPU comm. time | Overall time |
|---|---|---|---|
| 5.0:1 (4630:0920) | 6 | 722 | 1587 |
| 4.5:1 (4540:1010) | 6 | 671 | 1584 |
| 4.0:1 (4440:1110) | 6 | 542 | 1543 |

| 3.5:1 (4320:1230) | 7 | 313 | 1480 |
|---|---|---|---|
| 3.0:1 (4160:1390) | 10 | 222 | 1471 |
| 2.5:1 (3970:1580) | 103 | 77 | 1502 |
| 2.0:1 (3700:1850) | 394 | 58 | 1724 |
| 1.5:1 (3330:2220) | 878 | 54 | 2055 |
| 1.0:1 (2780:2770) | 1500 | 46 | 2531 |

to be equal as the load of a node in the multi-node configuration. Thus, when the number of nodes increases in the multi-node environment, the corresponding load on the single node decreases. The benchmark continued for 100 seconds and the calculated radial steps were measured. As illustrated by Fig. 5.65, the flat region in the XFLAT performance was still observable.

Since, the flat region appeared only when the number of nodes increased beyond 16 nodes (*i.e.* the performance did not improve by increasing the number of nodes to 18, 20, 22, *etc.*). As a result, the performance benchmarks should be continued for the single node load beyond the 16-node benchmark. Thus, on a single compute node (single CPU and single MIC) a load equal to the load of a single node in the 18node, 20-node, 22-node configurations was placed in order to study the performance. In addition, unlike the previous benchmarks the MIC:CPU load ratio also varied to study the effect of the processors idle time on the overall performance. Each benchmark continued for 100 seconds and the spent time at synchronization points were measured (Although, there was no inter-node MPI messages involved, either Table 5.18: XFLAT performance benchmark on Stampede (using 5550 × 10 × 100 neutrino beams) to find more refined MIC:CPU load ratio range. The first column shows the load ratios of MIC to CPU (number of beams on each side). The second and third columns show the communication time on MIC, CPU, respectively. The fourth column illustrates the overall run time.

| MIC:CPU (beams#) | MIC comm. time | CPU comm. time | Overall time |
|---|---|---|---|
| 3.4:1 (4289:1261) | 6 | 347 | 1485 |

| 3.3:1 (4260:1290) | 7 | 322 | 1489 |
| 3.2:1 (4229:1321) | 6 | 281 | 1472 |
| 3.1:1 (4196:1354) | 7 | 280 | 1499 |
| 3.0:1 (4163:1387) | 8 | 242 | 1492 |
| 2.9:1 (4127:1423) | 23 | 163 | 1441 |
| 2.8:1 (4090:1460) | 34 | 143 | 1456 |
| 2.7:1 (4050:1500) | 53 | 121 | 1473 |
| 2.6:1 (4008:1542) | 108 | 67 | 1524 |

the CPU or the MIC could reach the synchronization points sooner, thus there may be idle time).

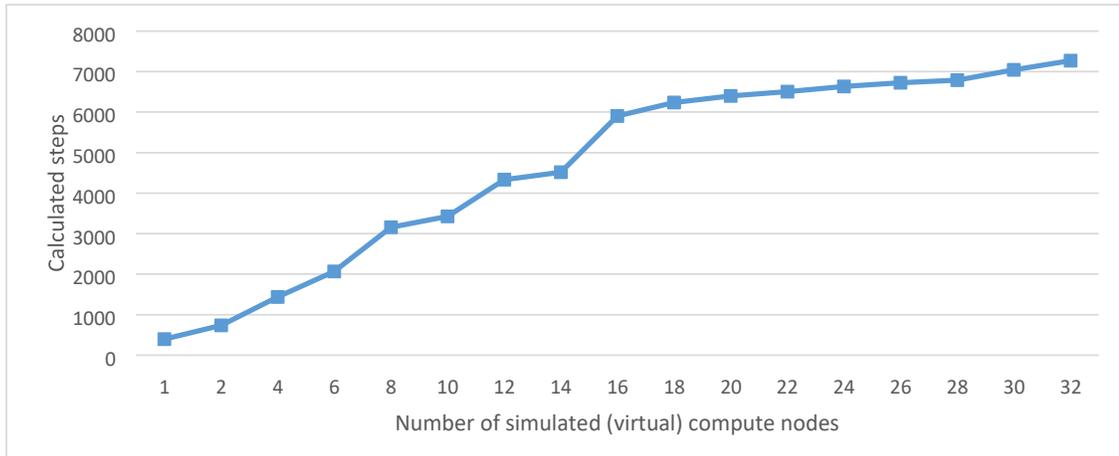

Figure 5.65: Stampede single-node (one CPU and one MIC) simulation of multinode benchmark. The horizontal line indicates that the load on the single node was virtually equal to the load of a node in the multi-node configuration. The number of calculated radial steps was measure for about 100 seconds.

Table 5.19: XFLAT performance benchmark on Stampede (using 12000 × 10 × 100 neutrino beams) to find more refined MIC:CPU load ratio range. The first column shows the load ratios of MIC to CPU (number of beams on each side). The second and third columns show the communication time on MIC, CPU, respectively. The fourth column illustrates the overall run time.

| MIC:CPU (beams#) | MIC comm. time | CPU comm. time | Overall time |
| --- | --- | --- | --- |
| 3.4:1 (9273:2727) | 0.08 | 6.28 | 30.8 |
| 3.3:1 (9209:2791) | 0.08 | 5 | 30.1 |
| 3.2:1 (9143:2857) | 0.08 | 4.9 | 30.6 |
| 3.1:1 (9073:2927) | 0.09 | 4.5 | 30.7 |
| 3.0:1 (9000:3000) | .3 | 3.2 | 30 |
| 2.9:1 (8923:3077) | .7 | 2.2 | 29.8 |
| 2.8:1 (8842:3158) | 1 | 2 | 30.3 |
| 2.7:1 (8757:3243) | 2 | 1.1 | 30.2 |
| 2.6:1 (8667:3333) | 3 | 1.1 | 31.3 |

Fig. 5.66 illustrates the idle time for the MPI functions. The time of the MPI synchronization points were recorded for the MIC and the CPU separately. The total load was set to 555 × 10 × 100 neutrino beams, which was the load similar to the load on a node in the 18-node configuration. Likewise, Fig. 5.67 and Fig. 5.68 illustrate the benchmark results of the single node by using 500 × 10 × 100 and 455 × 10 × 100 neutrino beams (which are equal to the load of a single node in the 20-node and 22-node environments), respectively. As observable, all the results show normal behavior, *i.e.*, by changing the load ratio, the timing on one side increased and on the other side decreased. Therefore, the MPI overhead cannot explain the previously observer behaviors in the performance trend.

The other factor that can potentially affect the XFLAT performance was the processors' internal load imbalance. Modern processors, especially Xeon Phi, are equipped with many cores and threads. Thus, the way that neutrino beams are distributed on threads may affect the overall performance of a processor. Hence, further benchmarks on multi-node environment were required. Furthermore, the multi-node system should be equipped with one type of processors (homogeneous).

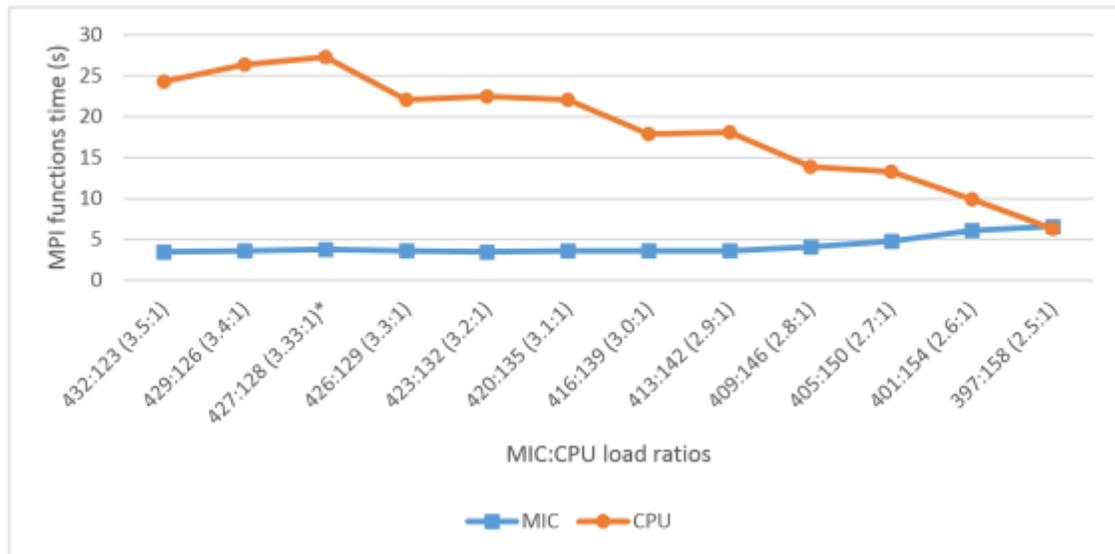

Figure 5.66: Illustration of the performance of a single-node in the simulated 18-node run (single CPU and single MIC) for various MIC:CPU load ratios. The wasted time at MPI synchronization functions were measured separately for the CPU and MIC. The calculated radial steps were measure for a 100 seconds run. The point market with ∗ indicates the number of beams divided by number of threads on the MIC is integer.

## 5.3.6 The distribution of neutrino beams on threads

Since the number of available hardware threads on the Stampede's Xeon Phis is 244 and the number of threads on its CPU is 8, an issue related to load distribution may arise when the load is imbalanced on the Xeon Phi's threads. When the number of neutrino beams is not divisible by the number of available threads, a few threads may end up receiving more tasks for processing, thus it causes the load imbalance issue. Furthermore, the load imbalance on threads may be related to the observed flat region in the XFLAT heterogeneous benchmarks. In order to further investigate the issue, a CPU-only and a MIC-only benchmarks were required. XFLAT should be able to scale well on both homogeneous and heterogeneous environments, thus it should be able to scale satisfactory in environments equipped with only one type of processor.

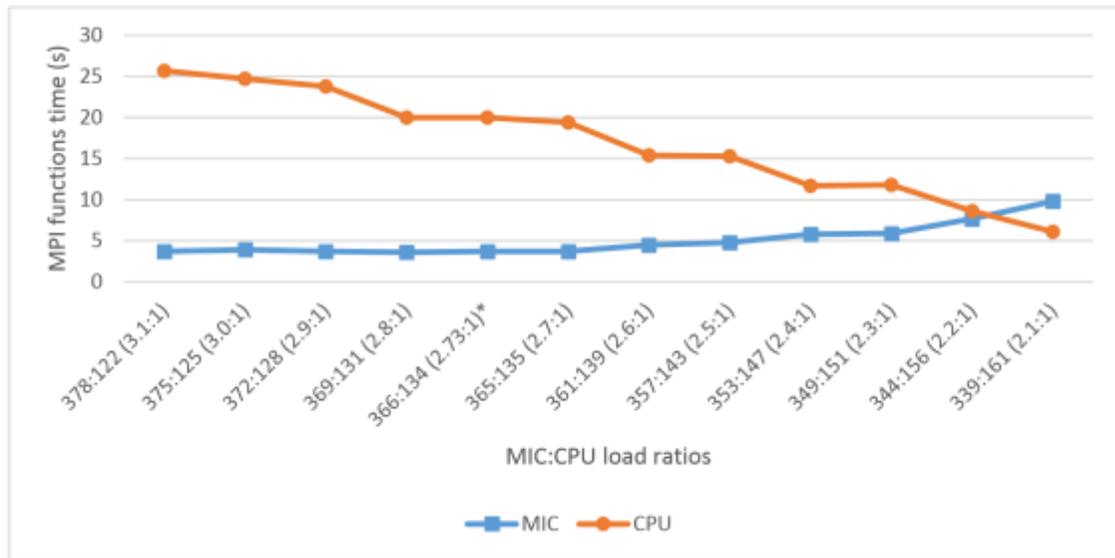

Figure 5.67: Illustration of the performance of a single-node in the simulated 20-node run (single CPU and single MIC) for various MIC:CPU load ratios. The wasted time at MPI synchronization functions were measured separately for the CPU and MIC. The calculated radial steps were measure for a 100 seconds run. The point market with ∗ indicates the number of beams divided by number of threads on the MIC is integer.

Other important benchmarks were the multi-node CPU-only and MIC-only benchmarks. The benchmarks utilized two CPUs or two MICs on Stampede's node. The problem size was set to $1000 \times 10 \times 100$ and the calculated radial steps for about 100 seconds period were measured. The results of the CPU and MIC benchmarks are depicted by Fig. 5.69. For the CPU-only benchmark, the code scaled linearly, however, for the MIC-only benchmark the flat region was still noticeable.

The flat region only appeared when the number of compute nodes increased beyond a particular number. Since, for a particular problem size by increasing the number of nodes, the number of neutrino beams per processor decreases, the issue may be related to the number of available neutrino beams per hardware thread. In fact, with a simple calculation it can be shown that the reason behind the flat region in the Xeon Phi's performance result was due to the load imbalance. For example, as depicted in Fig. 5.69, when the number of compute nodes was 20, the load on a single MIC was equal to 250 zenith angle beams. Since this number of beams were going to be distributed over 244 hardware threads, all of the threads performed calculations on 1 beam except only 6 threads to which 2 beams were assigned (250 mode 244 = 6). Hence, at the end of every iteration, all of the threads except 6, had to remain idle waiting for the last 6 threads. Obviously, it resulted in a performance hit, thus the performance results remained steady. On the other hand, when the number of compute nodes was 22, the load on each MIC was equal to 227 beams, thus the computational load on all of the threads were nearly the same. There was no need for threads to wait for only a few threads to complete their tasks.

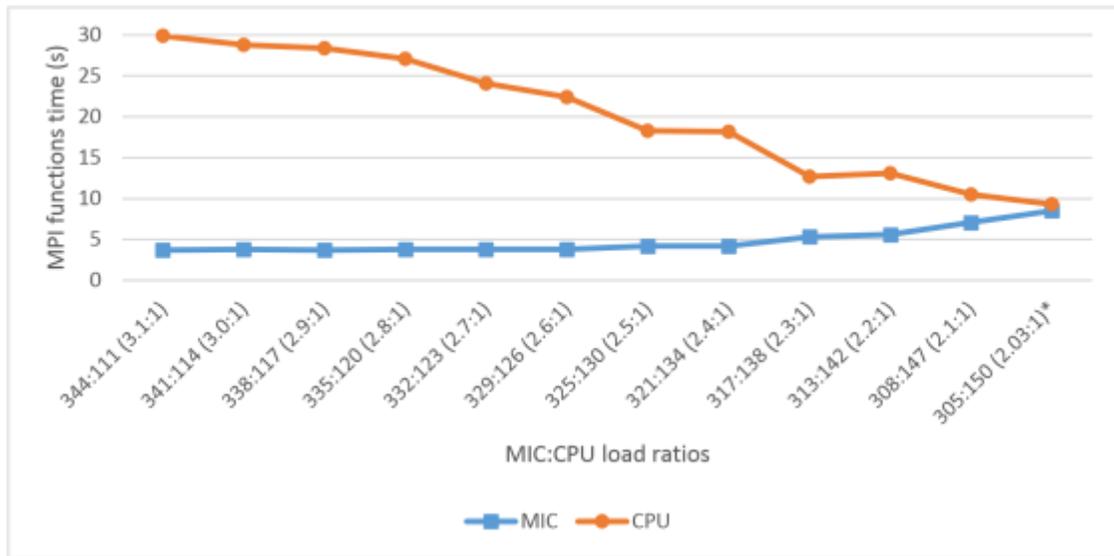

Figure 5.68: Illustration of the performance of a single-node in the simulated 22-node run (single CPU and single MIC) for various MIC:CPU load ratios. The wasted time at MPI synchronization functions were measured separately for the CPU and MIC. The calculated radial steps were measure for a 100 seconds run. The point market with ∗ indicates the number of beams divided by number of threads on the MIC is integer.

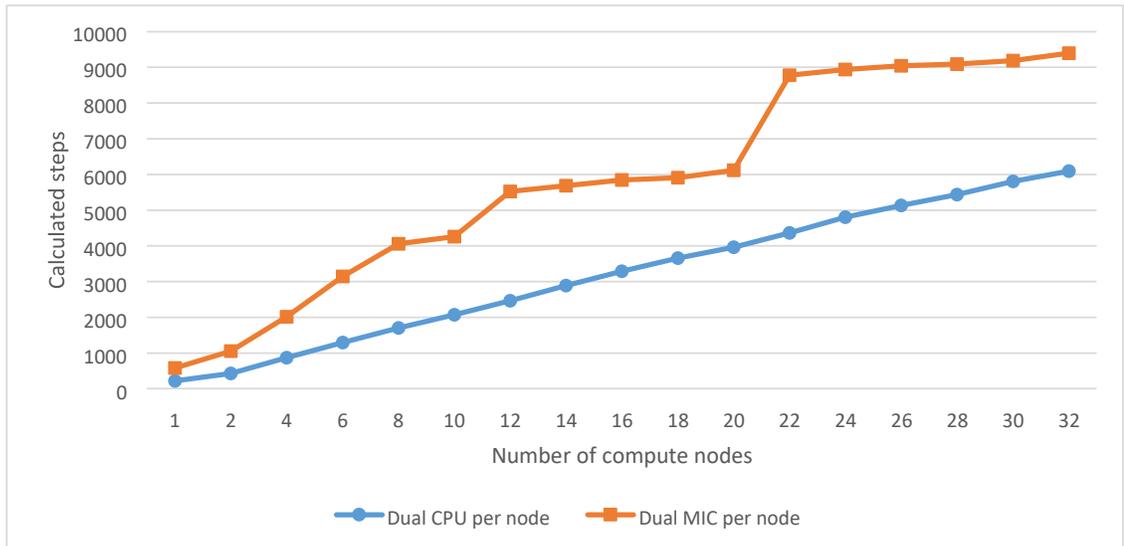

Figure 5.69: Multi-node homogeneous (CPU-only and MIC-only) benchmarks on Stampede. The blue line shows that the performance of the CPU scaled virtually linear by increasing the number of nodes. However, the orange line, illustrates that the performance of the MIC scaled non-linearly. The problem size was set to 1000× 10×100 and the calculated radial steps for about 100 seconds period were measured. Each node equipped with two CPUs and two MICs.

Nevertheless, if the reason behind the flat region was truly due to the load distribution, it must be repeatable for other benchmarks with different configurations and various loads as well. Therefore, another benchmark on single-CPU and single-MIC configuration was performed in which it simulated the multi-node benchmarks by adjusting the load ratio on processors based on the load of a node on multi-node environment. As a result, there was no inter-node communications. In addition, the new benchmark employed different problem size (3000 × 30 × 100 beams) with the

2.9:1 MIC to CPU load ratio. The number of calculated radial steps were measured for about 100 seconds.

As shown in Fig. 5.70, the flat behavior is observable between benchmarks with 6 to 10 nodes. For instance, for the 8-node run, there were 375 zenith beams on one node for which 279 were allocated on the Xeon Phi. Consequently, a very low number of threads (35 out of 244 threads) were allocated twice more beams, thus all other threads had to wait for them to complete their tasks for every iteration. As a result, there were no improvement in performance since the benchmark's result remained virtually identical to the 6-node run. On the other hand, for the 10-node run there were 223 zenith beam allocated on the Xeon Phi. This indicates that nearly all of the available threads had identical amount of beams for calculation, thus there were no waiting time for majority of threads per loop iteration. Consequently, all threads completed their tasks simultaneously that resulted in a significant performance gain in comparison to the 8-node run. As mentioned earlier, by going beyond 10 nodes, the number of available beams on Xeon Phi decreased and became too low that resulted in not utilizing the entire processor. Therefore, since the number of beams was less than the number of available threads on the Xeon Phi, the flat region after the 10-node run was due to insufficient load on the co-processor. The insufficient load resulted in permanent idle time for many threads of the co-processor.

The neutrino beams distribution issue can happen on both CPU and MIC. However, since there are large number of cores and threads on the MIC, the effect of load imbalance may affect a large number of threads. Therefore, even one thread can keep hundreds of threads in the waiting state, by forcing them to remain idle at every synchronization point. For instance, as depicted by Fig. 5.71, on the left side all of the hardware threads has equal amount of task, *i.e.* three tasks per thread. If each task takes $t$ seconds for completion, the Xeon Phi completes all the tasks in $3 * t$ seconds. Alternatively, if only a few threads has one more task (the right side), the Xeon Phi completes all of its tasks in $4 * t$ seconds per iteration, no matter how many threads have incomplete tasks. Thus, the rest of the threads have to remain idle and wait for the first two threads to complete their tasks.

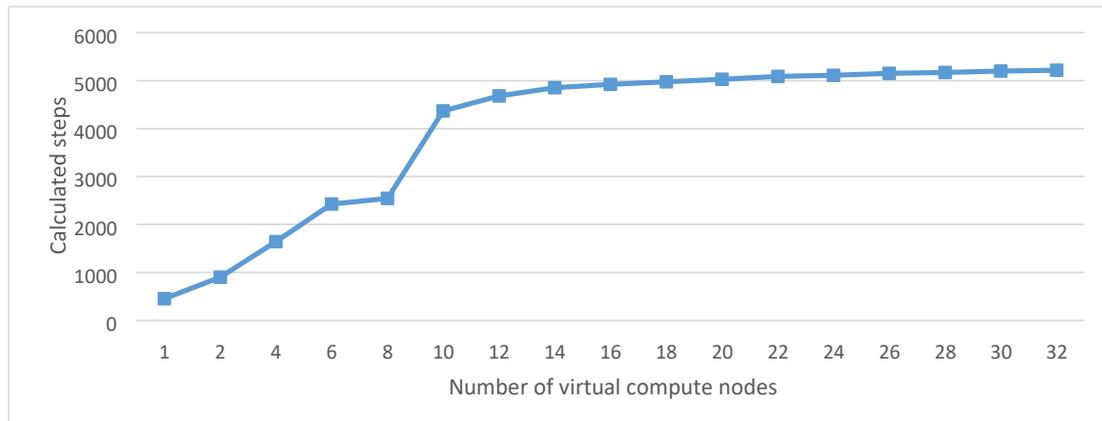

Figure 5.70: The simulation of multi-node benchmark on single node (one CPU and one MIC). At each point, the load on the node was adjusted to be equal to the load on a node in multi-node configuration. The benchmark employed different problem size (3000 × 30 × 100 beams) and the 2.9:1 MIC to CPU load ratio. The number of calculated radial steps were measured for about 100 seconds.

As a result, the reason for the observed flat regions in the result of the homogeneous runs on MIC as well as the heterogeneous runs, was due to the load imbalance issue. Moreover, it may happen when the load is not enough to be distributed equally on all threads. Hence, it is recommended that the number of nodes is chosen in the way that all of threads on every MIC, receive nearly equal amount of tasks.

### 5.3.7 Sweet spot location and MPI timing

The other performance section that was required to be analyzed was the location of the sweet spot. On heterogeneous environments, the location of the sweet spot (the optimum performance point) is directly related to the load ratios between processors.

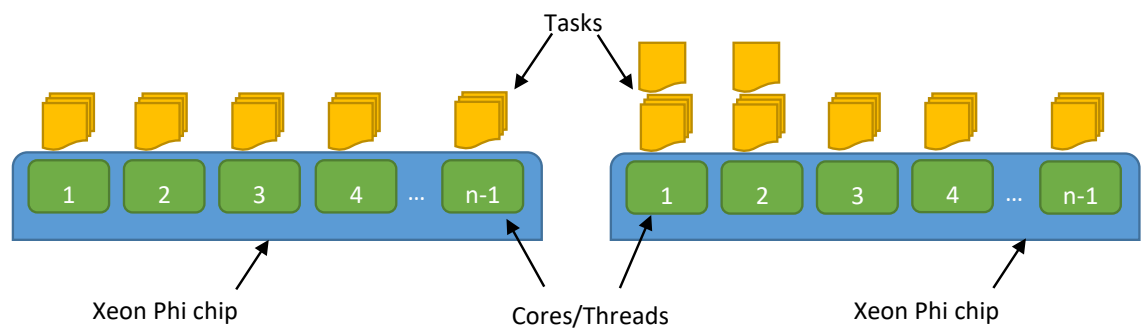

Figure 5.71: The effect of Xeon Phi threads load imbalance on the overall performance. On the left side, the load on the Xeon Phi (shown in blue) is balanced, thus each core/thread (shown in green) received equal amount of tasks (shown in yellow). They complete their tasks virtually at the same time. On the right side, only a few threads (the first two threads) received more tasks, thus for every iteration the rest of the threads must remain idle and wait for the first two threads to complete their tasks.

On both sides of the sweet spot, the timing increases for one processor and decreases for the other processor. The sweet spot, is the location when the two sides change their trends, *i.e.*, the crossing point in the performance plots. Thus, the crossing point is the point when the MPI timings on the CPU and MIC sides change their trends. For example, on the MIC side after decreasing, the timing starts to increase and on the CPU side the timing starts to decrease after increasing). As illustrated by Fig. 5.66, Fig. 5.67, and Fig. 5.68 the location of the crossing points for the MIC and CPU was not fixed and always changed by varying the load ratios. Therefore, the location of the crossing point was not fixed even when the inter-node communications
were eliminated.

In addition, it was expected that the location of the sweet spot, to be at the same point as the MPI timing crossing point (which is the point when the waiting time on both sides is minimized). Nevertheless, the two locations were not always located at the same point. For example, the XFLAT overall run time for the 5550 × 10 × 100 configuration and for various MIC:CPU load ratios is depicted by Fig. 5.72 and the MPI timing of each processor is depicted by Fig. 5.73. It anticipated that by changing the load ratio and placing less load on the MIC, the MPI timing on the MIC decreases until it reaches the crossing point, afterwards it anticipated that the timing remained steady at its minimum. The behavior is expected since at the MPI timing crossing point, processes on both the CPU and MIC reach the synchronization point virtually at the same time. Therefore, after passing the sweet spot, due to load imbalance one of the processes should always arrive earlier than the other process. Consequently, the late process should not remain idle, thus the MPI timing for the process is expected to remain minimized.

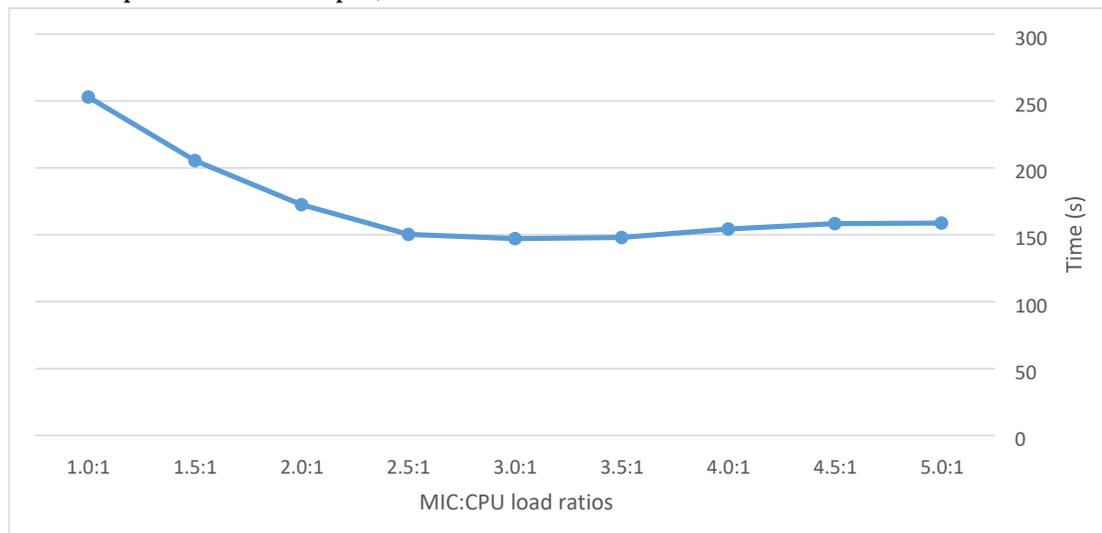

Figure 5.72: XFLAT overall run time on Stampede as the function of the MIC to CPU load ratios. The problem size was set to 5550 × 10 × 100, the code ran for 1k radial steps and the running time were measured in seconds.

However, the expected behavior was not observable all the time (see Fig. 5.73). One of the reasons can be due to the fact that in XFLAT the outer evolution loop contains several communication points per iteration and each communication point employs different MPI function. Although the message size is fixed and not the function of the problem size, invoking different MPI synchronization per iteration may affect the location of the sweet spot. Within the XFLAT evolution loop, there is one MPI broadcast, four MPI summation reductions, and one MPI maximization reduction invokes per loop iteration. In order to study the effect of each MPI invoke on the performance, the previous benchmark should have been repeated with less number of MPI calls (by eliminating some of the MPI invokes) to observe the behavior of the code and the sweet spot.

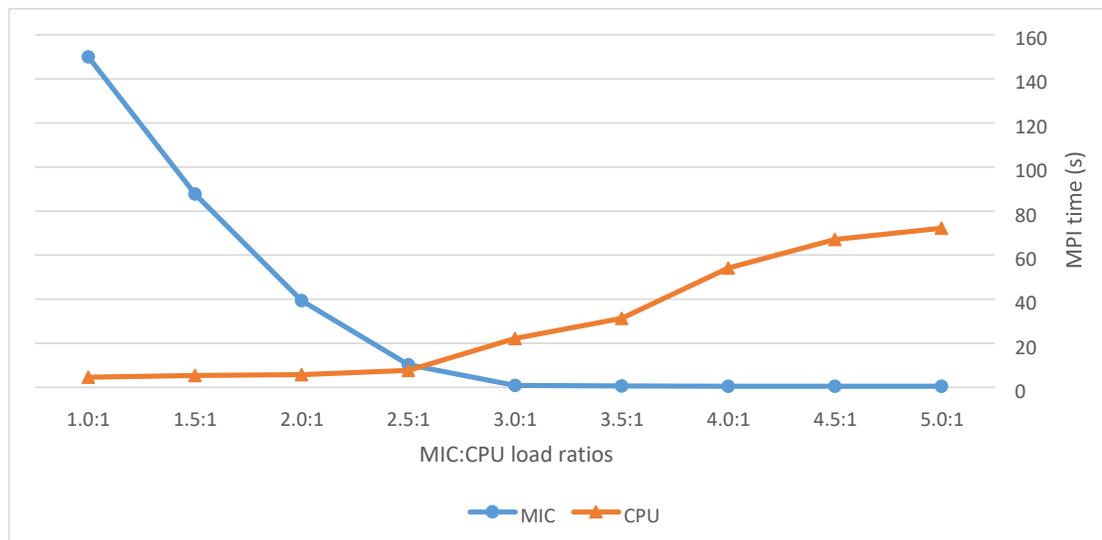

Figure 5.73: The measured MPI communications time for XFLAT on Stampede as the function of MIC to CPU load ratios. The problem size was set to 5550×10×100, the code continued to run for 1k radial steps and the running time were measured in seconds.

For the first benchmark, the rest of the MPI calls were eliminated except the first MPI broadcast call. Thus, the only remaining communication point was the single broadcast from the root process to all processes. The result of the first MPI invoke timing is shown in Tab. 5.20. The first row shows the MIC:CPU load ratios, the second, third, and fourth rows show the MIC time, the CPU time, and the overall Table 5.20: XFLAT overall time for various MIC:CPU load ratios when a single MPI broadcast presented per loop iteration. The first row shows the MIC:CPU load ratios, the second, third, and fourth rows show the MIC time, the CPU time, and the overall time, respectively. The highlighted cells show fixed time for each processor.

| MIC:CPU | 1.0:1 | 1.5:1 | 2.0:1 | 2.5:1 | 3.0:1 | 3.5:1 | 4.0:1 | 4.5:1 | 5.0:1 |
|---|---|---|---|---|---|---|---|---|---|
| MIC | 147.6 | 93.3 | 35.5 | 4.3 | 0.1 | 0.1 | 0.1 | 0.1 | 0.1 |
| CPU | 0.02 | 0.02 | 0.02 | 0.02 | 19.9 | 32.6 | 49.3 | 63.4 | 67.5 |
| Overall | 250.3 | 201.1 | 167.4 | 143.4 | 145.7 | 144.4 | 149.7 | 154.8 | 153.2 |

Table 5.21: XFLAT overall time for various MIC:CPU load ratios (refined range) when a single MPI broadcast presented per loop iteration. The first row shows the MIC:CPU load ratios, the second, third, and fourth rows show the MIC time, the CPU time, and the overall time, respectively. The highlighted cells show fixed time for each processor.

| MIC:CPU | 2.6:1 | 2.7:1 | 2.8:1 | 2.9:1 | 3.0:1 | 3.1:1 | 3.2:1 | 3.3:1 | 3.4:1 |
|---|---|---|---|---|---|---|---|---|---|
| MIC | 0.4 | 0.1 | 0.1 | 0.1 | 0.1 | 0.1 | 0.1 | 0.1 | 0.1 |
| CPU | 0.02 | 4.1 | 4.8 | 11.1 | 22 | 23.5 | 23.6 | 27.8 | 31.4 |
| Overall | 140.7 | 140 | 141.4 | 140.1 | 147.5 | 148.5 | 143 | 145.3 | 145.6 |

time, respectively. As observable the location of the MPI crossing point was in [2.5–3.0] load ratios range. Clearly, by increasing the load ratio, the MPI timing on the MIC reached its minimum and did not change afterwards, thus indicates there was no idle time on the MIC side. On the other hand, by decreasing the load ratio, the MPI timing on the CPU was minimized, thus shows the MIC had reached the communication point sooner and there was no idle time on the CPU side. In order to find the more precise sweet spot, the benchmark was repeated for [2.6–3.4] range. As perceptible in Tab. 5.21, the sweet spot was located at the load ratio crossing
point between 2.6:1 to 2.7:1.

Consequently, in the next set of benchmarks, all of the MPI invokes were excluded except the only MPI maximum reduction function. The results were shown in Tab. 5.22, and in Tab. 5.23 for a more refined load ratios span. For the single MPI Table 5.22: XFLAT overall time for various MIC:CPU load ratios when a single MPI reduction presented per loop iteration. The first row shows the MIC:CPU load ratios, the second, third, and fourth rows show the MIC time, the CPU time, and the overall time, respectively. The highlighted cells show fixed time for each processor.

| MIC:CPU | 1.0:1 | 1.5:1 | 2.0:1 | 2.5:1 | 3.0:1 | 3.5:1 | 4.0:1 | 4.5:1 | 5.0:1 |
|---|---|---|---|---|---|---|---|---|---|
| MIC | 153.3 | 88.9 | 37.9 | 5 | 0.1 | 0.1 | 0.1 | 0.1 | 0.1 |
| CPU | 0.25 | 0.32 | 0.37 | 0.4 | 19.8 | 35.7 | 51.4 | 66.5 | 68.3 |
| Overall | 252.2 | 206.2 | 168.9 | 144.1 | 146 | 147.7 | 153.1 | 158.7 | 152 |

Table 5.23: XFLAT overall time for various MIC:CPU load ratios (refined range) when a single MPI reduction presented per loop iteration. The first row shows the MIC:CPU load ratios, the second, third, and fourth rows show the MIC time, the CPU time, and the overall time, respectively. The highlighted cells show fixed time for each processor.

| MIC:CPU | 2.6:1 | 2.7:1 | 2.8:1 | 2.9:1 | 3.0:1 | 3.1:1 | 3.2:1 | 3.3:1 | 3.4:1 |
|---|---|---|---|---|---|---|---|---|---|
| MIC | 0.8 | 0.15 | 0.15 | 0.15 | 0.15 | 0.15 | 0.15 | 0.15 | 0.15 |

| CPU | 2.7 | 5.5 | 12.4 | 11.4 | 16.8 | 23.5 | 26.1 | 29.5 | 33.8 |
|---|---|---|---|---|---|---|---|---|---|
| Overall | 143 | 142 | 145.2 | 142 | 146.6 | 147.1 | 146.5 | 146.7 | 148.2 |

reduction, the sweet spot location was around in [2.6:1–2.7:1] range. Once again, after reaching the MPI timing crossing point, the timing remained constant on one side at its minimum similar to the previous results.

For the next set of benchmarks, since there are four identical MPI reductions (using sum operator) in the evolution loop, only those invokes were presented in the evolution loop. The results are shown in Tab. 5.24 and Tab. 5.25. As it is apparent, unlike the previous results, the location of the MPI crossing point was shifted to [2.7:1–2.8:1] range. In addition, the optimum point was shifted and this time located around the 2.9:1 point. The only differences between the previous benchmarks and this benchmark was presenting several identical MPI invokes per iteration. Therefore, by presenting multiple synchronization points (even identical) per iteration, the location of the MPI timing crossing point as well as the sweet spot Table 5.24: XFLAT overall time for various MIC:CPU load ratios when four MPI reductions presented per loop iteration. The first row shows the MIC:CPU load ratios, the second, third, and fourth rows show the MIC time, the CPU time, and the overall time, respectively. The highlighted cells show fixed time for each processor.

| MIC:CPU | 1.0:1 | 1.5:1 | 2.0:1 | 2.5:1 | 3.0:1 | 3.5:1 | 4.0:1 | 4.5:1 | 5.0:1 |
|---|---|---|---|---|---|---|---|---|---|
| MIC | 146.8 | 93.8 | 36.6 | 5.4 | 0.5 | 0.5 | 0.5 | 0.5 | 0.5 |
| CPU | 0.47 | 0.48 | 0.5 | 2.6 | 17.9 | 32.3 | 52.2 | 67.8 | 73.1 |
| Overall | 250.9 | 210.8 | 168 | 146.8 | 143.3 | 147.9 | 152.6 | 159.4 | 156.3 |

Table 5.25: XFLAT overall time for various MIC:CPU load ratios (refined range) when four MPI reduction presented per loop iteration. The first row shows the MIC:CPU load ratios, the second, third, and fourth rows show the MIC time, the CPU time, and the overall time, respectively. The highlighted cells show fixed time for each processor.

| MIC:CPU | 2.6:1 | 2.7:1 | 2.8:1 | 2.9:1 | 3.0:1 | 3.1:1 | 3.2:1 | 3.3:1 | 3.4:1 |
|---|---|---|---|---|---|---|---|---|---|
| MIC | 4.8 | 2 | 0.51 | 0.51 | 0.51 | 0.51 | 0.51 | 0.51 | 0.51 |
| CPU | 2.4 | 7.92 | 13.7 | 14 | 26.1 | 27.2 | 26 | 30.1 | 29.1 |
| Overall | 143.8 | 143.7 | 145.8 | 141.4 | 151.9 | 150.1 | 146 | 147.3 | 149 |

may be changed.

Although, there are four identical MPI invokes per iteration, the code sections between two MPI invokes is not identical, thus the performance of each section on the MIC and CPU may vary. For instance, some of the code sections may be single threaded, thus the CPU may perform faster in those regions, and in other sections the MIC may be faster that result in higher idle time for the MIC at the end the section. The combination of those different code sections and different MPI communications, may result in shifting the sweet spot as well as the location of the MPI crossing point.

In order to investigate the behavior of the code between the two MPI communication points, the XFLAT main loop was partitioned such that each MPI function and its preceding code section were isolated and benchmarked separately (see Fig. 5.3.7). while (conditions) {

    … //Code section before the MPI broadcast MPI_Bcast(…);

    … //Code section before the first MPI reduction MPI_Allreduce(…, MPI_SUM, …);

    … //Code section before the second MPI reduction MPI_Allreduce(…, MPI_SUM, …);

    … //Code section before the third MPI reduction MPI_Allreduce(…, MPI_SUM, …);

    … //Code section before the fourth MPI reduction MPI_Allreduce(…, MPI_SUM, …);

    … //Code section before the last MPI reduction MPI_Allreduce(…, MPI_MAX, …);

```
    … //Rest of the code
}
```

Figure 5.74: Overall structure of XFLAT evolution loop and the relative location of the MPI invokes.

All of the following benchmarks were perform on one of the compute nodes on Stampede, the total load was set to $5550 \times 10 \times 100$ beams. Each of the code section was benchmarked separately for various MIC:CPU load ratios. In all the following benchmarks, the code ran for 1k radial steps and the timing results were measured in second.

Since the first MPI communication point within the evolution loop is the MPI broadcast function, in the first benchmark the code prior to the broadcast was included in the loop as well as the last section of the code after the last MPI invoke within the loop (see Fig. 5.3.7). Hence, by altering the MIC:CPU load ratio, it was while (conditions) {

```
    … // Code section before the MPI broadcast

    MPI_Bcast(…);

    /* The rest of the code is excluded from the benchmark */
    /* //Code section before the first MPI reduction          */
    /* MPI_Allreduce(…, MPI_SUM, …);                           */
    /* //Code section before the second MPI reduction         */
    /* MPI_Allreduce(…, MPI_SUM, …);                           */
    /* //Code section before the third MPI reduction          */
    /* MPI_Allreduce(…, MPI_SUM, …);                           */
    /* //Code section before the fourth MPI reduction         */
    /* MPI_Allreduce(…, MPI_SUM, …);                           */
    /* //Code section before the last MPI reduction           */
    /* MPI_Allreduce(…, MPI_MAX, …);                           */
    … // the last part after the last MPI
```

```
}
```

Figure 5.75: The evolution loop structure after isolating the sections associated to the first MPI invokes.

possible to study the behavior and the performance of this code section on each processor.

The load ratio range for the MIC:CPU ratio was chosen to be similar to the ranges of the previous benchmarks. The benchmark's results for the first isolated section is shown in Tab. 5.26. As noticeable, the MPI time on the CPU side was always higher than the MIC side. This implies that the CPU was faster in performing this section, thus the process on the MIC always arrived afterwards. Therefore, the load on the CPU for the range of the benchmarked load ratios was lower than the optimum load ratio.

Similar to the previous benchmark, the next benchmark was performed for the
Table 5.26: XFLAT timing when the first section of the code including the MPI broadcast was benchmarked. The first row shows the MIC:CPU load ratios, the second, third, and fourth rows show the MIC time, the CPU time, and the overall time, respectively.

| MIC:CPU | 2.2:1 | 2.3:1 | 2.4:1 | 2.5:1 | 2.6:1 | 2.7:1 | 2.8:1 | 2.9:1 | 3.0:1 |
|---------|-------|-------|-------|-------|-------|-------|-------|-------|-------|
| MIC | 0.01 | 0.01 | 0.01 | 0.01 | 0.01 | 0.01 | 0.01 | 0.01 | 0.01 |
| CPU | 0.33 | 0.33 | 0.33 | 0.31 | 0.33 | 0.33 | 0.32 | 0.33 | 0.32 |
| Overall | 0.33 | 0.33 | 0.33 | 0.32 | 0.34 | 0.34 | 0.33 | 0.34 | 0.33 |

Table 5.27: XFLAT timing when the second section of the code including the first MPI reduction was benchmarked. The first row shows the MIC:CPU load ratios, the second, third, and fourth rows show the MIC time, the CPU time, and the overall time, respectively.

| MIC:CPU | 2.1:1 | 2.2:1 | 2.3:1 | 2.4:1 | 2.5:1 | 2.6:1 | 2.7:1 | 2.8:1 | 2.9:1 |
|---------|-------|-------|-------|-------|-------|-------|-------|-------|-------|
| MIC | 0.12 | 0.12 | 0.12 | 0.12 | 0.12 | 0.12 | 0.12 | 0.12 | 0.12 |
| CPU | 5.5 | 5.5 | 5.6 | 5.6 | 5.7 | 5.7 | 5.6 | 5.4 | 4.8 |
| Overall | 5.7 | 5.7 | 5.8 | 5.8 | 5.9 | 5.9 | 5.8 | 5.6 | 4.9 |

first MPI reduction invoke as well as the code section prior to the invoke. As shown in Tab. 5.27 and Fig. 5.76, there was no MPI timing crossing point for this range, since the CPU was faster in performing this section. Thus, for this MPI invoke and its associated code section, the MIC always arrived at the communication point after the CPU.

The following results are related to the four similar MPI reduction invokes with their associated code sections. Each section was benchmarked separately in order to study the performance and behavior of each section of the code. The results of the second, third, and fourth MPI reduction invokes (and their associated code) are shown in Tab. 5.28, Tab. 5.29, and Tab. 5.30 as well as Fig. 5.77, Fig. 5.78, and Fig. 5.79, respectively. As it is evident, the crossing point location changed between the benchmakrs. In addition, after the crossing point the timing on one side remained at its minimum. The observable difference among the results indicates that the performance of the code on the MIC and CPU varies between code sections. Although the communications are similar for each MPI invokes, by varying the load ratio, the optimum point for each code section is different. Thus, the performance of the code varies for each section.

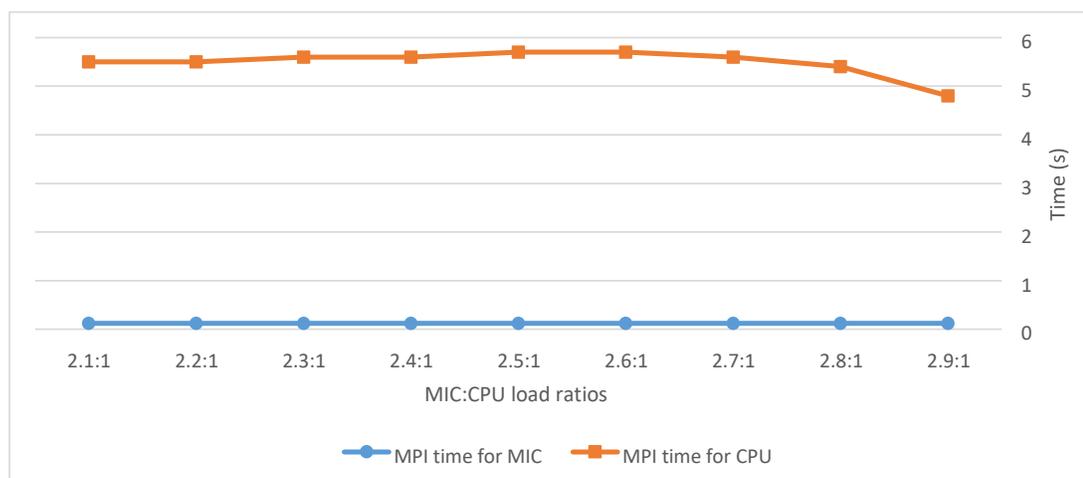

Figure 5.76: XFLAT timing behavior when the second section of the code including the first MPI reduction was benchmarked. The blue curve shows the measured time of MPI invokes on the MIC side, and the orange curve shows the measured time of MPI invokes on the CPU side.

Finally, the performance of the last MPI reduction invoke and its associated code was benchmarked as well. Unlike the previous MPI reductions, the last reduction invoke (using max operator) calculates the maximum of error values among nodes. The result of this section is shown in Tab. 5.31 and Fig. 5.80. It can be seen that similar to the previous reduction invokes, prior to reaching the crossing point the MPI time for the MIC was steady, and after passing the crossing point, on the right side of the point, the MPI time for CPU was stationary. In addition, the location of the crossing point was shifted toward higher load ratios, which indicates that the performance of this section on the MIC and CPU was not identical to the previous Table 5.28: XFLAT timing when the third section of the code including the second MPI reduction was benchmarked. The first row shows the MIC:CPU load ratios, the second, third, and fourth rows show the MIC time, the CPU time, and the overall time, respectively.

| MIC:CPU | 2.1:1 | 2.2:1 | 2.3:1 | 2.4:1 | 2.5:1 | 2.6:1 | 2.7:1 | 2.8:1 | 2.9:1 |
|---------|-------|-------|-------|-------|-------|-------|-------|-------|-------|
| MIC | 7 | 8.3 | 4.8 | 3.3 | 1.3 | 1.2 | .12 | .12 | .12 |
| CPU | .3 | .3 | .3 | .3 | .3 | .3 | 5.1 | 3.22 | 6.7 |
| Overall | 53.7 | 51.8 | 50.4 | 48.7 | 47.7 | 47.3 | 49.8 | 46.8 | 51.8 |

reduction sections.

Since, there are four MPI summation reduction invokes with the same message size within the evolution loop, and since the majority of computations are located within this section, the entire code block containing those invokes was combined together and benchmarked as well. As shown in Fig. 5.81, unlike the previous result, after passing the MPI timing crossing point, the MPI time still decreased on both sides. Therefore, the asymmetry and differences in the instructions of each code section may eventually shift the location of the sweet spot as well as the MPI timing crossing point.

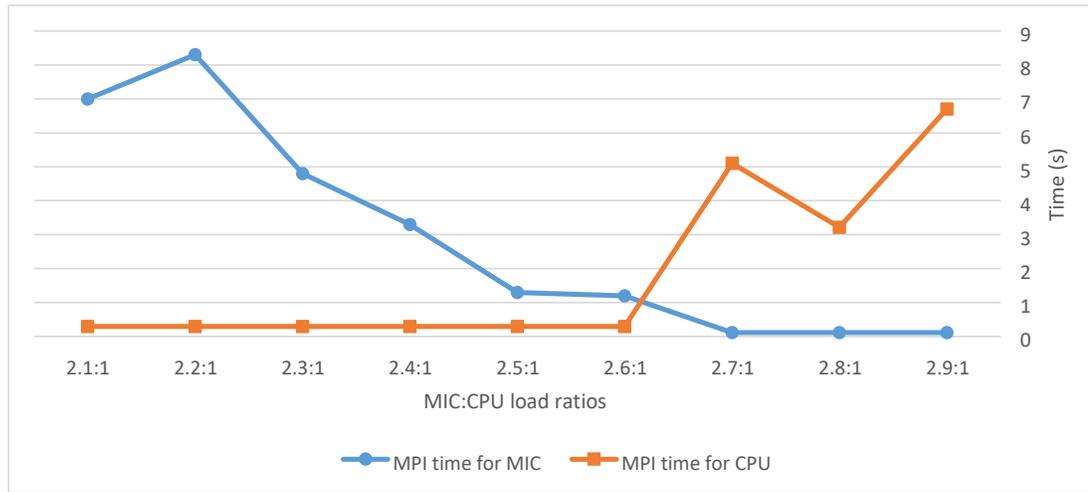

Figure 5.77: XFLAT timing behavior when the third section of the code including the second MPI reduction was benchmarked. The blue curve shows the measured time of MPI invokes on the MIC side, and the orange curve shows the measured time of MPI invokes on the CPU side.

Table 5.29: XFLAT timing when the fourth section of the code including the third MPI reduction was benchmarked. The first row shows the MIC:CPU load ratios, the second, third, and fourth rows show the MIC time, the CPU time, and the overall time, respectively.

| MIC:CPU | 2.1:1 | 2.2:1 | 2.3:1 | 2.4:1 | 2.5:1 | 2.6:1 | 2.7:1 | 2.8:1 | 2.9:1 |
|---------|-------|-------|-------|-------|-------|-------|-------|-------|-------|
| MIC | 8.3 | 6.8 | 5.9 | 2.1 | 2.4 | 1.5 | .4 | .12 | .12 |
| CPU | .3 | .3 | .3 | .3 | .3 | .3 | .4 | 2.2 | 2.1 |
| Overall | 35.9 | 34.5 | 34.1 | 32.6 | 31.8 | 31.0 | 30.1 | 31.0 | 30.2 |

It can be concluded that the location of the sweet spot, which shows the optimum point of the performance, can be located at a different point than the MPI timing crossing point. The reason for this behavior is due to the fact that the performance reduction was benchmarked. The first row shows the MIC:CPU load ratios, the second, third, and fourth rows show the MIC time, the CPU time, and the overall time, respectively.

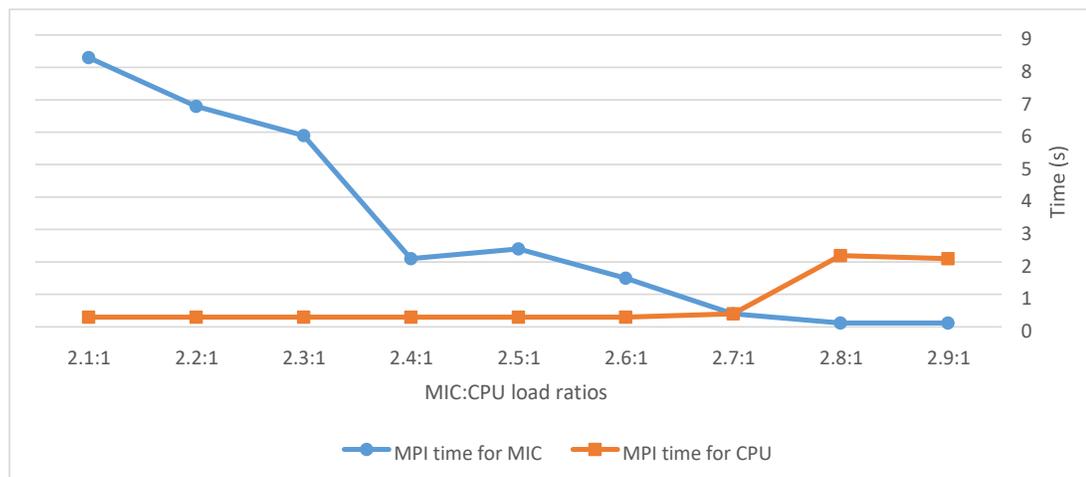

Figure 5.78: XFLAT timing behavior when the fourth section of the code including the third MPI reduction was benchmarked. The blue curve shows the measured time Table 5.30: XFLAT timing when the fifth section of the code including the fourth MPI of MPI invokes on the MIC side, and the orange curve shows the measured time of MPI invokes on the CPU side.

| MIC:CPU | 2.1:1 | 2.2:1 | 2.3:1 | 2.4:1 | 2.5:1 | 2.6:1 | 2.7:1 | 2.8:1 | 2.9:1 |
|---------|-------|-------|-------|-------|-------|-------|-------|-------|-------|
| MIC | 2. | .13 | .5 | .12 | .12 | .12 | .12 | .12 | .12 |
| CPU | .3 | .9 | .4 | 2.6 | 2.7 | 4.0 | 4.1 | 5.2 | 5.7 |
| Overall | 26.9 | 26.2 | 25.1 | 26.5 | 26.1 | 26.6 | 26.3 | 26.6 | 26.6 |

of the evolution loop is the function of the performance of sections that are separated by MPI synchronization points. On the MIC, the performance of some sections may be higher than that of the CPU, on the other hand, the performance of the other sections may be higher on the CPU than MIC. Therefore, the combination of those sections on both sides can affect the final position of the sweet spots as well as the MPI timing crossing point significantly. This will answer the previously unknown

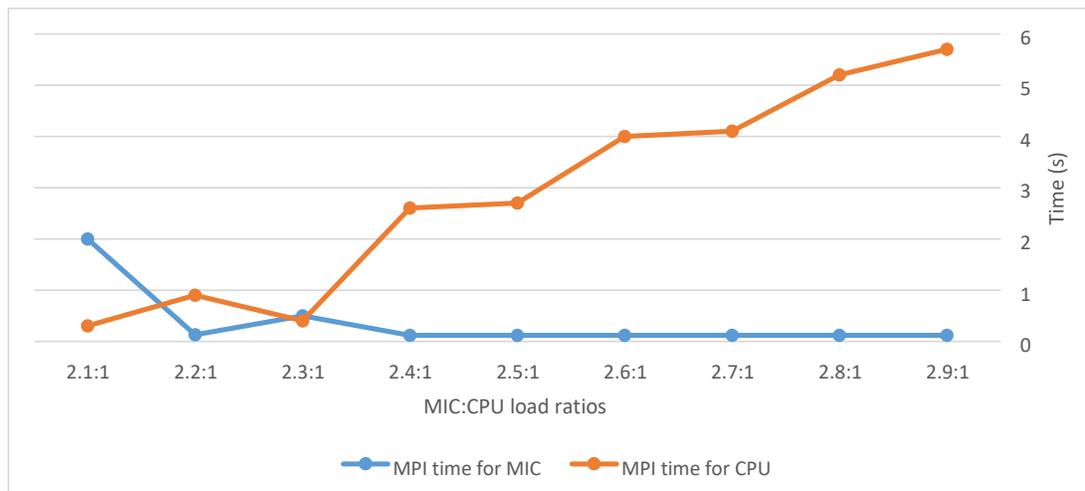

Figure 5.79: XFLAT timing behavior when the fifth section of the code including the fourth MPI reduction was benchmarked. The blue curve shows the measured time Table 5.31: XFLAT timing when the last section of the code including the last MPI reduction was benchmarked. The first row shows the MIC:CPU load ratios, the of MPI invokes on the MIC side, and the orange curve shows the measured time of MPI invokes on the CPU side.

second, third, and fourth rows show the MIC time, the CPU time, and the overall time, respectively.

| MIC:CPU | 2.4:1 | 2.5:1 | 2.6:1 | 2.7:1 | 2.8:1 | 2.9:1 | 3.0:1 | 3.1:1 | 3.2:1 |
|---------|-------|-------|-------|-------|-------|-------|-------|-------|-------|
| MIC | 8.61 | 8.11 | 6.22 | 4.54 | 3.15 | 2.59 | .95 | .15 | .15 |
| CPU | 0.3 | 0.3 | 0.3 | 0.3 | 0.3 | 0.3 | 0.3 | .64 | 1.67 |
| Overall | 39.5 | 39.1 | 37.4 | 36.2 | 35.1 | 34.4 | 33.6 | 33.0 | 33.2 |

reason of the difference between the location of the MPI timing crossing point and the overall optimum point. In addition, it explains the reason that the MPI timing did not reach its minimum immediately after passing the crossing point on the MIC or CPU side.

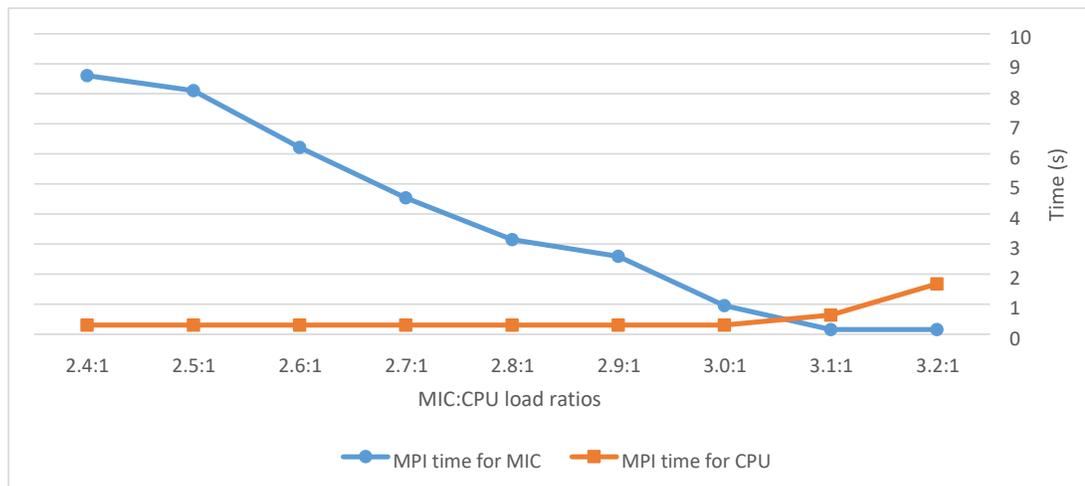

Figure 5.80: XFLAT timing behavior when the last section of the code including the last MPI reduction was benchmarked. The blue curve shows the measured time

of MPI invokes on the MIC side, and the orange curve shows the measured time of MPI invokes on the CPU side.

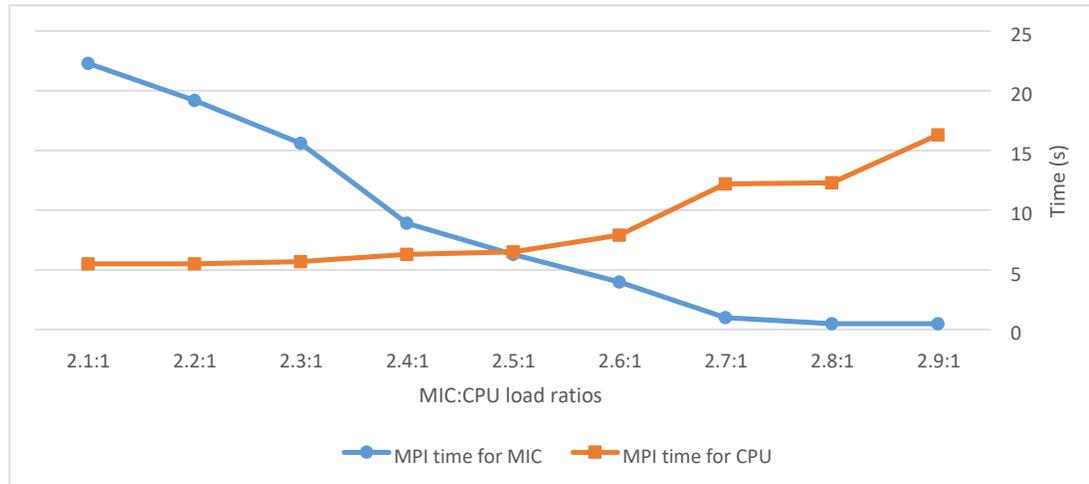

Figure 5.81: Illustration of the performance for the combined code section of the four MPI summation reductions for various MIC:CPU load ratios.

## 5.3.8 Merging MPI functions

From the previous results, it can be concluded that if the number of MPI synchronization points within the evolution loop can be reduced, there may be improvement in the overall performance of the loop. In order to minimize the number of communication points as well as the waiting time between compute nodes for each processor, the main evolution loop is required to be reordered. Consequently, by reducing the waiting time at synchronization points, the processing time and the efficiency may be improved. One approach to achieve this goal is to reduce the number of MPI synchronization points by merging two of the similar MPI _Allreduce() invokes together. By merging two MPI invokes together, instead of having six MPI invokes per loop iteration (one MPI Bcast(), four MPI Allreduce(SUM), and one MPI Allreduce(MAX)) there can be five MPI invokes per iteration. Nevertheless, not every MPI reductions can be merged together, since the code section before each one invoke may depend on the result of the previous reduction invoke. The original neutrino evolution loop arrangement is shown in List.5.1.

Listing 5.1: Illustration of the first version of the evolution's loop algorithm

```
while ( termination _conditions )
{
    //   setting   some   flags
    //   ...
    //   broadcasting   the   termination   conditions   from   the   root
    MPI _Bcast ( ... ) ;

    //===================== S1 ========================
    //   calculating   angle   bins
    //   derivative   of   length   for   each   angle   bins ,   etc .
    //   ...
    //   calculating   the   Hamiltonian H0 and
    //   exchanging   the   partial   result   of H0 between   nodes
        MPI _Allreduce ( ... , H0, MPI SUM,      ... ) ;
     //------------------- End of       S1 ----------------------

    //===================== S2 ========================
    //   evolving   neutrino   beams from   the   current   radial   point
    //   to   the   middle   point   ( half   step   size )   using   the H0
    //   ...
    // evolving neutrino beams from the current radial point // to the next point ( f
    u l l step ) using the calculated H0
    //   ...
    //   calculating   the   Hamiltonian H1 and
    //   exchanging   the   partial   result   of H1 between   nodes
        MPI _Allreduce ( ... , H1, MPI SUM,      ... ) ;
```

```
//------------------- End of        S2 ---------------------

//==================== S3 ========================
//  evolving    neutrino    beams from    the    current    radial    point
//  to  the  next  point  (full  step  size)  using   the H1
//  ...
//   calculating    the    average   of  the  two    previously    evolved
//   neutrino    beams
//  ...
//   calculating   the    Hamiltonian H2 and
//   exchanging   the   partial    result    of H2 between    nodes
     MPI_Allreduce ( ... , H2, MPI SUM,        ... ) ;
   //------------------- End of        S3 ---------------------

//==================== S4 ========================
// evolving neutrino beams from the middle point // to the next point (
half step size ) using the H2
//  ...
//   calculating   the    Hamiltonian H3
//   exchanging   the   partial    result    of H3 between    nodes
     MPI_Allreduce ( ... , H3, MPI SUM,       ... ) ;
   //------------------- End of        S4 ---------------------

//==================== S5 ========================
//  evolving    neutrino    beams from    the    current    radial    point
//  to  the  next  point  (full  step  size)  using   the H3
```

```
//  …
//   calculating   the   average   of   the   two   previously   evolved
//   neutrino   beams
//  …
//   calculating    the maximum error   as  well  as  finding
//   the     global maximum error among   a l l   the  beams
   MPI _Allreduce ( . . . , MPI MAX,    . . . ) ;
   //−−−−−−−−−−−−−−−−−− End of    S5 −−−−−−−−−−−−−−−−−−−−

//   checking   for   any   necessary   I/O
//  …
//   adjusting   the   radius   for   the   next   iteration
//  …
}
```

It can be observed that the third MPI reduction section did not depend on the code between the second and third MPI reductions, thus the computation part for the Hamiltonian $H2$ could be moved to the line after the computation of the Hamiltonian $H1$. Therefore, the neutrino evolution loop was reordered in a new way as shown in

List.5.2.

Listing 5.2: Illustration of the second version of the evolution's loop algorithm

```
while ( termination _conditions )
{
    //   setting   some   flags
    //   . . .
    //   broadcasting   the   termination   conditions   from   the   root
    MPI _Bcast ( . . . ) ;

    //====================== S1 =========================
    //   calculating   angle   bins
    //   derivative   of   length   for   each   angle   bins ,   etc .
    //   . . .
    //   calculating   the   Hamiltonian H0 and
    //   exchanging   the   partial   result   of H0 between   nodes
        MPI _Allreduce ( . . . , H0, MPI SUM,       . . . ) ;
      //-------------------- End of     S1 ----------------------

    //====================== S2 =========================
    //   evolving   neutrino   beams from   the   current   radial   point
    // to the middle point ( half step size ) using the H0 // . . .
    //   evolving   neutrino   beams from   the   current   radial   point
    //   to   the   next   point   ( f u l l   step )   using   the H0
```

```
//  ...
//    calculating     the     Hamiltonian H1 and
//    exchanging    the    partial    result    of H1 between    nodes
     MPI_Allreduce ( ... , H1, MPI SUM,      ... ) ;
//    calculating     the     Hamiltonian H2 and
//    exchanging    the    partial    result    of H2 between    nodes
     MPI_Allreduce ( ... , H2, MPI SUM,      ... ) ;
  //-------------------- End of      S2 ----------------------

//===================== S3 ========================
//   evolving    neutrino    beams from    the    current    radial    point
//   to    the    next    point    (full    step    size )    using    the H1
//  ...
//    calculating    the    average    of the two    previously    evolved
//    neutrino    beams
```

```
//   . . .
// evolving neutrino beams from the middle point // to the next point (
half step size ) using the H2
//   . . .
//   calculating    the    Hamiltonian H3
//   exchanging   the    partial    result    of H3 between    nodes
     MPI _Allreduce ( . . . , H3, MPI SUM,       . . . ) ;
   //-------------------- End of      S3 ----------------------

//====================== S4 =========================
//   evolving    neutrino    beams from    the    current    radial    point
//   to   the   next   point   ( f u l l   step   size )   using    the H3
//   . . .
//   calculating    the   average   of   the   two    previously    evolved
//   neutrino    beams
//   . . .
//   calculating      the maximum error    as   well   as   finding
//   the       global maximum error among     a l l   the   beams
     MPI _Allreduce ( . . . , MPI MAX,       . . . ) ;
   //-------------------- End of      S4 ----------------------

//   checking   for   any   necessary   I/O
//   . . .
//   adjusting   the   radius   for   the   next   iteration
//   . . .
}
```

The last step to restructuring the evolution algorithm was to merge the procedures that were responsible for computing the Hamiltonian $H1$ and the Hamiltonian $H2$ as a single fused method. As a result, the two MPI reductions were fused together as a single MPI function invoke. Although the message size in the fused reduction invoke was twice lengthier than the previous invokes (since the partial results for the Hamiltonian $H1$ and the Hamiltonian $H2$ were exchanged in one message), the number of data exchanging points in the loop was one less than the previous algorithms. Hence, the performance of the code was expected to improve on the heterogeneous environment, since the wasted time at the MPI exchange point, which may be due to the imperfect load balancing, was reduced. The new restructured evolution loop is shown in List.5.3.

Listing 5.3: Illustration of the third version of the evolution's loop algorithm

```
while ( termination _conditions )
{
     //    setting    some    flags
     //   . . .
     //    broadcasting    the    termination    conditions    from    the    root
     MPI _Bcast ( . . . ) ;

     //====================== S1 =========================
     //    calculating    angle    bins
     //    derivative    of    length    for    each    angle    bins ,    etc .
     //   . . .
     //    calculating    the    Hamiltonian H0 and
     //    exchanging    the    partial    result    of H0 between    nodes
        MPI _Allreduce ( . . . , H0, MPI SUM,        . . . ) ;
      //-------------------- End of        S1 ----------------------
     //====================== S2 =========================
     //    evolving    neutrino    beams from    the    current    radial    point
     //      to    the    middle point   ( half   step    size )   using   the H0 //
      . . .
     // evolving neutrino beams from the current radial point // to the next point ( f
     u l l step ) using the calculated H0
     //   . . .
     //    calculating    the    Hamiltonian H1 and H2 and
     //    exchanging    the    partial    result    of H1 and H2 together
      // between    nodes
       MPI _Allreduce ( . . . ,     H1 _H2 , MPI SUM,    . . . ) ;
```

```
//-------------------- End of      S2 ----------------------

//===================== S3 ========================
//   evolving    neutrino    beams from    the    current    radial    point
//   to   the   next   point  (full   step   size )   using    the H1
//   ...
//    calculating    the    average    of  the  two    previously    evolved
//   neutrino    beams
//   ...
// evolving neutrino beams from the middle point // to the next point (
half step size ) using the H2
//   ...
//    calculating    the    Hamiltonian H3
//   exchanging    the   partial    result    of H3 between    nodes
    MPI_Allreduce ( ... , H3, MPI SUM,      ... ) ;
 //-------------------- End of      S3 ----------------------
```

```
//====================== S4 =========================
//   evolving    neutrino    beams from   the   current    radial     point
//  to   the   next   point  ( full   step   size )   using    the H3
//  ...
//   calculating    the   average   of   the two    previously     evolved
//   neutrino    beams
//  ...
//   calculating      the maximum error    as   well   as   finding
//   the      global maximum error among    all   the   beams
   MPI _Allreduce ( ... , MPI MAX,     ... ) ;
 //-------------------- End of       S4 ----------------------

//   checking   for   any    necessary   I/O
//  ...
//   adjusting   the   radius   for   the   next   iteration
//  ...
}
```

Several multi-node benchmarks were required in order to study and analyze the performance of the code in the old and the new algorithms. In all of the following benchmarks on Stampede, the MIC:CPU load ratio was set at 3:1, thus the load on the MIC was fixed at $600 \times 10 \times 100$ ($\Theta$ *angles* $\times \Phi$ *angles* $\times$ *Energy bins*) and the load on the CPU was fixed at $200 \times 10 \times 100$ ($\Theta$ *angles* $\times \Phi$ *angles* $\times$ *Energy bins*). The benchmarks were repeated for various number of nodes and continued for 10k radial steps. The maximum MPI time of MICs, maximum MPI time of CPUs and the overall timing were measured.

The first benchmark employed one CPU and one MIC per compute node, and second benchmark utilized two CPUs and two MICs per compute node. Furthermore, each set of benchmarks were repeated on 16 and 32 nodes. As a result, the minimum number of MPI tasks was 32 (16 tasks on 16 CPUs and 16 tasks on 16 MICs) and the maximum number of MPI tasks was 128 MPI (64 tasks on 32 dual-CPU and 64 tasks on 32 dual-MIC nodes).

In Fig. 5.82 (top), the performance of the old style algorithm (version 1) is depicted. As visible, by increasing the number of compute nodes the overall timing slightly surged. It is due to the inter-node communications overhead, since by increasing the number of nodes, the amount of inter-node messages increases.

The timing of the reordered algorithm (version 2) was virtually identical to the version 1 algorithm, as shown in Fig. 5.82 (middle). Therefore, the reordering of the MPI functions did not boost the performance of the evolution algorithm.

Surprisingly, as shown in Fig. 5.82 (bottom), when two of the MPI reductions were fused together in the new version of the algorithm (version 3), the overall timing was significantly jumped for only the dual processor per node benchmarks. The issue with the performance may be related to the restructuring of the code. Moreover, it may be due to the message size in the version 3 algorithm for which the MPI message size was doubled. Thus, the MPI communications might be slower due to the lengthier messages. In order to examine and understand whether or not the behavior was related to the loop structure, the neutrino evolution loop was decomposed into different sections and each one was benchmarked separately. Hence, any of the differences between benchmarks could be tracked down to the exact line
of the code.

The neutrino evolution loop is decomposed into multiple sections by MPI synchronization points, thus similar to previous benchmarks, each section ending with a synchronization point was isolated and benchmarked separately. The following plots depict the performance of the each isolated section of the neutrino evolution loop.

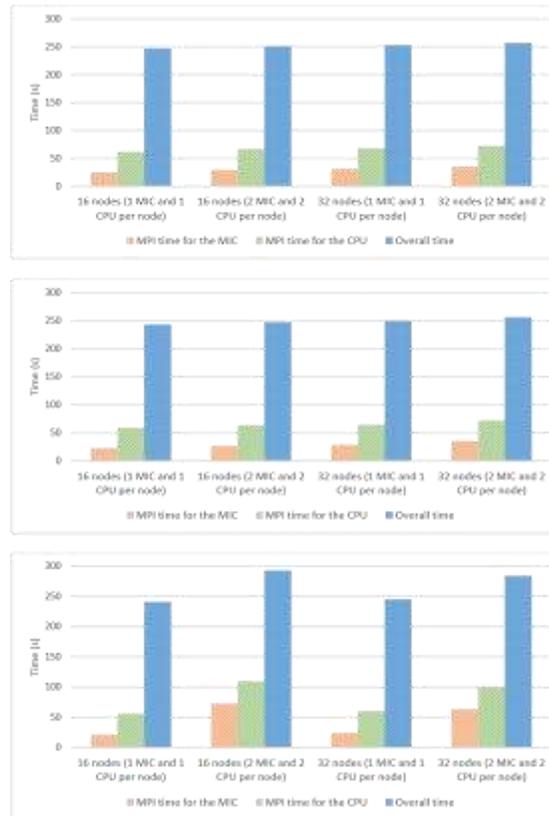

Figure 5.82: Performance of the old style algorithm (top panel), the reordered algorithm (middle panel), and the new style algorithm (bottom panel). Six MPI function calls per loop iteration. The red bar shows the maximum MPI time among the MICs' processes, the green bar shows the maximum MPI time among the CPUs' processes, and the blue bar shows the overall run time in second.

Since, each section ended with a synchronization point via MPI, only the timing of the codes associated before each MPI invoke was measured. Fig. 5.83 illustrates the performance of the section S1, which was identical for all of the three versions of the evolution algorithm. As observable, the amount of computations for this section was low, and there was no significant difference between the results.

The benchmarks for the next sections were performed similar to the first section. Fig. 5.84 shows the performance of the S2 section for the algorithms version 1 (top),

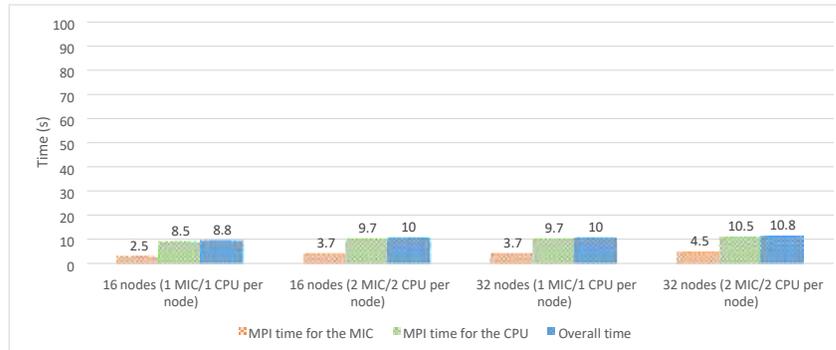

Figure 5.83: Performance of the section S1 of all algorithms. The red bar shows the maximum MPI time among the MICs' processes, the green bar shows the maximum MPI time among the CPUs' processes, and the blue bar shows the overall run time in second.

2 (middle), and 3 (bottom), respectively.

As illustrated by Fig. 5.84 (bottom panel), the MPI time of the dual-MIC benchmarks were jumped by a significant amount that resulted in increasing the overall run time. This section may be the root cause of the performance difference, however, the rest of the code sections had to be benchmarked as well in order to verify that the issue is only due to this section of the loop. Continuing the benchmarks for the next sections, as depicted in Fig. 5.85 and Fig. 5.86, the performance of the S2 and S3 sections for the old algorithm (version 1) are shown respectively. No significant performance differences was observed for these sections.

The algorithm version 2 and 3 had one more code section that needed to be benchmarked. The performance of the code section before the last MPI reduction (S3) for the algorithm version 2 and version 3 is illustrated by Fig. 5.87. Since the code inside this section was identical between the two algorithms, the result was equivalent for both approaches. No significant performance dissimilarity was observed for the last code section.

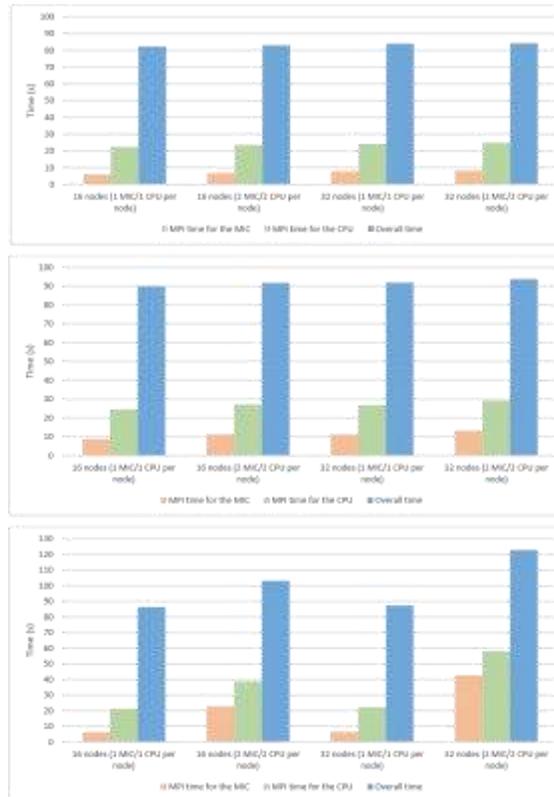

Figure 5.84: Performance of the section S2 of the algorithm version 1 (top), version 2 (middle), and version 3 (bottom). The red bar shows the maximum MPI time among the MICs' processes, the green bar shows the maximum MPI time among the CPUs' processes, and the blue bar shows the overall run time in second.

Finally, the performance of section S4 is depicted by Fig. 5.88. All of the algorithms shared the same code for this section, thus the performances were identical for them. Once more, no significant differences between the results were observed.

In the previous plots, the performance gap was only appeared when the section in which the fused MPI calls was exploited, *i.e.*, the algorithm version 3 (see Fig. 5.84). In addition, it was only observed for the benchmarks that utilized two co-processors per node. In that section of the code the only major difference between the three algorithms was the MPI message size in which the message size for the algorithm

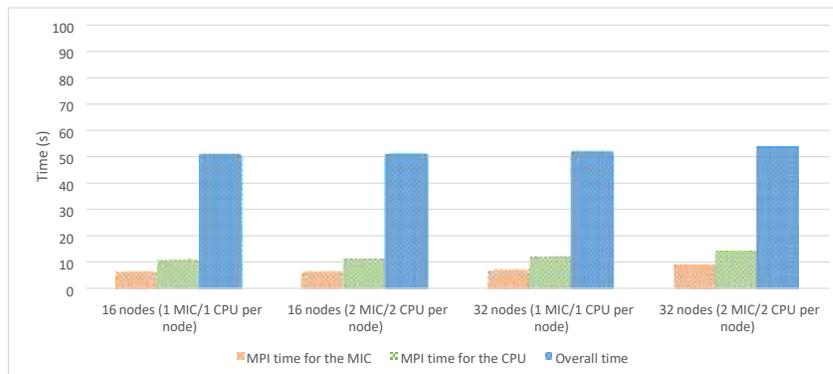

Figure 5.85: Performance of the section S3 of the algorithm version 1. The red bar shows the maximum MPI time among the MICs' processes, the green bar shows the maximum MPI time among the CPUs' processes, and the blue bar shows the overall run time in second.

version 3 was doubled in size.

Since, the performance hit only happened when the size of MPI messages increased, there may be an issue related to the message size. In order to further investigate the issue, the MPI message size was artificially doubled and quadrupled for both algorithm version 1 and version 3 (the performance of version 1 and 2 were identical), using dummy arrays. As a result, the MPI messages were padded with data for which the calculations were not dependent on, however, they increased the overall MPI message size. Since, the only difference between the algorithms at the communication point was the inter-node message size, the results of the new benchmarks were expected to differ by changing the message size. The benchmarks were repeated on 2, 4, 8, 16, and 32 compute nodes. As shown in Fig. 5.89, by doubling the message size, the performance of the algorithm version 3 remained steady, with no observable drop in the performance. Furthermore, quadrupling the message size
did not change the performance trend as well (Fig. 5.90).

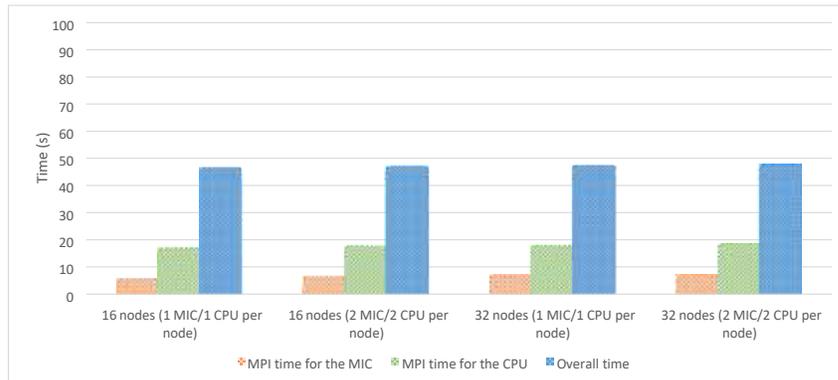

Figure 5.86: Performance of the section S4 of the algorithm version 1. The red bar shows the maximum MPI time among the MICs' processes, the green bar shows the maximum MPI time among the CPUs' processes, and the blue bar shows the overall run time in second.

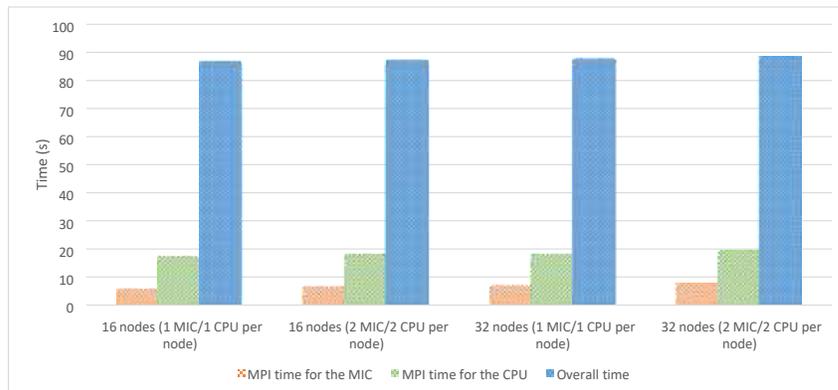

Figure 5.87: Performance of the section S3 of the algorithm version 2 and version 3. The red bar shows the maximum MPI time among the MICs' processes, the green bar shows the maximum MPI time among the CPUs' processes, and the blue bar shows the overall run time in second.

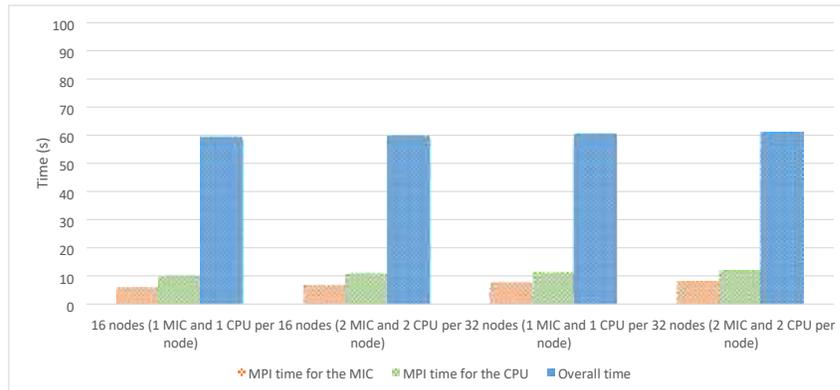

Figure 5.88: Performance of the last section before the last reduction for all of the algorithms. The red bar shows the maximum MPI time among the MICs' processes, the green bar shows the maximum MPI time among the CPUs' processes, and the blue bar shows the overall run time in second.

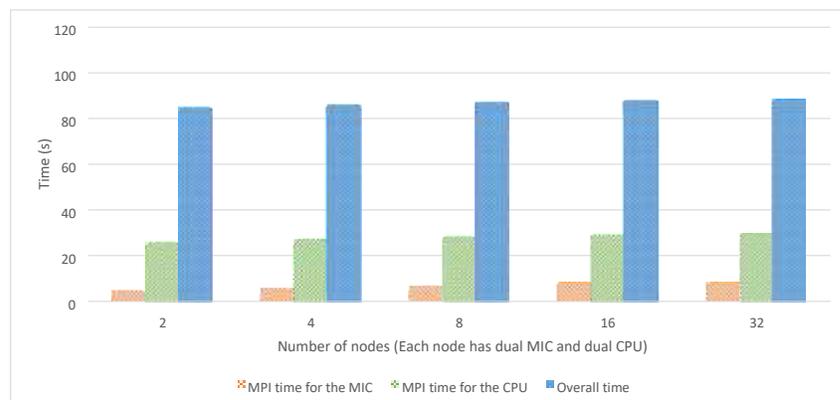

Figure 5.89: Performance of the section S2 of the algorithm version 3 using doubled MPI message size. The red bar shows the maximum MPI time among the MICs' processes, the green bar shows the maximum MPI time among the CPUs' processes, and the blue bar shows the overall run time in second.

Similar benchmarks were repeated for the algorithm version 1, first by doubling the MPI message size and next by quadrupling the message size. The performance of the benchmarks using the doubled message size is illustrated by Fig. 5.91. As noticeable, by increasing the number of node, the overall timing started to surge by at least 50% similar to the benchmark of the algorithm version 3. The communication MPI message size was identical to the algorithm version 3 (its regular message size) for which the same behavior were observed. The results of the benchmarks using quadrupled message size are depicted in Fig. 5.92. As recognizable, by quadrupling the message size the overall timing backed to the expected normal behavior. Hence,

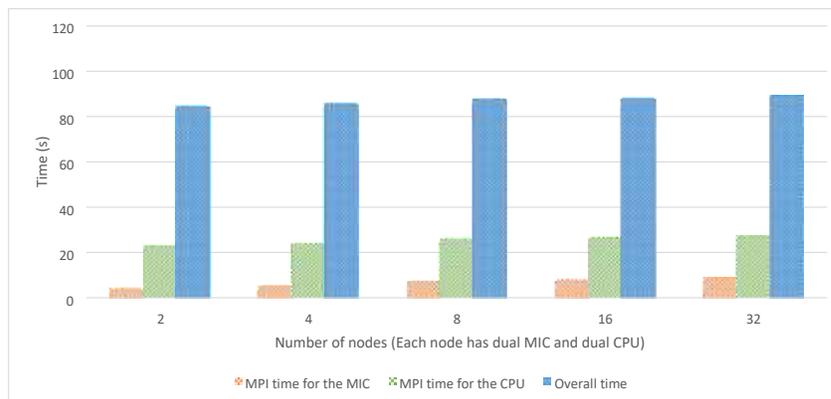

Figure 5.90: Performance of the section S2 of the algorithm version 3 using quadrupled MPI message size. The red bar shows the maximum MPI time among the MICs' processes, the green bar shows the maximum MPI time among the CPUs' processes, and the blue bar shows the overall run time in second.

the issue only happened with a particular MPI message size for both algorithms.

At this point it can be concluded that the poor performance of the code on the dual-MIC nodes only occurred for a specific MPI message size, and it was not the function of the utilized instructions before the MPI invoke nor the number of employed compute nodes. Furthermore, from the previous plots it is understandable that the issue was not due to the code structure or a bug in XFLAT. Since the issue only happened for a specific MPI message size and the strange behavior backed to normal when the message size changed, it may be due to the other factors.

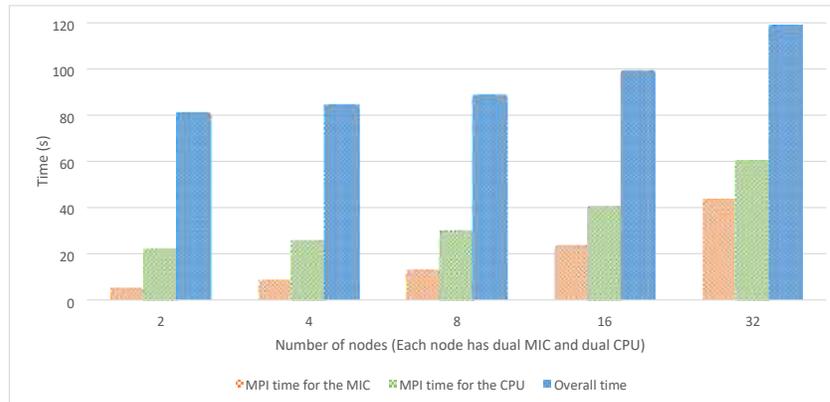

Figure 5.91: Performance of the section S2 of the algorithm version 1 using doubled MPI message size. The red bar shows the maximum MPI time among the MICs' processes, the green bar shows the maximum MPI time among the CPUs' processes, and the blue bar shows the overall run time in second.

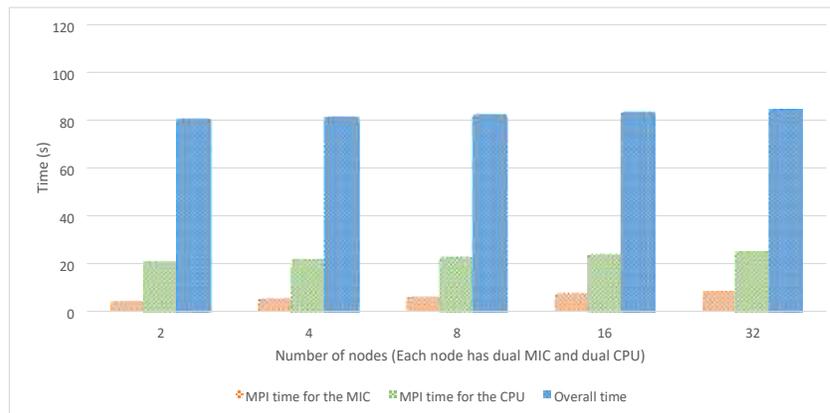

Figure 5.92: Performance of the section S2 of the algorithm version 1 using quadrupled MPI message size. The red bar shows the maximum MPI time among the MICs' processes, the green bar shows the maximum MPI time among the CPUs' processes, and the blue bar shows the overall run time in second.

One possibility for this issue was that it may be due to a bug in the Intel C++ Compiler or Intel MPI library. For the previous benchmarks the Intel C++ Compiler's version was 13.1.1.163 and the Intel MPI library version was 4.1.0.030 were employed. After upgrading the toolkit to the recently available packages (the compiler version 14.0.1 and the MPI library version 4.1.3.049) the benchmarks were repeated by utilizing similar load for the algorithm version 1 and 3 (the load per each CPU and MIC was fixed). The number of nodes varied from 16 to 128 compute nodes. The performance of the algorithm version 1 and version 3 (one CPU and one MIC per node) are depicted by Fig. 5.93 and Fig. 5.94, respectively. Moreover, Fig. 5.95 and Fig. 5.96 illustrate the performance of the algorithm version 1 and version 3 when two CPUs and two MICs were employed per node, respectively.

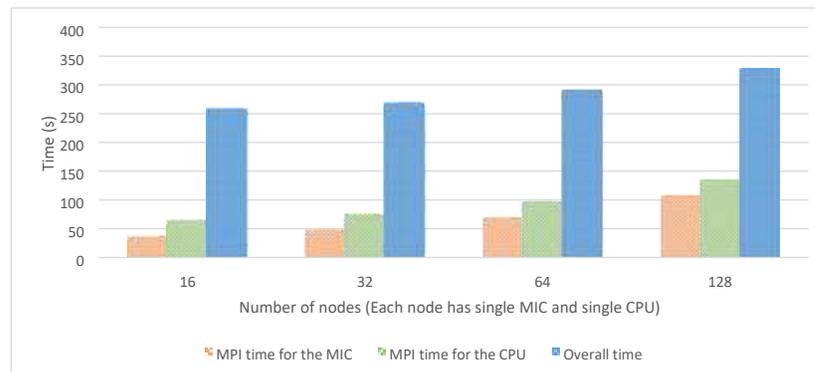

Figure 5.93: Performance the old algorithm (version 1) utilizing newer MPI library (one CPU and one MIC per node) on different number of employed nodes. The red bar shows the maximum MPI time among the MICs' processes, the green bar shows the maximum MPI time among the CPUs' processes, and the blue bar shows the overall run time in second.

As apparent from the results, there was no unexpected jump in MPI timing and the behavior of the single-processor and dual-processor benchmarks were similar.

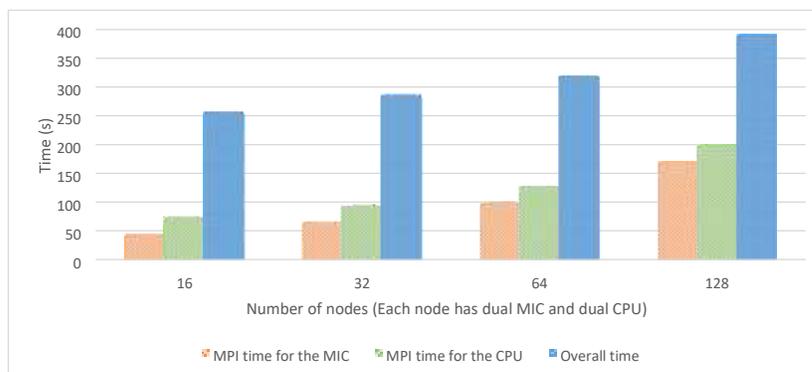

Figure 5.94: Performance the old algorithm (version 1) using newer MPI library (two CPUs and two MICs per node) on different number of employed nodes. The red bar shows the maximum MPI time among the MICs' processes, the green bar shows the maximum MPI time among the CPUs' processes, and the blue bar shows the overall run time in second.

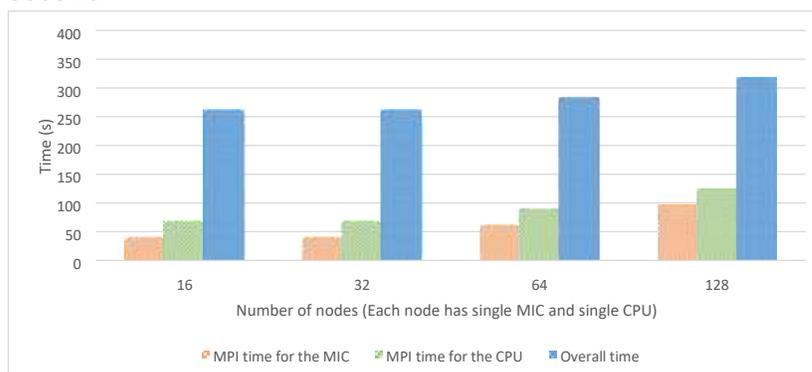

Figure 5.95: Performance the new algorithm (version 3) using newer MPI library (one CPU and one MIC per node) on different number of employed nodes. The red bar shows the maximum MPI time among the MICs' processes, the green bar shows the maximum MPI time among the CPUs' processes, and the blue bar shows the overall run time in second.

Consequently, the issue cleared completely after upgrading the libraries. It turned out that the issue was related to a bug in the Intel MPI library 4.1.0.030, thus by upgrading the library to version 4.1.3.049 the issue resolved and never happened

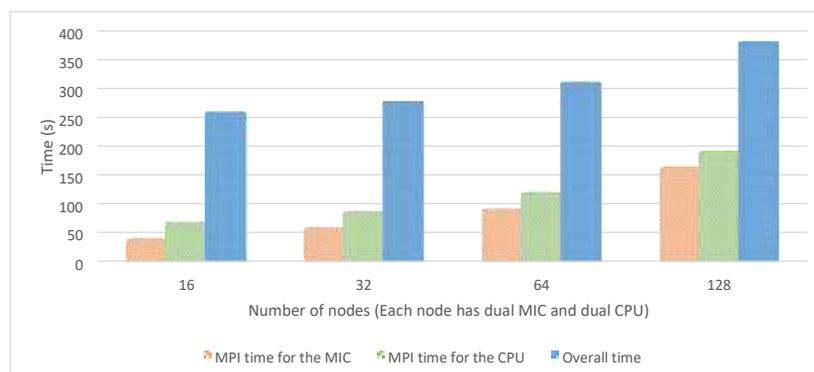

Figure 5.96: Performance the new algorithm (version 3) using newer MPI library (two CPUs and two MICs per node) on different number of employed nodes. The red bar shows the maximum MPI time among the MICs' processes, the green bar shows the maximum MPI time among the CPUs' processes, and the blue bar shows the overall run time in second.

again.

From the previous results, it can be concluded the XFLAT performance scales satisfactory by increasing the number of employed nodes. Even by employing 512 total tasks (128 compute nodes, each equipped with two CPUs and two MICs, thus four tasks per node), the XFLAT timing did not increase dramatically. In addition, merging two MPI messages together only improved the overall performance by a few percents. Therefore, XFLAT is not bounded by inter-node communication nor by the number of employed nodes and can scales satisfactory with hundreds of tasks.

### 5.3.9 XFLAT auto-benchmarking code

As a result, in all of the previous benchmarks, the prediction of the sweet spot for XFLAT on heterogeneous multi-node environments is not straight forward. On a heterogeneous environment, different processor types with different clock frequencies and multiple inter-node and intra-node buses each with different speed and latency, can make the prediction of the location of the sweet spot complex. In addition, different memory hierarchies with different architectures and speed, utilizing different I/O routes for different processor, and the overall code structures and algorithms, can make the prediction of the sweet spot even more complicated. Thus, on a heterogeneous environment, without the knowledge of the optimum point for the load ratios among processors, it is not possible to launch the optimum configuration that result in the best performance. Furthremore, that problem can become more complicated when the number of optimum nodes is unknown. For instance, Fig. 5.97 illustrates the performance of XFLAT on dual-CPU and dual-MIC nodes for three MIC:CPU load ratios when the number of nodes varies from 1 to 24 nodes. The problem size was set to $10000 \times 10 \times 100$, the run time continued for about 100 seconds and the number of computed radial steps were measured. As evident, from 8 to 12 nodes, there was no significant performance gain, thus utilizing only 8 nodes was the best choice. On the other hand, when the number of nodes was 14, the best load ratio for which the highest performance was achievable was 2:1. By going beyond 16 nodes, practically there was no performance gain, thus utilizing more nodes could only waste resources and no noticeable performance improvement.

As a result, the problem of finding the sweet spot can be describe as the search in a 2D space of the load ratios and number of nodes, in order to find the optimum point of configuration. For the reason that the optimum point varies by changing the problem size (as well as by employing different modules), before launching the production run, XFLAT can search the 2D space on a single compute node to find and report the near optimum regions. The benchmark contains two sections. The first section attempts to guess the near optimum computational capability of the CPU and MIC by launching a pre-defined problem size on each processor. After finding the near-optimum load ratios, XFLAT starts to simulate the load on a single node in multi-node configurations by varying the load on a CPU and MIC. XFLAT receives a range for the number of compute nodes (the minimum and maximum number of feasible compute nodes) from the configuration file. Afterwards, XFLAT calculates the load on a CPU and MIC for a given number of compute nodes, then starts the benchmark using the actual neutrino evolution code. For each particular load on the benchmark node, XFLAT varies the MIC to CPU load ratio and repeats the benchmark by running the evolution code (see 5.3.9). Hence, iteratively XFLAT can search the compute nodes range and the load ratios space in order to find the sweet spot for a given problem size and for a particular physics module.

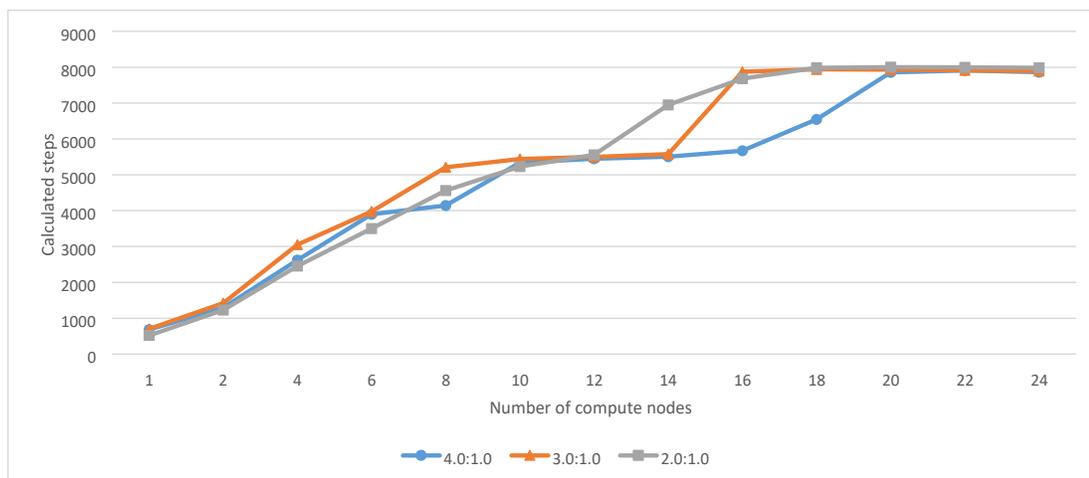

Figure 5.97: XFLAT performance over a range of compute nodes. The problem size was 10000 × 10 × 100, the run time continued for 100 seconds and the number of computed radial steps were measured. The blue curve shows XFLAT performance

when the MIC:CPU ratio is 4, the orange curve shows XFLAT performance when the MIC:CPU ratio is 3, and the grey curve shows XFLAT performance when the MIC:CPU ratio is 2.

```
// Benchmark each processor using a pre-defined problem size result =
Benchmark();

// Search the 2D space iteratively for (i =
minNumNode : maxNumNode)
{ for (j = minLoadRatios : maxLoadRatios)
     { results[i][j] = Benchmark();
     }
}
// Report the results cout <<
results
```

Figure 5.98: The benchmark for searching the 2D space of load ratios and compute nodes in order to find the sweet spot for a particular problem size.

# Chapter 6

# Summary and future work

In this work we have developed an astrophysical simulation code, XFLAT, to study neutrino oscillations in supernovae. Because of its modular design, XFLAT can be easily expanded to investigate neutrino oscillations in various geometries and physical
environments.

We have designed XFLAT to utilize all three major levels of parallelism which are available to modern supercomputers, i.e. the multi-node/device parallelism at the top level, the multi-core parallelism within a single computing node/device, and the vectorization or SIMD within a single core. In implementing the three levels of parallelism, we have chosen to use open standards to make the code portable. Specifically, we used MPI for the multi-node/device parallelism and OpenMP for both the multi-node/device parallelism and the vectorization within a single core. In order to efficiently utilize the SIMD hardware in the CPU and Xeon Phi, we have adopted the Structure-of-Array scheme for the low level module. But for upper level modules we used the Array-of-Structure scheme to make the code more modular and easy to maintain. The design of XFLAT make it suitable to run on both the CPU and the Intel Xeon Phi accelerator the latter of which is based on the Intel Many *Chapter 6. Summary and future work*

Core Architecture (MIC).

We have studied the performance of XFLAT with various configurations and in many scenarios on the Stamped supercomputer as well as two local testbed workstations. We find that, in the best scenarios, XFLAT can perform about 3 times as fast on the first-generation Xeon Phi as on an 8-core Xeon CPU, and about 4× speedup can be achieved on Xeon Phi as compared to an 6-core Xeon CPU. Because the Xeon Phi can be installed as a PCIe extension card to a compatible workstation, our work suggests that the Xeon Phi can be a low-cost choice to dramatically increase the performance of existing computers or computer clusters.

In our study we have found that it can be a great challenge to maintain the load balance in a heterogeneous environment where both the CPU and Xeon Phi are employed. This is because the many-core architecture of the Xeon Phi can support up to about 240 threads per device. There can be a significant drop in the performance of the Xeon Phi when the number of jobs on the device is slightly more than a multiple of the number of its hardware threads.

In our study we have also found that the I/O performance of the first-generation Xeon Phi was very poor. To avoid this I/O penalty we have implemented an indirect I/O module for the Xeon Phi in which the output data from the Xeon Phi is redirected to the CPU. This I/O bottleneck may be fixed in the next-generation Xeon Phi.

The recent work by [Raffelt et al., 2013, Duan and Shalgar, 2015, Abbar and Duan, 2015] has shown that the directional, spatial and time symmetries employed in the 3dimension neutrino Bulb model (with 1 spatial and 2 momentum dimensions) could be broken spontaneously. As a result, simulations of neutrino oscillations in full 7-dimension supernova models (with 1 temporal, 3 spatial and 3 momentum dimensions) must be performed in order to find out the real impacts of neutrino oscillations on supernova physics. This paradigm shift implies an increase of several *Chapter 6. Summary and future work* orders of magnitude in the computation intensity. The development of XFLAT is a first attempt towards this direction. In the coming years the XFLAT project will be expanded to include more realistic physical models to simulate the fascinating phenomenon of neutrino oscillations in various physical environments!

# Appendix A

# XFLAT Documentations

## A.1    Compilation and Build Instructions

XFLAT is a command line application intended to study neutrino flavor oscillations in supernovae environments. The code is C++ implementation with the hybrid architecture that exploits SIMD, OpenMP and MPI for performance acceleration. It is capable to be run on heterogeneous supercomputers and can utilize both traditional CPUs and the newer Intel Many Integrated Core Architecture or Intel MIC (Xeon Phi).

The code contains several modules that can be swapped in or out from the build using the provided switches via the Makefile. In addition to the modules, other features can be switched on or off from the Makefile as well. Features such as the usage of SIMD, OpenMP, and MPI are controllable from the Makefile. Turning off optimizations can help the debugging process. The following code features can be switched on or off from the second line of the provided Makefile starting with CXXFLAGS, and using -D switch:

*A.1. COMPILATION AND BUILD INSTRUCTIONS*

SIMD # when defined the code will use SIMD instructions
OMP # when defined the code will use OpenMP threads
MMPI # when defined the code will use MPI

In addition there are multiple modules that can be changed from the same line of the Makefile. These modules can be categorized as follow: Geometry related modules:

SA # Single Angle supernova module
MA # Multi Angle supernova module
MAA # Multi Azimutal Angle extended supernova module
CLN # Cylindrical module
PNT # Plane module
LIN # Multi-source line module IO

related modules:

IOF # performs file IO for each node
IOFI # performs indirect IO in which MIC sends its data to CPU first

An example of the line with the usage of IOF and MAA modules which utilizes SIMD, OpenMP and MPI is shown as below:

CXXFLAGS = -O3 -openmp -DIOF -DMAA -DSIMD -DOMP -DMMPI

The code employs the NetCDF(either version 3 or 4) library for its current IO modules, however, if the NetCDF4 is used, HDF5 library is required as well.

In order to build binary for CPU the following commands must be issued from console:

$ cp Makefile.cpu Makefile
$ make all

 OR

$ make all -f Makefile.cpu

Likewise, in order to build the code for the Xeon Phi the following commands must
be issued:

$ cp Makefile.mic Makefile
$ make all

 OR

$ make all -f Makefile.mic

Consequently, the CPU binary will be called XFLAT.cpu and the Xeon Phi binary is
XFLAT.mic. Please note, if the OpenMP feature is switched on, the -openmp flag also
must be added to the compiler flags and if the MPI feature is switched on inside
the Makefile, a few MPI scripts are required so as to run it on multiple nodes.

In order to optimize the code for any Intel CPU, one has to add -xHOST to the
compiler flags set in the Makefile. Likewise, in order to build the code for MIC the -
mmic flag should be added to the compiler flags set.

## A.2 Source Code Directories and Files

There are two directories that contains the header files (include/) and the source files (src/). The header directory contains the following header files:

**Fenergy.h:** Contains function declarations which are responsible for energy spec-
tra calculations. These functions are called from NBeam module.

**Fio.h:** Holds function declarations which are responsible for dumping data onto files as well as other I/O related tasks that depends on the settings, different source files can implement those functions.

**Global.h:** Encloses global settings and constants. These settings affects the general behavior of XFLAT.

**Matt.h:** Encapsulates the matter profile functions and its related settings. A run may or may not utilize this module.

**NBeam.h:** Encloses Neutrino Beam class declaration. This class contains all the variables and functions for calculating neutrino beam interaction and evolution. Each neutrino beam in the system must instantiate this class. This module serves as the lower layer module for other upper layer modules.

**NBGroup.h:** This module encapsulates all the upper layer modules such as the numerical, Physics, and I/O modules.

**Nmr.h:** Contains numerical related function declaration. Those functions are responsible for solving the numerical algorithm.

**Parser.h:** Encloses the parser class declaration. This class is responsible for taking a config.txt file, parse it, and put the extracted values to the corresponding variables.

**Phy.h:** Holds physics and geometry related function declaration. Depends on the geometry, different source files can implement those functions.

**Util.h:** Holds auxiliary functions and variables declarations. These functions may be utilized from the other modules.

The source directory contains the following C++ files:

**main.cpp:** Holds the main function of the application (the entry point of the application).

**Fenergy.cpp:** Implements the energy spectra functions and its related procedures.

**IO f.cpp:** One of the implementation of Fio.h header. In this source file, every node dumps its data directly to its own file.

**IO fi.cpp:** One of the implementation of Fio.h header. Due to limited I/O capability of the current generation of Intel MIC, Xeon Phi nodes send their data to corresponding CPUs first, the CPU dumps both their own data and the Phis data to file.

**Matt.cpp:** Implements matter profile functions. Depends on the configuration, the functions may or may not be called in a run.

**NBeam.cpp:** Implements Neutrino Beam class. The class encapsulates the wavefunctions of a neutrino's beam.

**NBGroup.cpp:** Encapsulates upper modules. Contains functions for initializing and finalizing other modules.

**Nmr.cpp:** One of the upper layer modules. Implements the neutrino evolution numerical algorithm.

**Parser.cpp:** Implements the Parser class. It is responsible for parsing the configuration file.

**Phy CLN.cpp:** This is one of the several Physics modules. This module implements the cylindrical geometry.

**Phy MA.cpp:** This module implements supernova bulb model (multi-zenith angle geometry).

**Phy MAA.cpp:** This module contains the extended supernova physics. It implements the multi-zenith and multi-azimuthal angle geometry.

**Phy PLN.cpp:** This module implements the multi-zenith and multi-azimuthal plain geometry.

**Phy SA.cpp:** Contains the implementation of the single-angle supernova model. Due to the geometry limitations this module cannot utilize more than a single processor's core.

**Util.cpp:** This modules holds many general variables and are used in several other modules. It also implements several auxiliary functions.

## A.3   The config.txt file

In order to run XFLAT, a configuration file has to pass via command line argument to the program. This file contains several settings related to the behavior of the application and many other values for initializing program's variables.

Each keyword in the file must be starts in a new line. Keywords are constant and cannot be change unless the corresponding keyword in the `Parser.cpp` class implementation is changed accordingly. Thus, each keyword contains a constant character string and ends in a = and after a white space its value is stored:

`Keyword1= value1`

Comments can be added to the configuration file as well. They starts with # and continues until reaching the new line character. Therefore, they can be added after the value of a keyword, or can be added in a separate line:

# comment1

Keyword1= value1 # extra comment

Please note that if an expected variable is not initialized in the configuration file, its initial value will be undefined.

The current keywords can be categorized in two different categories: Those that control the general behavior of the application (mostly related to the I/O tasks) and those from which the physics related variables are initialized.

General keywords are listed as follow:

dumpMode= This keyword expects an integer from which the way data dump onto a file is set. The values must be read in binary mode. Putting 0 (also 0 in binary mode) for the value means no dumping data. Putting 1 (also 1 in binary mode) is the first mode which means dumping a whole snapshot in which all the wave-function values are written onto the file. The value 2 (10 in binary mode) means dumping only the weighted average over energy bins onto file. The next independent value can be 4 (100 in binary mode) and so on. Note that in this way values can be combined together as binary flags. Therefore, putting 3 (11 in binary mode) means performing I/O in both the first mode (01 in binary) and second mode (10 in binary).

filePrefix= This keyword takes a string as the general name for files. In the application some other strings may attach to it as well. For example when the data dump mode is 1, the "Snapshot" string is also attach to it. Moreover, if multiple files is generated, a counter number starts from 0 is also attached to the file's name afterwards. Finally, for the MPI runtime, each node will attach its own MPI id at the end of the file (separated from the rest of string with ' ').

newFile _step= Takes an integer value indicating that after a [articular number of I/O task, a new file has to be generated. In this way it is possible to prevent creating a huge file.

sync step= Indicates that after a specific number of data dumping iterations, the file has to be synchronized with the disk. In this way it is possible to prevent data loss due to application crash.

r _step1= If for each radial iteration in the evolution loop, data are to be dumped on file, the performance will drop dramatically. In addition, there is no need to perform I/O task for every iteration in the loop as the difference between the two consistent radial steps is normally negligible. Therefore, the value (float) of this keyword indicates that only after advancing a particular distance in Km one snapshot is dumped to file. Hence, there is a radial distance equals to this keyword's value between each snapshot. This keyword only controls the I/O mode of 1, for the second I/O mode the keyword is r step2. As a result, the frequency of saving data can be
controlled independently for each I/O mode.

t _step1= This keyword is similar to the previous keyword.        The I/O is only performed after a particular seconds. Similar to the previous keyword, it only controls
the first I/O mode. For the second I/O mode the keyword is t step2.

itr _step1= Similar to the previous keywords, the I/O task is allowed to be performed after a particular radial iterations. It only affects the behavior of the first I/O mode. For the second I/O mode the keyword is itr step2.

start _beam=, end _beam= These two keywords indicates the starting and ending indices of neutrino beams that the current node has to perform computations on. For instance, the distribution of 1000 beams over two identical nodes can be done in this way: for the first node the start beam= 0 and end _beam= 500 and for the second node start _beam= 500 and end beam= 1000. Thus the 500th beam is the first beam on the second node. The distribution of beams over nodes depends on the Physics module. For instance, for the bulb model, the neutrino beams along the zenith angle are distributed over nodes, however, for the other modules, depends on the geometry the distribution can be different. If the value of these two keywords is a negative number, in order to find distribution range of neutrino beams on each compute node, prior to distributing the load over nodes a benchmark code is performed on each node and based on the computational capability of each node, the starting and ending beam's indices are defined.

multiNodeBench= If the value is non zero, it indicates that the multi-node benchmarking should be on. The multi-node benchmarking is the benchmark to find the optimum MIC to CPU load ratios among multiple nodes.

minNodes=, maxNodes= These two values indicate the minimum and maximum number of desired nodes on which the search for the optimum number of nodes should
be performed.

hasMatter= Indicates whether or not the matter profile is included into the run. Zero means the matter profile is excluded, and non-zero values indicates the matter profile is included into the run.

Tn= The total execution time of the application in second. Note that the time for memory allocations and initilizations is not included, thus this time is only the allowable time for executing the main neutrino evolution loop.

Ts= The total number of iteration for the main evolution loop. The program will be finished after reaching the Tn seconds or after performing Ts radial iteration.

Keywords related to physics are listed as follow:

eps0= The error tolerance that indicates the maximum allowable error between two computed wave-functions.

ksi= A float value between 0 and 1 for controlling the adaptive step size behavior. It is a safety factor to ensure success on the next try.

dm2= The value of the neutrino mass-squared difference. It is positive for the normal mass hierarchy and negative for the inverted mass hierarchy.

theta= The vacuum mixing angle.

R0= The starting radius in Km.

Rn= The final radius in Km.

dr= The initial $\Delta r$ value in Km. Normally less than 1 Km.

max _dr= The maximum possible value for 'dr'. The higher values will be trimmed to this value.

E0= The starting point of the energy spectra in MeV.

E1= The ending point of the energy spectra in MeV.

Abins= The number of angle bins along zenith direction. The value is always greater or equal to one.

Pbins= The number of azimuth angle bins. The value is always greater or equal to one.

Ebins= The number of energy bins over the range of the energy spectra. The value is always greater or equal to one.

SPoints= The number of emission surface points, for the multi emission points systems.

Flvs= The number of neutrino flavors in the system.

Ye= The electron fraction or the net number of electrons per baryon. nb0=

The baryon density at the neutrino sphere.

Rv= The neutrino sphere radius.

Mns= The mass of the neutron star in solar mass unit. gs=

The statistical weight in relativistic particles.

S= The entropy per baryon.

hNS= The scale height.

Lve=, Lv e=, Lvt=, Lv t= The energy luminosity for electron, anti-electron, tau, and anti-tau neutrinos in erg/s.

Tve=, Tv _e=, Tvt=, Tv _t= The neutrino temperature for electron, anti-electron, tau, and anti-tau neutrinos in MeV.

eta _ve=, eta _v e=, eta _vt=, eta v t= The degeneracy parameter for electron, anti-electron, tau, and anti-tau neutrinos.

# A.4     Methods and Variables of Modules

Here the role of each method and variable within modules is described.

## A.4.1       Neutrino Beam module's methods (NBeam.h/NBeam.cpp)

This module is one of the lower layer modules. It contains the NBeam class, which holds arrays of wave-function. Each element of those arrays represent a neutrino in a particular energy bin. The NBeam module has the following functions and class:

void init(int flavors, int ebins) First, this function calls the energy spectra module's initialization method. Next, depends on the number of flavors, it allocated several arrays for storing energy bins' values. The size of each array is determined by the number of energy bins. Afterwards, it fills up the energy arrays based on the normalized value of the energy spectra. In addition, it calculates and stores the vacuum Hamiltonian for each bin.

void freemem() This method frees up all the arrays that were allocated in the initialization function.

inline void upd _nu coef(const double *restrict nu, const double *restrict anu, const double n cf, const double an _cf, double *restrict ret) throw()

This inlined function calculates the difference of the multiplication of the energy spectra functions and the density matrices between a neutrino particle and its antiparticle. In addition, each density matrix is weighted by its pre-computed coefficient as follow: $\varrho'_{\nu_\alpha} f_{\nu_\alpha}(E') \frac{L_{\nu_\alpha}}{\langle E_{\nu_\alpha} \rangle} - \varrho'^*_{\bar\nu_\alpha} f_{\bar\nu_\alpha}(E') \frac{L_{\bar\nu_\alpha}}{\langle E_{\bar\nu_\alpha} \rangle}$ class NBeam This class represent a single neutrino beam that contains arrays of wavefunctions expanding over a range of energy spectra. There are several variables and methods inside this class:

- The constructor: The argument is the index that represent the type of the particle (electron neutrino, anti-electron neutrino, etc.) for the beam:

  NBeam(int prtc);

- Energy spectra function's setter and getter: Two functions are available to receive the value of energy spectra function. The first method receives an index of energy bin and returns the value of the energy function based on the received index. The second method returns the pointer to the array of energy bins. The pointer can be used to set or get each energy bins' value separately:

inline double Fv(int e) const; inline double*

Fv() const;

- Wave-function's components setter and getter: Sometimes it requires that the wavefunction's components be accessible based on an index (*i.e.* 0 for the real part of the first number, 1 for the imaginary part of the first number, 2 for the real part of the second number, 3 for the imaginary part of the second number, etc.), therefore the following methods are provided to make wavefunction's components accessible based on an index number: inline const double* psi(int cmpn) const throw(); inline double* psi(int cmpn) throw();

- Energy bins' setters and getters: These set of functions return the pointer to the components of wavefunction arrays. The pointer can be used to set or get each energy bins' value. The following functions are available for the current version of NBeam class: inline double* Ar(); inline double* Ai(); inline double* Br(); inline double* Bi(); inline const double* Ar() const; inline const double* Ai() const; inline const double* Br() const; inline const double* Bi() const;

- General wavefunction's setters and getters: These set of functions can set the value of each energy bin or return the current value of each energy bin. For each wave-function component, there has to be at least one setter and getter functions. The argument of each method is the index of the queried energy bin. The following functions are available for the current version of NBeam class:

inline double& Ar(const int e);

> inline double& Ai(const int e); inline double&
>
> Br(const int e); inline double& Bi(const int e);
>
> inline const double& Ar(const int e); inline
>
> const double& Ai(const int e); inline const
>
> double& Br(const int e); inline const double&
>
> Bi(const int e);

- Density matrix: The density matrix is calculated from the wavefunction. For instance, a wavefunction with two complex components has a two by two complex density matrix. Yet, since the second row in the matrix can be constructed from the first row, this method only returns the computed first row of the density matrix. The returned value is in a four-element array. There are two possible ways to compute the density matrix from a wavefunction. The first one is by passing an energy bin index, and the other is by passing the wavefunction components as argument:

> inline void density(const int ebin, Res t ret) const throw(); inline void density(const
>
> double ar, const double ai, const double br, const double bi, Res t ret) const
>
> throw();

- Neutrinos' evolution: After computing the Hamiltonian, a method is required to evolve the current wavefunctions using the Hamiltonian. There are two methods that can be used for the neutrinos' evolution. The first function, takes an energy bin index, delta-radius, the Hamiltonian, and returns the components of the evolved wavefunction. The second method takes delta-radius, the Hamiltonian, the components of the current wavefunction, and returns the components of the new wave-function:

inline void U( const int e, const double dr, const double h r0, const double h _i0, const double h r1, const double h i1, const double n _ar, const double n ai, const double n br, const double n bi) throw(); inline void U( const double dr, const double h _r0, const double h _i0, const double h r1, const double h i1, const double a r, const double a _i, const double b r, const double b i, double& n _ar, double& n _ai, double& n br, double& n _bi) const throw();

- Summation over energy bins: The summation over all energy bins is necessary in order to calculate the Hamiltonian, thus a function is provided to calculate the sum and store inside the class: void calcHSum() throw();

- Energy bins summation: This method returns the previously computed summation over energy bins. The returned value is in a four-element array that is the first row of the summation matrix:

  void getHSum(Res _t ret) const throw();

- Neutrino's Beam Evolution: These set of functions receive another neutrino's beam and based on the neutrino-neutrino background Hamiltonian, matter potential, and neutrinos' mass difference term, evolves the wavefunctions:

  void evolveBinsAvgErr( const NBeam& beam, const int ptc idx, const double dr, const double *restrict hvv, const double hmatt, NBeam& beamAvg, NBeam& beamErr ) throw(); void evolveBinsHvvAvg( const NBeam& beam, const int ptc idx, const double dr, const double *restrict hvv, const double hmatt, NBeam& beamAvg ) throw(); void evolveBinsAvg( const NBeam& beam, const int ptc idx, const double dr, const double *restrict hvv, const double hmatt, NBeam& beamAvg ) throw(); void evolveBinsAvg( const int ptc idx, const double dr, const double *restrict hvv, const double hmatt, NBeam& beamAvg ) throw(); void evolveBinsHvv( const

NBeam& beam, const int ptc idx, const double dr, const double *restrict hvv, const double hmatt ) throw(); void evolveBins( const NBeam& beam, const int ptc idx, const double dr, const double *restrict hvv, const double hmatt ) throw(); void evolveBins( const int ptc _idx, const double dr, const double *REST hvv, const double hmatt ) throw();

Other than the neutrinos' evolution, several of the methods perform other task such as calculating the summation of energy bins for the background Hamiltonian or taking the average between two neutrino beams or calculating the maximum error between two neutrino beams. Since function fusing can increase the overall performance, it is recommended to use the fused version of the function to perform more tasks on the same data. The particle index (0 for electron neutrino, 1 for anti-electron neutrino, 2 for mu neutrino, 3 for antimu neutrino, *etc.*), the delta-radius, the neutrino-neutrino background and the matter potential are passed as arguments too.

- Average of two beams: The average function takes a neutrino beam's array as an argument, then calculates the average between the argument and the current object's beam, and replace the current beam with the result:

  void addAvg( const NBeam& beam ) throw();

- Find the maximum error between two beams' wavefunctions: In order to detect whether or not the current delta-radius is appropriate for calculations, the maximum error between two neutrino beams is computed. Two functions are privuded. The first one that takes a neutrino beam and calculates and returns the maximum error over all energy bins. The second function is similar to the first one but also computes the summation over energy bins which can be used in future for the Hamiltonian computations:

double calcErr( const NBeam& beam ) const throw(); double

calcErrHvv( const NBeam& beam ) throw();

- Maximum calculate error return: If the maximum error is computed in one of the fused function before, this function only returns it: double& getErr() throw();

## A.4.2    Numeric module's methods (Nmr.h/Nmr.cpp)

This module is responsible for the numeric algorithm, and can be replaced with the other numerical modules using different algorithms. There are several functions in this module that have to be implemented:

int init(int len) This function is the first function to be called within this module. It takes an integer argument that is the length of the neutrino beams (number of trajectory beams multiplied by number of particles). Afterwards, it allocates the memory for the entire neutrino beam's arrays. The number of arrays may be varied and depends on the algorithm. Next, it calls the initBeam() function of the physics module for each of the allocated arrays, in order to initialize them.

void freemem() Calls the freeBeam() function of the physics module, afterwards performs the neutrino beams' deallocation.

int evolutionLoop() throw() This function contains the main neutrinos' evolution loop. In fact, it implements the numerical algorithm. If MPI is enabled, there are several points that nodes communicate and exchange data. At the first point, within the main evolution loop, the master node sends the termination condition, current radius, and delta-radius to all of the nodes. Next, this function calculates the radial advancement based on the number of middle points. Then, it continues by calculating the matter density profile. Afterwards, the Hamiltonian computations are performed. The current algorithm continues with the neutrino beams evolution, error calculations, and if the program is in the multi-node mode, all the nodes exchange maximum local errors in order to find maximum global error. Next, if the maximum error is less that a predefined error threshold an IO module's function is called to perform any necessary IO operations. Finally, it continues the next iteration.

### A.4.3    Physics modules' methods (Phy.h/Phy *.cpp)

Currently, there are several physics related module each having different geometry and physics. All of them have to implement at least all the methods within the header file (Phy.h). These general methods are as the following:

void init() The Initilization method of the physics module that performs memory allocations in a specific order and arrays initialization.

void freemem() Deallocates all the allocated memories for this module.

int beamLen() Returns the length of the current neutrino beam's array. It can be used from other modules to find out about the neutrino beams' length for memory allocation purposes (*e.g.* $\Theta \times \Phi \times$ *num of particles*)

int getDim() Returns the number of dimensions for the data related to the plugged-in physics module. This can be useful from the I/O modules in order to format the data for the NetCDF file (*e.g.* 6 = [r, theta, phi, num of particles, wavefunction's components, E])

void getDimInfo(std::string str[]) Returns an array containing strings for the names of each dimension (*e.g.* 6 = ["r" "theta", "phi", "prtcl", "comp", "ebin"]) size t*startDim() Returns an array of starting point of each data dimension.

To be used by the I/O module (*e.g.* [current radius, 0, 0, 0, 0, 0])

size t* countDim() Returns an array containing the length of each data dimension. This can determine the size of the current snapshot. To be used by the I/O module (*e.g.* [1, theta bins num, phi bins num, num of particles, num of components, num of energyBins])

int& startBeamIdx() Returns the starting beam number for the dimension that is distributed over nodes. The value may vary on each node (*e.g.* 0 for the first node, 500 for the second node, 1000 for the third node, 2000 for the fourth node)

int& endBeamIdx() Returns the length of the dimension that is distributed over nodes. The value may vary on each node (*e.g.* 500 for the first node, 1000 for the second node, 2000 for the third node, 3000 for the fourth node)

int firstDimLen() Returns the size of the dimension on which the problem size is distributed over nodes (*e.g.* depending on the module can be theta bins, or phi bins, *etc.*)

void initBeam(NBeam* beam) Receives an array of NBeam objects and initializes each object's internal arrays (Particles and Ebins) accordingly by calling each the constructor.

void freeBeam(NBeam* beam) Receives an array of NBeam objects and calls the deconstructor of each object.

void calcAngleBins(const double r, const int step num) The The functionality depends on the module but normally calculates and caches the cosine bins at the current radius *r* and for different step numbers (current point, mid-point, full-point). Depends on the module may calculates other angle bins as well.

void calcDeltaLs(const double dr, const int cur _pnt, const int s pnt, const int e pnt) Calculates and caches $\delta l$ for each angle bins. The 'dr' parameter is the radius difference, the 'cur _pnt' is the point at which the $\delta l$ is calculated and the last two points are the points at which the average is calculated. (*i.e.* $dl[cur\, pnt] = dr/(.5 * (cos[s\, pnt] + cos[e\, pnt])$ ) void newHvv(double*& hvv) Allocates memory for an array of Hamiltonians.

void deleteHvv(double*& hvv) Deletes the allocated memory for Hamiltonian array.

void avgBeam(const NBeam *restrict ibeam, NBeam *restrict obeam) This function only calculates the average of the two input neutrino beam arrays and store the result into the second beam.

The following methods, receive an array of NBeam objects, evolve and save them into an output array of NBeam objects. They may receive the matter profile's value and Hamiltonian array hvv. In Addition, they may perform other tasks including the calculation of partial summation for neutrino-neutrino background over energy bins, or calculating the average of two NBeam arrays and store them into the last parameter:

- void evolve(const nbm::NBeam *restrict ibeam, const int pnt, const double *restrict hvv, const double hmatt, nbm::NBeam *restrict obeam) throw();

- void evolveHvv(const nbm::NBeam *restrict ibeam, const int pnt, const double *restrict hvv, const double hmatt, nbm::NBeam *restrict obeam) throw();

- void evolveAvg(const nbm::NBeam *restrict ibeam, const int pnt, const double *restrict hvv, const double hmatt, nbm::NBeam *restrict obeam, nbm::NBeam *restrict obeamAvg) throw();

- void evolveHvvAvg(const nbm::NBeam *restrict ibeam, const int pnt, const double *restrict hvv, const double hmatt, nbm::NBeam *restrict obeam, nbm::NBeam *restrict obeamAvg) throw();

- void evolveAvgErr(const nbm::NBeam *restrict ibeam, const int pnt, const double *restrict hvv, const double hmatt, nbm::NBeam *restrict obeam, nbm::NBeam *restrict obeamAvg, nbm::NBeam *restrict obeamErr) throw();

- void evolveAvg(const int pnt, const double *restrict hvv, const double hmatt, nbm::NBeam *restrict iobeam, nbm::NBeam *restrict obeamAvg) throw();

## A.4.4    I/O module's methods (Fio.h/IO f.cpp, IO fi.cpp)

Currently, there are two implemented I/O modules in XFLAT. When the first I/O module (IO f.cpp) is employed, each node will dump data onto its own file using the NetCDF. However, since the I/O performance is very poor on Xeon Phi and may cause a serious bottleneck on heterogeneous environments, another module is provided (IO fi.cpp) that indirectly sends the Xeon Phi's data to the corresponding CPU. Therefore, CPU is responsible to write down its own data as well as the Xeon Phi's data onto files. Here are the public methods for the I/O modules:

void init(int file counter=0) Initializes the I/O module. In addition, it receives a counter number which indicates the number of generated files so far. The counter is provided since sometimes it is not possible to store all the data onto a single file. Thus, this function again can be called from the I/O module for initializing another file.

void freemem() Deallocates all the allocated memories for this module.

void fillInitData(NBeam* nubeam) It is possible to resume the code using the previously generated data file. In that case, this function receives an array of NBeam objects in order to initialize the neutrinos' state function by using the provided data. The data file is the second argument that is passed from the console to the code.

void dumpToFile( const NBeam *restrict nubeam, const int itr, const double r ) This is the main function for saving data. It receives an array of NBeam objects as well as the current iteration number and radius. If the current iteration or radius have reached a predefined thresholds, it stores the data onto file.

# A.5 The Dependency of Functions in XFLAT

This section describes the function call hierarchy in XFLAT.

NBGroup Module:

Listing A.1: NBGroup::init()

```
\\**********************************************************
\\ Initialization void init ()
{
\\\ Performs benchmarks for heterogeneous multi−node runs
      node _benchmark ();
\\\ I n i t i a l i z e other modules nbm: : init ( Util : : Flvs ()
      , Util : : Ebins ()); phy : : init (); fio : : init ();
      matt : : init ();
}
```

Listing A.2: NBGroup::particleLoop()

```
\\**********************************************************  \\ Calls    the
neutrino      evolution      loop from the numeric module
void particleLoop ()
{ nmr : : evolutionLoop ();
}
```

Listing A.3: NBGroup::freeme()

```
\\*********************************************************
\\    Finalization    void
freemem ()
{
\\\ Free up memories of each module
     fio : : freemem ();
     phy : : freemem ();
}
```

Numeric Module:

Listing A.4: Nmr::evolutionloop()

```
\\*********************************************************
\\ The main neutrino evolution loop int evolutionLoop
()
{
\\\ Allocates         Hamiltonian arrays
     phy : : newHvv(h ); while ( . .
. )
{
 \\\  Get the matter values
     matt : : getHm ( . . . ) ;

 \\\  Calculates             angles dependent values per each bin
     phy : : calcAngleBins ( . . . ) ;

 \\\  Calculates       dl from dr
     phy : : calcDeltaLs ( . . . ) ;

 \\\  Calculates           the Hamiltonian at each point
     phy : : calc _Hvv ( . . . ) ;
```

```
\\\  Evolves neutrino beams with the            calculated      Hamiltonian
     phy : : evolve ( . . . ) ;
\\\  Calculates       the average    flavor    states    between two beams
     phy : : avgBeam ( . . . ) ;
\\\  Conditionally        saves the     results    onto   f i l e
     fio : : dumpToFile ( . . . ) ;
}
\\\  Deallocates          the memory of Hamiltonian arrays
     phy : : deleteHvv ( . . . ) ;
}
```

NBeam Module:

Listing A.5: NBeam::evolveBins()

```
\\****************************************************
\\ Computes the        result   of    the neutrino     evolution    per bin
\\ To be called from the Physics module
void NBeam: : evolveBins ( . . . )
{
. . .
}
```

Listing A.6: NBeam::calcErr()

```
\\****************************************************
\\ Calculates            the maximum error between two NBeam objects
\\ To be called from the Physics module
double NBeam: : calcErr ( . . . )
```

```
{
...
}
```

Listing A.7: NBeam::addAvg()

```
\\****************************************************
\\ Adds two NBeam flavor       states    together ,    afterwards    saves
\\ the     result    onto    ' this '    object
void NBeam: : addAvg ( . . . )
{
...
}
```

Listing A.8: NBeam::calcESum()

```
\\****************************************************
\\ Calculates        the summation of     flavor     states
\\ The loop      is     over energy bins        within the NBeam object
void NBeam: : calcESum ()
{
...
}
```

Physics Module:

Listing A.9: Phy::evolve()

```
\\********************************************************
\\ Loops over       all     the neutrino beams
\\ calls      the NBeam: : evolveBin ()    function      per each angle beam void
evolve ( . . . )
{
      for ( angle     : ANGLE BEAMS)
          NBeam: : evolveBins ( . . . ) ;
}
```

Listing A.10: Phy::avgBeam()

```
\\******************************************************** \\ Loops over
      all     the neutrino beams , calls   the addAvg()
void avgBeam ( . . . )
{
      for ( angle     : ANGLE BEAMS)
          NBeam: : addAvg ( . . . ) ;
}
```

Listing A.11: Phy::calc Hvv()

```
\\********************************************************
\\ Calculates  the Hamiltonian void calc _Hvv
( . . . )
{
\\\ Calculates         the angle dependent     results   of the     partial
\\\ Hamiltonian getHvv _partial (
      . . . ) ;

\\\  In multi−node env .       exchange the    results   between   all   nodes
          MPI _Allreduce ( . . . ) ;
```

```
\\\  Loops over    all   angles   to compute the    final   Hamiltonian
     for ( angle      : ANGLE BEAMS)
          getHvv ( . . . ) ;
}
```

I/O Module:

Listing A.12: IOf::fillInitData()

```
\\***********************************************************
\\ Fills          wavefunctions by the       flavor       states from a previous
\\ run ( loaded from a          f i l e )
void fillInitData ( . . . )
{
     for (bm : NEUTRINO BEAMS)
          memcpy(&NBeam[bm] . psi , &data );
}
```

Listing A.13: IOf::dumpToFile()

```
\\***********************************************************
\\ Based on the verbose mode decides which function                to    call
\\ for performing     different      writes void
dumpToFile ( . . . )
{
. . .
}
```

# Appendix B

# Kernels

In this appendix the majority of the developed codes for the benchmarked kernels are provided. The codes can be compiled with the Intel C++ Compilers, although the gcc compiler should be able to build most of them.

## B.1    Raw Performance Benchmarks of the Xeon Phi

Code for the benchmarking the Xeon Phi was similar to [Intel, 2013]. The following code was used to benchmark MADD (multiply and addition) as well as transcendental
functions throughput.

The compiler flags for the CPU was as follow:

icpc -O3 -openmp bench.c -xHOST

The compiler flags for the Xeon Phi was as follow:

icpc -O3 -openmp bench.c -mmic
> Listing B.1: Benchmark for MADD and transcendental functions

```
#include <cstdio>
#include <cstdlib >
#include <cstring>
#include <omp.h>
#include <sys/time .h>
#include <cmath>

// dtime
//
//   returns   the   current   wall   clock   time
//
double dtime ()
{
     double tseconds = 0.0; struct timeval mytime ; gettimeofday(&mytime ,(
     struct timezone *)0); tseconds = (double)(mytime . tv _sec+mytime . tv _usec
     *1.0e−6);
       return( tseconds      );
}

#define FLOPS ARRAY SIZE (1024*1024)
#define MAXFLOPS ITERS 1000000
#define LOOPCOUNT 4096

// number    of  f l o a t  pt  ops  per    calculation
#define FLOPSPERCALC 2
#define REAL double
```

```
   //    define some arrays – // make sure they
      are 64 byte    aligned
// for   best   cache   access
REAL fa [FLOPS ARRAY SIZE]attribute(( align (64)));
REAL fb [FLOPS ARRAY SIZE]            __attribute __ (( align (64)));

int main( int argc , char *argv [ ]            )
{ int i , j , k ; int numthreads ; double tstart
    , tstop , ttime ; double gflops = 0.0;
     REAL a=.05;

    //
     // i n i t i a l i z e   the   compute   arrays
    //
    //

#pragma omp parallel
#pragma omp master numthreads = omp get num
     threads ();           _   _     _

     printf (" Initializing \r\n" ); #pragma omp
     parallel for for ( i =0; i<FLOPS ARRAY SIZE;
      i++)
    {

          fa [ i ] = (REAL) i * 0.1;
          fb [ i ] = (REAL) i * 0.2;
```

```
} printf ("Starting Compute on %d threads\r\n" ,numthreads );

tstart = dtime ();
```

```
// scale   the   calculation   across   threads   requested   need to
// set   environment          variables OMP NUM THREADS and KMP AFFINITY
    #pragma omp parallel for private( j , k)

  for   ( i =0; i<numthreads ;        i++)
  {
        // each     thread  will  work   it's own array     section
        //  calc  offset  into  the   right    section
        int offset = i *LOOPCOUNT;

                                              calculations
        //    loop many times   to  get  lots   of
        for ( j =0; j<MAXFLOPS ITERS;        j++)
        {
            //   scale  1 st   array and add      inthe   2nd array
            #pragma  omp  simd  for   (k=0;
            k<LOOPCOUNT; k++)
            {
                    /// FMADD benchmark
                              fa [ k+offset ] = a*fa [ k+offset ] + fb [ k+offset ] ;

                    ///    sin ()/ cos ()    benchmark
                    // fa [ k+offset ] = sin ( fa [ k+offset ] ) ;
```

```
                    // fb [ k+o f f s e t ] = cos ( fb [ k+o f f s e t ] ) ;

                     /// exp ()    benchmark
                         // fa [ k+o f f s e t ] = exp ( fa [ k+o f f s e t ]*.000001);
                         // fb [ k+o f f s e t ] = exp ( fb [ k+o f f s e t ]*.000003);

              }
          }
      }
     tstop = dtime ();
      // # of    gigaflops    we   just     calculated
       gflops = (double)(      1.0e−9*numthreads*LOOPCOUNT*
                           MAXFLOPS ITERS*FLOPSPERCALC);

     // elasped time ttime = tstop −
     tstart ;

     //
     //   Print   the    results
     //
     if (( ttime ) > 0.0)
     { printf ("GFlops=%e , Secs=%e , GFlops per sec=%e\r\n" ,
                       gflops ,    ttime ,     gflops /ttime );
     }
     return( 0 );
}
}
```

# B.2    SoA and AoS Benchmarks

Listing B.2: Structure of Arrays

```
class Data{
public : double ar [] , ai [] , br [] , bi [ ] ;
    ...
    void init (double x) {
        #pragma omp simd
            for ( int j = 0;        j < SIZE ; ++j ) {
                ///    vectorized    computations
            }
    }
};

int main( int argc , char** argv ) {
    ...
    #pragma omp parallel for
        for ( int i = 0;        i < LEN; ++i ) {
            ...
            for ( int k = 0; k < L; ++k ) {
                ///  c a l l   array      with LEN length ,

            }
    }
    ...
}                                       in   p a r a l l e l
```

Listing B.3: Array of Structures

```
class Data
{ public
:
      double ar ,    ai , br ,   bi ;
      . . .
      void init (double x)
      {
          ///   serial      computations on  class ' members
      }
};

int main( int argc , char** argv )
{
      . . .
      #pragma omp parallel for
         for ( int i = 0;         i < LEN * SIZE ; ++i )
      {
          . . .
          for ( int k = 0; k < L; ++k)
          {
              ///  c a l l   array       with LEN*SIZE length ,
              data [ index ] . init ( . . . ) ; }
      }
}                                                    p a r a l l e
                                                  in   l
```

## B.3        Function Arguments and Their Performance

Listing B.4: The effect of function arguments within vectorized loops

```
void func2 ( const double ar , const double ai ,
                const double br , const double bi ,
                    double& res _r , double& res _i      ) {
            /// Computations !
}

void func1 (const int e , double& res _r , double& res _i ) {
            /// Computations !
}

int main( int argc , char** argv )
{
      . . .
      for ( int z = 0; z < N; ++z)
      {
            #pragma omp simd
            for ( int e = 0; e < E; ++e)
            {
                  ///  Either  one of the  should    be commented out
                  func1 (e , mm[ e ] , nn[ e ] ) ;
                   func2 ( ar [ e ] ,   ai [ e ] ,   br [ e ] ,     bi [ e ] , mm[ e ] , nn[ e ] ) ;
            }
      }
}
```

# B.4    OpenMP Parallel Loops

OpenMP loops can be written in various ways that may affect the performance.

Listing B.5: First approach for parallelizing a region via a parallel region that encloses everything and single regions within the loop.

```
int main() {
     . . .
    #pragma omp parallel {
            for ( int i = 0;          i < N; ++i ) {

              #pragma omp single { p[ i
                  /SIZE ] += . . . q [ i
                  /SIZE ] += . . .
            }
            #pragma omp for
              for ( int j = 0;          j < SIZE ; ++j )  {
                c [ j ] += . . . b[ j
                ] += . . . a [ j ] +=
                      . . .
            }

            #pragma omp single { r [ i
                  /SIZE ] −= . . . s [ i
                  /SIZE ] −= . . .
            }
            #pragma omp for
              for ( int j = 0;          j < SIZE ; ++j )
                z [ j ] += . . .                          {
```

```
        y [ j ] += . . . x [ j
              ] += . . .
    }

    #pragma omp single { m[ i
         /SIZE ]  −= . . . q [ i
         /SIZE ] += . . .
    }
    #pragma omp for
       for ( int j = 0;        j < SIZE ; ++j )     {
```

```
                    d[ j ] −= . . . e [ j
                     ] −= . . . f [ j ]
                        −= . . .
            }

            #pragma omp single { n[ i
                /SIZE ] += . . . s [ i
                /SIZE ] −= . . .
            }
            #pragma omp for
                for ( int j = 0;         j < SIZE ; ++j )
                    u[ j ] −= . . . v [ j
                    ] −= . . . w[ j ] −=
                            . . .
            }
        }
    }
}                                               {
```

Listing B.6: Second approach for parallelizing a region via a parallel region inside the main loop that encloses everything.

```
int   main() {
    . . .
    for ( int i = 0;            i < N; ++i ) {

        #pragma omp parallel {

            #pragma omp single { p[ i
                /SIZE ] += . . . q [ i
                /SIZE ] += . . .
            }
            #pragma omp for
                for ( int j = 0;           j < SIZE ; ++j )  {
                    c [ j ] += . . . b[ j
                    ] += . . . a [ j ] +=
                    . . .
                }

            #pragma omp single { r [ i
                /SIZE ] −= . . . s [ i
                /SIZE ] −= . . .
            }
            #pragma omp for
                for ( int j = 0;           j < SIZE ; ++j )
                    z [ j ] += . . . y [ j
                    ] += . . . x [ j ] +=
                    . . .                               {
```

```
                }

                #pragma omp single { m[ i
                    /SIZE ]  −= . . . q  [ i
                    /SIZE ] += . . .
                }
                #pragma omp for
                    for ( int j = 0;          j < SIZE ; ++j )      {
                        d[ j ] −= . . . e [ j
                         ] −= . . . f [ j ]
                            −= . . .
                }

                #pragma omp single { n[ i
                    /SIZE ]  += . . . s  [ i
                    /SIZE ] −= . . .
                }
                #pragma omp for
                    for ( int j = 0;          j < SIZE ; ++j )
                        u[ j ] −= . . . v [ j
                        ] −= . . . w[ j ] −=
                                . . .
                }
            }
        }
}                                              {
```

Listing B.7: Third approach for parallelizing a region via a parallel region that encloses everything and single regions within the loop. Threads at the end of parallel for loop does not wait for the other threads.

```
int main() {

    . . .

    #pragma omp parallel {

            for ( int i = 0;        i < N; ++i ) {

              #pragma omp single { p[ i
                  /SIZE ] += . . . q [ i
                  /SIZE ] += . . .
            }
                #pragma omp for nowait
                  for ( int j = 0;         j < SIZE ; ++j )  {
                    c [ j ] += . . . b[ j
                    ] += . . . a [ j ] +=
                          . . .
            }

              #pragma omp single { r [ i
                  /SIZE ] −= . . . s [ i
                  /SIZE ] −= . . .
            }
                #pragma omp for nowait
                  for ( int j = 0;         j < SIZE ; ++j )
                    z [ j ] += . . . y [ j
                        ] += . . .                          {
```

```
                x [ j ] += . . .
        }

        #pragma omp single { m[ i
            /SIZE ] −= . . . q [ i
            /SIZE ] += . . .
        }
          #pragma omp for nowait
           for ( int j = 0;        j < SIZE ; ++j )      {
              d[ j ] −= . . . e [ j
               ] −= . . . f [ j ]
                  −= . . .
        }

        #pragma omp single { n[ i
            /SIZE ] += . . . s [ i
            /SIZE ] −= . . .
        }
          #pragma omp for nowait
           for ( int j = 0;        j < SIZE ; ++j )
              u[ j ] −= . . . v [ j
              ] −= . . . w[ j ] −=
                  . . .
        }
      }
    }
}                                                {
```

Listing B.8: Fourth approach for parallelizing a region via separated parallel for regions.

```
int   main()
{
     . . .
     for ( int i = 0;            i < N; ++i )
     {

          p[ i /SIZE ] += . . . q [ i
          /SIZE ] += . . .

          #pragma omp parallel for
              for ( int j = 0;          j < SIZE ; ++j )
          {
              c [ j ] += . . . b[ j
              ] += . . . a [ j ] +=
              . . .
          }

          r [ i /SIZE ] −= . . . s [ i
          /SIZE ] −= . . .

          #pragma omp parallel for
              for ( int j = 0;          j < SIZE ; ++j )
          {
              z [ j ] += . . . y [ j
              ] += . . . x [ j ] +=
              . . .
```

```
        }

        m[ i /SIZE ] -= . . . q [ i
        /SIZE ] += . . .

        #pragma omp parallel for
            for ( int j = 0;          j < SIZE ; ++j )
        {
            d[ j ] -= . . . e [ j
            ] -= . . . f [ j ]
            -= . . .
        }

        n[ i /SIZE ] += . . . s [ i
        /SIZE ] -= . . .

        #pragma omp parallel for
            for ( int j = 0;          j < SIZE ; ++j )
        {
            u[ j ] -= . . . v [ j
            ] -= . . . w[ j ] -=
            . . .
        }

    }
}
```

## B.5 Benchmarks of the I/O Loops

Listing B.9: Saving data via NetCDF within nested loops.

```
void IO(NBeam* beams)
{
    . . .
        for ( int tet = 0;          tet < thetas ; ++tet )
    {
        start [1] = tet ;
            for ( int phi = 0;        phi < phis ; ++phi )
        {
            start [2] = phi ;
            for ( int p = 0; p < P; ++p )
            {
                start [3] = p;
                    for ( int c = 0; c < C; ++c )
                {
                    start [4] = c ;
                    NCRUN(      nc put-vara _double ( nc1id ,        psid ,
                            start , count ,
                                beams [( tet*phis+phi )*P+p ] . psi (c )));
                }
            }
        }
    }
    . . .
}
```

Listing B.10: Saving data via NetCDF within a single loop.

```
void IO(NBeam* beams)
{
    ...
        for ( int i = 0;          i < thetas*phis*P; ++i )
    {
        ///    Calculates    the    proper    indecis

                -  sizet i0 = i / ( phis*P); sizet
        j0  =  -  i % ( phis*P); sizet i1 = j0
        /  P;  -  sizet i2 = j0 % P;
                -

          start [1] = i0 ;       start [2] = i1 ;       start [3] = i2 ;

        for ( int c = 0; c < C; ++c)
        {
            start [4] = c ;
            NCRUN( nc put vara _double ( nc1id , psid , start , count
                ,
                beams [ i ] . psi (c )) ); }
    }
    ...
}
```

Listing B.11: Saving data via NetCDF after completion of a single loop. Within the loop data may be extracted from NBeam objects and store onto a buffer.

```
void IO(NBeam* beams)
{
     . . .
         for ( int i = 0;           i < thetas*phis*P; ++i )
     {
                   - sizet i0 = i / ( phis*P); sizet
          j0  =  -  i % ( phis*P); sizet i1 = j0
          /  P;  -  sizet i2 = j0 % P;
                   -
          for ( int c = 0; c < C; ++c)
          {
               for ( int e = 0; e < ebins ; ++e)
                              buffer [( i *C+c)* ebins+e ] = beams [ i ] . psi (c )[ e ] ;

          }
     }
     NCRUN(     nc put vara _double (       nc1id ,    psid ,
                                       start ,    count ,    buffer )    );
     . . .
}
```